\pdfoutput=1
\documentclass[
preprint,
superscriptaddress,
amsmath,amssymb,
aps,
prx,
]{revtex4-2}

\usepackage{hyperref}% add hypertext capabilities
 \hypersetup{
     colorlinks=true,
     linkcolor=blue,
     filecolor=magenta,      
     urlcolor=cyan
     }
 \usepackage{rotating}
 \usepackage{graphicx}% Include figure files
 \usepackage{dcolumn}% Align table columns on decimal point
 \usepackage{bm}% bold math
 \usepackage[T1]{fontenc}

 \usepackage{multirow}
 \usepackage{tabularx}
 \usepackage{booktabs}
 \usepackage{xcolor}

\makeatletter
\renewcommand{\figurename}{Figure}

\newcommand{\team}[1]{\MakeUppercase{#1}}

\usepackage{algorithm,algorithmic}
\newlength\myindent
\newcommand\bindent{%
	\begingroup
	\setlength{\itemindent}{\myindent}
	\addtolength{\algorithmicindent}{\myindent}
}
\newcommand\eindent{\endgroup}
 
\usepackage{multirow}
\usepackage{graphicx}% Include figure files

\def\bibsection{\section*{\refname}} 
\renewcommand\refname{References}

\usepackage[section]{placeins}

\begin{document}

\title{Objective comparison of methods to decode anomalous diffusion}

\author{Gorka Mu\~noz-Gil}%
\affiliation{ICFO -- Institut de Ci\`encies Fot\`oniques, The Barcelona Institute of Science and Technology, Av. Carl Friedrich Gauss 3, 08860 Castelldefels (Barcelona), Spain}
\author{Giovanni Volpe}
\email{giovanni.volpe@physics.gu.se}
\affiliation{Department of Physics, University of Gothenburg, Origov{\"a}gen 6B, SE-41296 Gothenburg, Sweden}
\author{Miguel Angel Garcia-March}
\affiliation{Instituto Universitario de Matem\'{a}tica Pura y Aplicada, Universitat Polit\`{e}cnica de Val\`{e}ncia, Spain}
\author{Erez Aghion}
\affiliation{Max Planck Institute for the Physics of Complex Systems, N{\"o}thnitzer Stra{\ss}e 38, DE-01187 Dresden, Germany}
\author{Aykut Argun}
\affiliation{Department of Physics, University of Gothenburg, Origov{\"a}gen 6B, SE-41296 Gothenburg, Sweden}
\author{Chang Beom Hong}
\affiliation{Department of Physics, Pohang University of Science and Technology, Pohang 37673, Korea}
\author{Tom Bland}
\affiliation{The Francis Crick Institute, 1 Midland Road, London, NW1 1AT, UK}
\author{Stefano Bo}
\affiliation{Max Planck Institute for the Physics of Complex Systems, N{\"o}thnitzer Stra{\ss}e 38, DE-01187 Dresden, Germany}
\author{J. Alberto Conejero}
\affiliation{Instituto Universitario de Matem\'{a}tica Pura y Aplicada, Universitat Polit\`{e}cnica de Val\`{e}ncia, Spain}
\author{Nicol\'{a}s Firbas}
\affiliation{Instituto Universitario de Matem\'{a}tica Pura y Aplicada, Universitat Polit\`{e}cnica de Val\`{e}ncia, Spain}
\author{\`{O}scar Garibo i Orts}
\affiliation{Instituto Universitario de Matem\'{a}tica Pura y Aplicada, Universitat Polit\`{e}cnica de Val\`{e}ncia, Spain}
\author{Alessia Gentili}
\affiliation{Department of Chemistry, University College London, 20 Gordon Street, London WC1H 0AJ, UK}
\author{Zihan Huang}
\affiliation{School of Physics and Electronics, Hunan University, Changsha 410082, China}
\author{Jae-Hyung Jeon}
\affiliation{Department of Physics, Pohang University of Science and Technology, Pohang 37673, Korea}
\author{H\'{e}l\`{e}ne Kabbech}
\affiliation{Department of Cell Biology, Erasmus MC, Rotterdam, the Netherlands}
\author{Yeongjin Kim}
\affiliation{Department of Physics, Pohang University of Science and Technology, Pohang 37673, Korea}
\author{Patrycja Kowalek}
\affiliation{Faculty of Pure and Applied Mathematics, Hugo Steinhaus Center, Wroc\l{}aw University of Science and Technology, Wroc\l{}aw, Poland}
\author{Diego Krapf}
\affiliation{Department of Electrical and Computer Engineering, Colorado State University, Fort Collins, Colorado 80523, USA}
\author{Hanna Loch-Olszewska}
\affiliation{Faculty of Pure and Applied Mathematics, Hugo Steinhaus Center, Wroc\l{}aw University of Science and Technology, Wroc\l{}aw, Poland}
\author{Michael A. Lomholt}
\affiliation{PhyLife, Department of Physics, Chemistry and Pharmacy, University of Southern Denmark, DK-5230 Odense M, Denmark}
\author{Jean-Baptiste Masson}
\affiliation{Institut Pasteur, Decision and Bayesian Computation lab, Paris}
\author{Philipp G. Meyer}
\affiliation{Max Planck Institute for the Physics of Complex Systems, N{\"o}thnitzer Stra{\ss}e 38, DE-01187 Dresden, Germany}
\author{Seongyu Park}
\affiliation{Department of Physics, Pohang University of Science and Technology, Pohang 37673, Korea}
\author{Borja Requena}
\affiliation{ICFO -- Institut de Ci\`encies Fot\`oniques, The Barcelona Institute of Science and Technology, Av. Carl Friedrich Gauss 3, 08860 Castelldefels (Barcelona), Spain}
\author{Ihor Smal}
\affiliation{Department of Cell Biology, Erasmus MC, Rotterdam, the Netherlands}
\author{Taegeun Song}
\affiliation{Department of Physics, Pohang University of Science and Technology, Pohang 37673, Korea}
\affiliation{Center for AI and Natural Sciences, Korea Institute for Advanced Study, Seoul, Korea}
\affiliation{Department of Data Information and Physics, Kongju National University, Kongju 32588, Korea}
\author{Janusz Szwabi\'{n}ski}
\affiliation{Faculty of Pure and Applied Mathematics, Hugo Steinhaus Center, Wroc\l{}aw University of Science and Technology, Wroc\l{}aw, Poland}
\author{Samudrajit Thapa}
\affiliation{Institute for Physics \& Astronomy, University of Potsdam, Karl-Liebknecht-Str 24/25, D-14476 Potsdam-Golm, Germany}
\affiliation{Sackler Center for Computational Molecular and Materials Science, Tel Aviv University, Tel Aviv 69978, Israel}
\affiliation{School of Mechanical Engineering, Tel Aviv University, Tel Aviv 69978, Israel}
\author{Hippolyte Verdier}
\affiliation{Institut Pasteur, Decision and Bayesian Computation lab, Paris, France}
\author{Giorgio Volpe}
\affiliation{Department of Chemistry, University College London, 20 Gordon Street, London WC1H 0AJ, UK}
\author{Artur Widera}
\affiliation{Department of Physics and Research Center OPTIMAS, Technische Universität Kaiserslautern, 67663 Kaiserslautern, Germany}
\author{Maciej Lewenstein}
\affiliation{ICFO -- Institut de Ci\`encies Fot\`oniques, The Barcelona Institute of Science and Technology, Av. Carl Friedrich Gauss 3, 08860 Castelldefels (Barcelona), Spain}
\affiliation{ICREA, Pg. Llu\'is Companys 23, 08010 Barcelona, Spain}
\author{Ralf Metzler}
\affiliation{Institute for Physics \& Astronomy, University of Potsdam, Karl-Liebknecht-Str 24/25, D-14476 Potsdam-Golm, Germany}

\author{Carlo Manzo}
\email{carlo.manzo@uvic.cat}
\affiliation{Facultat de Ci\`encies i Tecnologia, Universitat de Vic -- Universitat Central de Catalunya (UVic-UCC), C. de la Laura,13, 08500 Vic, Spain}
\affiliation{ICFO -- Institut de Ci\`encies Fot\`oniques, The Barcelona Institute of Science and Technology, Av. Carl Friedrich Gauss 3, 08860 Castelldefels (Barcelona), Spain}

\date{\today}% It is always \today, today,

\begin{abstract}
Deviations from Brownian motion leading to anomalous diffusion are found in transport dynamics from quantum physics to life sciences. 
The characterization of anomalous diffusion from the measurement of an individual trajectory is a challenging task, which traditionally relies on calculating the trajectory mean squared displacement. 
However, this approach breaks down for cases of practical interest,  e.g.,  short or noisy trajectories, heterogeneous behaviour, or non-ergodic processes. Recently, several new approaches have been proposed, mostly building on the ongoing machine-learning revolution.
To perform an objective comparison of methods, we gathered the community and organized an open competition, the Anomalous Diffusion challenge (AnDi). Participating teams applied their algorithms to a commonly-defined dataset including diverse conditions.
Although no single method performed best across all scenarios, machine-learning-based approaches achieved superior performance for all tasks. The discussion of the challenge results provides practical advice for users and a benchmark for developers.
\end{abstract}


\maketitle

\section*{Comments}
This is the author’s version of the article published in Nature Communications under a Creative Commons Attribution 4.0 International License (CC BY 4.0). The final published version is available at \hyperlink{https://doi.org/10.1038/s41467-021-26320-w}{https://doi.org/10.1038/s41467-021-26320-w}.

\section*{Introduction \label{sec:intro}}

\begin{figure*}[!ht]
	\includegraphics[width=0.7\textwidth]{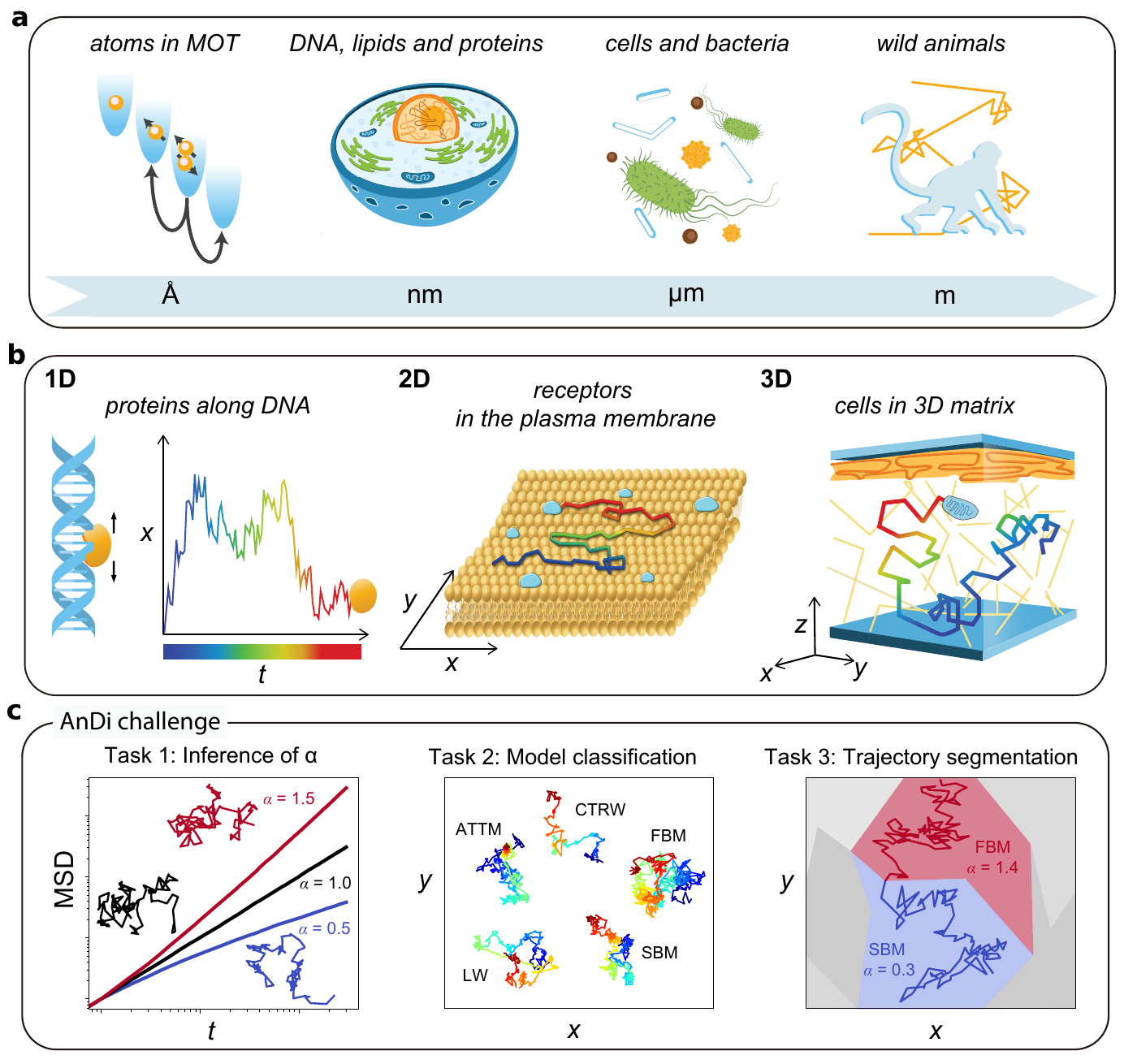}%
	\caption{\label{fig:summary_andi} \textbf{The AnDi challenge tasks and datasets.} \textbf{a}, Random walks, characterized by an erratic change of an observable, occur at all length and time scales in a variety of systems. Examples are provided by atoms in magneto-optical traps; the diffusion of cellular components, such as DNA, proteins, lipids, and organelles; the motion of bacteria and cells; and animals foraging and mating. \textbf{b}, Trajectories of tracers in  spaces of different dimensionality: 1D, Proteins sliding along DNA fragments; 2D, receptors diffusing in the plasma membrane; 3D, cells migrating in a 3-dimensional matrix. The color code of the trajectories represents time. \textbf{c}, The challenge tasks. Task 1~--~Inference of the anomalous diffusion exponent. Representative trajectories and corresponding MSD for diffusive ($\alpha = 1$, black lines), subdiffusive ($0<\alpha<1$, blue lines), and superdiffusive ($1<\alpha<2$, red lines) motion. Task 2~--~Classification of the  underlying anomalous diffusion model. Representative trajectories for continuous-time random walk (CTRW), fractional Brownian motion (FBM), L\'{e}vy walk (LW), annealed transient time motion (ATTM), and scaled Brownian motion (SBM). Different diffusion models produce subtle changes.	Details of the models are described in the text and in \protect\hyperlink{sec:methods_h}{Methods}, ``Theoretical models''. Task 3~--~Segmentation and characterization of a trajectory with changepoint. Trajectory switching diffusion model and/or exponent as a result of diffusion in spatially-heterogeneous environment, represented by the colored patches.}
\end{figure*}

The random walk~\cite{pearson1905problem} is a mathematical model ubiquitously employed at all scales in a variety of scientific fields, including physics, chemistry, biology, ecology, psychology, economics, sociology, and computer science (Fig.~\ref{fig:summary_andi}a)~\cite{klafter2011first, hughes1995random}. Random walks are characterized by an erratic change of an observable over time (e.g., position, temperature, or stock price, Fig.~\ref{fig:summary_andi}b). The archetypal example of a random walk is Brownian motion, which describes the movement of a microscopic particle in a fluid as a consequence of thermal forces ~\cite{metzler2014anomalous}. 

The space explored by random walkers over time is commonly measured by the mean squared displacement (MSD), which grows linearly in time for Brownian walkers (${\rm MSD} \propto t$) \cite{metzler2014anomalous}. Deviations from such a linear behavior displaying an asymptotic power-law dependence (${\rm MSD} \propto t^\alpha$) have been observed in several fields and are generally referred to as anomalous diffusion~\cite{metzler2014anomalous}: subdiffusion for $0<\alpha<1$, and superdiffusion for $\alpha>1$ (as particular cases, $\alpha=0$ corresponds to immobile trajectories, $\alpha=1$ to Brownian motion, and $\alpha=2$ to ballistic motion).
The left panel in Fig.~\ref{fig:summary_andi}c shows some examples of MSDs for Brownian (black line), subdiffusive (blue line), and superdiffusive (red line) motion together with the corresponding trajectories in 2D.
For example, anomalous diffusion occurs in the diffusion of lipids and receptors in the cell membrane~\cite{krapf2015mechanisms}, in the transport of molecules within the cytosol~\cite{sabri2020elucidating} and the nucleus~\cite{di2018anomalous}, in the foraging and mating strategies of animals~\cite{humphries2012foraging}, in sleep-wake transitions during sleep~\cite{lo2002dynamics}, and in the fluctuations of the stock market~\cite{plerou2000economic}.

The recurrent observation of anomalous diffusion has driven an important theoretical effort to understand and mathematically describe its underlying mechanisms. This effort has provided a palette of microscopic models characterized by different spatial (step length) and temporal (step duration) random distributions, both with and without long-range correlations~\cite{metzler2014anomalous}. Important models for the interpretation of experimental results are continuous-time random walk (CTRW)~\cite{scher1975anomalous}, fractional Brownian motion (FBM)~\cite{mandelbrot1968fractional}, L\'{e}vy walk (LW)~\cite{klafter1994levy}, annealed transient time motion (ATTM)~\cite{massignan2014nonergodic}, and scaled Brownian motion (SBM)~\cite{lim2002self} (some sample trajectories are shown in the central panel of  Fig.~\ref{fig:summary_andi}c, see \hyperlink{sec:methods_h}{Methods}, ``Theoretical models'').

In typical experiments aimed at understanding diffusion, the available data consists of trajectories of a tracer, such as a molecule in a cell, a stock price in the stock market, a foraging animal in its environment. The aim is to extract from these trajectories information about properties of the tracer and of the medium where its motion takes place, namely to infer the anomalous diffusion exponent $\alpha$, to determine the underlying diffusion model and, finally, to determine whether these properties change over time and space. 

The first crucial step to characterize the tracer's motion is the determination of the anomalous diffusion exponent $\alpha$ (Task 1, Fig.~\ref{fig:summary_andi}c). It is typically estimated by fitting the  MSD to a power law~\cite{kepten2015guidelines}. Traditionally, the MSD is defined as the ensemble average over a group of tracers (EA-MSD, Equation~\eqref{eq:eamsd}), in analogy to the solution to Fick's second law for the spreading of a bunch of particles in a homogeneous medium~\cite{metzler2014anomalous}. 
When long tracks are available, the MSD can be instead obtained as a time average from the trajectory of a single tracer (TA-MSD, Equation~\eqref{eq:tamsd}). 
While seemingly a straightforward procedure, determining $\alpha$ from the MSD can introduce significant errors and biases: i) the accuracy of the estimation depends on fluctuations, which can only be reduced by increasing the number of tracers (for EA-MSD) or the length of the trajectory (for TA-MSD), which is often not possible because of practical constraints; 
ii) the value of $\alpha$ is biased by noise, such as the localization precision of experimental trajectories~\cite{chenouard2014objective}, which needs to be estimated independently to introduce a proper correction~\cite{martin2002apparent, kepten2015guidelines}; iii) while for a stationary motion in a homogeneous medium, EA-MSD and TA-MSD have the same exponent, several systems are intrinsically heterogeneous and non-stationary~\cite{weigel2011ergodic, manzo2015weak}, which can lead to non-ergodicity (i.e., the non-equivalence of time and ensemble averages). Typically, the exponent $\alpha$ of the EA-MSD characterizes the physical properties of the systems (e.g., the trapping time distribution in CTRW or the time-dependence of diffusivity in SBM). However, in several non-ergodic systems, the TA-MSD shows a linear behavior with respect to the timelag in the long time limit even when $\alpha \neq 1$~\cite{metzler2014anomalous}; iv) the behavior of the MSD at short times or timelags might differ from its asymptotic limit~\cite{metzler2014anomalous}, thus long trajectories are required for the correct estimation of $\alpha$.

The second critical issue is to determine the underlying diffusion model (Task 2, Fig.~\ref{fig:summary_andi}c), which is related to its driving physical mechanism. Here, difficulties arise because the calculation of the MSD is not very informative, since different models provide curves with the same scaling exponent. Other statistical parameters have been proposed for this task and algorithms based on the combination of several estimators allow to distinguish between pairs of models~\cite{magdziarz2009fractional, meroz2013test, chen2017anomalous, schwarzl2017quantifying}, but there is no general consensus on how to unambiguously determine the underlying diffusion model from a trajectory.

The third issue is to determine whether the properties of the motion of a given tracer change over time~\cite{manzo2015weak, weron2017ergodicity, sabri2020elucidating, yamamoto2021universal} (Task 3, Fig.~\ref{fig:summary_andi}c). This can be both the result of heterogeneity in the environment (e.g., patches with different viscosity on a cellular membrane) or of time-varying properties of the tracer (e.g., different activation states of a molecular motor). In these cases, the determination of $\alpha$ and of the underlying diffusion model must be combined with a segmentation of the trajectory to identify fragments with homogeneous characteristics. Several methods have been proposed for the segmentation of time traces~\cite{truong2020selective}, mostly based on changes in diffusion constant, velocity, or diffusion mode (e.g., immobile, random, directed)~\cite{yin2018detection, vega2018multistep, akimoto2017detection,arts2019particle}. Only recently, attempts have been made to determine changepoints with respect to a switch in $\alpha$~\cite{weron2017ergodicity,sikora2017elucidating,bo2019measurement} and diffusion model~\cite{lanoiselee2017unraveling}. Until now, a systematic assessment of changepoint detection methods for anomalous diffusion has not been performed.

In recent years, advances in fluorescence techniques have greatly increased the availability of high-precision trajectories of single molecules in living systems~\cite{manzo2015review}, producing an increasing drive to develop methods for quantifying anomalous diffusion~\cite{thapa2018bayesian, burnecki2015estimating, weron2017ergodicity,sikora2017elucidating, kepten2015guidelines, krapf2019spectral, thapa2020leveraging}. Furthermore, the recent blossoming of machine learning has promoted the accessibility of new powerful tools for data analysis~\cite{cichos2020machine} and further widened the palette of available methods~\cite{munoz2020single, granik2019single, bo2019measurement, kowalek2019classification}. Some of the novel approaches have already delivered new insights into anomalous diffusion in different scenarios~\cite{jamali2021anomalous,munoz2020phase, cherstvy2019non}.

This recent increase of available methods performing similar tasks requires an objective assessment on a common reference dataset to define the state of the art and guide end-users in the optimal choice of characterization tools for each specific application. To assess the performance in quantifying anomalous diffusion, we have therefore run an open competition, the Anomalous Diffusion (AnDi) Challenge, divided in three different tasks: anomalous exponent inference, model classification, and trajectory segmentation, each for 1D, 2D, and 3D trajectories. The performance of submitted methods was assessed with common metrics on simulated datasets with trajectory length and signal-to-noise level reproducing realistic experimental conditions (\hyperlink{sec:methods_h}{Methods}, ``Structure of the datasets''). The submitted methods were also compared on the blind analysis of experimental trajectories (\hyperlink{sec:SI-note3_h}{Supplementary Note 2}). Although several experiments provide 2D and 3D trajectories, we first present and discuss in detail the results obtained for the 1D trajectories. This choice is driven by the fact that the 1D-case is conceptually easier to understand, thus complex methods are in general first developed in 1D and then extended to multidimensional space, as testified by the larger participation for this dimension.  Thus, it allows us to assess the performance of a larger set of methods including those that might eventually be extended to 2D and 3D. We follow the same rationale when describing the physical models and their simulation.

\section*{Results} \label{sec:results} 

\subsection*{Competition design}

The challenge consisted of three tasks: Task 1 (T1)~--~inference of the anomalous diffusion exponent $\alpha$; Task 2 (T2)~--~classification of the underlying diffusion model; Task 3 (T3)~--~trajectory segmentation (Fig.~\ref{fig:summary_andi}c and \hyperlink{sec:methods_h}{Methods}, ``Organization of the challenge''). The aim of the last task was to identify the changepoint  within a trajectory switching $\alpha$ and diffusion model, as well as to determine the exponent and model for the identified segments. Each task was further divided into three subtasks corresponding to the trajectory dimensions (1D, 2D, and 3D, Fig.~\ref{fig:summary_andi}b), totaling 9 independent subtasks. Participants could choose to submit predictions for any combination of subtasks. For the competition, we let developers build and use their own tools to provide predictions for the common dataset. While this choice limited the methods assessed to those provided by the community, it ensured that those algorithms were properly applied. Datasets were generated as described in \hyperlink{sec:methods_h}{Methods}, ``Structure of the datasets'' and ``Theoretical models''. 

\subsection*{Challenge participants and performance evaluation}

We received submissions from 13 teams for T1, 14 teams for T2, and 4 teams for T3. One of the methods participating to T3 had results comparable with random predictions and was thus excluded from the discussion of the results. Basic information about methods used by participating teams can be found in \hyperlink{sec:methods_h}{Methods}, ``Challenge methods'', Table~\ref{tab:methods}, and \hyperlink{sec:SI-note1_h}{Supplementary Note 1}. A detailed description of each of the methods can be found in the referenced articles.

We investigated the performance of the methods submitted for each task separately using the metrics described in \hyperlink{sec:methods_h}{Methods}, ``Metrics''.  A summary of rankings for all tasks and methods is presented in Supplementary Fig.~\ref{fig:ranks_summary}. Full rankings for T1 and T2 in all dimensions are presented in Fig.~\ref{fig:task1}a-c and Fig.~\ref{fig:task2}a-c, respectively, together with representative information for the best-in-class methods for the 1D case  (Fig.~\ref{fig:task1}d-g and Fig.~\ref{fig:task2}d-g, respectively). The same analysis is presented in Supplementary Fig.~\ref{fig:T1_allD} and Supplementary Fig.~\ref{fig:T2_allD} for higher dimensions.  Results for T3 in 1D are shown in Fig.~\ref{fig:task3}a-c, together with representative information for the best-in-class methods (Fig.~\ref{fig:task3}e-f). Results for all dimensions are presented in Figs.~\ref{fig:task3}d-e and Supplementary Fig.~\ref{fig:T3_allD}.

\subsection*{Task 1: Inference of the anomalous diffusion exponent}
\label{sec:task1}

\begin{figure*}[!ht]
	\includegraphics[width=0.9\textwidth]{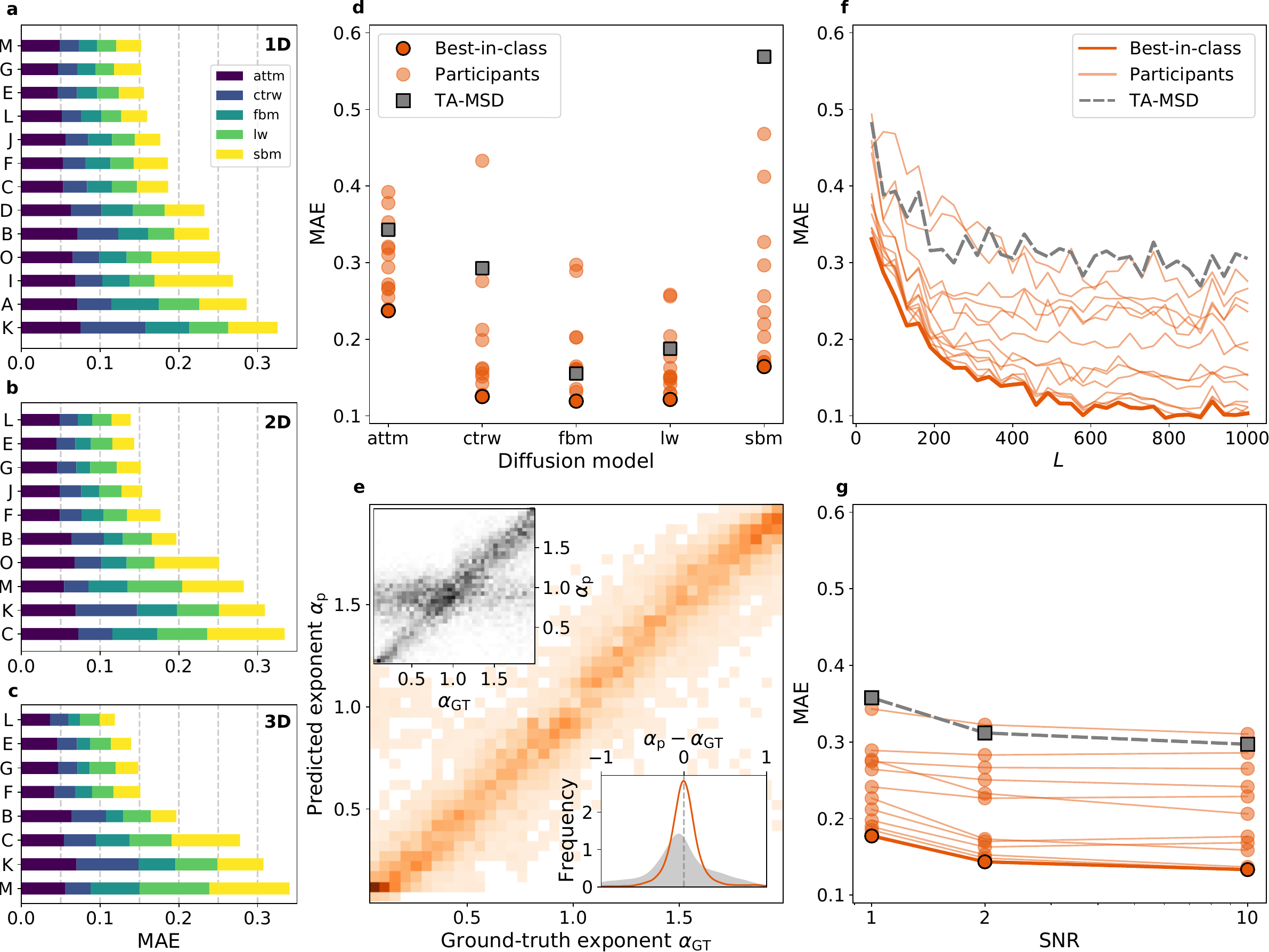}%
	\caption{\label{fig:task1} \textbf{Challenge results for Task 1: inference of $\alpha$}. 
		\textbf{a-c}, Final leaderboards according to the ${\rm MAE}$ obtained by participants for 1D (\textbf{a}), 2D (\textbf{b}), and 3D (\textbf{c}). The colors represent the relative contributions to the overall mean absolute error (${\rm MAE}$) calculated for each underlying diffusion model and normalized such that the sum of all contributions gives the value of the same metric calculated over the whole dataset.
		\textbf{d}, ${\rm MAE}$ obtained by participating teams as a function of the diffusion model for 1D trajectories. \textbf{e}, Probability distribution of the predicted vs ground-truth anomalous diffusion exponent for the best-in-class team in 1D (team \team{m}). Insets: (top left)  Probability distribution of the predicted vs ground-truth anomalous diffusion exponent for the baseline method (${\rm TA-MSD}$). (bottom right) Frequency of the bias between predicted and ground-truth anomalous diffusion exponent for the best-in-class team  (team \team{m}, orange line) and the baseline method (${\rm TA-MSD}$, gray area) in 1D. 
		\textbf{f}, ${\rm MAE}$ obtained by participating teams as a function of the trajectory length in 1D. 
		\textbf{g}, ${\rm MAE}$ obtained by participating teams as a function of the SNR in 1D.
		All results for T1 in 1D, 2D and 3D are provided in Supplementary Figs.~\ref{fig:T1_allD}, \ref{fig:T1_allMAE}-\ref{fig:T1_3D_all2dhist}, \ref{fig:T1_1D_bias}-\ref{fig:T1_3D_bias}. 
	}
\end{figure*}

The inference of the exponent $\alpha$ is the most popular way to quantify anomalous diffusion and 13 teams participated in T1 of the AnDi Challenge (Fig.~\ref{fig:task1}a-c). We observed a rather large spread of performances, but for each dimension we could identify a cluster of four top methods with comparable performance, scoring better than the rest. Three methods (\team{e}, \team{g}, and \team{l}) were consistently part of the top group in all dimensions. All top teams used machine-learning approaches: teams \team{e}, \team{g}, \team{j}, and \team{m} applied them to raw or simply pre-processed trajectories; teams \team{f} and \team{l} used statistical features as inputs.  All these methods, except \team{l} and \team{j}, were based on a length-specific training. 

Besides the overall ${\rm MAE}$, Fig.~\ref{fig:task1}a-c also shows the performance obtained for specific diffusion models (colors within bars) by all participating teams. In Fig.~\ref{fig:task1}d-g, the methods are compared with the simple fitting of the TA-MSD, used as a baseline method (\hyperlink{sec:methods_h}{Methods}, ``Alternative and baseline estimators''). Most methods perform better than TA-MSD. As expected, the fit of the TA-MSD shows better performance on ergodic (FBM) and ultra-weakly non-ergodic (LW) rather than on (weakly) non-ergodic models (CTRW, ATTM, and SBM), for which TA-MSD and EA-MSD have different scaling exponents (Fig.~\ref{fig:task1}d and Supplementary Fig.~\ref{fig:T1_allMAE}). Interestingly, the top performing methods do not suffer from this limitation and provide similar ${\rm MAE}$ for all the models, with exception of the ATTM (short ATTM trajectories might not undergo any change of diffusion coefficient and, therefore, the result is indistinguishable from pure Brownian motion, impacting the final performance).
As an example, in Fig.~\ref{fig:task1}e, we show a 2D histogram of the predicted exponent vs the ground truth for the best-in-class method (team \team{m}) and the TA-MSD (upper inset) in 1D. As most of the top-scoring methods (Supplementary Fig.~\ref{fig:T1_1D_all2dhist}, Supplementary Fig.~\ref{fig:T1_2D_all2dhist}, and Supplementary Fig.~\ref{fig:T1_3D_all2dhist}), the best-in-class method achieves similar performance over the whole range of $\alpha$, whereas TA-MSD has a lower accuracy for $\alpha\simeq 0.5$ to $1$. Obtaining precise predictions for $\alpha\simeq 1$ is particularly relevant, since the correct assessment of the exponent in this regime would further allow the discrimination between normal and anomalous diffusion. In addition, the method of team \team{m} (similarly to other top methods, (Supplementary Fig.~\ref{fig:T1_1D_bias}, Supplementary Fig.~\ref{fig:T1_2D_bias}, and Supplementary Fig.~\ref{fig:T1_3D_bias})) does not show any bias, whereas the TA-MSD systematically underestimates the value of $\alpha$ as a consequence of localization error~\cite{martin2002apparent, kepten2015guidelines} (Fig.~\ref{fig:task1}e, lower inset). 

In Fig.~\ref{fig:task1}f, we explore the effect of the trajectory length on the exponent prediction. As the trajectory length increases, the ${\rm MAE}$ rapidly decreases toward a value $\approx 0.1$ for the best performing methods. Thus, the ${\rm MAE}$ of machine-learning approaches features a striking improvement with respect to the nearly constant ${\rm MAE}$ of the TA-MSD, demonstrating the capability of machine learning to take advantage of the information contained in longer trajectories.

Last, we investigate the effect of the level of noise (Fig.~\ref{fig:task1}g). Even for SNR$=1$, i.e.,  when the standard deviation of the noise has the same amplitude as the displacement standard deviation, the top-performing methods show a greater than $2$-fold improvement in predicting $\alpha$ with respect to TA-MSD. Thus, while localization noise delays convergence of TA-MSD to its asymptotic behavior~\cite{kepten2015guidelines}, the top methods seem able to determine patterns associated to the correct exponent even from short-time behaviors, which is an ability particularly useful for many potential applications to the analysis of experimental data.

\subsection*{Task 2: Classification of the underlying diffusion model}
\label{sec:task2}

\begin{figure*}[!ht]
	\includegraphics[width=0.9\textwidth]{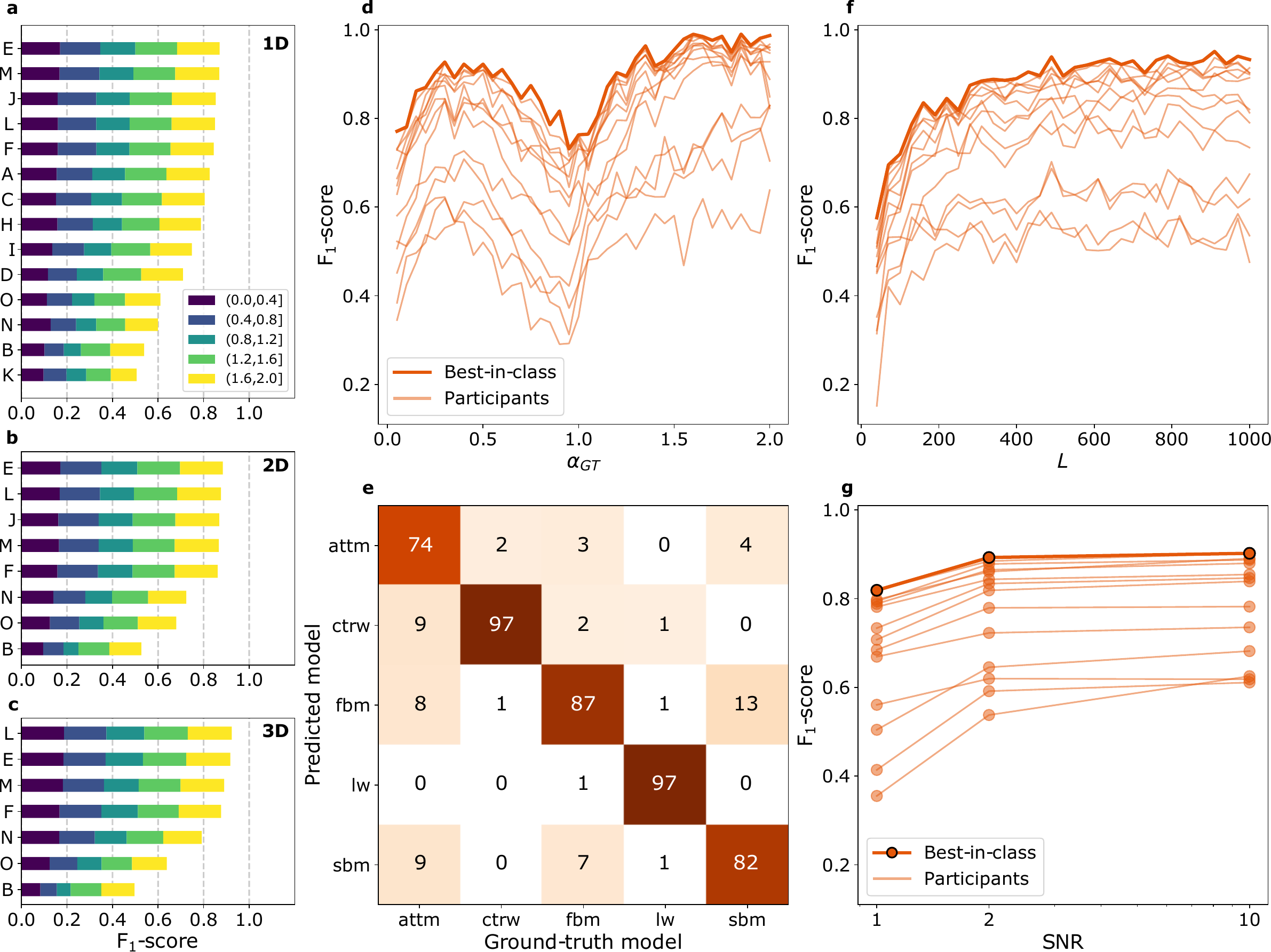}%
	\caption{\label{fig:task2}  \textbf{Challenge results for Task 2: diffusion model classification}. 
		\textbf{a-c}, Final leaderboards according to the ${\rm F_1}$-score obtained by participants for 1D (\textbf{a}), 2D (\textbf{b}), and 3D (\textbf{c}). The colors represent the relative contributions to the overall ${\rm F_1}$-score calculated for different ranges of anomalous diffusion exponents and normalized such that the sum of all contributions gives the value of the same metric calculated over the whole dataset. 
		\textbf{d}, ${\rm F_1}$-score obtained by participating teams as a function of the anomalous diffusion exponent for 1D trajectories.
		\textbf{e}, Confusion matrix for the predictions of the best-in-class team in 1D (team \team{e}). Numbers in matrix cells represent the number of correctly and incorrectly classified trajectories for each ground-truth model as percentages of the number of trajectories of the corresponding ground-truth model (column-based normalization, so that their sum along the columns should add up to 100, with minor deviation due to rounding). Thus, the percentages of correctly classified observations can be thought of as class-wise recalls.
		\textbf{f}, ${\rm F_1}$-score obtained by participating teams as a function of the trajectory length in 1D. 
		\textbf{g}, ${\rm F_1}$-score obtained by participating teams as a function of the SNR in 1D.
		All results for T2 in 1D, 2D and 3D are provided in Supplementary Figs.~\ref{fig:T2_allD}, \ref{fig:T2_allF1}-\ref{fig:T2_3D_conf}, \ref{fig:T2_1D_ROC}-\ref{fig:rank_AUC}.
	}
\end{figure*}

We present the performance of the submitted methods to classify trajectories between the 5 diffusion models in Figs.~\ref{fig:task2} and~\ref{fig:T2_allD}.
For each dimension of this task, a different number of methods showed comparable performance (Fig.~\ref{fig:task2}a-c). For each dimension, we selected the 2 teams that achieved top scores. These top positions were occupied by three teams with machine-learning methods operating on raw trajectories (teams \team{e} and \team{m}) or features (team \team{l}). In general, the use of features as input to machine learning models seems to provide better results as the trajectory dimension increases.

We also dissect the results as a function of the exponent $\alpha$, as shown in Fig.~\ref{fig:task2}a-c (colors within the bars), and in more detail in Fig.~\ref{fig:task2}d for 1D, and in Supplementary Fig.~\ref{fig:T2_allF1} for all dimensions. For all methods, the worst performance is achieved for $\alpha\simeq 1$. This is expected because in this regime all models converge to pure Brownian motion and thus feature large similarity in their long-time statistical properties, even though their microscopic generative dynamics are different. A similar situation occurs for $\alpha\rightarrow 0$, a regime in which, independently of the underlying model, trajectories are nearly immobile and dominated by localization noise. Still, most of the methods show good predictive capability (${\rm F_1} \gtrsim 0.7$) even in these two regimes, since they probably learn to recognize details or patterns of the microscopic dynamics.
The confusion matrix of the best-in-class method (team \team{e}) for the 1D subtask  (Fig.~\ref{fig:task2}e) provides a representative view of the classification capabilities of these methods. Results obtained by other methods are shown in Supplementary Fig.~\ref{fig:T2_1D_conf}, Supplementary Fig.~\ref{fig:T2_2D_conf}, and Supplementary Fig.~\ref{fig:T2_3D_conf}. The best accuracy is obtained for CTRW and LW, for which the method of team \team{e} is able to identify their markedly different features. However, it shows a higher level of error when discriminating between Gaussian processes, such as FBM and SBM~\cite{thapa2020leveraging}. The worst performance is obtained for ATTM, whose trajectories display a large heterogeneity in diffusion coefficients and lack a characteristic timescale. Rather long trajectories (including at least a switch of diffusivity) are thus necessary to distinguish ATTM from the other models.

Similarly to what we observe for T1, the trajectory length and the presence of localization noise affect the performance of the methods, as shown in Figs.~\ref{fig:task2}f and \ref{fig:task2}g, respectively. Nevertheless, even for very short and noisy trajectories, the results obtained by the top methods display excellent accuracy (${\rm F_1} \approx 0.6$ to $0.8$), taking into account that the largest noise level severely hides the actual diffusive dynamics. 

\subsection*{Task 3: Segmentation of the trajectory}
\label{sec:task3}

Recently, several experimental studies have evidenced the occurrence of switching of diffusion model and $\alpha$ within individual trajectories~\cite{weron2017ergodicity,sabri2020elucidating}. 
However, methods to determine and analyze such changes are not established and widely employed yet.
Probably, for this reason, the participation to T3 was reduced as compared to T1 and T2, with two teams proposing machine-learning methods (RNNs for team \team{e} and CNN for team \team{j}), and team \team{b} using Bayesian inference. The methods taking part in T3 were specifically designed for the challenge and have not been tested on other time-dependent processes, e.g., such as those involving a continuous change of anomalous diffusion properties.

\begin{figure*}[!ht]
	\includegraphics[width=0.9\textwidth]{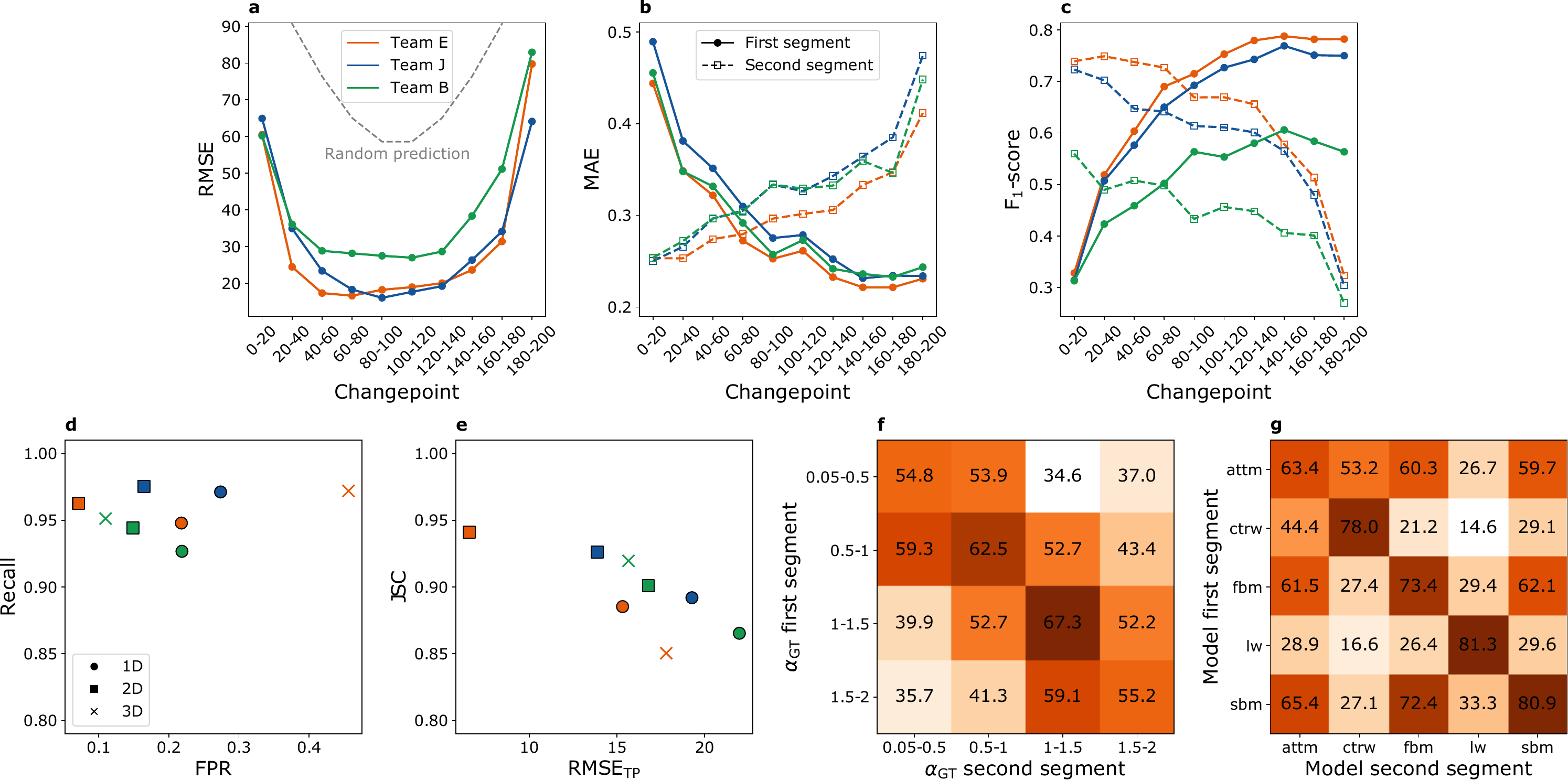}%
	\caption{\label{fig:task3} \textbf{Challenge results for Task 3: trajectory segmentation and characterization}. 
		\textbf{a}, Root mean square error (${\rm RMSE}$) as a function of the changepoint location along the trajectory for all teams in 1D (teams \team{e}, \team{j}, and \team{b}). Dashed lines represents the RMSE associated to a random prediction of the changepoint position. \textbf{b}, Corresponding mean absolute error (${\rm MAE}$) of the prediction of $\alpha$ and, \textbf{c}, ${\rm F_1}$-score for the classification of the diffusion model for the first (solid symbols/continuous line) and second segment (empty symbols/dashed line) as a function of the changepoint location along the trajectory.
		\textbf{d}, Plot of the recall vs the FPR for all participating teams. \textbf{e}, Plot of JSC vs RMSE$_{\rm{TP}}$ for all participating teams.
		For the calculation of the metrics in \textbf{d}-\textbf{e}, only trajectories presenting a changepoint at a distance larger than 20 points from the start/end points were considered as undergoing a switch. RMSE$_{\rm{TP}}$ was estimated only for true positive position pairs. Colors indicate teams, following the same color code as in \textbf{a}. 
		\textbf{f}-\textbf{g},  ${\rm RMSE}$ as a function of the anomalous diffusion exponent (\textbf{f}) and of the diffusion model (\textbf{g}) of the first and second segment for the best-in-class team in 1D (team \team{e}).
		All results for T3 in 1D, 2D and 3D are provided in Supplementary Fig.~\ref{fig:T3_allD}.
		}
\end{figure*}

The main objective of T3 is the precise assessment of the changepoint between two diffusive regimes, characterized by different diffusion models and anomalous diffusion exponents. As shown in Fig.~\ref{fig:task3}a, participants to this task achieved RMSE well below the one obtained from random predictions. The RMSE is heavily affected by the position of the changepoint, being minimum for changepoints located near the center of the trajectory. As described earlier, the performance for predictions of $\alpha$ and the diffusion model strongly depends on the trajectory length. In this task, they are thus correlated to the changepoint position, which sets the segment length. Therefore, the larger (smaller) the distance of the changepoint from the origin, the better (worse) the prediction for the first segment is and the worse (better) than for the second segment (Fig.~\ref{fig:task3}b-c).

For the challenge purposes, we simulated all trajectories as having a changepoint that could be located at any position, including the endpoints. In this view, the presence of a changepoint at one extreme was interpreted as a trajectory not having an ``actual'' changepoint. Similarly, participants were required to always provide a prediction for the changepoint position. In the case of not detecting a changepoint, the predicted position should have coincided either with the start or the end point of the trajectory, considered equivalent for this evaluation.  With this design, the RMSE simultaneously provides an evaluation of the localization precision as well as of its specificity. We also performed further analyses to independently assess the sensitivity and specificity of the participating methods and gain further insight into their performance.  Since Fig.~\ref{fig:task3}a-c show that it is challenging to estimate the changepoint when it is located very close to the trajectory start/end points, we considered trajectories with a changepoint within $\epsilon = 20$ points from the start/end as not having a changepoint. The same criterion was applied to the predictions provided by each method. Predictions/ground truth pairs located at $\epsilon<t<L-\epsilon$ were counted as true positives. Predictions/ground truth pairs located at $t \le \epsilon$ or $t \ge L-\epsilon$ were counted as true negatives. Mixed cases were considered as false positive or false negatives. Based on this classification, we could evaluate the recall (Equation~\eqref{eq:recall}), the false positive rate (Equation~\eqref{eq:fpr}), and the Jaccard similarity coefficient (Equation~\eqref{eq:Jaccard}). We also calculated the RMSE$_{\rm{TP}}$, defined as the RMSE obtained only for true positives.

The plot of the recall vs. the false positive rate (Fig.~\ref{fig:task3}d) shows that all submitted methods detect more than 92\% of the inner changepoints but present a rate of false positives larger than $\approx 10$\% and sometimes as high as $\approx 40$\%. We think that several factors might interplay to produce this behavior. As explained earlier, participants always provided a prediction for the changepoint position, the latter being equal to one of the trajectory endpoints if no changepoint was detected. In the latter analysis, our distance-based criterion relaxes this requirement to a distance $\epsilon = 20$ points from the endpoints.  Thus, the high false positive rate reflects the methods' limitations when dealing with changepoints close to the trajectory endpoints that, instead of being associated to no changepoint, are generally predicted to be more internal. Nevertheless, since the challenge metric does not explicitly account for false positive identifications, predicting an inner changepoint even when the odds of predicting a false positive are high might be a conservative choice to keep the RMSE low. In some case, this effect is produced by the choice of the architecture. For example, in 1D and 3D, team \team{e} applied a strategy based on the averaging of predictions obtained through different networks. In this way, they could reduce the RMSE even for changepoints close to the trajectory endpoints (Fig.~\ref{fig:task3}a), but it also led to a high rate of false positives (Fig.~\ref{fig:task3}d-e), associated with contrasting predictions of the networks (e.g., a very early changepoint and a very late changepoint), averaging into an internal point. 

In addition, we aimed at exploring the relationship between the overall detection performance and changepoint localization precision. As a measure of detection performance, we used the Jaccard similarity coefficient for binary classification (Equation~\eqref{eq:Jaccard}) that, with respect to the recall, further accounts for false positive detection.  The localization precision was instead estimated by RMSE$_{\rm{TP}}$ resulting from true positive identifications. The plot of the Jaccard similarity coefficient vs RMSE$_{\rm{TP}}$ (Fig.~\ref{fig:task3}e) shows that, despite the false positive rate,  all submitted methods show good overall detection performance and comparable precision (RMSE$_{\rm{TP}} = 10 \mbox{--} 20$ points). Interestingly, the performance of teams \team{b} and \team{j} improves with the dimensionality of the problem, consistently with the increase of information provided by the additional components of the motion. Team \team{e} also shows an improvement from 1D to 2D, in agreement with this explanation. The degradation of performance of team \team{e} in 3D can be ascribed to their approach to the problem through the independent training of three 1D networks, showing obvious limitations when applied to a diffusion model that is not the simple composition of 1D diffusion along orthogonal directions.

The combination of $\alpha$-exponents and diffusion models of the two segments is also expected to affect the changepoint localization precision. However, our dataset has a rich parameter space entangling several variables (anomalous model, $\alpha$, noise, changepoint location) and some imbalance since not all the models can have any value of $\alpha$. To highlight changes in RMSE due to a switch in $\alpha$ or in the diffusion model, we restricted the analysis to a subset of trajectories with a single noise level (SNR=10, Fig.~\ref{fig:task3}f,g). Unsurprisingly, the ${\rm RMSE}$ is minimal when there is a large change in $\alpha$, as between nearly immobile motion ($\alpha < 0.5$) to either superdiffusion ($1 \le \alpha < 1.5$) or directed or ballistic motion ($1.5 \le \alpha \le 2$) (Fig.~\ref{fig:task3}f). The worst case scenario is instead observed when both segments undergo mild sub- ($0.5 \le \alpha <  1$) or superdiffusion ($1 \le \alpha <  1.5$). The matrix shows a reasonable level of symmetry, considering the large heterogeneity of the dataset. However, in the presence of small changes of $\alpha$, such as between $0.05 \le \alpha <  0.5$ and $0.5 \le \alpha < 1$, or between $1 \le \alpha <  1.5$ and $1.5 \le \alpha \le  2$, the methods seem to detect changes involving an increase of $\alpha$ with better precision. 

This dependence is related in a nontrivial fashion to the change in ${\rm RMSE}$ observed as a function of diffusion models (Fig.~\ref{fig:task3}g). In fact, while FBM and SBM allow Brownian, sub- and superdiffusion, CTRW and ATTM do not allow superdiffusion, and LW does not allow subdiffusion. Changepoints associated with a switch of $\alpha$ but with no change of model are the most difficult to precisely locate. The smallest ${\rm RMSE}$ is observed when LW switches to CTRW. In contrast, models involving an abrupt (ATTM) or smooth change of diffusivity (SBM) are the most difficult to distinguish from the others.

\subsection*{Analysis of experimental data} \label{sec:experiments}

\begin{figure*}[h!]
	\includegraphics[width=0.9\textwidth]{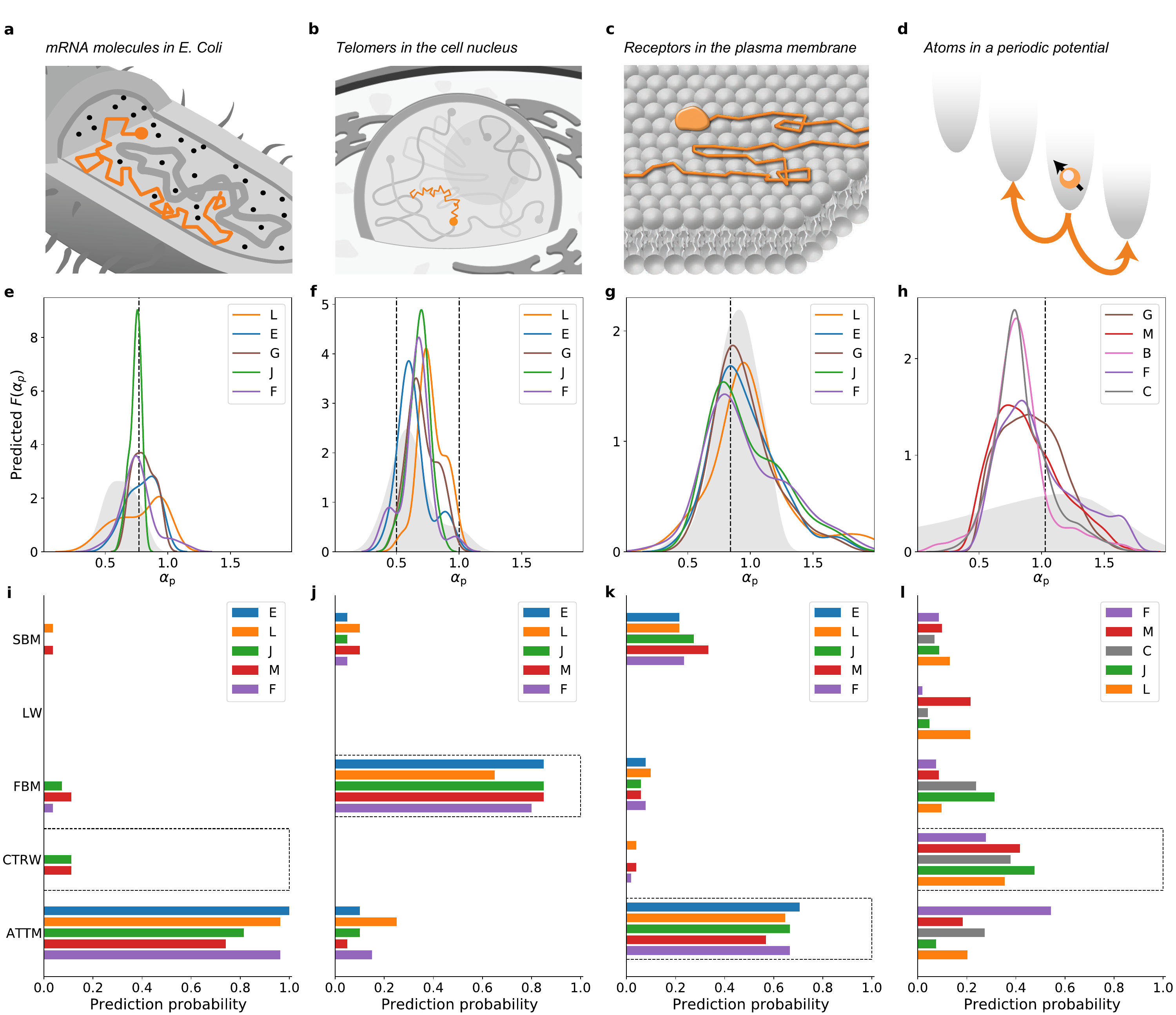}%
	\caption{\label{fig:experiments} \textbf{Analysis of experimental datasets}  \textbf{a-d}, Schematic representation of the experiments analyzed in the contest of the AnDi challenge: 2D motion of mRNA molecules inside live \emph{E. coli} cells (\cite{golding2006physical}, \textbf{a}); 2D motion of telomeres in the nucleus of mammalian cells (\cite{stadler2017non, krapf2019spectral}, \textbf{b}); 2D motion of biomolecular receptors moving on the membrane of mammalian cells (\cite{manzo2015weak}, \textbf{c}); and 1D motion of single atoms moving in a 1D periodic optical potential (\cite{kindermann2017nonergodic}, \textbf{d}).  \textbf{e-h}, Histograms of the estimation of the anomalous diffusion exponent $\alpha_{\rm{p}}$ predicted by top teams for trajectories from experimental datasets. Gray areas correspond to the results of baseline method TA-MSD. Dashed lines indicate the original estimations of $\alpha$ provided by Refs.~\cite{golding2006physical} (\textbf{e}), \cite{stadler2017non, krapf2019spectral} (\textbf{f}), \cite{manzo2015weak} (\textbf{g}), and \cite{kindermann2017nonergodic} (\textbf{h}). \textbf{i-l}, Histograms of the diffusion model predicted by top teams for trajectories from experimental datasets. Dashed boxes indicate the original classifications provided by Refs.~\cite{golding2006physical} (\textbf{i}), \cite{stadler2017non, krapf2019spectral} (\textbf{j}), \cite{manzo2015weak} (\textbf{k}), and \cite{kindermann2017nonergodic} (\textbf{l}). We show predictions obtained by the top 5 teams for the corresponding subtask. For the last dataset, we further selected the teams based on their performance on short ($L\approx10$) trajectories.
    All results for the analysis of the experimental data are presented in Supplementary Figs. \ref{fig:T1_GC}-\ref{fig:T2_WI}.
    }
\end{figure*}

The datasets provided to the participants for the scoring of the methods participating in T1 and T2 also included experimental trajectories of mRNA molecules in bacterial cells, telomeres in the cell nucleus, proteins in the cell membrane and cytoplasm, single atoms in an optical trap, and tracer particles in cell cytoplasm and stirring liquid, from previously published works. For these trajectories, no objective ground truth is available besides the interpretation given in the literature. Therefore, it is not possible to assess their absolute errors and they were not included in the scoring. However, we found it interesting to carry out a comparative analysis of the predictions blindly provided by the 5 top-scoring challenge participants in each task. Out of the whole dataset, we discuss the results for 4 representative experiments~\cite{golding2006physical, krapf2019spectral, manzo2015weak, stadler2017non, kindermann2017nonergodic} for the inference of $\alpha$ (Fig.~\ref{fig:experiments}a-d) and the classification of the underlying model (Fig.~\ref{fig:experiments}e-h). The results obtained by all methods are shown in Supplementary Figs.~\ref{fig:T1_GC}~--~~\ref{fig:T2_WI}.

The first dataset includes 2D trajectories of mRNA molecules inside live {\it E. coli} cells from the  work by Golding and Cox~\cite{golding2006physical} (Fig.~\ref{fig:experiments}a). Together with Ref.~\cite{caspi2000enhanced}, these data provide one of the first evidences of subdiffusion in cellular systems. These experiments have generated a lively discussion about their underlying diffusion model (mainly between FBM and CTRW) and ergodicity~\cite{he2008random,magdziarz2009fractional,magdziarz2011anomalous,molina2016fractional}.
All top-ranking methods provided distributions of exponents centered (median between $0.75$ and $0.81$) around the value estimated in the original publication ($\alpha=0.77$) with variable width (st. dev. between $0.04$ and $0.18$) (Fig.~\ref{fig:experiments}e). However, the methods agreed in classifying the large majority (between $74\%$ and $100\% $) of trajectories as ATTM (Fig.~\ref{fig:experiments}i). This classification confirms the occurrence of ergodicity breaking, since both CTRW and ATTM are compatible with non-ergodic behavior and both have power-law waiting-time distribution. The preference toward ATTM might arise because of its varying diffusivity that better accounts for heterogeneity due to the biological environment or to variable noise.

The second dataset of experiments includes 2D trajectories of telomeres in the nucleus of mammalian cells~\cite{stadler2017non, krapf2019spectral} (Fig.~\ref{fig:experiments}b). It was previously shown that their TA-MSD features a FBM-like subdiffusive scaling for short and intermediate times with a mean exponent  $\alpha \simeq 0.5$,  approaching a linear behavior ($\alpha \simeq 1$) at longer timescales~\cite{stadler2017non}. Also in this case, the classification methods largely agree and associate most of the trajectories to FBM (between $65\%$ and $85\% $) (Fig.~\ref{fig:experiments}j). However, the determination of the exponent often produces a bimodal distribution with median values between $0.61$ and $0.75$ (Fig.~\ref{fig:experiments}f). Likely, the methods are not able to pick up the crossovers between diffusion regimes and rather assign an average exponent to each trajectory. The analysis of these experiments deserves further methodological effort, since heterogeneous diffusion is emerging as a key feature of random motion in the biological environment~\cite{lanoiselee2018diffusion}.

The third dataset consists of 2D trajectories recorded for receptors diffusing in the plasma membrane of mammalian cells  (Fig.~\ref{fig:experiments}c). In the original work~\cite{manzo2015weak}, the TA-MSD was found to scale roughly linearly, whereas the EA-MSD showed subdiffusion with $\alpha \simeq 0.84$; this non-ergodicity was attributed to a temporal change of diffusivity and associated to ATTM. Once more, the classification methods largely confirmed previous results. A large percent of trajectories were attributed to the two models with time-dependent diffusion coefficient, namely the ATTM (between $57\%$ and $71\% $) and the SBM (between $22\%$ and $33\% $) (Fig.~\ref{fig:experiments}k). Moreover, inference methods consistently detected a large heterogeneity in $\alpha$, including both sub- and superdiffusion, with a slightly subdiffusive overall value, median between $0.86$ and $0.95$ (Fig.~\ref{fig:experiments}g), in agreement to the original study~\cite{manzo2015weak}.

To demonstrate the applicability of these methods beyond biological systems and at different spatio-temporal scales, we included a dataset with 1D trajectories obtained for single atoms moving in a 1D periodic potential and interacting with a near-resonant light field that acts as a thermal bath~\cite{kindermann2017nonergodic} (Fig.~\ref{fig:experiments}d). These data were originally interpreted as evidence of CTRW with $\alpha = 1$~\cite{kindermann2017nonergodic}. Subsequently, the CTRW was deduced from microscopic parameters reproducing the trajectories without free parameters~\cite{dechant2019continuous}. Because of the intrinsic complexity of this experiment, the trajectories were extremely short ($\approx 10$ datapoints), a regime that challenges the predictive power of any approach. Indeed, in this range of trajectory length all the methods showed rather large uncertainties on simulated data (Fig.~\ref{fig:task1}f and Fig.\ref{fig:task2}f). However, since the microscopic mechanisms are well known, we aimed at using these experiments as a benchmark to check the predictive limits of the different approaches for very short trajectory length in a real scenario. The top regression methods for such short trajectories in 1D provided distributions spread over a wide range of $\alpha$, with medians between $0.8$ and $0.91$ (Fig.~\ref{fig:experiments}h). The results of model classification were also less conclusive with respect to the previous cases, likely a consequence of having short trajectories and of having $\alpha \simeq 1$, a regime where detectable differences among models are reduced (as shown in Fig.~\ref{fig:task2}d). Predictions might also suffer from the lack of training data based on the microscopic model of Ref.~\cite{dechant2019continuous}, of which CTRW with $\alpha = 1$ is an approximation. Still, the CTRW was the most-likely model for 4 of the 5 top-scoring methods (between $28\%$ and $48\% $, Fig.~\ref{fig:experiments}l), thanks to the capability of these methods to extract information from the microscopic dynamics of the generative models and not only from the long-term properties of the trajectory and its MSD. 

The methods participating in T3 were not initially planned to be applied to the analysis of experimental data, due to the lack of trajectories featuring changes of diffusion models and/or anomalous diffusion exponent with availability of previous analysis for comparison. However, when applied to some of the experimental trajectories described above, they did not evidence a significant occurrence of changepoints, as expected. 

\section*{Discussion}

The results of the AnDi Challenge (T1) show that the choice of the analysis method strongly affects the accuracy in the determination of the anomalous diffusion exponent $\alpha$, in particular for more challenging conditions. Most of the methods outperform the conventional TA-MSD, even for long trajectories. For each dimension, we could identify a group of methods with comparable performance that greatly improve the precision of the anomalous diffusion exponent with respect to the baseline provided by the classical estimation of the MSD. These approaches were all based on machine learning, so we can infer that machine-learning-based methods can go beyond classical statistics, probably because they can extract from the trajectories of complex models some information that is not easily assessed by classical statistics.
Despite a little degradation of performance, top-ranking methods perform best also for short and noisy trajectories, as shown by the correlation between metrics calculated over a subset of trajectories ($L<200$, SNR$=1$) with respect to the same metrics obtained over the whole dataset (Supplementary Fig.~\ref{fig:noi_vs_norm}). This is a major improvement for trajectory analysis, since it enables collecting information from short and noisy tracks (e.g., those obtained by SPT PALM~\cite{manley2008high}) and from time segments of trajectories exhibiting heterogeneous behavior, without further averaging. However, the aspect that mostly boosts the overall performance is the ability to extract the anomalous diffusion exponent (an intrinsic ensemble property) for non-ergodic models from single trajectories (Fig.~\ref{fig:task1}d). Top-performing methods are capable of determining model properties usually obtained from ensemble averages or feature distributions from patterns present in single trajectories. It is quite remarkable that this is possible even in the presence of noise that is known to hide non-ergodic behavior in some classical estimators~\cite{jeon2013noisy} or with short trajectories that limit obtaining sufficient statistics for features such as the waiting-time distribution. This is a major limitation for approaches based on classical statistics (e.g., Bayesian inference) with models having several hidden variables that need to be systematically integrated.  
The availability of reliable methods to infer $\alpha$ will encourage researchers to further investigate the deviations from Brownian behavior that emerge in many experiments of interest, e.g., for biology and physics.

The AnDi challenge (T2) has led to the first concerted effort to develop methods able to classify individual trajectories among several mathematical models of diffusion. Machine-learning methods ranked top in the leader board and achieved an overall accuracy greater than $80\%$ at detecting the ground-truth diffusion models. The comparison of ${\rm F_1}$-score and AUC/ROC (Supplementary Fig.~\ref{fig:T2_1D_ROC}, Supplementary Fig.~\ref{fig:T2_2D_ROC}, Supplementary Fig.~\ref{fig:T2_3D_ROC}, and Supplementary Fig.~\ref{fig:rank_AUC}) shows that most of the methods are quite confident at providing the correct classification. However, a limitation of all these classification approaches is that they can only choose among the diffusion models provided in the training. To robustly extend model classification to actual experiments, it can be useful to further widen the palette of models (e.g., by using {\it ad hoc} models), include a none-of-the-above class, and/or to include some metric of the confidence of the estimation (e.g., by using an entropy measure calculated on the predictions of an ensemble of machine-learning models). 
Trajectory segmentation (T3 of the AnDi challenge) has been widely investigated when changes occur with respect to an estimator of the observable such as the mean or the variance~\cite{truong2020selective}. Determining changes of anomalous diffusion is a rather novel problem, triggered by recent experimental findings~\cite{weron2017ergodicity, sabri2020elucidating}. We kept the challenge design rather simple, with trajectories of fixed length featuring exactly one changepoint. Even in this simple condition, the wide parameter space made the problem rather challenging, limiting the participation to T3 to only 4 teams. Yet, the submitted results showed an interesting asymmetry: The changepoint localization precision seems not only to depend on the relative length of the segments but also on the changepoint location (Fig.~\ref{fig:task3}a), producing a lower RMSE for changepoints located at the beginning of the trajectory. Similarly, the methods show best performance in estimating $\alpha$ and diffusion model for the first segment (Fig.~\ref{fig:task3}b-c). We believe that this is at least partly a consequence of the inaccurate localization of the changepoint and the non-stationarity of some models. The inexact localization of the changepoint produces two spurious segments, altering the tail of the first segment and the initial point of the second by removing or adding spurious points. For non-stationary models, the initial point encloses information about the initiation of the physical process, thus improper segmentation impacts more severely the evaluation of the second segment~\cite{cherstvy2014ageing}. 

From the blind analysis of various experimental datasets, we observed that the top methods, although based on different principles, lead to very similar results. This is encouraging as it points to an objective underlying reality of the anomalous diffusion phenomena and its mechanisms, which can be measured experimentally and has now been underpinned by the results of the AnDi challenge. Importantly, the results provided by the challenge methods were also in line with the conclusions of previous studies~\cite{golding2006physical, krapf2019spectral, manzo2015weak, stadler2017non, kindermann2017nonergodic}, further reinforcing their reliability. Interestingly, while the original works required a combination of several estimators, including ensemble averages, the challenge methods were able to provide compatible predictions in a one-shot analysis and with no prior knowledge about the experimental conditions.
This is a particularly remarkable result, since the methods were not specifically trained to work with parameters used in experiments. In fact, experimental trajectories often show broad distributions of diffusion coefficients. In spite of a fixed localization error, this produces a non-uniform SNR with respect to our simulations. Also, experiments have different sampling rates with respect to the characteristic diffusion timescale. Accounting for the variability introduced by these effects during the training might improve the methods' prediction capability, further boosting their performance. 

The number of experiments producing individual random trajectories is steadily increasing, accompanied by the production of {\it ad hoc} analysis tools. The AnDi challenge gave the opportunity to obtain a first assessment of some of these tools, oriented at detecting anomalous diffusion.  In particular, we focused on methods quantifying deviation from pure Brownian behavior in terms of anomalous diffusion exponent and the underlying mathematical model. However, similar experiments are often analyzed following a more phenomenological approach, e.g., the classification of motion as diffusive, immobile, confined, or directed. Although the latter classification offers a more intuitive interpretation of random motion occurring in some systems, the models included in the challenge are strictly connected to these diffusion modalities. In fact, they allow a generalization of anomalous diffusion beyond the life sciences and include macroscopic natural and human processes, ranging from the foraging of animals to the spread of diseases, to trends in financial markets and climate records. 

Building on these considerations, we believe it is necessary to establish clear and unified guidelines to identify and report anomalous diffusion, in particular from  experiments, where the ground truth is not known. Possibilities in this sense might involve a list of key parameters to be quantified together with their respective confidence interval, e.g., based on the comparative use of multiple methods, involving both machine learning and classical statistics. The joint approach will allow to combine advantages from both worlds: while machine learning methods are becoming more available and powerful, they often operate as a black box; estimators based on classical statistics can thus help to provide deep insight on anomalous diffusion phenomena. 

The AnDi challenge gathered a large part of the community to trigger this discussion and collaborate on this unifying task. We hope this effort might be extended in the future to reach a larger consensus.
To this aim, we have built an interactive tool (\url{http://andi-challenge.org/interactive-tool/}) where datasets and results of the challenge are stored; new methods can undergo an automated benchmarking according to the challenge rules and compare their scores with those of other participants. In fact, since the conclusions of the challenge, several participants have already improved their scores. Therefore, the challenge is permanently open and performance improvements will be continuously updated on demand.

\hypertarget{sec:methods_h}{}
\section*{Methods \label{sec:methods_h}}

\subsection*{Organization of the challenge}

We ran the Anomalous Diffusion (AnDi) challenge as a time-limited competition from March 1, 2020, to November 1, 2020. The competition was hosted on the Codalab platform (\url{https://competitions.codalab.org/competitions/23601}) and divided in three phases (Development, Validation, and Challenge). The competition has later been converted to an open challenge, continuously accepting new submissions.  Datasets, methods, list of participants, and results of the AnDi Challenge are available at \url{http://andi-challenge.org}. Software for simulation and analysis is hosted on the competition GitHub repository \url{https://github.com/AnDiChallenge}.

\subsection*{Challenge methods}

Among the participants, we could distinguish fifteen substantially different approaches (Table~\ref{tab:methods} and \hyperlink{sec:SI-note1_h}{Supplementary Note 1}). We classify the approaches based on three different criteria, as detailed in Table~\ref{tab:methods}. First, we group methods based on the type of approach used, whether involving machine-learning or classical statistics. A large majority of methods are based on machine-learning architectures, such as recurrent neural networks (RNN), convolutional neural networks (CNN), gradient boosting machines, graph neural networks, extreme learning machine (ELM), or sequence learners. Other methods are based on statistical approaches, such as Bayesian inference, temporal scaling, and random interval spectral ensemble (RISE). 
A second grouping involves the type of input data used. Some methods employed feature engineering using classical statistics as an input, whereas other were simply fed raw trajectories. A further classification is based on whether methods required a specific training or model for different (ranges of) trajectory lengths (length-specific) or not. Several methods could be directly used or easily adapted to run multiple tasks. 

\subsection*{Structure of the datasets}

Simulated datasets were composed of synthetic trajectories generated according to five different mathematical models, both ergodic and non-ergodic: annealed transient time motion (ATTM, weakly non-ergodic), a motion with random changes of diffusion coefficient in time~\cite{massignan2014nonergodic}, continuous-time random walk (CTRW, weakly non-ergodic), a motion undergoing local trapping with a wide distribution of waiting times~\cite{scher1975anomalous}, fractional Brownian motion (FBM, ergodic), a motion with long-range correlated steps, often used to describe viscoelastic effects~\cite{mandelbrot1968fractional}, L\'{e}vy walk (LW, ultra-weakly non-ergodic), a motion displaying irregular jumps with constant velocity, often associated with animal foraging strategies~\cite{klafter1994levy},  and scaled Brownian motion (SBM, weakly non-ergodic), a motion whose diffusion coefficient features deterministic time-dependent changes~\cite{lim2002self}. We considered trajectories with anomalous diffusion exponents in the range $\alpha\in[0.05,2]$. Exponents were restricted to $\alpha\ge0.05$ because smaller exponents produce practically immobile trajectories. Note that CTRW and ATTM are strictly subdiffusive ($\alpha\leq 1$), LW is superdiffusive ($\alpha\geq 1$), FBM cannot have ballistic behavior ($\alpha < 2$), whereas SBM covers the whole exponent range. 

Each dataset contained $10^4$ trajectories of variable length. All trajectories were first generated with a length $L=1000$. For theoretical models providing trajectory sampling at irregular times (CTRW and LW), oversampling was used to obtain tracer coordinates at uniform times.  The trajectories were then standardized to have a unitary standard deviation $\sigma_{D}$ of the distribution of displacements over unit time.  To mimic experimental data, trajectories were corrupted with a finite localization precision. For this, a random number from a normal distribution $\mathcal{N}(0,\sigma_{\rm noise})$ was added to each trajectory coordinate.  Last, the displacements' standard deviation was scaled by a random number sampled from a normal distribution $\mathcal{N}(0,1)$ to include the effect of an effective diffusion coefficient (see Fig.~\ref{fig:summary_andi}a-c for exemplary trajectories in each dimension). 
Trajectories were thus cut to the desired length. For T1 and T2, trajectories were cut to lengths $L\in[10,1000]$, whereas for T3 all trajectories had length $L=200$. A different dataset was generated for each task to ensure the proper balance of the feature to be determined. Therefore, the dataset for T1 had a balanced distribution of anomalous exponents but not of diffusion models, whereas the dataset for T2 was balanced with respect to the diffusion models. For T3, trajectories were obtained by concatenating trajectories simulated for all models and exponents. Each trajectory had a random changepoint at a discrete index $t_{\rm GT}\in[1,199]$ corresponding to a change at least in one of the two features ($\alpha$ and diffusion model). An example of such kind of trajectories is presented in Fig.~\ref{fig:summary_andi}c.

Three levels of noise were used to corrupt trajectories, corresponding to $\sigma_{\rm{noise}} = 0.1$, $0.5$, $1$. The SNR was calculated as $\rm{SNR}=\sigma_{D}/\sigma_{\rm noise}$, where $\sigma_{D}$ is the standard deviation of the distribution of displacements over unit time. Due to the previous standardization, the SNR levels thus were $\rm{SNR}=1$, $2$, $10$. Trajectories in 2D and 3D were allowed to have different noise levels along different directions.
The overall SNR was calculated as the average of SNRs calculated along orthogonal directions. 

We developed the {\tt andi-datasets} Python package~\cite{andigithub} to allow participants to generate their own dataset (e.g., for training).  Examples of trajectories for various exponents and models are presented in Fig.~\ref{fig:summary_andi}c. Details about available functions can be found in the hosting repository \url{https://github.com/AnDiChallenge/ANDI_datasets}.

\subsection*{Theoretical models}

In this section, we present a brief introduction to the concepts of anomalous diffusion and ergodicity breaking. We provide theoretical insights about the anomalous diffusion models considered in the AnDi challenge, as well as the description of the pseudocode used for simulations in 1D.  Finally, we describe how to extend the algorithms to simulate the diffusion models in 2D and 3D, since for some models this is not simply obtained as the composition of motion along independent directions. The Python implementation of all the algorithms described below is available at \url{https://github.com/AnDiChallenge/ANDI_datasets} ~\cite{andigithub}.

\subsubsection*{Anomalous diffusion and ergodicity breaking \label{sec:weak}}

When analyzing trajectories, diffusion is typically quantified through the calculation of the mean squared displacement (MSD). The MSD grows linearly in time for Brownian walkers, $\mathrm{MSD}\sim t$, while it shows a power-law scaling for anomalous diffusion, $\mathrm{MSD}\sim t^\alpha$, where $\alpha$ is the anomalous diffusion exponent. In practice, the MSD can be calculated either by performing an ensemble average of the positions of a set of $N$ tracers,
\begin{equation}
	\mbox{EA-MSD}(t) = \frac{1}{N} \sum_{i=1}^N [{\bf x}_i(t)-{\bf x}_i(0)]^2,
	\label{eq:eamsd}
\end{equation}
or, for the trajectory of a single tracer, sampled at $L$ discrete times $t_i = i\Delta t$,  as a time-average:
\begin{equation}
	\mbox{TA-MSD}(\Delta \!=\! m \Delta t) \!=\! 
	\frac{1}{L - m}\!\!\!\sum_{i=1}^{L-m}\!
	\left[{\bf x}(t_i+m\Delta t)-{\bf x}(t_i)\right]^2.
	\label{eq:tamsd}
\end{equation}

In its most general definition, a process is considered ergodic if any single realization is able to explore all the possible configurations of the system. The impossibility of performing such an exploration is usually referred to as ergodicity breaking. For a (strong) non-ergodic process, the space of configurations is separated into mutually inaccessible domains, hence preventing its full exploration. If those domains are indeed accessible, but a single tracer is unable to visit them in a finite time, the process is instead defined as weakly non-ergodic~\cite{bouchaud1992weak}. In this case, a sufficiently large ensemble of tracers may indeed explore all possible configurations, hence producing a difference between ensemble and time averages.

In the context of anomalous diffusion, a system is said to show weak ergodicity breaking if the TA-MSD does not converge to EA-MSD in the infinite time limit~\cite{metzler2014anomalous}. Generally, while the EA-MSD still shows a power-law scaling, the TA-MSD scales linearly with the timelag~\cite{metzler2014anomalous}. Moreover, the value of the TA-MSD for different trajectories at a given timelag is a random variable, whose distribution can be analytically calculated for some diffusion models~\cite{barkai2012single}. 
One can then define the time and ensemble averaged TEA-MSD over a set of $N$ trajectories as
\begin{equation}
	\mbox{TEA-MSD}(\Delta) =\frac{1}{N}\sum_{i=1}^N 	\mbox{TA-MSD}(\Delta)_i,
\end{equation}
where $\mbox{TA-MSD}(\Delta)_i$ is the TA-MSD for the $i$-th trajectory. 
The so-called ergodicity breaking parameter (EB) \cite{he2008random} can be calculated as
\begin{equation}
\mbox{EB}=\langle \zeta^2 \rangle-1,
\end{equation}
where $\zeta=\mbox{TA-MSD}(\Delta)/  \mbox{TEA-MSD}(\Delta)$. The EB parameter, in the limit $\Delta/T\to 0$, is a widely used tool to quantify ergodicity breaking (here $T=L \Delta t$ represents the trajectory length). For ergodic diffusion, then $\mbox{EB}\rightarrow 0$, while any other value showcases a non-ergodic behavior. Processes like CTRW, ATTM and SBM show weak ergodicity breaking \cite{2005Bel,2007Rebenshtok,massignan2014nonergodic}, whereas Brownian motion and FBM are ergodic, though convergence of the EA-MSD to the TA-MSD may be slow for certain values of  the  anomalous exponent $\alpha$ \cite{2009DengErgodic}. Indeed, as discussed in \cite{schwarzl2017quantifying}, the ergodicity of FBM requires a careful analysis as a function of $\alpha$, and often other statistical measures are necessary to study ergodicity breaking. To find a technique to study short trajectories, it is important to note that, for  CTRW and ATTM, the TA-MSD shows a short-time linear behavior TA-MSD$\propto\Delta$ even for anomalous trajectories. This showcases one of the limitations of the fitting of the  TA-MSD to determine the anomalous diffusion exponent. For the case of LW, a different kind of ergodicity breaking named ultraweak can been identified, where time and ensemble averages only differ by a constant factor~\cite{2013Godec,2013Godecb}.

%%%%%%%%% CTRW %%%%%%%%%%%
\subsubsection*{Continuous time random walk \label{sec:ctrw}}

The continuous time random walk (CTRW) defines a large family of random walks with arbitrary displacement density for which the \textit{waiting time}, i.e., the time between subsequent steps, is a stochastic variable~\cite{scher1975anomalous}. Here, we consider a specific case of CTRW for which waiting times are sampled from a power-law distribution $\psi(t)\sim t^{-\sigma}$ and displacements are sampled from a Gaussian distribution with variance $D$ and zero mean. In such case, the anomalous diffusion exponent is $\alpha = \sigma-1$ (the $\mbox{ EA-MSD}= \langle {\bf x}(t)^2 \rangle\propto t^\alpha$). Since the waiting times are generated from a power law distribution, for $\sigma=2$ the $\mbox{EA-MSD}$ features Brownian diffusion with logarithmic corrections~\cite{klafter2011first}. For $\alpha=1$ one should instead use a Poisson density, or a fixed waiting time (i.e., the limit of a one-sided L\'{e}vy stable density in the limit $\alpha=1$). 

The algorithm used to simulate CTRW trajectories is described in Algorithm~\ref{alg:ctrw}. Notice that the variable $\tau$ stands for the total time at $i$-th iteration. Also notice that the output vector $\vec{x}$ corresponds to the position of the particle at the irregular times given by $\vec{t}$. 

\begin{algorithm}[H]
	\caption{Generate CTRW trajectory}
	\label{alg:ctrw}
	\begin{algorithmic}
		\STATE{\bfseries Input:}
		\bindent
		\STATE{length of the trajectory $T$}
		\STATE{anomalous exponent $\alpha$}
		\STATE{diffusion coefficient $D$}
		\eindent
		\STATE{\bfseries Define:}
		\bindent
		\STATE{$\vec{x} \rightarrow$ empty vector}
		\STATE{$\vec{t} \rightarrow$ empty vector}
		\STATE{$N(\mu, s) \rightarrow$ Gaussian random number generator with  mean $\mu$ and standard deviation $s$}
		\eindent
		\STATE{$i = 0; \ \tau = 0$}
		\WHILE {$\tau<T$}
		\STATE $t_i \leftarrow$  sample randomly from  $\psi(t)=t^{-\sigma}$
		\STATE $x_i \leftarrow x_{i-1} + N(0,\sqrt{D})$
		\STATE $\tau \leftarrow \tau + t_i$
		\STATE $i \leftarrow i+1$
		\ENDWHILE
		\STATE{\bfseries Return:} $\vec{x}, \ \vec{t}$
	\end{algorithmic}
\end{algorithm}

%%%%%%%%%%% FBM %%%%%%%%%%%
\subsubsection*{Fractional Brownian motion \label{sec:fbm}}

In fractional Brownian motion (FBM), $x(t)$  is a  Gaussian process with stationary increments. This process is symmetric, $\langle x(t)\rangle=0$, and importantly its EA-MSD scales as $\langle x(t)^2 \rangle=2K_{\rm{H}}t^{2H}$. Here,  $H$ is the Hurst exponent,  which is related to the anomalous diffusion exponent as $H = \alpha/2$~\cite{mandelbrot1968fractional, jeon2010fractional}. Also, the two-time correlation is $\langle x(t_1)x(t_2) \rangle=K_{\rm{H}}(t_1^{2H}+t_2^{2H}-|t_1-t_2|^{2H})$. 

FBM can also be introduced as a process arising from a generalized Langevin equation where the noise is non-white (aka fractional Gaussian noise, fGn). The fGn has a standard normal distribution with zero mean and power-law correlations:
\begin{align}
	\left< \xi_{\rm fGn}(t_1)\xi_{\rm fGn}(t_2)\right> & = 2K_{\rm H }H(2H-1)|t_1-t_2|^{2H-2} \nonumber \\
	&+ 4K_{\rm H }H|t_1-t_2|^{2H-1}\delta(t_1-t_2).
\end{align}
The FBM features two regimes: one  where the noise is  positively correlated ($ 1/2<H<1$, i.e., $1<\alpha<2$, superdiffusive) and one where the noise is negatively correlated ($ 0<H<1/2$, i.e., $0<\alpha<1$, subdiffusive). For $H=1/2$ ($\alpha = 1$) the noise is uncorrelated, hence the FBM converges to Brownian motion. 

For a $d$-dimensional FBM, the corresponding position vector has zero mean, $\langle {\bf x}(t)\rangle=0$, the EA-MSD is   $\langle {\bf x}(t)^2 \rangle=2dK_{\rm{H}}t^{2H}$, the autocorrelation is $\langle {\bf x}(t_1){\bf x}(t_2) \rangle=dK_{\rm{H}}(t_1^{2H}+t_2^{2H}-|t_1-t_2|^{2H})$,  and the fGN reads
\begin{align}
	\left< \xi_{\rm fGn,i}(t_1)\xi_{\rm fGn,j}(t_2)\right> & = 2K_{\rm H }H(2H-1)|t_1-t_2|^{2H-2}\delta_{ij} \nonumber \\
	&+ 4K_{\rm H }H|t_1-t_2|^{2H-1}\delta(t_1-t_2)\delta_{ij},
\end{align}
where $i,j$ in the subindex of the fGN denotes a different cartesian coordinate. 

Various numerical approaches have been proposed to solve the FBM generalized Langevin equation exactly. Here, we use the Davies-Harte method~\cite{davies1987} and the Hosking method~\cite{hosking1984modeling} via the {\tt FBM} Python package(\url{https://pypi.org/project/fbm/}). Details about the numerical implementations can be found in the associated references.
%%%%%%%%%%% LW %%%%%%%%%%%
\subsubsection*{L\'evy walk \label{sec:lw}}

The L\'evy walk (LW) is a particular case of CTRW. The time between steps is  irregular~\cite{klafter1994levy}, but,  in contrast to the CTRW considered here, the distribution of displacements for a LW is not Gaussian. We considered the case in which the flight times (i.e., the times between steps) are retrieved from the distribution $\psi(t)\sim t^{-\sigma-1}$. In one dimension,  the displacements are $\Delta x$ and the step length is $|\Delta x|$. The displacements   are correlated with the flight times such that the probability to move  a step  $\Delta x$ at time $t$ and stop at the new position to wait for a new random event to happen is  $\Psi(\Delta x , t) = \frac{1}{2}\delta(|\Delta x|-vt)\psi(t)$, where $v$ is the velocity. From here, one can show that the anomalous exponent is given by
\begin{equation}
	\alpha =
	\begin{cases}
		2       & \mbox{if} \ 0<\sigma<1 \\
		3-\sigma  & \mbox{if} \ 1<\sigma<2.
	\end{cases}
\end{equation}
The details of the numerical implementation for the LW are given in Algorithm~\ref{alg:lw}. Notice that we use a random number $r$, which can take values 0 or 1, to decide in which sense the step is performed. Also note that, as for the CTRWs, the output vectors $\vec{x},  \vec{t}$ represent irregularly sampled positions and times. 

\begin{algorithm}[H]
	\caption{Generate LW trajectory}
	\label{alg:lw}
	\begin{algorithmic}
		\STATE{\bfseries Input:}
		\bindent
		\STATE{length of the trajectory $T$}
		\STATE{anomalous exponent $\alpha$}
		\eindent
		\STATE{\bfseries Define:}
		\bindent
		\STATE{$\vec{x} \rightarrow$ empty vector}
		\STATE{$\vec{t} \rightarrow$ empty vector}
		\STATE{$v \rightarrow$ random number $\in (0,10]$}
		\eindent
		\STATE{$i = 0$}
		\WHILE {$\tau<T$}
		\STATE $t_i \leftarrow$ sample randomly from $\psi(t)\sim t^{-\sigma-1}$
		\STATE $x_i \leftarrow (-1)^rvt_i$, where random $r$  is 0 or 1 with equal probability.
		\STATE $\tau \leftarrow \tau + t_i$
		\STATE $i \leftarrow i+1$
		\ENDWHILE
		\STATE{\bfseries Return:} $\vec{x},  \vec{t}$
	\end{algorithmic}
\end{algorithm}

%%%%%%%%% ATTM %%%%%%%%%%%
\subsubsection*{Annealed transient time motion \label{sec:attm}}

The annealed transient time motion (ATTM) implements the motion of a Brownian particle whose diffusion coefficient varies in time~\cite{massignan2014nonergodic}. The tracer performs Brownian motion for a random time $t_1$ with a random diffusion coefficient $D_1$, then for $t_2$ with $D_2$, etc. The diffusion coefficients are sampled from a distribution such that $P(D)\sim D^{\sigma-1}$ with $\sigma>0$ as $D \rightarrow 0$ and that decays rapidly for large $D$. If the random times $t$ are sampled from a distribution with expected value $E[t|D] = D^{-\gamma}$, with $ \sigma < \gamma <\sigma+1 $, the anomalous diffusion exponent is $\alpha=\sigma/\gamma$ (corresponding to the subdiffusive \textit{regime I} of the model described in Ref.~\cite{massignan2014nonergodic}). Here, we consider that the distribution is a delta function, $P_t(t|D)\sim \delta (t-D^{-\gamma})$. Hence, the period of time $t_i$ in which the particle performs Brownian motion with a random diffusion coefficient $D_i$ is $t_i=D_i^{-\gamma}$, with $D_i$ extracted from the distribution described above.  The numerical implementation of the ATTM model is given in Algorithm~\ref{alg:attm}. Note that, in contrast to CTRW and LW, now the only output is $\vec{x}$ because the trajectory is already produced at regular time intervals of duration $\Delta t$.  
\begin{algorithm}[H]
	\caption{Generate ATTM trajectory}
	\begin{algorithmic}
		\label{alg:attm}
		\STATE{\bfseries Input:}
		\bindent
		\STATE{length of the trajectory $T$}
		\STATE{anomalous exponent $\alpha$}
		\STATE{sampling time $\Delta t$}
		\eindent
		\STATE{\bfseries Define:}
		\bindent
		\WHILE {$\sigma > \gamma$ and $\gamma > \sigma+1$}
		\STATE{$\sigma \leftarrow$ uniform random number $\in (0,3]$}
		\STATE{$\gamma = \sigma/\alpha$}
		\ENDWHILE
		\STATE{$\mbox{BM}(D,t,\Delta t) \rightarrow$ generates a Brownian motion trajectory of length $t$ with diffusion coefficient $D$, sampled at time intervals $\Delta t$}
		\STATE{$\vec{x} \rightarrow$ empty vector}
		\eindent
		\WHILE {$\tau<T$}
		\STATE $D_i \leftarrow$ sample randomly  from  $P(D)=D^{\sigma-1}$
		\STATE $t_i \leftarrow D_i^{-\gamma}$
		\STATE number of steps $N_i=\rm{round}(t_i/\Delta t)$
		\STATE $x_i,...,x_{i+N_i} \leftarrow \mbox{BM}(D_i,t_i,\Delta t)$
		\STATE $i \leftarrow i+N_i+1$
		\STATE $\tau=\tau+N_i\Delta t$
		\ENDWHILE
		\STATE{\bfseries Return:} $\vec{x}$
	\end{algorithmic}
\end{algorithm}

%%%%%%%%%%% SBM %%%%%%%%%%%%%%%%%%%%
\subsubsection*{Scaled Brownian motion \label{sec:sbm}}

The scaled Brownian motion (SBM) is a process described by the Langevin equation with a time-dependent diffusivity $K(t)$
\begin{equation}
	\frac{dx(t)}{dt}=\sqrt{2K(t)}\xi(t),
\end{equation}
where $\xi(t)$ is white Gaussian noise~\cite{lim2002self}. For the case in which $K(t)$ has a power-law dependence with respect to $t$ such that  $K(t)=\alpha K_\alpha t^{\alpha-1}$, the EA-MSD follows $\left< x^2(t) \right>_N\sim K_\alpha t^\alpha$ with $K_\alpha = \Gamma(1+\alpha)K_\alpha$. The numerical implementation of SBM is presented in Algorithm~\ref{alg:sbm}.

\begin{algorithm}[H]
	\caption{Generate SBM trajectory}
	\label{alg:sbm}
	\begin{algorithmic}
		\STATE{\bfseries Input:}
		\bindent
		\STATE{length of the trajectory $T$}
		\STATE{anomalous exponent $\alpha$}
		\eindent
		\STATE{\bfseries Define:}
		\bindent
		\STATE{{\tt erfcinv}$(\vec{a}) \rightarrow$ Inverse complementary {\rm erf} of $\vec{a}$ }
		\STATE{\textit{U}$(L) \rightarrow$ returns $L$ uniform random numbers $\in [0,1]$}
		\eindent
		\STATE{\bfseries Calculate:}
		\bindent
		\STATE{$\overrightarrow{\Delta x} \leftarrow (1^\alpha,2^\alpha,...,T^\alpha) - (0^\alpha,...,(T-1)^\alpha)$}
		\STATE{$\overrightarrow{\Delta x} \leftarrow 2\sqrt{2} U(L) \overrightarrow{\Delta x}$},
		\STATE{$\vec{x} \leftarrow \mbox{{\tt cumsum}}(\overrightarrow{\Delta x})$}.
		\eindent
		\STATE{\bfseries Return:} $\vec{x}$
	\end{algorithmic}
\end{algorithm}

\subsubsection*{Simulations in higher dimensions}

The algorithms presented above provide examples for the simulation of 1D trajectories. In order to maintain the properties of each anomalous diffusion model, extension to 2D and 3D was performed differently depending on the considered model. For ATTM, CTRW, FBM, and SBM in 2D, trajectories were obtained by the simple composition of (independent) motion performed over orthogonal axes. The same was done for FBM and SBM in 3D. For ATTM and CTRW (3D), and for LW (2D and 3D), waiting times and displacement lengths were sampled according to the recipe provided by each particular model in 1D. However, the displacement length was used to sets the radius of the circle (2D) or the sphere (3D) over which the tracer step ended up. The direction was randomly chosen to ensure the uniform sampling of the circle or the sphere, and coordinates along orthogonal axes were calculated accordingly.

\subsection*{Metrics}

We calculated several metrics to quantify the performance of the submitted methods with respect to the ground truth in the various tasks. Although only the most representative metrics were used to build the competition leaderboard, others were used to gain further insight about the methods. We further built an interactive tool (\url{http://andi-challenge.org/interactive-tool/}) for comparing method performance (Supplementary Fig.~\ref{fig:interactive_screenshot}). This application also provides a useful tool for developers to benchmark new methods.

\subsubsection*{Challenge metrics}

\begin{itemize}
\item{Mean absolute error (${\rm MAE}$).}
Methods were required to provide an accurate prediction for the anomalous diffusion exponent $\alpha$ for a single trajectory (T1) or for a part of a trajectory after segmentation (T3). Method performance was thus quantified by the ${\rm MAE}$ between the predicted value and the ground truth:
\begin{equation}
	\label{eq:mae}
	{\rm MAE} = \frac{1}{N} \sum_{i=1}^{N}{| \alpha_{i,\rm{p}} - \alpha_{i,\rm{GT}} |},
\end{equation}
where $N$ is the number of trajectories in the dataset, and $\alpha_{i,\rm{p}}$ and $\alpha_{i,\rm{GT}}$ represent the predicted and ground truth values of the anomalous exponent of the $i$-th trajectory, respectively.

\item{$F_1$-score.} 
For T2 and T3, the methods have to provide a score of the probability for a trajectory (or a segment) to be assigned to one of the five diffusion models. Predictions for which the highest probability value corresponded to the ground-truth model were identified as true positives. As a summary statistics for model classification, we used the ${\rm F_1}$-score.
For multi-class classification problems, scoring metrics such as precision, recall, and ${\rm F_1}$-score can be computed as a macro-average (which evaluates the metric independently for each class and then take the average, giving all classes the same weight), or as a micro-average (which aggregates the contributions of all classes to compute the average metric). Micro-averaging is generally preferable when class imbalance is present. Although the challenge was based on a balanced dataset with each class equally represented, we used a micro-averaged ${\rm F_1}$-score in order not to provide any hint to participants about the content of the dataset.
The micro-averaged ${\rm F_1}$-score was calculated as
\begin{equation}
	\label{eq:f1}
    F_1 = \frac{2\mbox{TP}}{2\mbox{TP}+\mbox{FP}+\mbox{FN}},
\end{equation}
where TP, FP, and FN represent true positives, false positives, and false negatives calculated over the whole dataset, respectively.

\item{Root mean square error (RMSE).} The trajectory segmentation problem in T3 requires the location of the point where a trajectory undergoes a change in anomalous diffusion. The most important consideration for a changepoint method is how accurately it localizes the changepoint itself. The quantification of this accuracy was performed through the RMSE between the predicted and ground truth position:
\begin{equation} 
	\label{eq:rmse}
	{\rm RMSE} = \sqrt{\frac{1}{N} \sum_{i=1}^{N} \Big(t_{i,{\rm p}} - t_{i,{\rm GT}} \Big)^2},
\end{equation}
where $t_{i,{\rm p}}$ and $t_{i,{\rm GT}}$ represent the predicted and ground truth values of the changepoint  position, respectively. Unlike for T1, where we used the MAE, in this case we opted for the RMSE. This quadratic metric gives a higher weight to large errors, thus penalizing methods that provide predictions very far from the ground truth.

\item{Mean reciprocal rank (MRR).}
For ranking purposes of T3, the precision in determining the changepoint position, the anomalous diffusion exponent $\alpha$, and the diffusion model were summarized into a single statistics for the overall method evaluation, given by the MRR: 
\begin{equation}
\label{eq:MRR}
	{\rm MRR} = \frac{1}{3}\cdot 
	\left(\frac{1}{{\rm rank}_{\rm MAE} } +\frac{1}{{\rm rank}_{F_1}}+\frac{1}{{\rm rank}_{\rm RMSE}}\right),
\end{equation}
where ${\rm rank}_{\rm MAE}$, ${\rm rank}_{\rm F_1}$, and ${\rm rank}_{\rm RMSE}$ correspond to the position in an ordered list based on the value of the corresponding metrics. For this task, ${\rm MAE}$ and ${\rm F_1}$-score  were calculated by treating each segment (before and after the predicted changepoint) as an individual trajectory and averaging the metrics obtained over the two segments.
\end{itemize}

\subsubsection*{Additional metrics}
Further statistics were used for the comparative analysis of the performance of the methods.
\begin{itemize}
\item{Anomalous exponent bias.}
For the determination of the anomalous diffusion exponent in T1 and T3, besides the accuracy, we further assessed whether the predicted value systematically differed from the ground truth. For this reason, we calculated the distribution of the difference between predicted and ground truth exponent (Supplementary Fig.~\ref{fig:T1_1D_bias}, Supplementary Fig.~\ref{fig:T1_2D_bias}, and Supplementary Fig.~\ref{fig:T1_3D_bias}), and estimated the bias $\theta$ as its expectation value:
\begin{equation}
	\label{eq:bias}
	\theta = \frac{1}{N} \sum_{i=1}^{N}{( \alpha_{i,\rm{p}} - \alpha_{i,\rm{GT}} )}.
\end{equation}
As shown in Fig.~\ref{fig:task1}, the estimation of the anomalous diffusion exponent from the fit of the TA-MSD shows a negative bias (i.e., the predicted exponent $\alpha_{\rm{p}}$ is systematically smaller than the ground truth exponent $\alpha_{\rm{GT}}$). Such effect is particularly important close to $\alpha_{\rm{GT}} = 1$ and is associated to the presence of localization error~\cite{martin2002apparent}. However, as shown in Supplementary Fig.~\ref{fig:T1_1D_bias}, Supplementary Fig.~\ref{fig:T1_2D_bias}, and Supplementary Fig.~\ref{fig:T1_3D_bias}, the top performing methods show little or no bias in their predictions.

\item{Receiver operating characteristic (ROC) curve and area under the curve (AUC).} 
The calculation of the ${\rm F_1}$-score assumes that a method outputs a discrete classifier (i.e., a unique choice for the diffusion model). However, many methods output continuous numbers associated to the probability of the input to belong to each class. Thus, these values assigned to each model contain more information about the performance of the classifier. This information can be summarized by the ROC curve and the corresponding AUC. The ROC curve reports the true positive rate (or sensitivity) versus  the false negative (one minus the specificity) for different levels of probability thresholds: if an input has a certain class probability above the threshold, it is considered to belong to such class. The AUC is given by the integral of the ROC curve and is equal to the probability that a classifier will rank a randomly chosen positive instance higher than a randomly chosen negative one. It thus provides a useful tool to compare the sensitivity and specificity of a given classifier.
In particular, being based on probability instead of class labels, ROC/AUC report how ``doubtful'' a method is about its choice of the model.
ROC curves for each class versus the others are shown in Supplementary Fig.~\ref{fig:T2_1D_ROC}, Supplementary Fig.~\ref{fig:T2_2D_ROC}, and Supplementary Fig.~\ref{fig:T2_3D_ROC}  for all teams.   Micro- (i.e., considering each class as a binary prediction) and macro-averaged (i.e., considering an equal weight for the classification of each label) ROC curves are also reported. The ROC/AUC analysis confirms that ATTM is the most problematic model to classify, whereas the best results are obtained for CTRW and LW.  The scatter plot of values of ${\rm F_1}$-score vs. micro-averaged AUC show a rather good correlation (Supplementary Fig.~\ref{fig:rank_AUC}), with the exception of a few models (teams \team{l}, \team{d} and \team{n}) that perform considerably better in terms of ${\rm F_1}$-score.

\item{Recall, false positive rate, Jaccard similarity coefficient, and RMSE$_{\rm{TP}}$.} 
For the assessment of the changepoint localization error in T3, we followed two different evaluation approaches. For the challenge evaluation, we simply quantified the ${\rm RMSE}$. Trajectories showing no changepoint were considered as having a dummy changepoint either at index $1$ or $199$. However, to get a better understanding of methods' performance, we also considered an alternative analysis. For this, trajectories with ground truth and predicted changepoints within a distance $\epsilon = 20$ from the start/end points were considered as not having a changepoint. We thus considered four cases:
\begin{itemize}
    \item predicted and ground-truth positions located at $\epsilon<t<L-\epsilon$, counted as true positives (TP);
    \item predicted and ground-truth positions located at $t\le\epsilon$ or $t\ge L-\epsilon$, counted as true negatives (TN);
    \item the predicted position located at $\epsilon<t<L-\epsilon$ but the ground-truth located at either $t\le\epsilon$ or $t\ge L-\epsilon$, counted as false positive (FP);
\item the predicted position located at either $t\le\epsilon$ or $t\ge L-\epsilon$.    but the ground-truth located at $\epsilon<t<L-\epsilon$, counted as false negative (FN).
\end{itemize}
Based on this classification, we evaluated the recall (also known as sensitivity):
\begin{equation}
{\rm recall}=  \frac{{\rm TP}}{{\rm TP}+{\rm FN}};    \label{eq:recall}
\end{equation}
the false positive rate:
\begin{equation}
{\rm FPR}=  \frac{{\rm FP}}{{\rm FP}+{\rm TN}};
\label{eq:fpr}
\end{equation}
and the Jaccard similarity coefficient (JSC) for binary classification:
\begin{equation}
{\rm JSC}=  \frac{{\rm TP}}{{\rm TP}+{\rm FP}+{\rm FN}}.
    \label{eq:Jaccard}
\end{equation}
We also calculated the RMSE$_{\rm{TP}}$, corresponding to the RMSE obtained only for prediction/ground-truth pairs classified as true positives.  
\end{itemize}

\subsection*{Alternative and baseline estimators}
\subsubsection*{Inference of the anomalous diffusion exponent}
Several classical statistical methods have been employed to characterize anomalous diffusion from single trajectories and quantify the anomalous diffusion exponent. Many of them rely on the analysis of the EA-MSD or TA-MSD presented in Eqs.~\eqref{eq:eamsd} and~\eqref{eq:tamsd}.

We developed a simple tool to perform the estimation of the anomalous exponent to establish a performance baseline for T1 of the challenge. The code calculates the TA-MSD and performs a linear fit of its logarithm with respect to the logarithm of the timelag for the first $k$ datapoints, where $k$ is the maximum between $10$ and the $10\%$ of the trajectory length. The anomalous diffusion exponent is thus obtained as the slope of the straight line. This criterion has been shown to provide reliable results for the fitting of TA-MSD for Brownian diffusion~\cite{michalet2010mean}. Although the choice of a different timescale or the use of an independently calculated localization precision can produce better results~\cite{kepten2015guidelines}, we intentionally limited the code to a simple fitting algorithm with a straightforward criterion for the choice of the number of data points to fit. As shown in Fig.~\ref{fig:task1}d, for ergodic models (FBM), such a simple fit produces results comparable with the best methods. In addition, as it can be observed using the interactive tool (\url{http://andi-challenge.org/interactive-tool/}), the estimation of $\alpha$ through the fit of the TA-MSD even outscores the other methods for ergodic models (FBM) at the highest SNR (e.g., SNR$=10$). The code is available at \url{https://github.com/AnDiChallenge/ANDI_datasets} ~\cite{andigithub}.

For the sake of completeness, we would like to mention other statistical approaches not considered in the challenge that can be used to tackle T1. Besides the ${\rm MSD}$, another popular methodology for the quantification of the anomalous diffusion exponent is the moment scaling spectrum $\rm{ MSS}$~\cite{ferrari2001strongly, sbalzarini2005feature}. $\rm{MSS}$ considers several high-order moments of the displacement distribution to obtain their scaling exponents and use them to calculate the slope of the exponent curve versus the moment order, which is found proportional to $\alpha$.

The anomalous diffusion exponent is strictly linked to specific characteristics of the diffusion model, thus it can also be obtained by means of their quantification~\cite{metzler2014anomalous}. However, this approach require the knowledge (or an educated guess) of the diffusion model. If, in addition, distributions of the associated quantities can be obtained, then anomalous diffusion exponent can be estimated through their fitting.  For instance, for CTRW, the anomalous diffusion exponent can be extracted by fitting the waiting time distribution $\psi(t)$~\cite{weigel2011ergodic}; for the ATTM, by fitting the distribution of diffusion coefficients or transit times~\cite{manzo2015weak};  or, for a L\`evy walk from the flight time or step length distribution~\cite{ariel2015swarming}.

\subsubsection*{Classification of the underlying diffusion model}
Even though the problem of associating a trajectory to an underlying diffusion model has been long investigated, there is still no clear general consensus on how to unambiguously determine the underlying physical mechanism from a trajectory. To the best of our knowledge, model classification is generally performed using a combination of multiple estimators and further corroborated by a comparison with the corresponding analysis of simulated data.  
Several statistical parameters have been proposed in this sense. Algorithms based on multiple estimators can allow to distinguish between pairs of models~\cite{magdziarz2009fractional, meroz2013test, chen2017anomalous}. Some of the proposed approaches are based on estimating trajectory statistical features to determine ergodicity~\cite{magdziarz2009fractional, he2008random} and Gaussianity~\cite{slkezak2019codifference}, and thus restrict the number of possible models. 
Lastly, the velocity autocorrelation function~\cite{burov2011single} and the power spectral density~\cite{krapf2019spectral} have been shown to have model-dependent fingerprints for some diffusion models. 
However, none of these method can be directly used to classify the trajectories as required for T2. First attempts to provide a direct and generalized classification have been proposed only recently~\cite{ thapa2018bayesian,kowalek2019classification, munoz2020single} and the developing teams have participated in the challenge. Therefore, we decided not to provide any baseline estimation for this task.

\subsubsection*{Trajectory segmentation}

Although a few methods have been recently developed for the detection of trajectory changepoints with respect to a switch in $\alpha$~\cite{weron2017ergodicity,sikora2017elucidating,bo2019measurement} and diffusion model~\cite{lanoiselee2017unraveling}, there is no consensus on a well-established method that can be used as a baseline for T3. 
Limited to the the changepoint detection part, we thus decided to compare methods' performance with the results of a random prediction, as shown in dashed lines in Fig.~\ref{fig:task3}a and Supplementary Fig.~\ref{fig:T3_allD}. For this, we simply calculate the RMSE for selecting a random point on a trajectory having a changepoint at $t_{\rm GT}$. The error associated to such a random prediction is not uniform, since it depends on the changepoint position $t_{\rm GT}$ along the trajectory, as well as on the trajectory length $L$. The random predictor RMSE$_{\rm random}$ can thus be calculated as the RMSE for a trajectory with a changepoint at position $t_{\rm GT}$ and random predictions $t$ of the changepoint drawn from a uniform distribution in the range $[0,L]$
\begin{widetext}
 \begin{equation}
 {\rm RMSE}_{\rm random}(t_{\rm GT}) =\sqrt{ \frac{1}{L} \int_{0}^L   \Big( t - t_{\rm GT} \Big)^2  dt \vphantom{\frac{t_{\rm GT}^3+\Big(L-t_{\rm GT} \Big)^3}{3L  }} } = \sqrt{ \frac{t_{\rm GT}^3+\Big(L-t_{\rm GT} \Big)^3}{3L  } \vphantom{\frac{1}{L} \int_{0}^L   \Big( t - t_{\rm GT} \Big)^2 }},
 \end{equation}
 \end{widetext}
where $L$ is the trajectory length.

\section*{Data availability}

The simulated data used in this study are available for download at the competition website \url{http://andi-challenge.org/challenge2020/}. Ground-truth for datasets used in the first phase of the competition for training are also available.

\section*{Code availability}

All software used for the Challenge is available at \url{https://github.com/AnDiChallenge}. The code of the {\tt andi-datasets} package~\cite{andigithub} used to generate the competition datasets is available at \url{https://github.com/AnDiChallenge/ANDI_datasets}.

%\bibliography{biblio.bib}

\begin{acknowledgments}
The authors would like to thank: Paula Kowalek for the graphical illustrations; Matthias Weiss and Maria Garcia-Parajo for sharing experimental data;  Daniel Adam for help with compiling the data of single-atom trajectories.
G.M.-G., B.R., and M.L. acknowledge support from ERC AdG NOQIA, Agencia Estatal de Investigación ``Severo Ochoa'' Center of Excellence CEX2019-000910-S, Plan National FIDEUA PID2019-106901GB-I00/10.13039 / 501100011033, FPI), Fundació Privada Cellex, Fundació Mir-Puig, and from Generalitat de Catalunya (AGAUR Grant No. 2017 SGR 1341, CERCA program, QuantumCAT U16-011424, co-funded by ERDF Operational Program of Catalonia 2014-2020), MINECO-EU QUANTERA MAQS (funded by State Research Agency (AEI) PCI2019-111828-2 / 10.13039/501100011033), EU Horizon 2020 FET-OPEN OPTOLogic (Grant No 899794), and the National Science Centre, Poland-Symfonia Grant No. 2016/20/W/ST4/00314. 
Giov.V. and A.A. acknowledge funding from ERC StG ComplexSwimmers (Grant No. 677511) and from the Knut and Alice Wallenberg Foundation.
M.A.G.-M. acknowledges funding from the Spanish Ministry of Education and Vocational Training (MEFP) through the Beatriz Galindo program 2018 (BEAGAL18/00203).
R.M. acknowledges DFG grant ME 1535/12-1.
Gior.V. and A.G. acknowledge sponsorship for this work by the U.S. Office of Naval Research Global (Award No. N62909-18-1-2170).
Z.H. acknowledges funding from the Fundamental Research Funds for the Central Universities.
J.-H.J. acknowledges NRF grants 2020R1A2C4002490 and 2017K1A1A2013241. 
T.B. acknowledges support by the Francis Crick Institute, which receives its core funding from Cancer Research UK (FC001086), the UK Medical Research Council (FC001086), and the Wellcome Trust (FC001086), and thanks Nate Goehring for supervision and acquisition of funding.
This research was funded in whole, or in part, by the Wellcome Trust (FC001086). For the purpose of Open Access, the author has applied a CC BY public copyright license to any Author Accepted Manuscript version arising from this submission.
J.A.C. acknowledges support from the ALBATROSS project (National Plan for Scientific and Technical Research and Innovation 2017-2020, No. PID2019-104978RB-I00).
P.K, H.L.-O. and J.S. were funded by the Polish National Science Centre (NCN-DFG Beethoven Grant No. 2016/23/G/ST1/04083) and acknowledge the support by the Wroclaw Centre for Networking and Supercomputing (calculations were performed using their BEM computing cluster).
S.T. acknowledges the Deutscher Akademischer Austauschdienst for PhD Scholarship (DAAD Program ID 57214224).
H.K. and I.S. acknowledge funding from the Dutch Research Council (NWO) through the GENOMETRACK project of the Building Blocks of Life research program (Project No. 737.016.014).
C.M. acknowledges funding from FEDER/Ministerio de Ciencia, Innovaci\'{o}n y Universidades -- Agencia Estatal de Investigaci\'{o}n through the ``Ram\'{o}n y Cajal'' program 2015 (Grant No. RYC-2015-17896), and the ``Programa Estatal de I+D+i Orientada a los Retos de la Sociedad'' (Grant No. BFU2017-85693-R); from the Generalitat de Catalunya (AGAUR Grant No. 2017SGR940). C.M. also acknowledges the support of NVIDIA Corporation with the donation of the Titan Xp GPU and funding from the PO FEDER of Catalonia 2014-2020 (project PECT Osona Transformaci\'{o} Social, Ref. 001-P-000382).
\end{acknowledgments}

\section*{Author contributions}
C.M. conceived the study. C.M., G.M.-G., Giov.V., M.A.G.-M, M.L., and R.M. organized the challenge and the corresponding workshop. G.M-G. designed and implemented the software for data generation and comparison of results. G.M.-G. generated the data and ground truth used in all challenge phases.  G.M.-G. and C.M. verified the files submitted by the participants and performed the scoring of all methods. G.M.-G., C.M., Giov.V., and M.A.G.-M. analyzed the results. The methods were designed, implemented, run, and described by the participating teams: team \team{a}: B.R., G.M.-G.; team \team{b}: S.T., M.L., J.-H.J., S.P., Y.K.; team \team{c}: J.-B.M., H.V.; team \team{d}: T.S., C.B.H., J.-H.J.; team \team{e}: A.A., S.B.; team \team{f}: H.K., I.S.; team \team{g}: Z.H.; team \team{h}: N.F., J.A.C., O.G.O.; team \team{i}: C.M.; team \team{j}: T.B.; team \team{k}: E.A., P.G.M.; team \team{l}: Gior.V., A.G.; team \team{m}: O.G.O., J.A.C.; and teams \team{n,o}: H.L.-O., P.K., J.S. D.K., C.M., and A.W. provided experimental datasets. The article was written by C.M., G.M.-G., Giov.V., and M.A.G.-M. with input from all authors.

\section*{Competing interest} 

The authors declare no competing interests.

%%%%% FIGURES and TABLES %%%%%%%%%
\clearpage
\onecolumngrid
\section*{Figures and Tables}

\clearpage

\begin{sidewaystable}
\centering
			\begin{tabular}{@{}cllccccc@{}}
				\textbf{Label}&
				\multicolumn{1}{c}{\textbf{Team name}} &
				\multicolumn{1}{c}{\textbf{Method}} &
				\textbf{Class} &
				\textbf{Input} &
				\textbf{Tasks} &
				\textbf{$L$-specific} 
				\\ \hline\hline
				\team{a} & Anomalous Unicorns & Ensemble of CNN and RNN~\cite{munoz2020phase, wolpert1992stacked}    & ML     & Traj          & T1(1D), T2(1D)     & No       \\
				\team{b} & BIT                & Bayesian inference~\cite{krog2018bayesian, park2021bayesian}         & Stat  & Traj          & All              & No      \\
				\team{c} & DecBayComp         & Graph neural networks~\cite{verdier2021learning}     & ML     & Traj + Feat & T1, T2(1D, 2D)     & No    \\
				\team{d} & DeepSPT            & ResNet + XGBoost~\cite{he2016deep,chen2016guestrin}           & ML     & Traj + Feat & T1(1D), T2(1D)     & No       \\
				\team{e} & eduN               & RNN + Dense NN~\cite{argun2021classification}             & ML       & Traj          & All              & Yes      \\
				\team{f} & Erasmus MC         & bi-LSTM + Dense NN~\cite{arts2019particle}         & ML       & Feat        & T1, T2 & Yes      \\
				\team{g} & HNU                & LSTM~\cite{li2021wavenet}                       & ML       & Traj          & T1          & Yes      \\
				\team{h} & NOA                & CNN + bi-LSTM~\cite{donahue2015long}              & ML       & Traj          & T1(1D)            & No       \\
				\team{i} & QUBI               & ELM~\cite{manzo2021extreme}                        & ML       & Feat       & T1(1D), T2(1D)     & No       \\
				\team{j} & FCI                & CNN~\cite{bai2018empirical, granik2019single}                        & ML       & Traj          & T1(1D, 2D), T2(1D, 2D), T3(1D, 2D)   & No       \\
				\team{k} & TSA                & Scaling analysis and feature engineering~\cite{aghion2021moses}  & Stat    & Feat        & T1, T2(1D)   & No    \\
				\team{l} & UCL                & Feature engineering + NN~\cite{gentili2021characterization}  & ML    & Feat        & T1, T2  & No       \\
				\team{m} & UPV-MAT            & CNN + bi-LSTM~\cite{garibo2021efficient}              & ML      & Traj          & T1, T2  & Yes    \\
				\multirow{2}{*}{\team{n}} &
				\multirow{2}{*}{Wust ML A} &	
				1D: RISE + forest classifier~\cite{Lines2018} &
				\multirow{2}{*}{ML}&
				\multirow{2}{*}{Feat} &
				\multirow{2}{*}{T2} &
				\multirow{2}{*}{No} 
				\\
				&                      & 2D, 3D: MrSEQL + logistic reg.~\cite{LeNguyen2019}   &    &      &                           &        & \\
				\team{o} & Wust ML B          & Gradient boosting regression + classifier~\cite{kowalek2019classification, janczura2020classification, loch2020impact}  &  ML & Feat & T1(1D, 2D), T2     & No       \\
			\end{tabular}

	\caption{\label{tab:methods} \textbf{Participating teams and summary of methods.} See \protect\hyperlink{sec:SI-note1_h}{Supplementary Note 1} for further details on these methods.  Methods were classified based on the type of approach (as machine learning (ML), or classical statistics (Stat));  their input data (as raw/lightly preprocessed trajectories (Traj), or features (Feat)); and their training procedure (as length-specific ($L$-specific, Yes), or not (No)).}

\end{sidewaystable}

\clearpage

\setcounter{figure}{0}
\setcounter{table}{0}
\renewcommand{\figurename}{Supplementary Figure}
\renewcommand{\thefigure}{S\arabic{figure}}
\renewcommand{\thetable}{S\arabic{table}}
\makeatletter
\renewcommand{\theHfigure}{S\arabic{figure}}
\renewcommand{\theHtable}{S\arabic{table}}
\makeatother

\def\bibsection{\section*{\refname}} 
\renewcommand\refname{Supplementary References}

\section*{Supplementary Figures and Tables}
\hypertarget{sec:SI_h}{}
% screenshot of interactive app
\begin{figure}[!ht]
	\includegraphics[width=\textwidth]{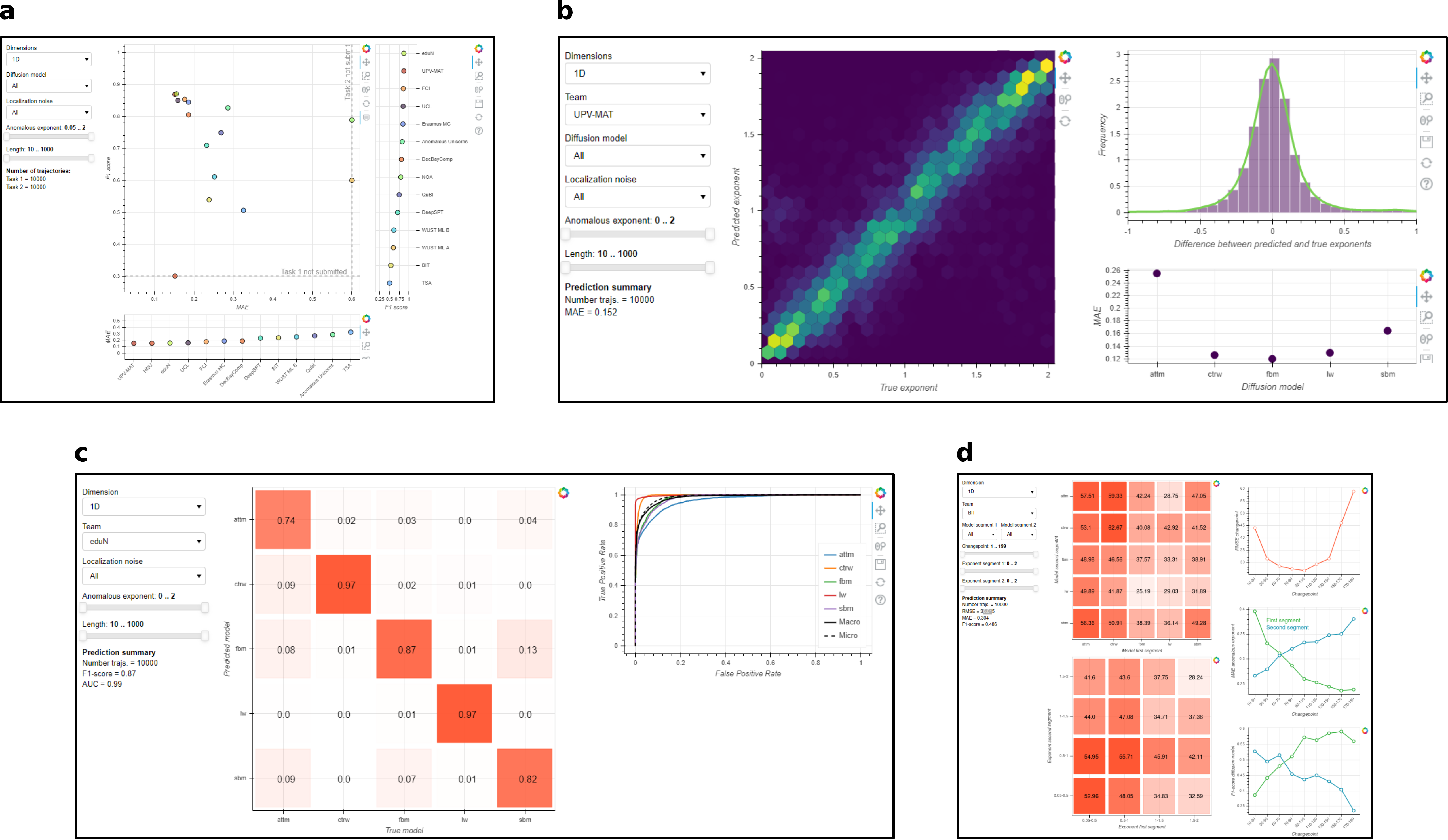}
	\caption{\label{fig:interactive_screenshot} \textbf{Screenshots of the interactive tool for performance comparison.} \textbf{a}, Summary of the results obtained for T1 and T2 according to corresponding challenge metrics. Hovering on each symbol reveals team name and scores. \textbf{b-d}, Plots of the metrics and estimators used to assess methods' performance for T1 (\textbf{b}), for T2 (\textbf{c}), and for T3 (\textbf{d}). For each task, plots can be displayed for user-selected subsets of the datasets. Sliders and buttons allow data selection based on task dimension, team, trajectory length, noise, $\alpha$, diffusion model, or changepoint position. The interactive tool is available at \url{http://andi-challenge.org/interactive-tool/}.}

\end{figure}
% summary T1/T2/T3
\newpage
\begin{figure}[!ht]
	\includegraphics[width=\textwidth]{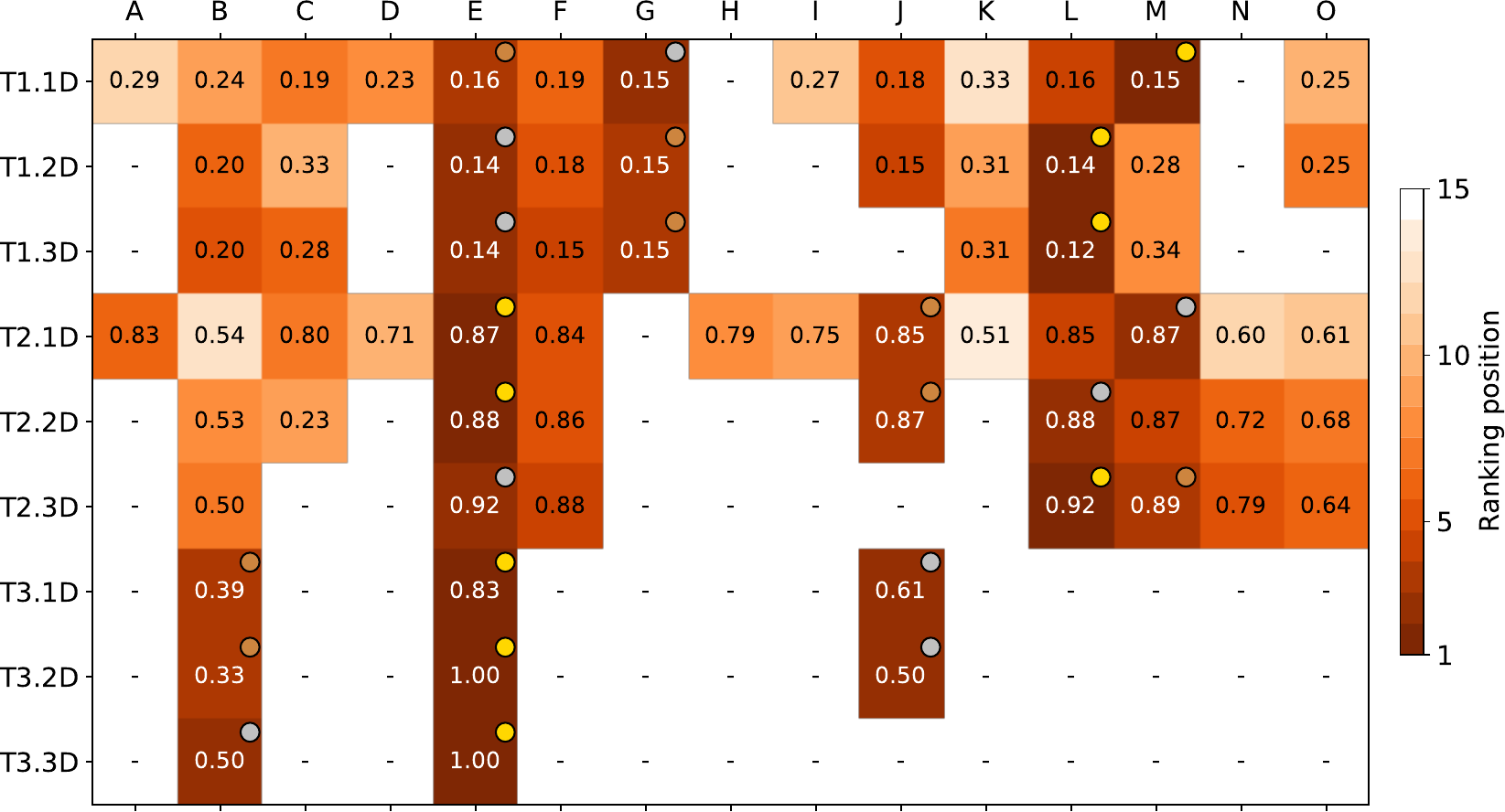}
	\caption{\label{fig:ranks_summary} \textbf{General ranking of the AnDi challenge.} Performance heatmap representing the value of the challenge metrics obtained by each team (\team{A} to \team{O}) for each task and dimension (T1.1D to T3.3D). The color code represents the relative position in the subtask leaderboard (the darker the color, the higher the rank). Top three teams of every subtask are labeled with a colored circle representing a medal  (first~--~gold, second~--~silver, third~--~bronze).}
\end{figure}
%
% summary T1 allD
\newpage
\begin{figure}[h]
	\includegraphics[width=\textwidth]{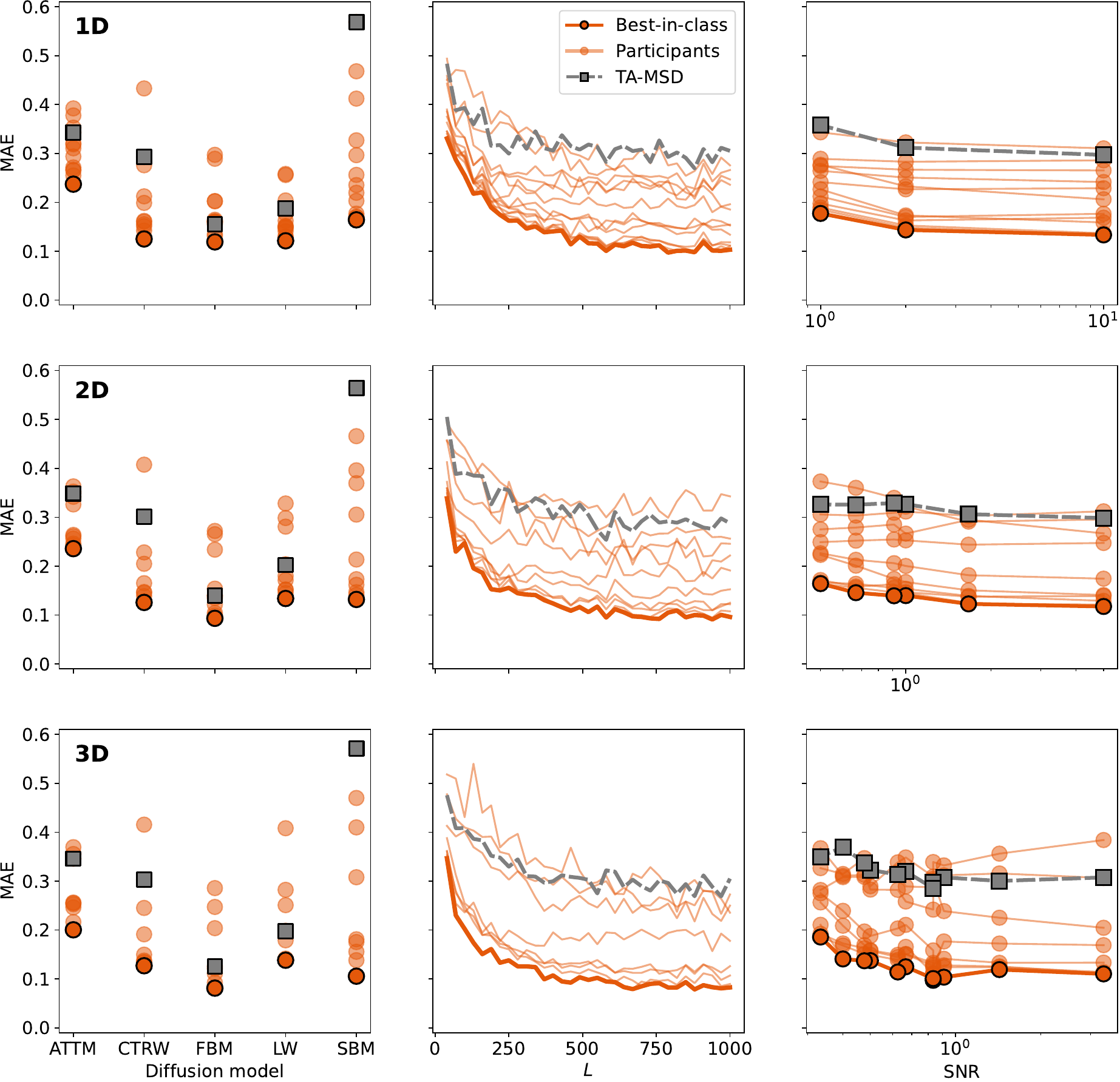}
	\caption{\label{fig:T1_allD}  \textbf{Comparison of method performance for T1.} ${\rm MAE}$ for all the submitted methods as a function of the diffusion model (left column), trajectory length (middle column), and ${\rm SNR}$ (right column).
	Rows show results obtained for different trajectory dimensions (from top to bottom, 1D, 2D, and 3D).}
\end{figure}
%
% summary T2 allD
\newpage
\begin{figure}[h]
	\includegraphics[width=\textwidth]{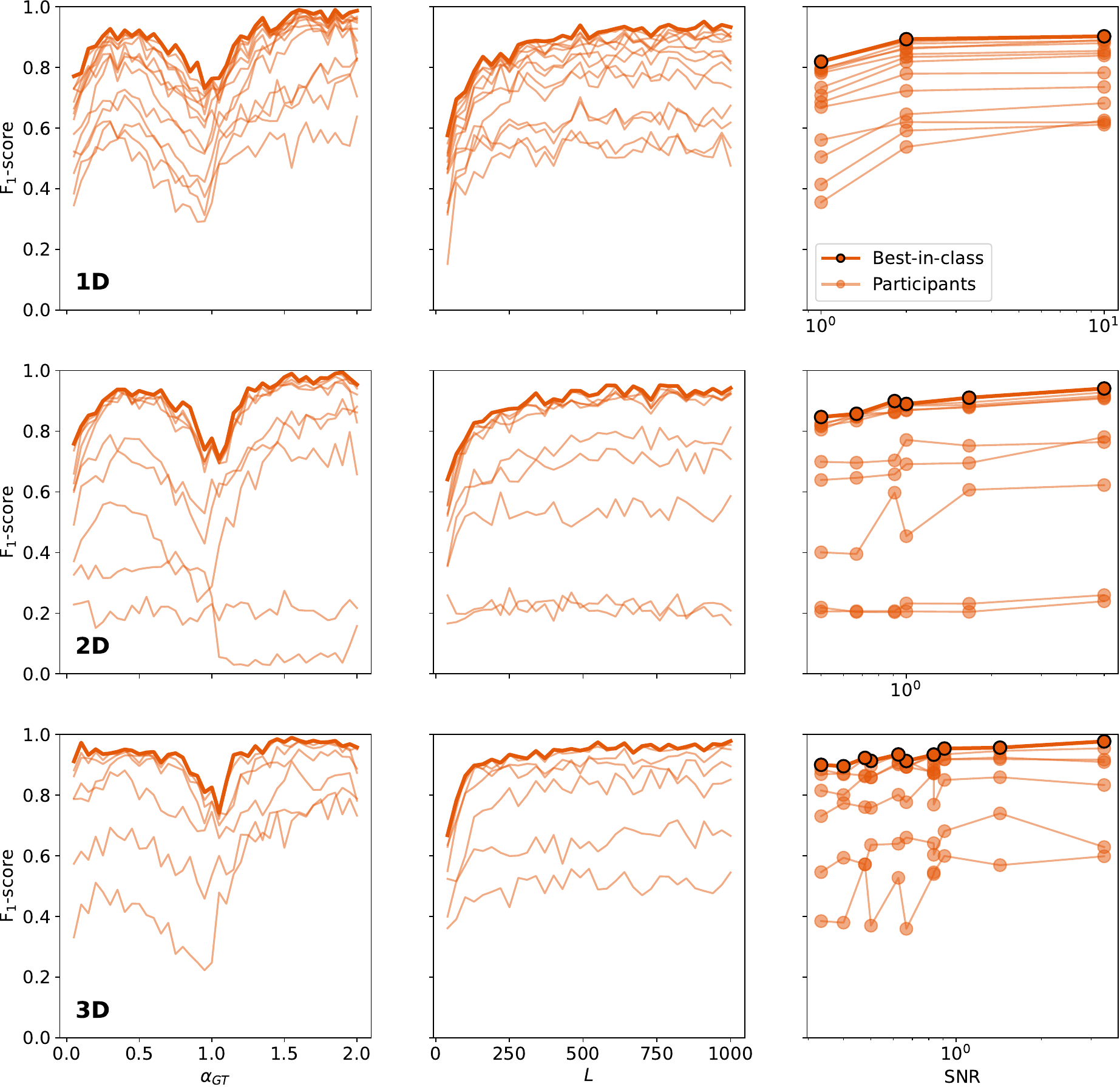}
	\caption{\label{fig:T2_allD}  \textbf{Comparison of method performance for T2.} ${\rm F_1}$-score for all the submitted methods as a function of $\alpha_{\rm{GT}}$ (left column), trajectory length (middle column), and ${\rm SNR}$ (right column). Rows show results obtained for different trajectory dimensions (from top to bottom, 1D, 2D, and 3D).}
\end{figure}
%
% summary T3 allD
\newpage
\begin{figure}[h]
	\includegraphics[width=\textwidth]{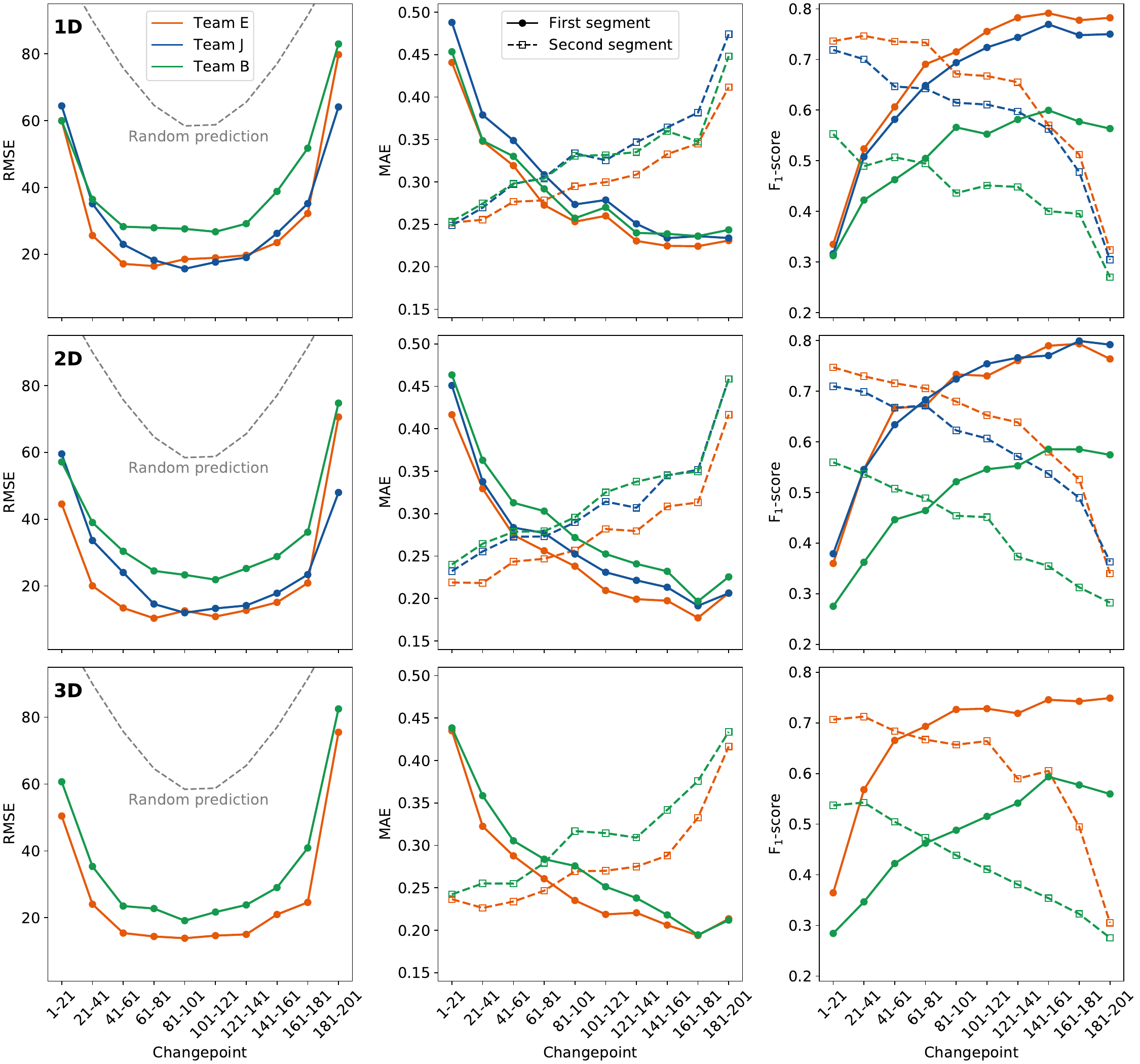}
	\caption{\label{fig:T3_allD} \textbf{Comparison of method performance for T3.} 
	$\rm{RMSE}$ for changepoint localization as a function of the changepoint position (left column), ${\rm MAE}$ for the prediction of $\alpha_{\rm{GT}}$ of the first (solid) and second segment (dashed) as a function of the changepoint position (middle column), and ${\rm F_1}$-score for classification of the diffusion model of the first (solid) and second segment (dashed) as a function of the changepoint position (right column). Rows show results obtained for different trajectory dimensions (from top to bottom, 1D, 2D, and 3D).}
\end{figure}

%
% T1 allD MAE per model
\newpage
\begin{figure}[h]
	\includegraphics[width=\textwidth]{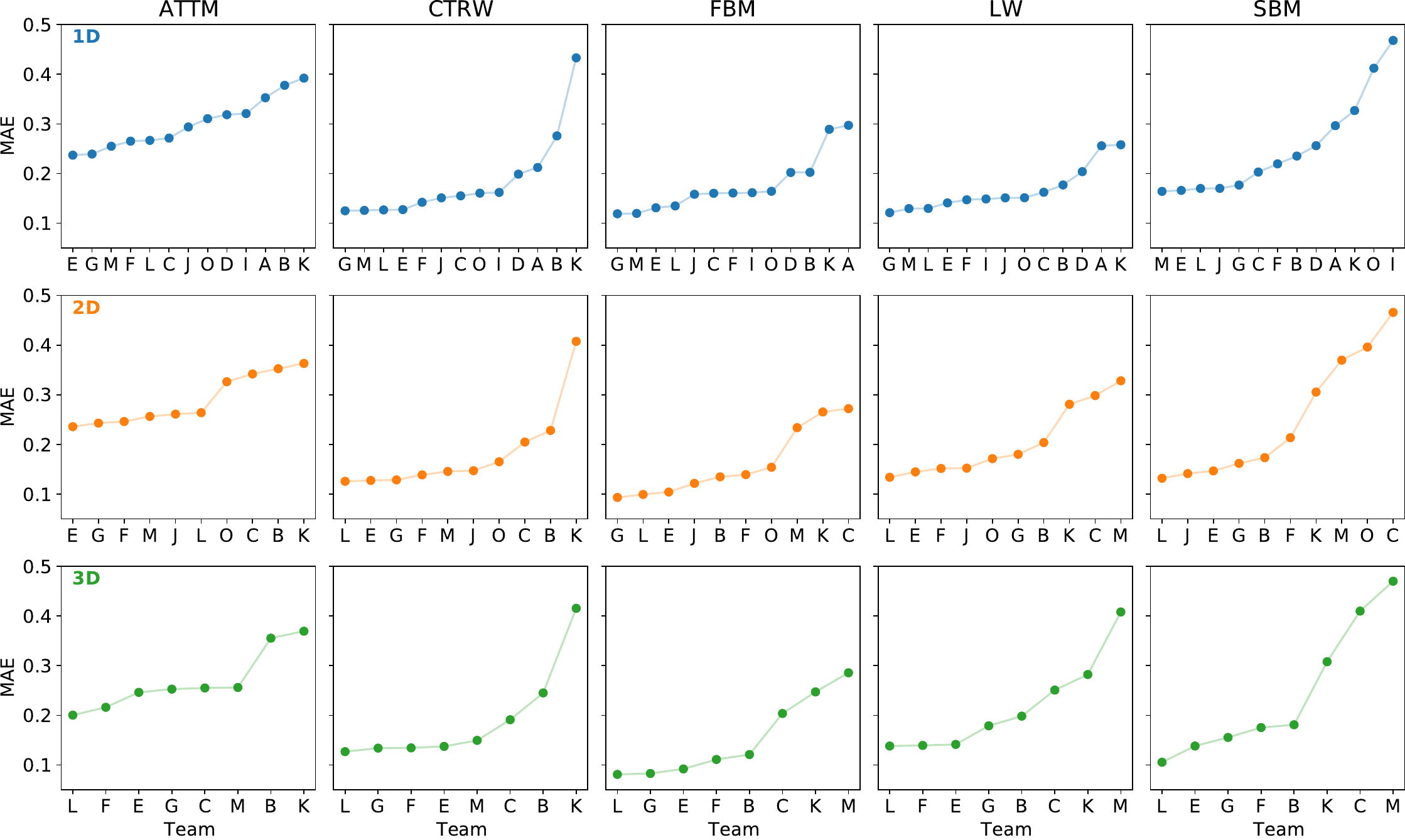}
	\caption{\label{fig:T1_allMAE} \textbf{T1 leaderboard per diffusion model.} ${\rm MAE}$ for the prediction of $\alpha_{\rm{GT}}$ obtained by submitted methods for each of the five diffusion model (columns). Rows show results obtained for different trajectory dimensions (from top to bottom, 1D, 2D, and 3D). Teams are ordered according to to their ranking in the leaderboard based on the ${\rm MAE}$ value.}
\end{figure}
%
% T1 1D 2dhist all teams
\newpage
\begin{figure}[h]
	\includegraphics[width=\textwidth]{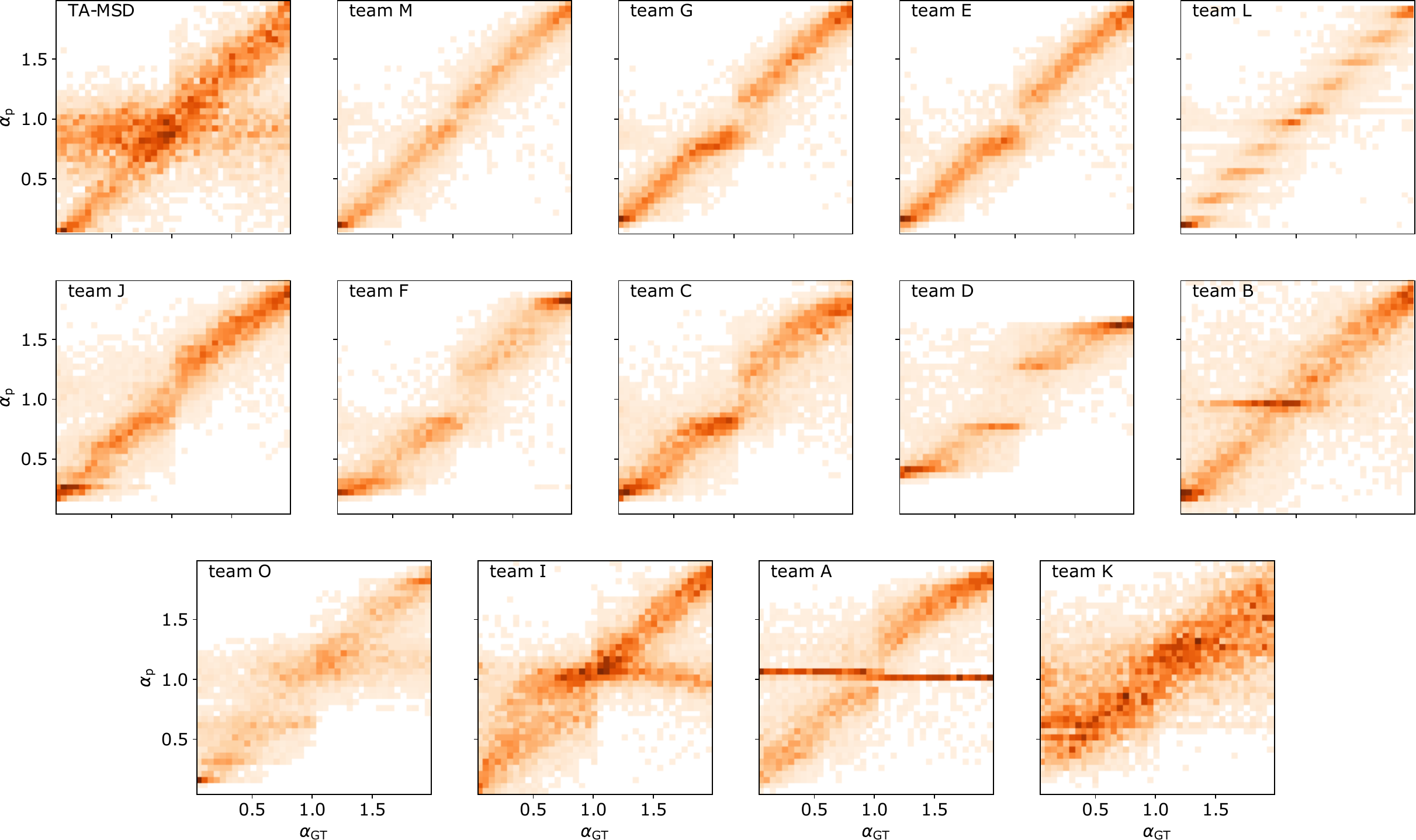} \caption{\label{fig:T1_1D_all2dhist} \textbf{T1.1D methods' performance.} 2D histograms of the ground truth ($\alpha_{\rm{GT}}$) vs the predicted exponent ($\alpha_{\rm{p}}$) for all the submitted methods for T1.1D. Teams are ordered according to to their ranking in the leaderboard.}
\end{figure}
%
% T1 2D 2dhist all teams
\newpage
\begin{figure}[h]
	\includegraphics[width=\textwidth]{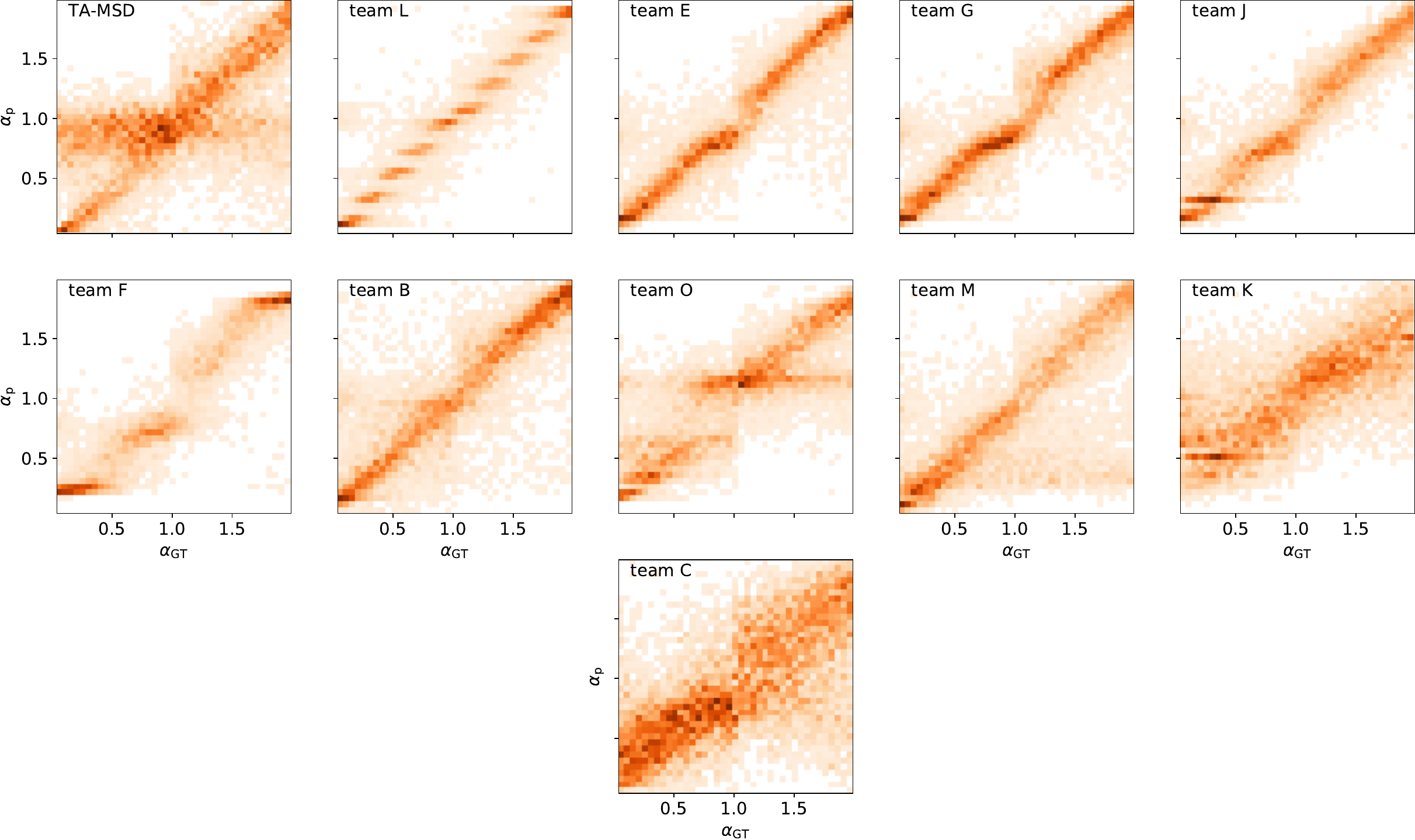}	\caption{\label{fig:T1_2D_all2dhist} \textbf{T1.2D methods' performance.} 2D histograms of the ground truth ($\alpha_{\rm{GT}}$) vs the predicted exponent ($\alpha_{\rm{p}}$) for all the submitted methods for T1.2D. Teams are ordered according to to their ranking in the leaderboard.}
\end{figure}
%
% T1 3D 2dhist all teams
\newpage
\begin{figure}[h]
	\includegraphics[width=\textwidth]{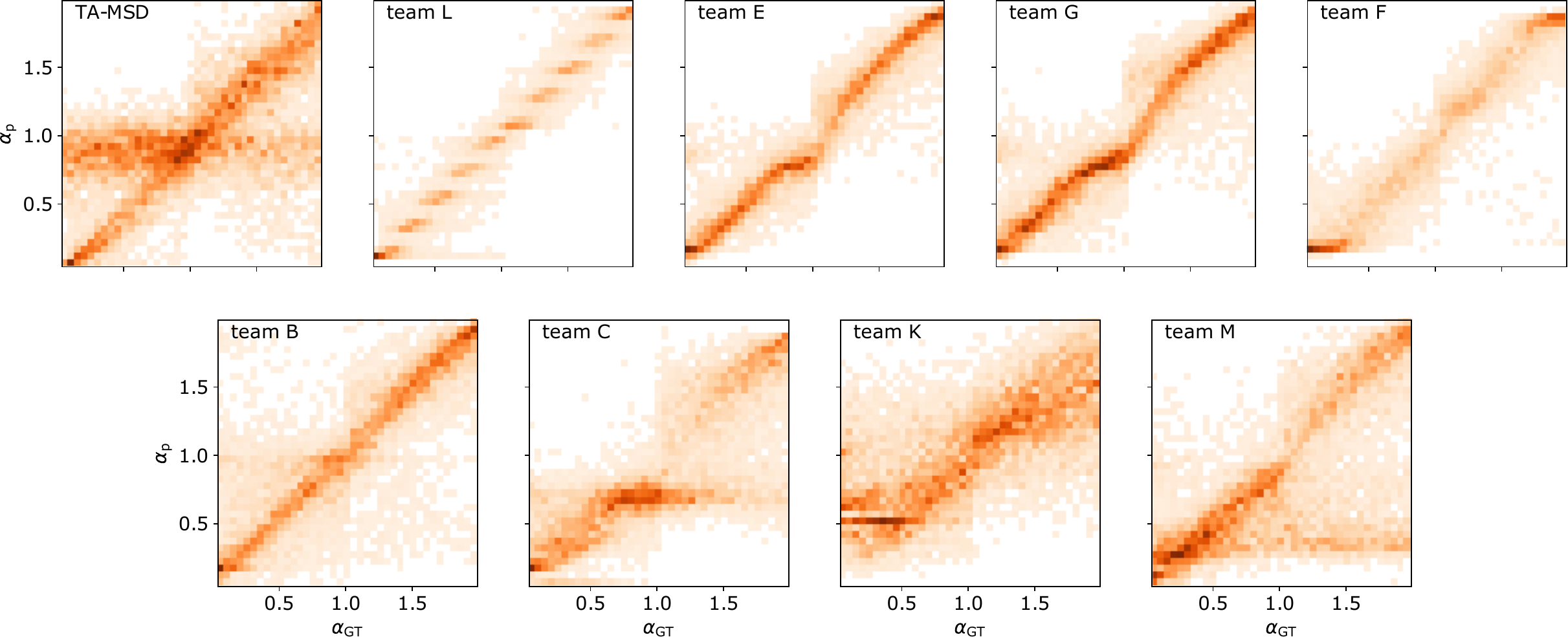}	\caption{\label{fig:T1_3D_all2dhist}  \textbf{T1.3D methods' performance.} 2D histograms of the ground truth ($\alpha_{\rm{GT}}$) vs the predicted exponent ($\alpha_{\rm{p}}$) for all the submitted methods for T1.3D. Teams are ordered according to to their ranking in the leaderboard.}
\end{figure}
%
% T2 allD F1 per alphas
\newpage
\begin{figure}[h]
	\includegraphics[width=\textwidth]{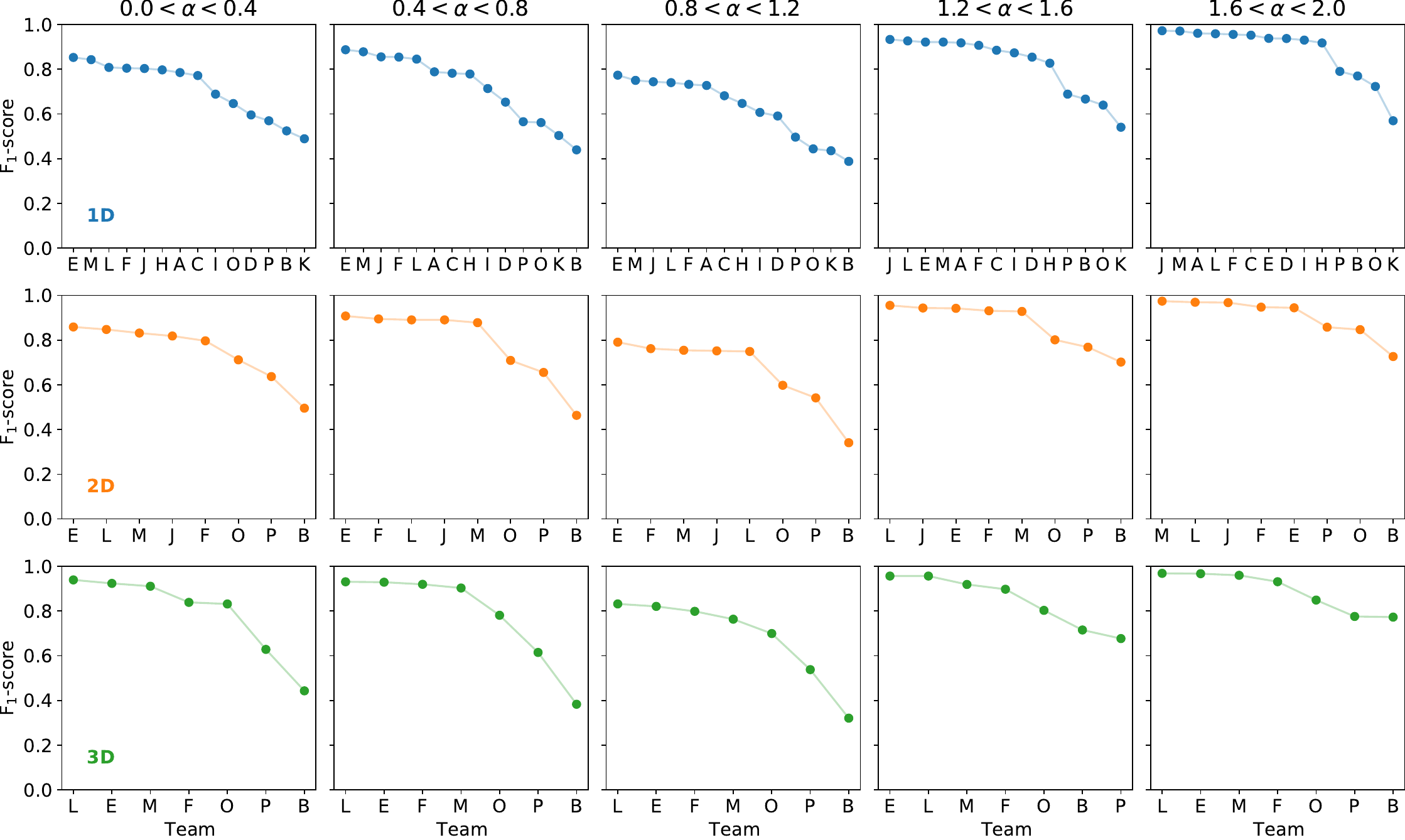}
	\caption{\label{fig:T2_allF1} \textbf{T2 leaderboard per range of $\alpha_{\rm{GT}}$.} ${\rm F_1}$-score for the prediction of the diffusion model obtained by submitted methods for five ranges of $\alpha_{\rm{GT}}$ (columns). Rows show results obtained for different trajectory dimensions (from top to bottom, 1D, 2D, and 3D). Teams are ordered according to to their ranking in the leaderboard based on the ${\rm F_1}$-score value.}
\end{figure}
%
% T2 1D Confusion matrix all teams
\newpage
\begin{figure}[h]
	\includegraphics[width=\textwidth]{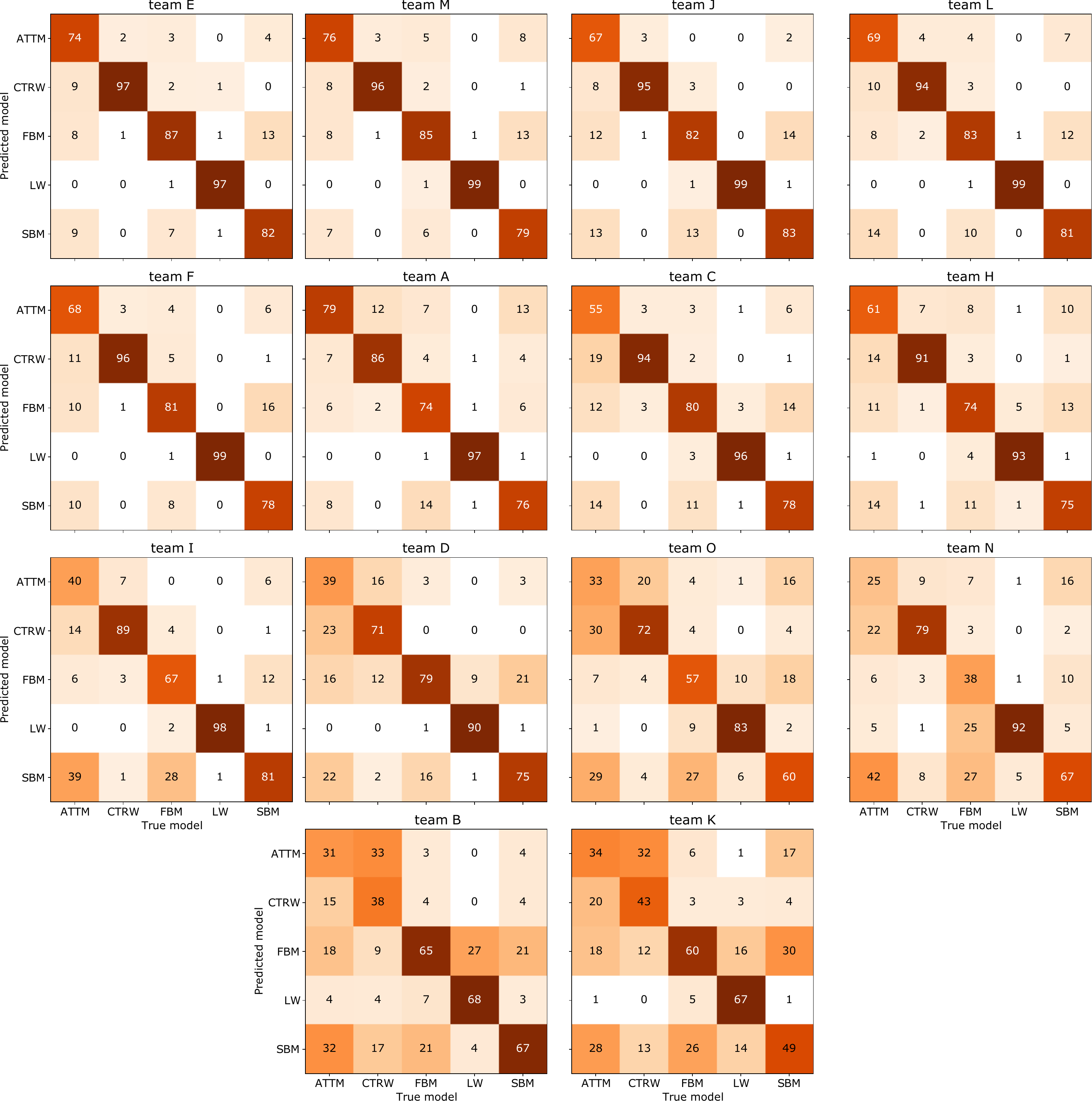}
	\caption{\label{fig:T2_1D_conf} \textbf{T2.1D methods' performance.}
	 Confusion matrix of the ground truth model vs the predicted model for all the submitted methods for T2.1D. Teams are ordered according to to their ranking in the leaderboard. Numbers in matrix cells represent the number of correctly and incorrectly classified trajectories for each ground-truth model as percentages of the number of trajectories of the corresponding ground-truth model (column-based normalization). Thus, the percentages of correctly classified observations can be thought of as class-wise recalls.}
\end{figure}
%
% T2 2D Confusion matrix all teams
\newpage
\begin{figure}[h]
	\includegraphics[width=\textwidth]{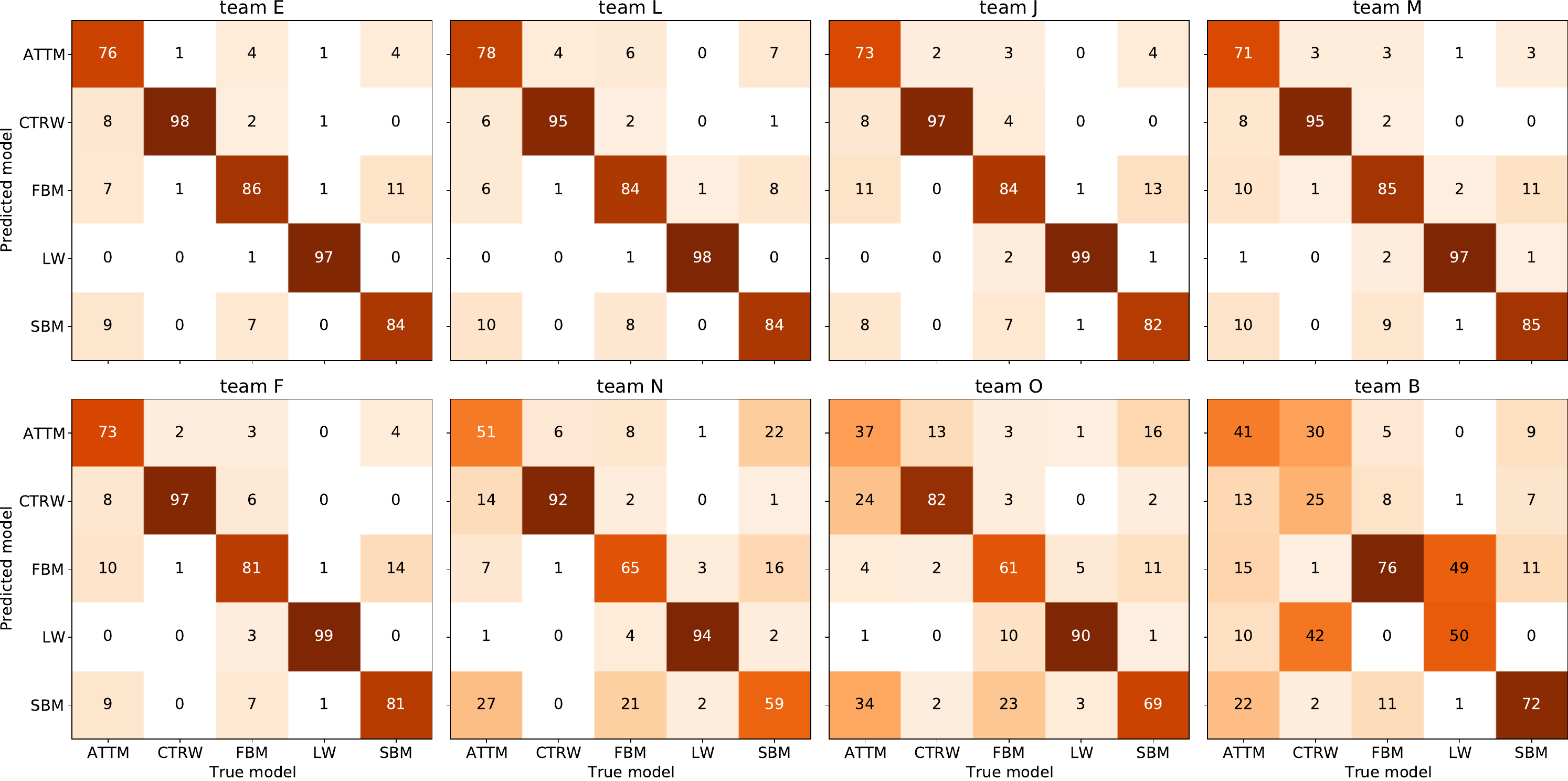}
	\caption{\label{fig:T2_2D_conf} \textbf{T2.2D methods' performance.}
	 Confusion matrix of the ground truth model vs the predicted model for all the submitted methods for T2.2D. Teams are ordered according to to their ranking in the leaderboard. Numbers in matrix cells represent the number of correctly and incorrectly classified trajectories for each ground-truth model as percentages of the number of trajectories of the corresponding ground-truth model (column-based normalization). Thus, the percentages of correctly classified observations can be thought of as class-wise recalls.}
\end{figure}
%
% T2 3D Confusion matrix all teams
\newpage
\begin{figure}[h]
	\includegraphics[width=\textwidth]{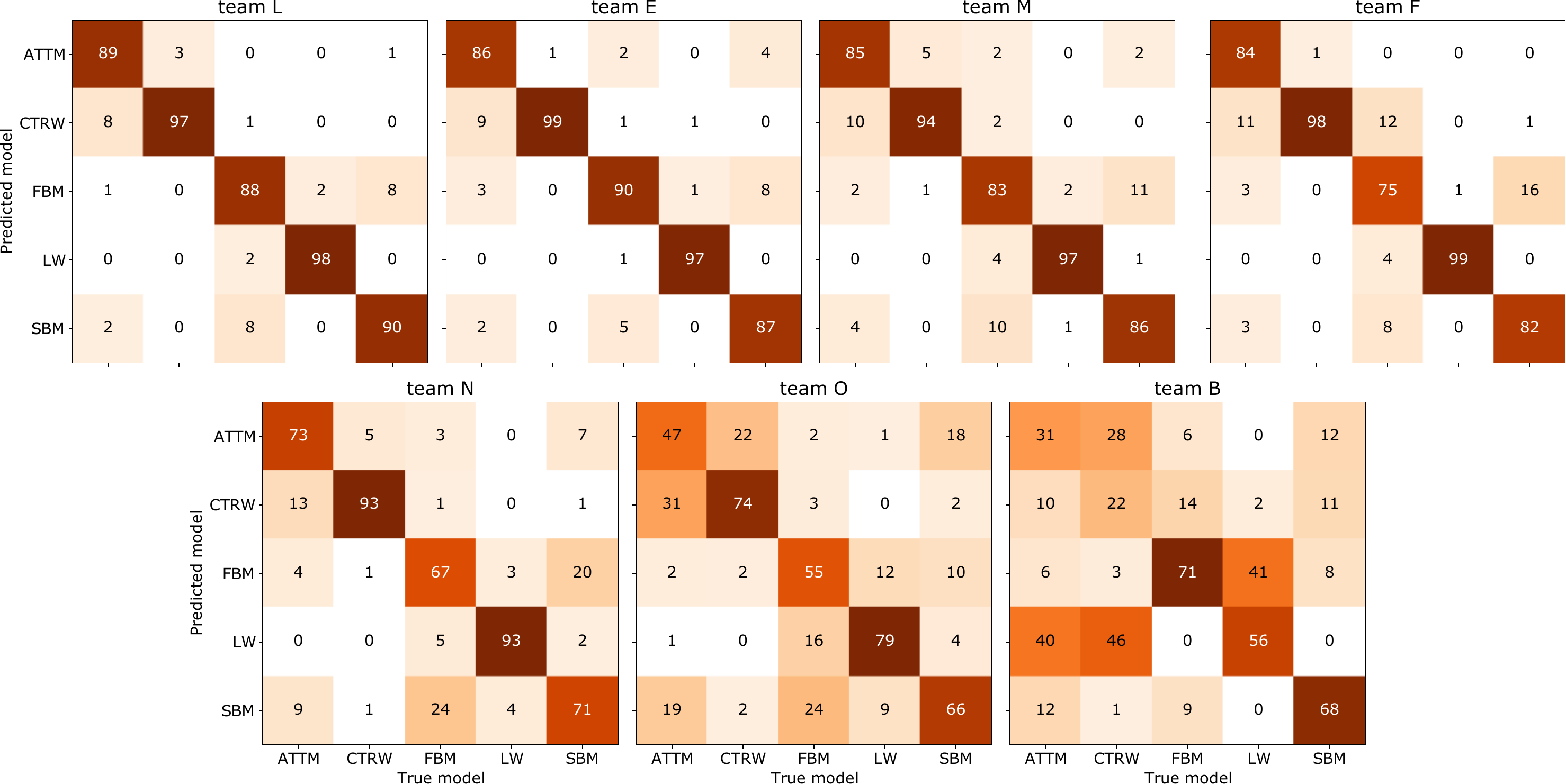}
	\caption{\label{fig:T2_3D_conf} \textbf{T2.3D methods' performance.}
	 Confusion matrix of the ground truth model vs the predicted model for all the submitted methods for T2.3D. Teams are ordered according to to their ranking in the leaderboard. Numbers in matrix cells represent the number of correctly and incorrectly classified trajectories for each ground-truth model as percentages of the number of trajectories of the corresponding ground-truth model (column-based normalization). Thus, the percentages of correctly classified observations can be thought of as class-wise recalls.}
\end{figure}
%
% Bias T1 1D
\newpage
\begin{figure}[h]
	\includegraphics[width=\textwidth]{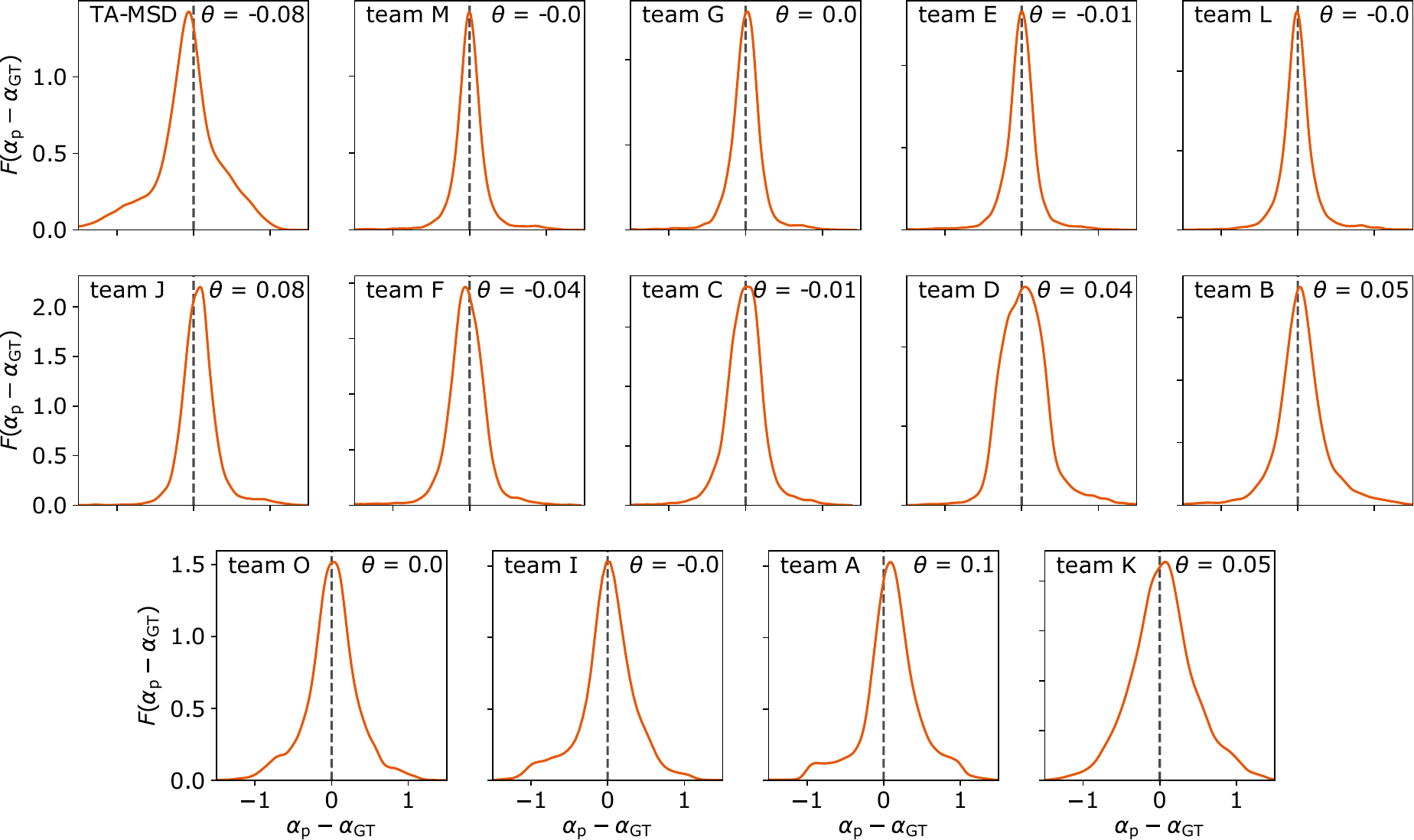}
	\caption{\label{fig:T1_1D_bias} \textbf{T1.1D prediction bias.} Empirical probability distributions of the difference between the predicted ($\alpha_p$) and the ground-truth exponent ($\alpha_{GT}$) for every method participating in T1.1D. The expectation value of the bias $\theta$ is reported in the plot. A dashed line representing the zero value is included as a guide-to-the-eye. Teams are ordered according to to their ranking in the leaderboard.}
\end{figure}
%
% Bias T1 2D
\newpage
\begin{figure}[h]
	\includegraphics[width=\textwidth]{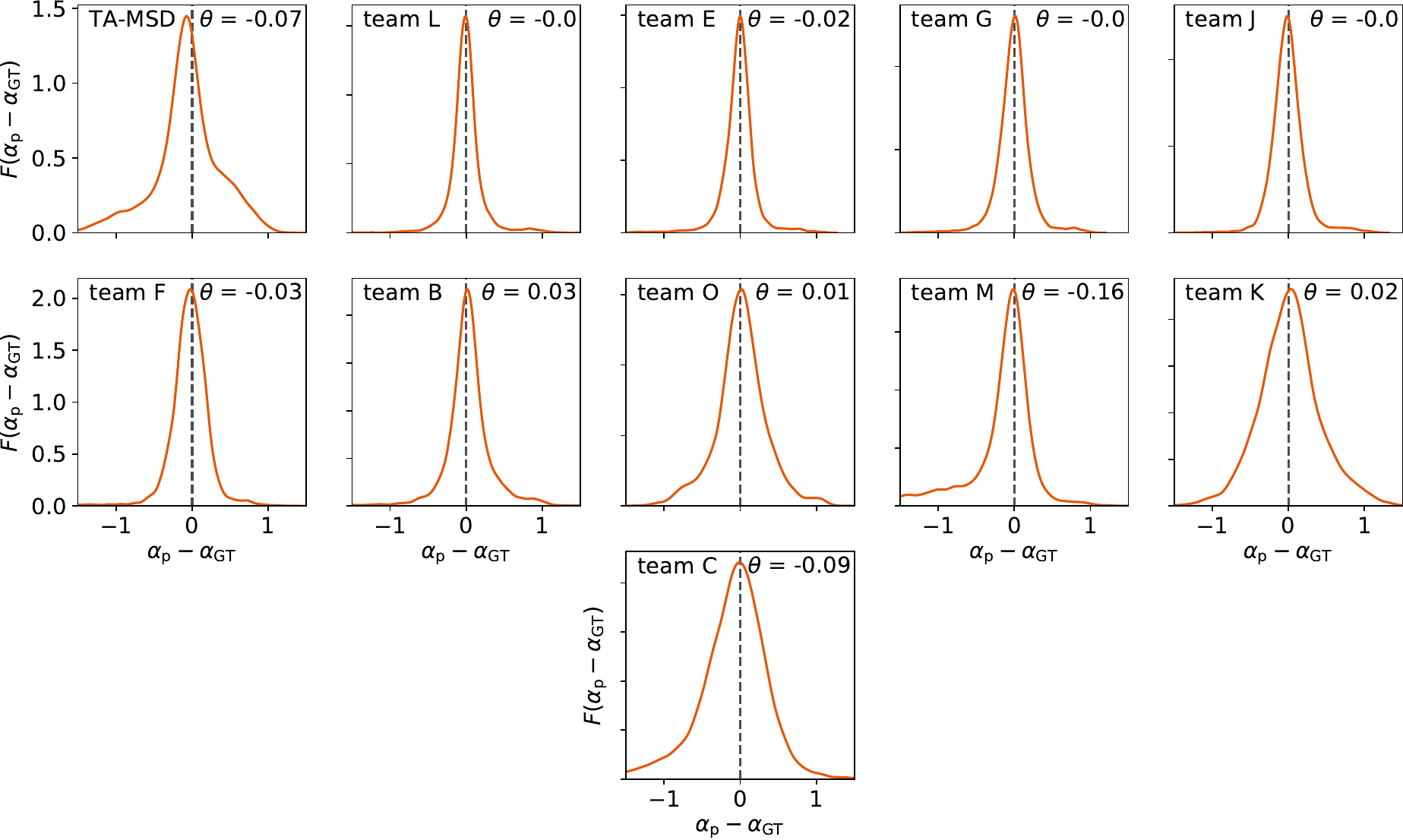}
	\caption{\label{fig:T1_2D_bias} \textbf{T1.2D prediction bias.} Empirical probability distributions of the difference between the predicted ($\alpha_p$) and the ground-truth true exponent ($\alpha_{GT}$) for every method participating in T1.2D. The expectation value of the bias $\theta$ is reported in the plot. A dashed line representing the zero value is included as a guide-to-the-eye. Teams are ordered according to to their ranking in the leaderboard.}
\end{figure}
%
% Bias T1 3D
\newpage
\begin{figure}[h]
	\includegraphics[width=\textwidth]{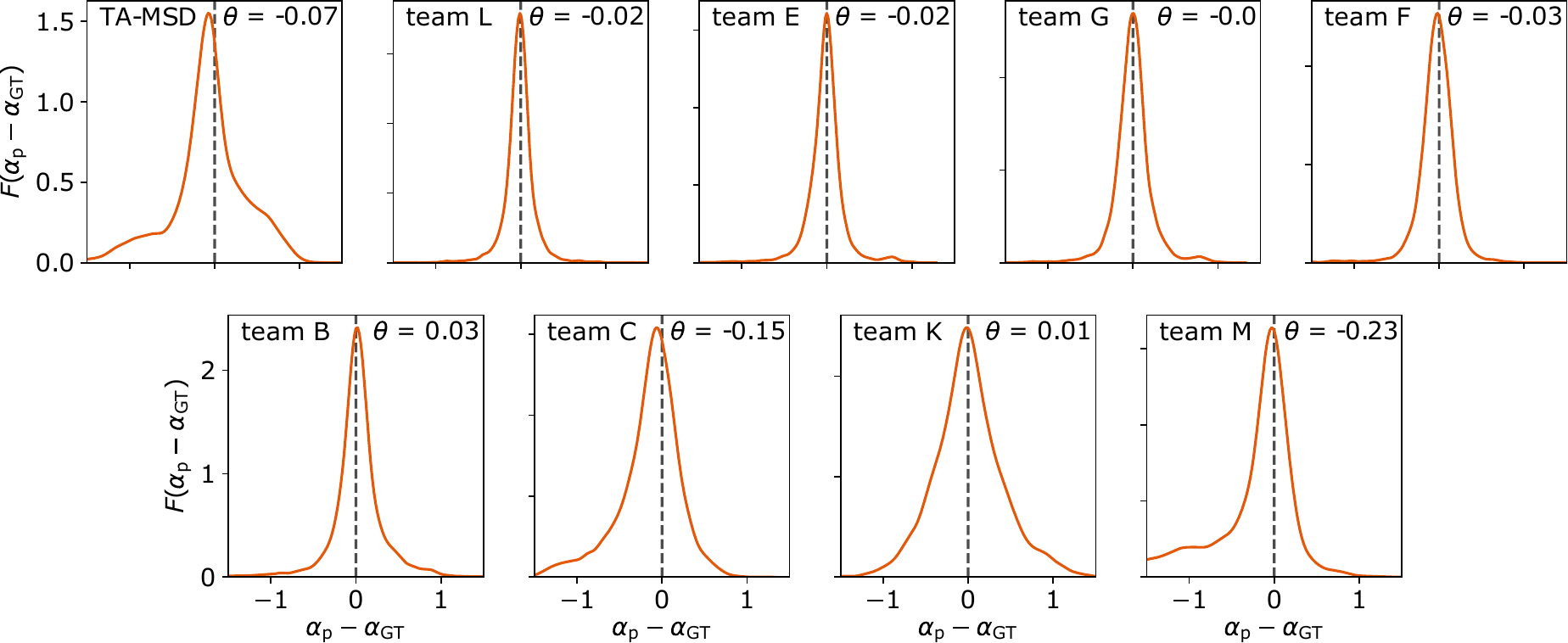}
	\caption{\label{fig:T1_3D_bias} \textbf{T1.3D prediction bias.} Empirical probability distributions of the difference between the predicted ($\alpha_p$) and the ground-truth exponent ($\alpha_{GT}$) for every method participating in T1.3D. The expectation value of the bias $\theta$ is reported in the plot. A dashed line representing the zero value is included as a guide-to-the-eye. Teams are ordered according to to their ranking in the leaderboard.}
\end{figure}
%
% T2 1D ROC/AUC all teams
\newpage
\begin{figure}[h]
	\includegraphics[width=\textwidth]{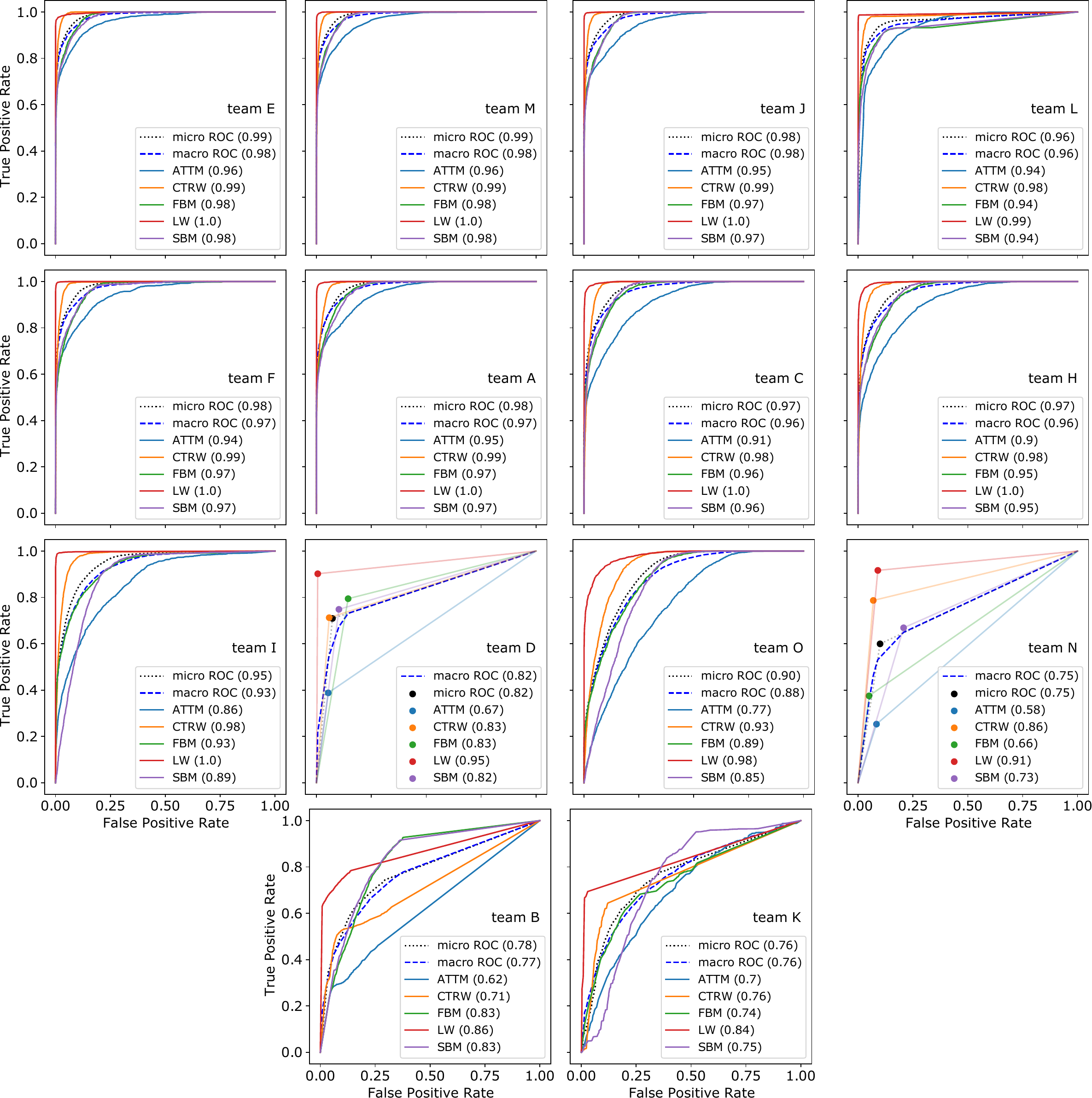}
	\caption{\label{fig:T2_1D_ROC} \textbf{T2.1D ROC curves.} ROC curves obtained for each diffusion model, plus micro- and macro-average, for all the methods participating in T2.1D. AUC values are reported in the legend. Teams are ordered according to to their ranking in the leaderboard.}
\end{figure}
%
% T2 2D ROC/AUC all teams
\newpage
\begin{figure}[h]
	\includegraphics[width=\textwidth]{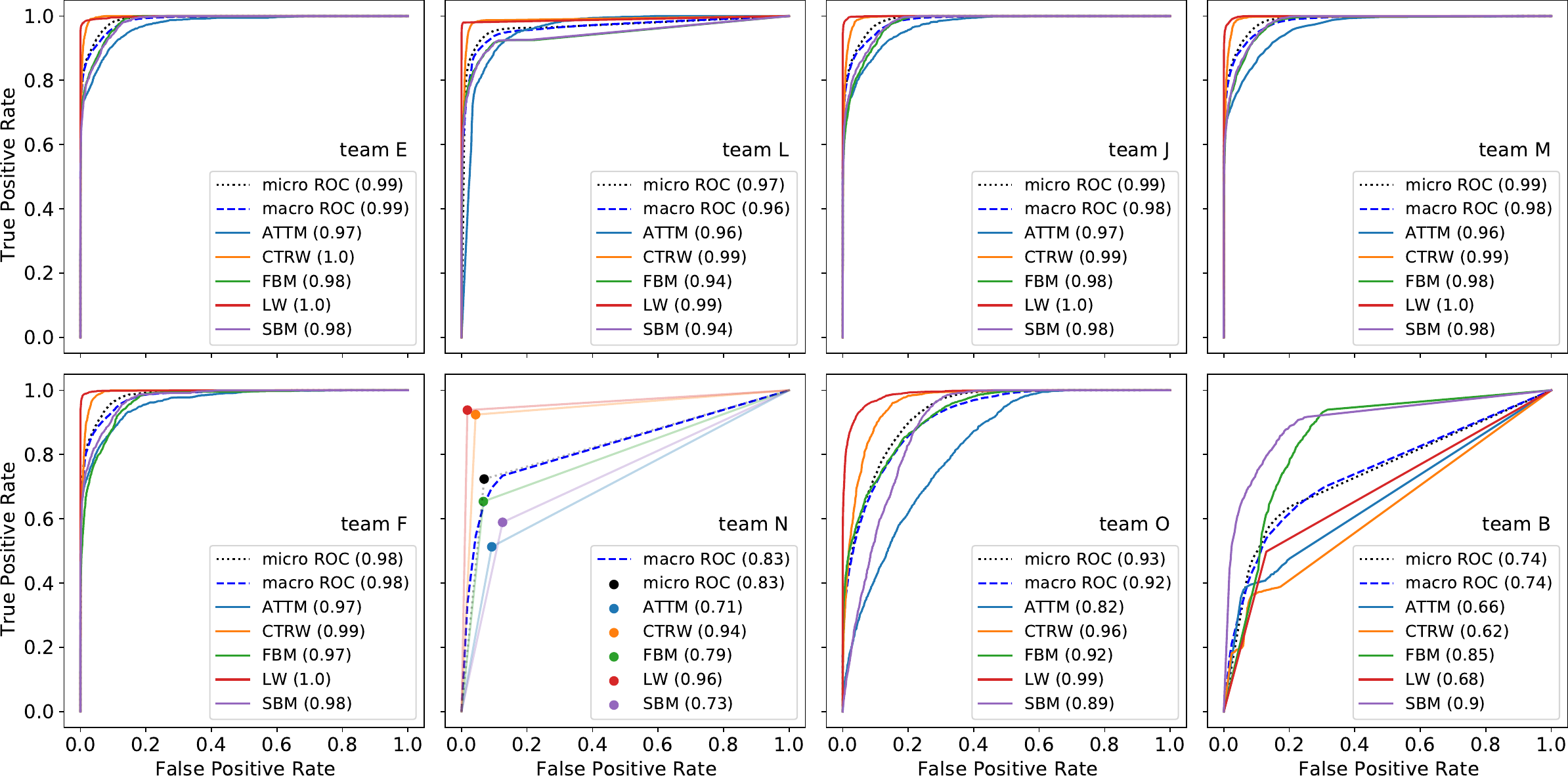}
	\caption{\label{fig:T2_2D_ROC} \textbf{T2.2D ROC curves.} ROC curves obtained for each diffusion model, plus micro- and macro-average, for all the methods participating in T2.2D. AUC values are reported in the legend. Teams are ordered according to to their ranking in the leaderboard.}
\end{figure}
%
% T2 3D ROC/AUC all teams
\newpage
\begin{figure}[h]
	\includegraphics[width=\textwidth]{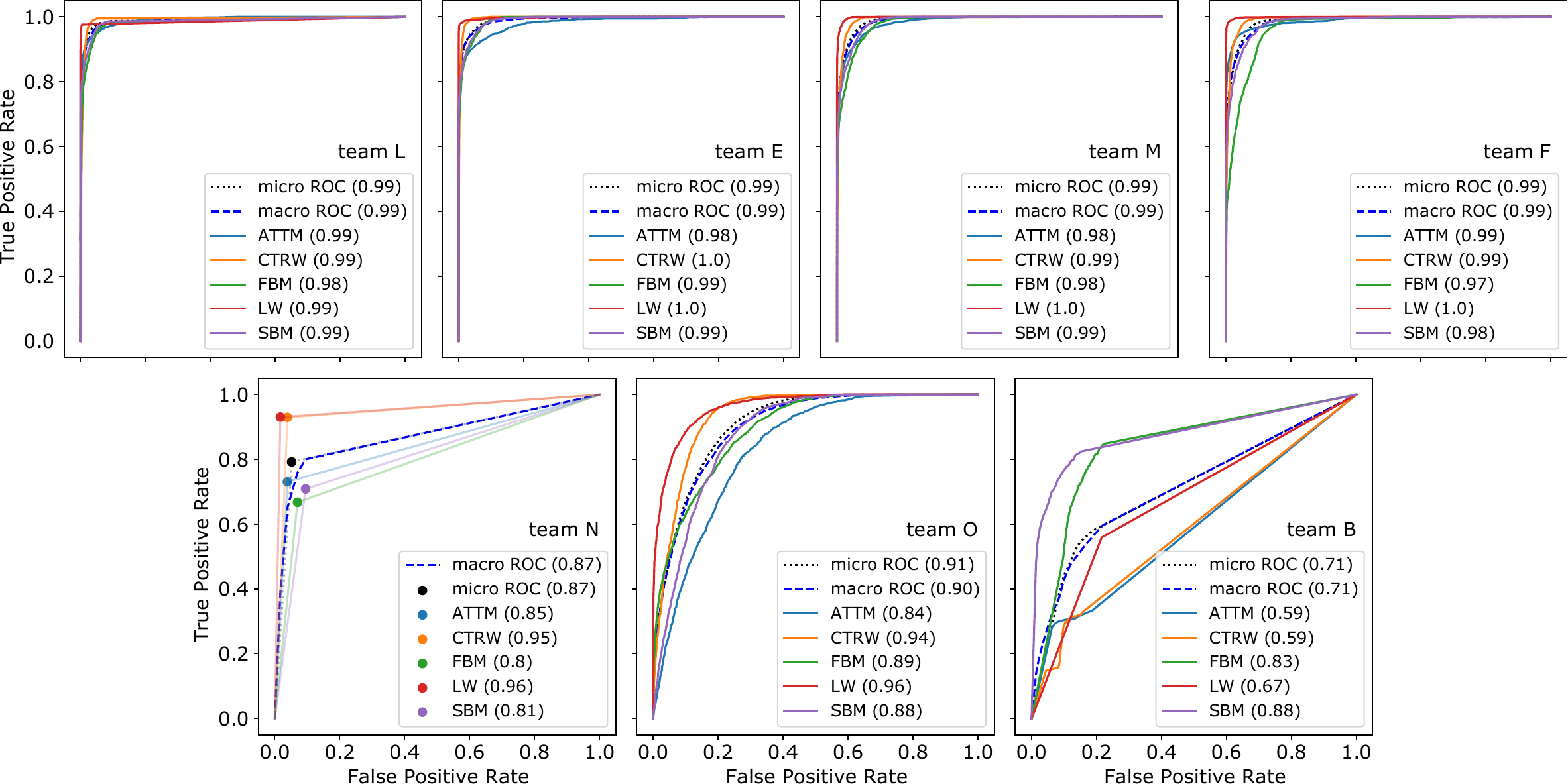}
	\caption{\label{fig:T2_3D_ROC} \textbf{T2.3D ROC curves.} ROC curves obtained for each diffusion model, plus micro- and macro-average, for all the methods participating in T2.3D. AUC values are reported in the legend. Teams are ordered according to to their ranking in the leaderboard.}
\end{figure}
%
% Ranking AUC
\clearpage
\newpage
\begin{figure}[h]
	\includegraphics[width=\textwidth]{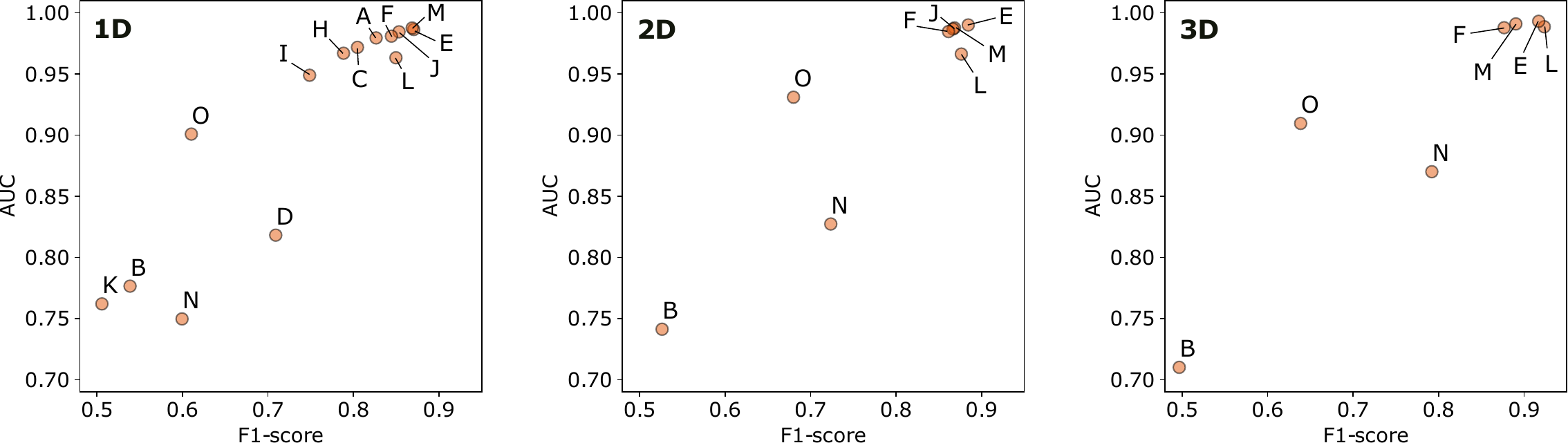}
	\caption{\label{fig:rank_AUC} \textbf{AUC vs ${\rm F_1}$-score for T2.} Scatter plot of the micro-averaged AUC vs the ${\rm F_1}$-score for all methods participating in T2.}
\end{figure}
%
%
% experiments
%%%% TASK 1 
\newpage
\begin{figure}[h]
	\includegraphics[width=\textwidth]{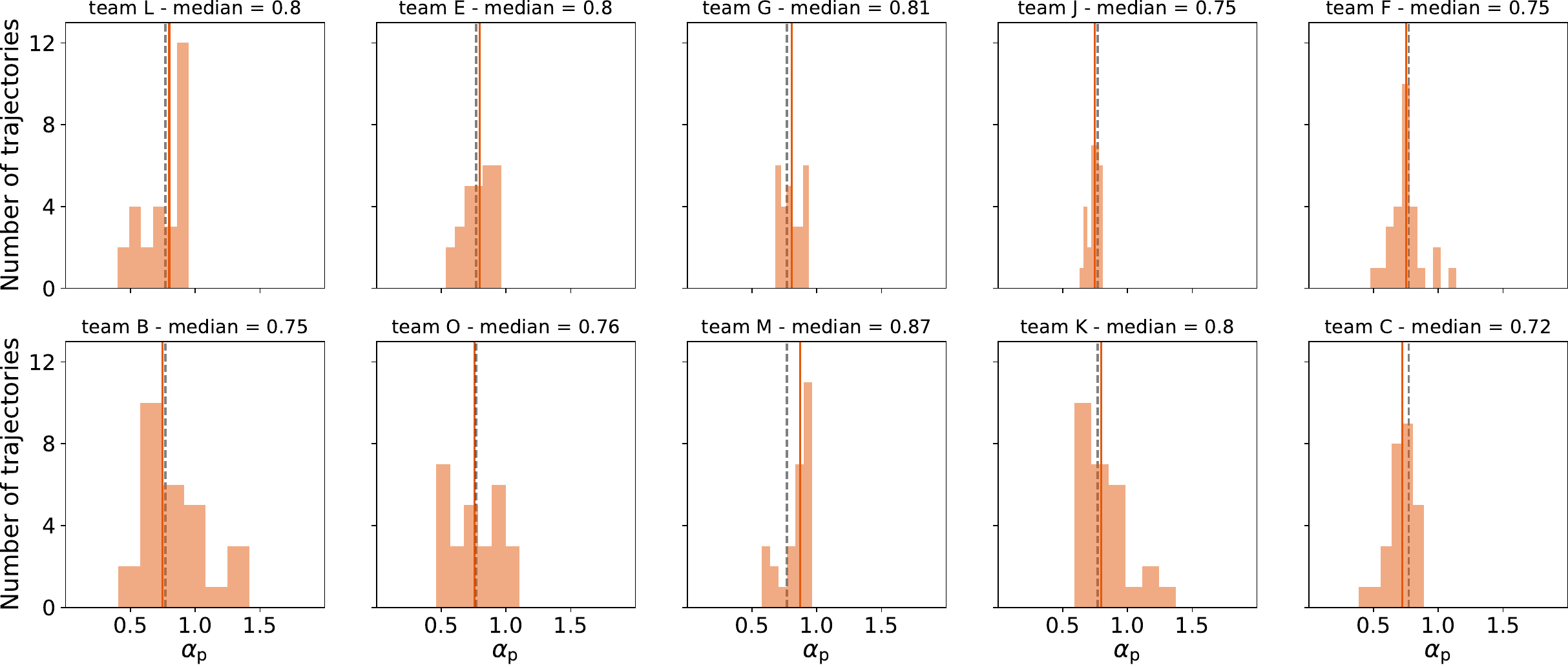}
	\caption{\label{fig:T1_GC} \textbf{Prediction of anomalous diffusion exponent for experimental trajectories from Ref.~\cite{golding2006physicalSI}.} Histogram of the anomalous diffusion exponent $\alpha_p$ predicted by all the methods participating in T1.2D. The continuous line represents the median value of $\alpha_p$. The dashed line indicates  the  original  estimation  of $\alpha$ provided  by  Ref.~\cite{golding2006physicalSI}.}
\end{figure}
\newpage
\begin{figure}[h]
	\includegraphics[width=\textwidth]{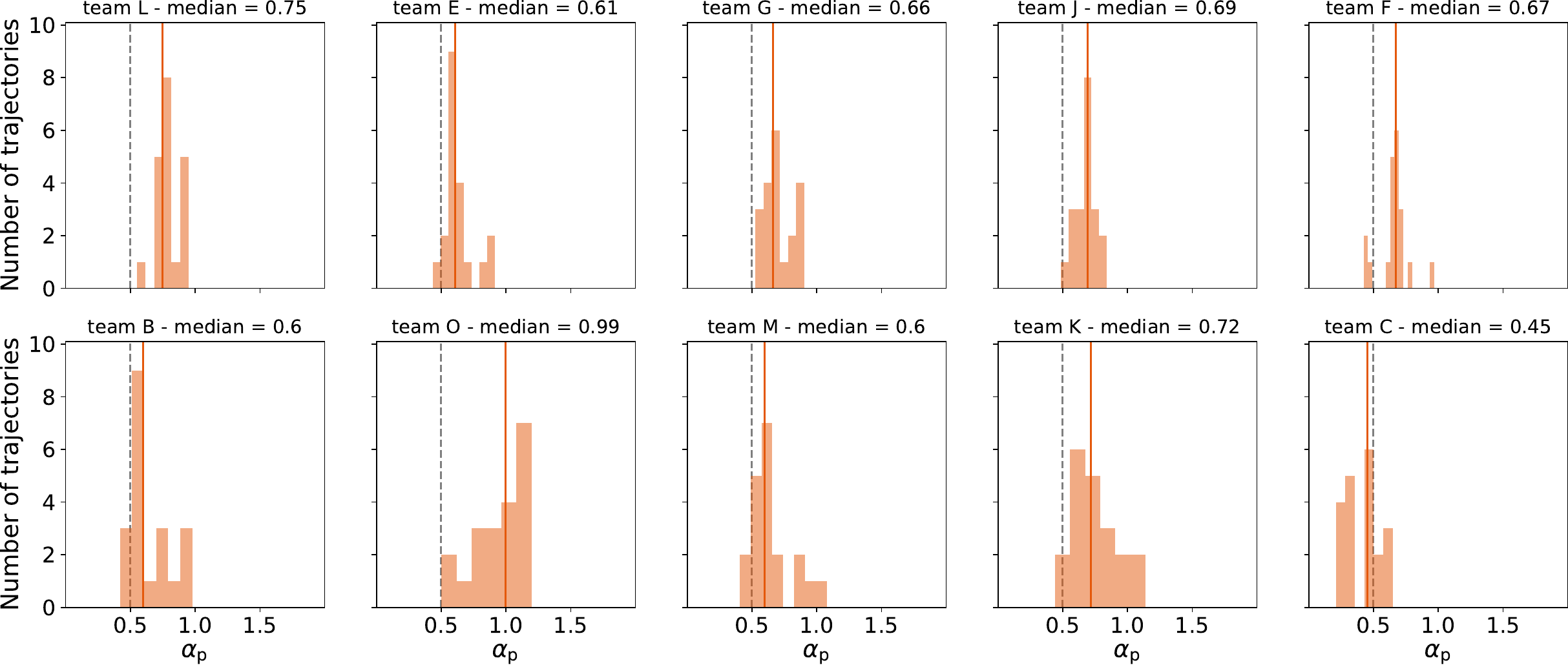}
	\caption{\label{fig:T1_WC} \textbf{Prediction of anomalous diffusion exponent for experimental trajectories from Ref.~\cite{stadler2017nonSI, krapf2019spectralSI}.} Histogram of the anomalous diffusion exponent $\alpha_p$ predicted by all the methods participating in T1.2D. The continuous line represents the median value of $\alpha_p$. The dashed lines indicate  the  original  estimation  of $\alpha$ provided  by  Refs~\cite{stadler2017nonSI, krapf2019spectralSI}.}
\end{figure}
\newpage
\begin{figure}[h]
	\includegraphics[width=\textwidth]{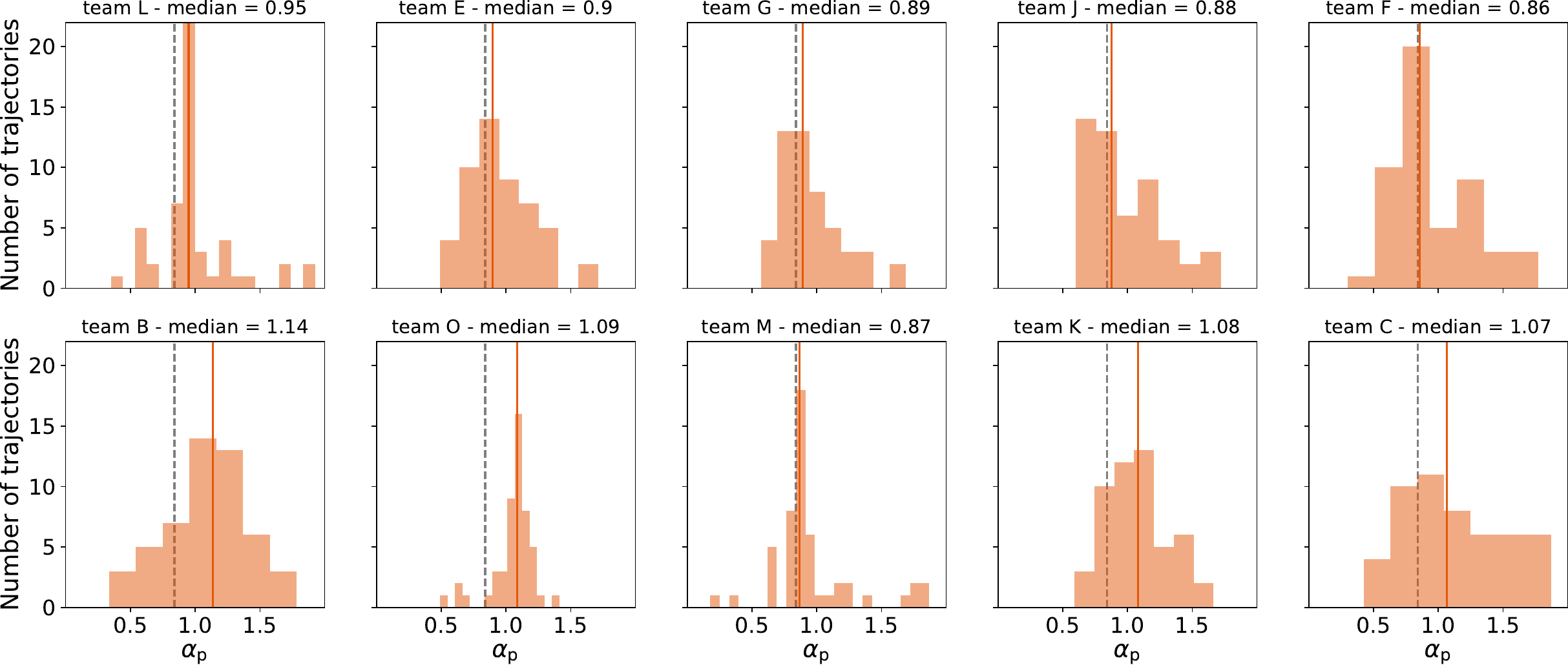}
	\caption{\label{fig:T1_M} 
	\textbf{Prediction of anomalous diffusion exponent for experimental trajectories from Ref.~\cite{manzo2015weakSI}.} Histogram of the anomalous diffusion exponent $\alpha_p$ predicted by all the methods participating in T1.2D. The continuous line represents the median value of $\alpha_p$. The dashed line indicates  the  original  estimation  of $\alpha$ provided  by  Ref.~\cite{manzo2015weakSI}.}
\end{figure}
\newpage
\begin{figure}[h]
	\includegraphics[width=\textwidth]{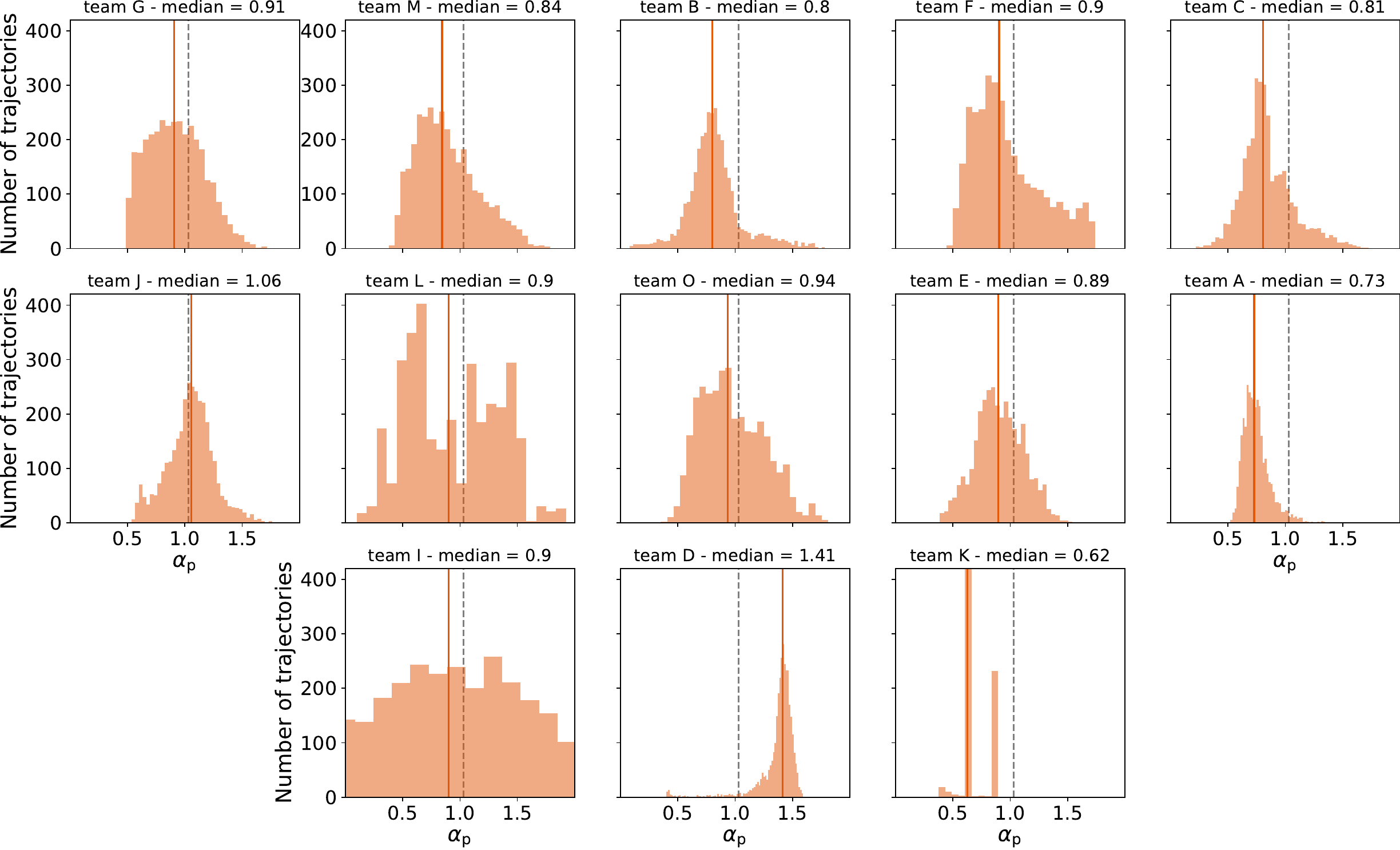}
	\caption{\label{fig:T1_WI} 	\textbf{Prediction of anomalous diffusion exponent for experimental trajectories from Ref.~\cite{kindermann2017nonergodicSI}.} Histogram of the anomalous diffusion exponent $\alpha_p$ predicted by all the methods participating in T1.1D. The continuous line represents the median value of $\alpha_p$. The dashed line indicates  the  original  estimation  of $\alpha$ provided  by  Ref.~\cite{kindermann2017nonergodicSI}.}
\end{figure}
% experiments
%%%% TASK 2
%
\newpage
\begin{figure}[h]
	\includegraphics[width=\textwidth]{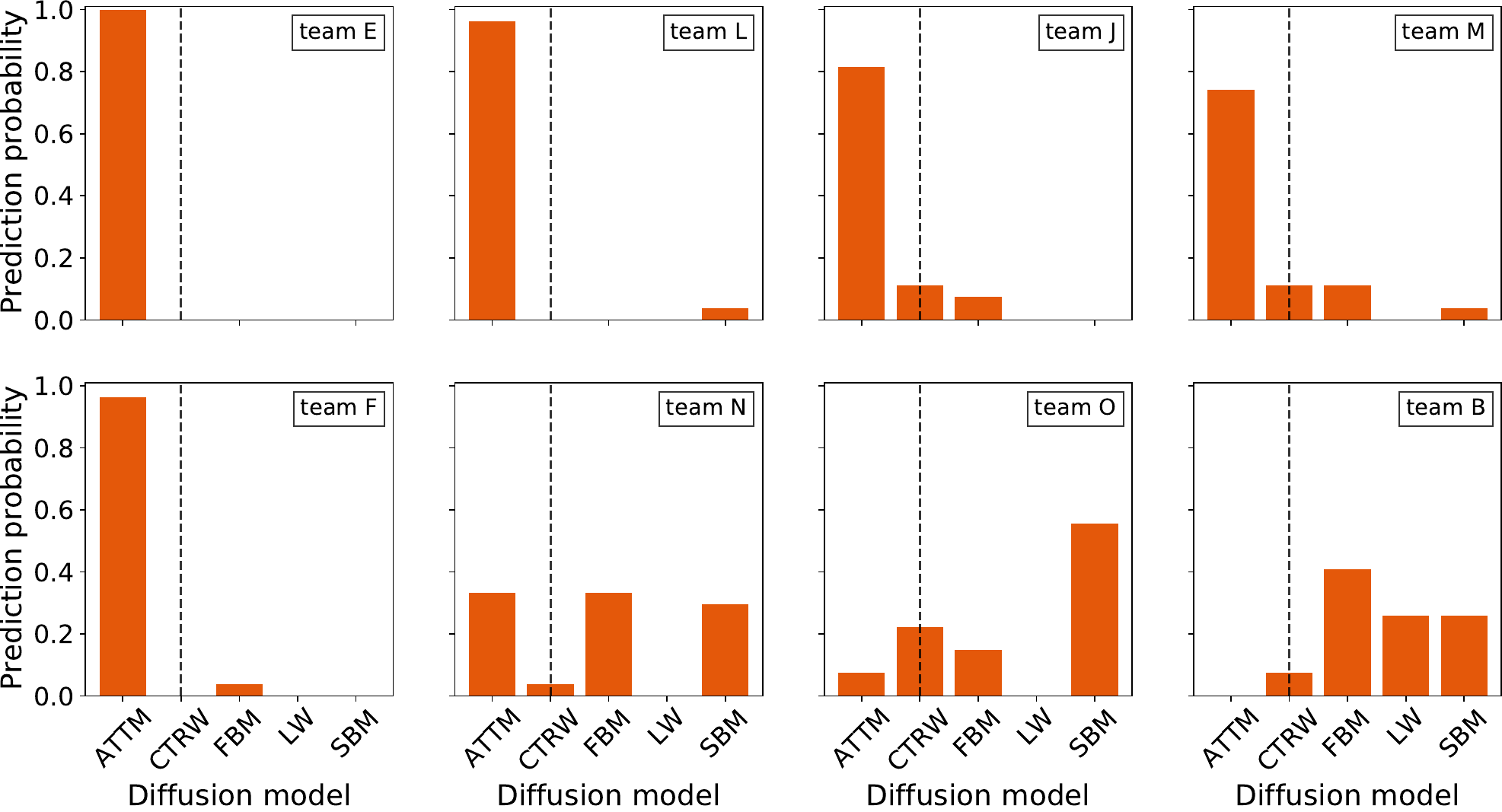}
	\caption{\label{fig:T2_GC} \textbf{Prediction of diffusion model for experimental trajectories from Ref.~\cite{golding2006physicalSI}.} Bar plot of the trajectory classification probability for the five anomalous diffusion model as predicted by all the methods participating in T2.2D. The dashed line indicates the original prediction of diffusion model provided by Ref.~\cite{golding2006physicalSI}.}
\end{figure}
\newpage
\begin{figure}[h]
	\includegraphics[width=\textwidth]{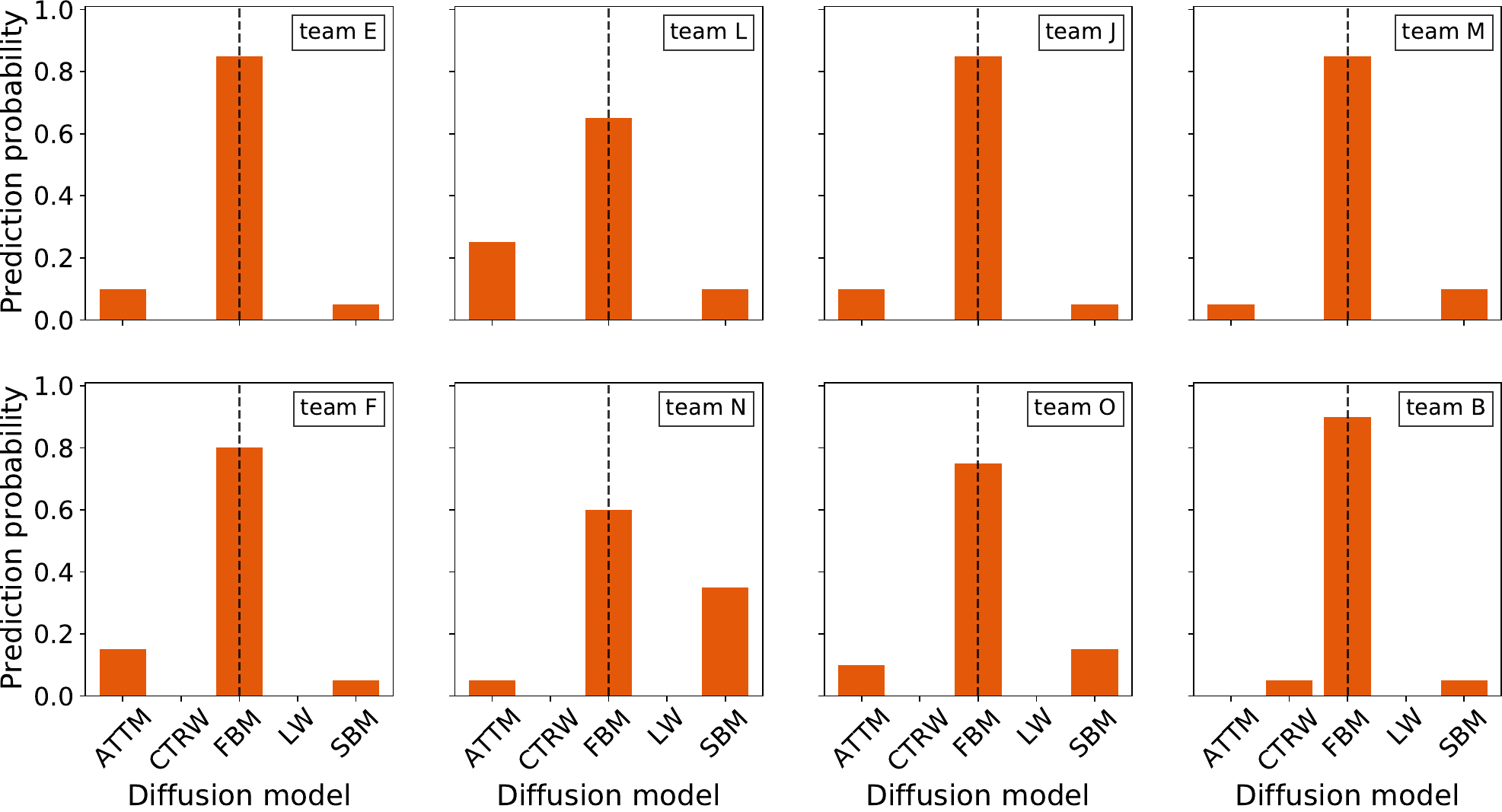}
	\caption{\label{fig:T2_WC} \textbf{Prediction of diffusion model for experimental trajectories from Refs.~\cite{stadler2017nonSI, krapf2019spectralSI}.} Bar plot of the trajectory classification probability for the five anomalous diffusion model as predicted by all the methods participating in T2.2D. The dashed line indicate the original  prediction of diffusion model provided by Refs~\cite{stadler2017nonSI, krapf2019spectralSI}.}
\end{figure}
\newpage
\begin{figure}[h]
	\includegraphics[width=\textwidth]{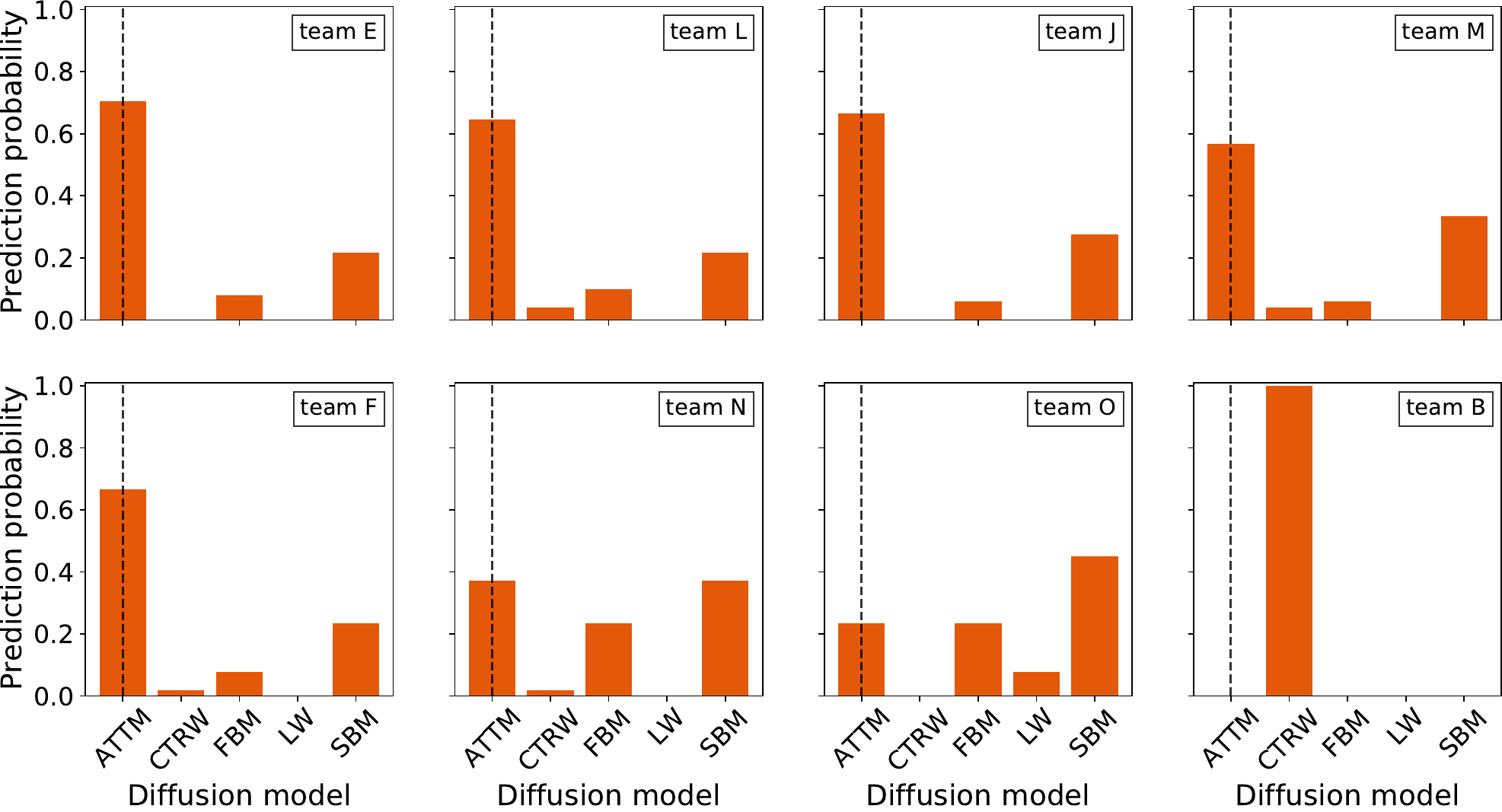}
	\caption{\label{fig:T2_M} \textbf{Prediction of diffusion model for experimental trajectories from Ref.~\cite{manzo2015weakSI}.} Bar plot of the trajectory classification probability for the five anomalous diffusion model as predicted by all the methods participating in T2.2D. The dashed line indicates the original prediction of diffusion model provided by Ref.~\cite{manzo2015weakSI}.}
\end{figure}
\newpage
\begin{figure}[h]
	\includegraphics[width=\textwidth]{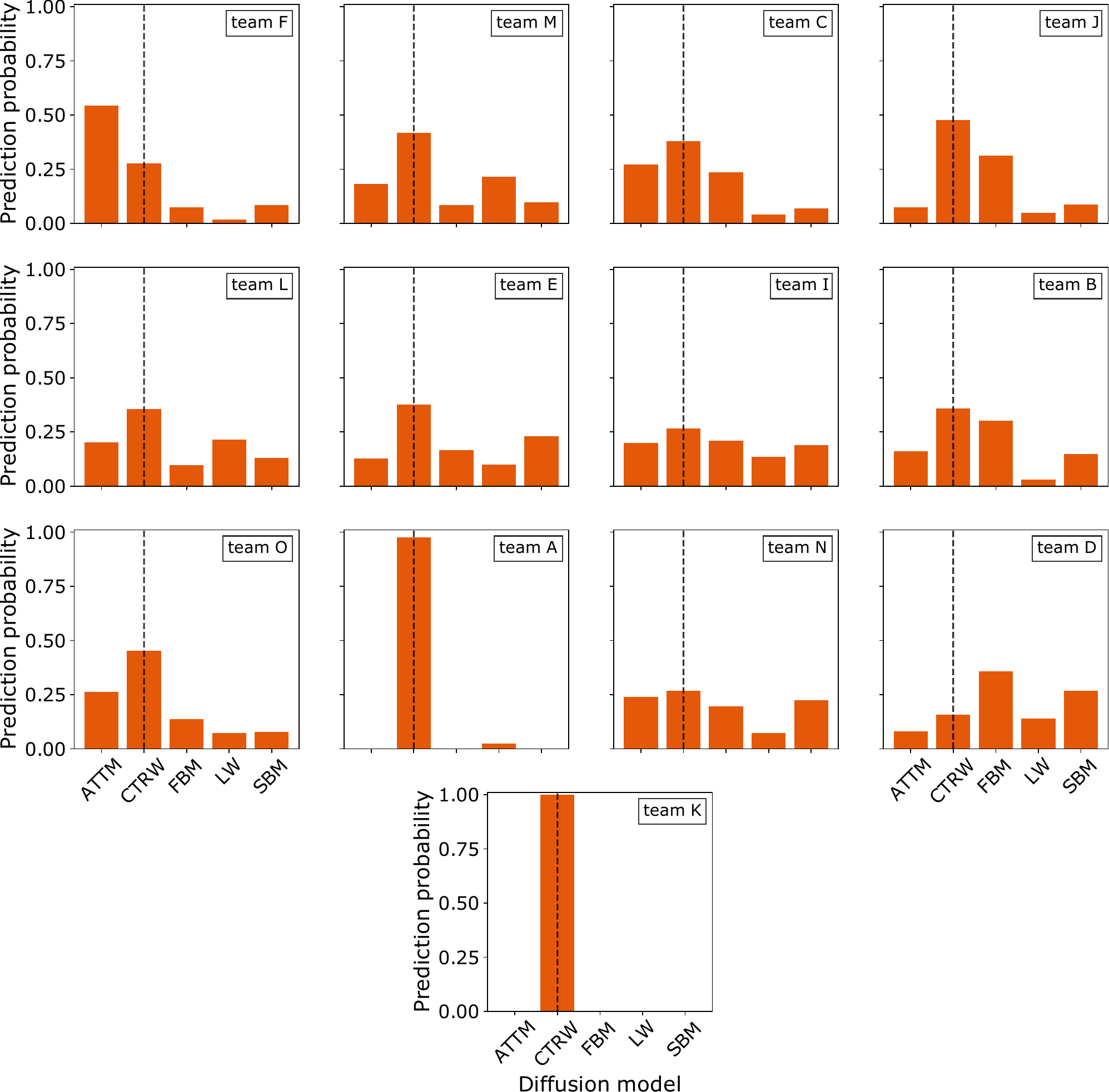}
	\caption{\label{fig:T2_WI} \textbf{Prediction of diffusion model for experimental trajectories from Ref.~\cite{kindermann2017nonergodicSI}.} Bar plot of the trajectory classification probability for the five anomalous diffusion model as predicted by all the methods participating in T2.1D. The dashed line indicates the original prediction of diffusion model provided by Ref.~\cite{kindermann2017nonergodicSI}.}
\end{figure}
\newpage
\begin{figure}[h]
	\includegraphics[width=\textwidth]{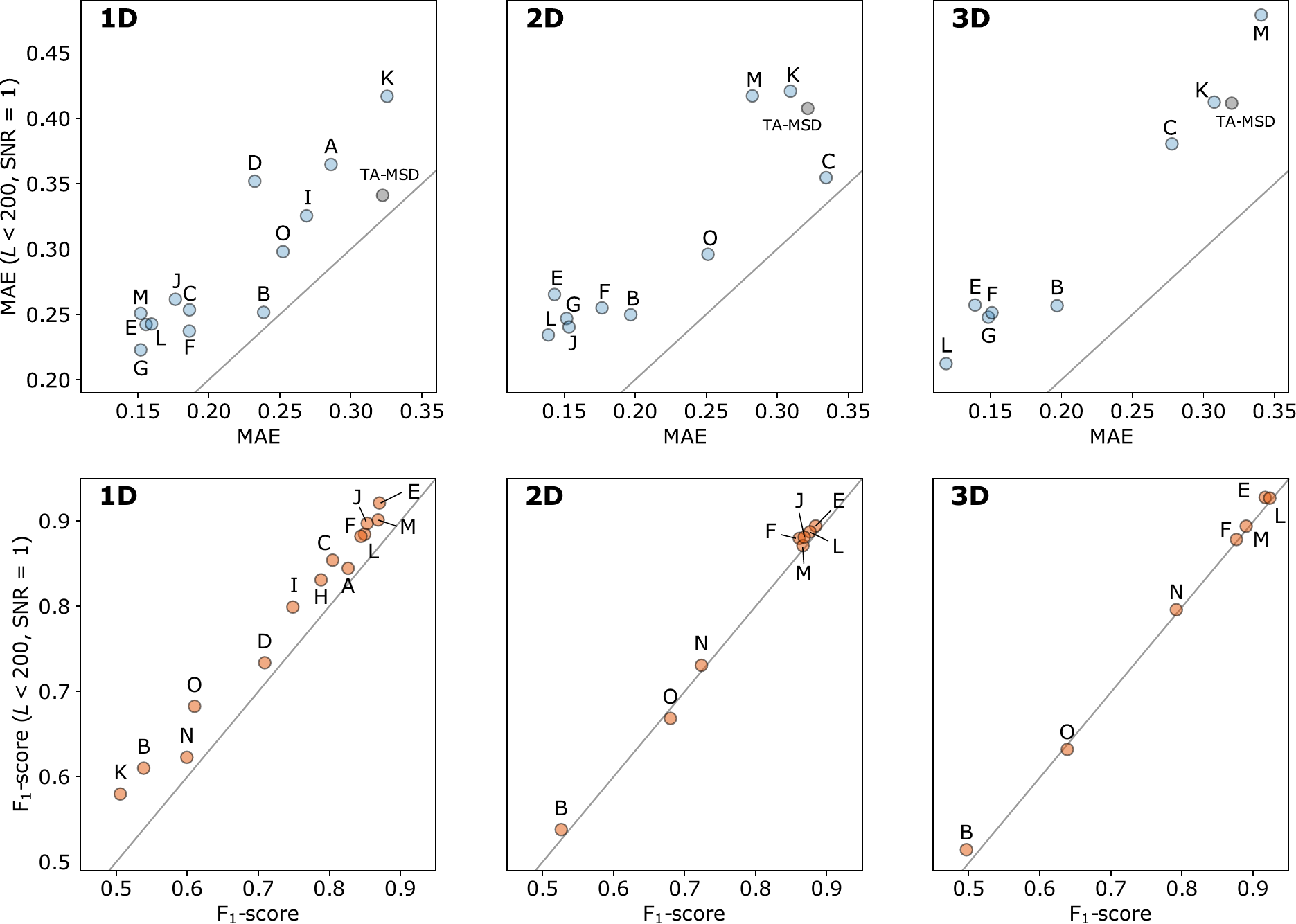}
	\caption{\label{fig:noi_vs_norm} \textbf{Metrics for short and noisy trajectories vs whole dataset.} Scatter plots of challenge metrics obtained over a subset of short and noisy trajectories ($L<200$, SNR$=1$) vs those obtained for the whole dataset for T1 (MAE, upper panels) and T2 (${\rm F_1}$-score, lower panels) in all the dimensions. Lines correspond to $y=x$, indicating equivalent performance on both datasets.}
\end{figure}

\clearpage
\linespread{1.0}\selectfont\centering
\hypertarget{sec:SI-note1_h}{}
\section*{Supplementary note 1: List of teams participating to the Challenge}

\begin{table}[!ht]
\linespread{1.0}\selectfont\centering
    \begin{tabular}{p{0.15\linewidth}  p{0.85\linewidth}}
    \hline
     \multicolumn{2}{l}{
Team \team{a}: \emph{Anomalous Unicorns}}     \\ \hline\hline
Contact: &  Borja Requena  \newline ICFO--The Institute of Photonic Sciences \newline Castelldefels (Barcelona), Spain \\ 
Reference: & Based on Refs.~\cite{munoz2020phaseSI, wolpert1992stackedSI} \\
Method: & HYDRAS (RNN + CNN) \\ 
Platform: & Python \\ 
Open-access: & \url{https://github.com/BorjaRequena/AnDi-unicorns} \newline \url{https://github.com/AnDiChallenge/AnDi2020_TeamA_AnomalousUnicorns}\\
Description: & Hydras are architectures that have a set of independent feature extractors (heads) that process the input trajectories. These all converge into a final set of fully connected layers (body) that process the output of the heads to perform inference. The feature extractors can be anything capable of processing trajectories of arbitrary lengths, such as recurrent neural networks (RNNs), convolutional neural networks (CNNs) or, even, other hydras. For T2, we have taken an ensemble of ten bi-headed hydras built with an RNN and a CNN as feature extractors. For T1, the resulting model is another ensemble of hydras that builds upon the result from T2. The resulting hydras have six heads: a hydra from T1 and five expert bi-headed hydras (RNN+CNN) that are trained to predict the anomalous exponent of a single diffusion model exclusively. This way, the body receives the output from all the model-specific feature extractors together with the opinion of the classifier. Each head is trained independently and then, in order to build the hydra, their weights are frozen while the body is trained. Finally, after a few epochs of body training, the head weights are unfrozen, and the entire hydra is trained with different learning rates: heads are trained with a much lower learning rate than the body. The entire source code can be found in the GitHub repository together with some examples. \\
Tasks: & T1.1D, T2.1D\\
\hline 
\end{tabular}
\end{table}

\begin{table}[!ht]
\linespread{1.0}\selectfont\centering
    \begin{tabular}{p{0.15\linewidth}  p{0.85\linewidth}}
    \hline
     \multicolumn{2}{l}{
Team \team{b}: \emph{BIT}}     \\ \hline\hline
Contact: &  Michael A. Lomholt  \newline PhyLife, Department of Physics, Chemistry and Pharmacy,
University of Southern Denmark \newline Odense M, Denmark\\ 
Reference: & \cite{krog2018bayesianSI, park2021bayesianSI}\\
Method: & Bayesian inference \\ 
Platform: & Matlab\\ 
Open-access: & \url{https://github.com/mlomholt/andi} \newline \url{https://github.com/AnDiChallenge/AnDi2020_TeamB_BIT}\\
Description: & Bayesian inference using annealed importance sampling to sample from the posterior distribution. We attempted to use Bayes theorem to calculate the posterior probability distributions for the models and parameters. The likelihood functions, and to a large extent also the priors, could be derived from the descriptions and codes provided by the organizers. Effective Bayesian inference could be achieved for the SBM and FBM~\cite{krog2018bayesianSI} models. However, the need to integrate out hidden waiting times impaired effective inference for ATTM, CTRW and LW. For ATTM and CTRW, we attempted to integrate out the waiting times together with the model parameters using Monte Carlo techniques. For LW, in 1D we used the forward algorithm on a hidden Markov model (but without including measurement noise)~\cite{park2021bayesianSI}, while in 2D and 3D we used a goodness-of-fit test after inference with the other four models to exclude them, followed by a fit to the TA-MSD to obtain the anomalous diffusion exponent of the LW. \\
Tasks: & All \\
\hline 
\end{tabular}
\end{table}

\begin{table}[!ht]
\linespread{1.0}\selectfont\centering
    \begin{tabular}{p{0.15\linewidth}  p{0.85\linewidth}}
    \hline
     \multicolumn{2}{l}{
Team \team{c}: \emph{DecBayComp}}     \\ \hline\hline
Contact: &  Jean-Baptiste Masson  \newline Institut Pasteur, Decision and Bayesian Computation lab \newline Paris, France\\ 
Reference: & \cite{verdier2021learningSI} \\
Method: & Gratin: graphs on trajectories for inference \\ 
Platform: & Python\\ 
Open-access: & \url{https://github.com/DecBayComp/gratin} \newline \url{https://github.com/AnDiChallenge/AnDi2020_TeamC_DecBayComp}\\
Description: & First, each trajectory is turned into a graph, where nodes are the positions and edges connect positions following a pattern based on their time difference. Then, features computed from normalized positions are attached to nodes (e.g., cumulative distance covered since origin, distance to origin, maximal step size since origin). These graphs are then passed as input to a graph convolution module (graph neural network), which outputs, for each trajectory, a latent representation in a high-dimensional space. This fixed-size latent vector is then passed as input to task-specific modules, which can predict the anomalous exponent or the random walk type. Several output modules can be trained at the same time, using the same graph convolution module, by summing task-specific losses. The model can receive trajectories of any size as inputs. The high-dimensional latent representation of trajectories can be projected down to a 2D space for visualization and provides interesting insights regarding the information extracted by the model (see details in Ref.~\cite{verdier2021learningSI}).\\
Tasks: & T1.1D, T2.1D \\
\hline
\end{tabular}
\end{table}

\begin{table}[!ht]
\linespread{1.0}\selectfont\centering
    \begin{tabular}{p{0.15\linewidth}  p{0.85\linewidth}}
    \hline
     \multicolumn{2}{l}{
Team \team{d}: \emph{DeepSPT}}     \\ \hline\hline
Contact: &  Taegeun Song  \newline  Center for AI and Natural Sciences, Korea Institute for Advanced Study \newline Seoul, Korea\\
Reference: & Based on Refs.~\cite{he2016deepSI,chen2016guestrinSI}\\
Method: & ResNet-MLP + XGBoost \\ 
Platform: & Python\\ 
Open-access: & \url{https://github.com/TaegeunSONG/DeepSPT} \newline \url{https://github.com/AnDiChallenge/AnDi2020_TeamD_DeepSPT}\\
Description: & 
We build our machine in the context of ensembles and hybrid structures. The applied preprocessing consists of three steps: 1) the noise is reduced by a 3-points moving average, 2) length of input trajectories are re-scaled to 100 points by a spline interpolation, and 3) the trajectories are normalized to the range$[0,1]$. First, we prepare each normalized trajectory and extract user-defined features from the trajectory as an input for the ensemble modules. Then, we construct an ensemble of ten identical modules based on residual net (ResNet)~\cite{he2016deepSI} and multi-layer perceptron (MLP). The ResNet input is the normalized trajectory and the following MLP receives both an output of the ResNet and the prepared features. Finally, the ten outputs from the ResNet-MLP module are analyzed by a scalable tree boosting system (XGBoost)~\cite{chen2016guestrinSI}.\\
Tasks: & T1.1D, T2.1D \\
\hline
\end{tabular}
\end{table}

\begin{table}[!ht] 
\linespread{0.85}\selectfont\centering
    \begin{tabular}{p{0.15\linewidth}  p{0.85\linewidth}}
    \hline
     \multicolumn{2}{l}{
Team \team{e}: \emph{eduN}}     \\ \hline\hline
Contact: &  Stefano Bo  \newline Max Planck Institute for the Physics of Complex Systems \newline Dresden, Germany\\ 
Reference: & \cite{argun2021classificationSI}\\
Method: & RANDI (LSTM + dense NN) \\ 
Platform: & Python\\ 
Open-access: & \url{https://github.com/booste/andi_for_organizers} \newline \url{https://github.com/AnDiChallenge/AnDi2020_TeamE_eduN}\\
Description: & The method is based on recurrent neural networks (RNN). The RNN used in all tasks share the same basic architecture and differ only in the last layer or two. All the RNN have two long short-term memory (LSTM) layers (of dimension 250 and 50, respectively). For inference tasks (T1 and T3) the last output of the second LSTM layer is directly connected to the output layer. For classification tasks (T2 and T3), the last output of the second LSTM layer is followed by a dense layer including 20 nodes, which is then connected to the five dimensional output layer (representing each model with softmax activation).

We train multiple RNN that specialize in analyzing trajectories of a certain length. When presented with a trajectory of length $l$, we use the predictions of the two RNN trained on the nearest lengths (one  on longer trajectories of length $L_+$ and one on shorter ones of length $L_-$) and weigh them according to their distance from $l$. %, $L_+$ and $L_-$.
For T1, we train 14 RNN for different lengths in 1D and 9 RNN for different lengths in 2D. For T2, we train 6 RNN for different lengths in 1D and 4 RNN for different lengths in 2D. In T3 all trajectories have the same length; we train 4 RNN: the first RNN to classify the model of the first segment, the second RNN to classify the model of the second segment, and two inference RNN; each inference RNN predicts the switching time, first exponent and the second exponent and their predictions are then averaged. We follow the same approach in 2D (but there we use a single RNN for the inference). 
We do not train RNN on 3D trajectories. For 3D data, we take projections on lower dimensions and use RNN trained on 2D and 1D data and average their outputs.

All RNN are trained using $3\times 10^6$  trajectories that are generated using {\tt andi-datasets} package~\cite{andigithubSI}. To avoid overtraining, we split these trajectories in 30 datasets (each containing $10^5$ trajectories) which are successively presented to the RNN. We use the first dataset to train for $5$ epochs splitting it in batches of size $32$. We then switch to another dataset, split it in batches of size $128$ and train for $4$ epochs. We repeat this procedure for $3$ other datasets. We iterate the procedure using $5$ datasets split into batches of size $512$ each considered for $3$ epochs and finally use $20$ datasets split into batches of size $2048$ for $2$ epochs each. For memory reasons, we did not use the batches of size 2048 for trajectories containing large amounts of measurement, such as long or high-dimensional trajectories. We use recurrent dropout (20\%) in both LSTM layers.

We preprocess the input data as follows: 1) We take the increment values of the trajectory. 2) We normalize the increments in a way that they have zero mean and unitary standard deviation for each trajectory. 
3) To optimize the training, we re-shape the input trajectories into shorter trajectories of higher dimensions. For example, for the inference of 1D trajectories of length $225$, the $224$ increments are split into $56$ blocks of dimension $4$, $b_j= [\Delta x_{4j}, \Delta x_{4j+1},\Delta x_{4j+2}, \Delta x_{4j+3}]$  with $j=0, \ldots 55$. The chosen block size varies according to the trajectory length and dimension.\\
Tasks: & All \\
\hline
\end{tabular}
\end{table}

\begin{table}[!ht]
\linespread{0.75}\selectfont\centering
    \begin{tabular}{p{0.15\linewidth}  p{0.85\linewidth}}
    \hline
     \multicolumn{2}{l}{
Team \team{f}: \emph{Erasmus MC}}     \\ \hline\hline
Contact: &  H\'{e}l\`{e}ne Kabbech   \newline Erasmus MC, Department of Cell Biology \newline Rotterdam, The Netherlands\\ 
Reference: & Based on Ref.~\cite{arts2019particleSI}\\
Method: & FEST\\ 
Platform: & Python\\ 
Open-access: & \url{https://github.com/hkabbech/FEST_AnDiChallenge} \newline \url{https://github.com/AnDiChallenge/AnDi2020_TeamF_ErasmusMC}\\
Description: & The Feature Extraction Stack long short-term memory (FEST) method was used to solve T1 and T2 and was applied to one-, two- and three-dimensional trajectory data. This method is divided in two parts: i) measurement of features at each point along the trajectories, and ii) training of a neural network consisting of a stack of bidirectional long short-term memory (LSTM) and fully connected (``Dense'') layers~\cite{goodfellow2016deepSI}.

The following features were computed: the displacements $\Delta \textbf{r}_n(t)=(\Delta \textbf{x}_n(t), \Delta \textbf{y}_n(t), \Delta \textbf{z}_n(t))$ of a particle between time $t$ and $t+n$ (which is the difference between two particle positions $\textbf{r}_t$ and $\textbf{r}_{t+n}$, where $\textbf{r}_t=(x_t, y_t, z_t)$ and $n\geq 1$) and the distances $d_{n}(t) = \sqrt{\Delta \textbf{x}_{n}(t)^2 + \Delta \textbf{y}_{n}(t)^2 + \Delta \textbf{z}_{n}(t)^2}$. The features for 1D and 2D cases were similarly defined. Subsequently, a mean of distances between time $t-p$ and $t+p$, $\overline{d_{n,p}}(t)$, was calculated as
$\overline{d_{n,p}}(t) = \frac{1}{2p+1}\sum_{k=t-p}^{t+p} d_{n}(k)$, where $p\geq 1$. All the mentioned features characterize how fast particles move. To gain information on the direction of motion, for 2D and 3D cases, the angles $\theta_n(t)$ between two displacement vectors $\Delta \textbf{r}_n(t)$ and $\Delta \textbf{r}_n(t-n)$ were computed. 

The number of features that were used as input to the neural network depended greatly on the number of dimensions. For 1D case, only displacements could be computed, therefor we used $\Delta \textbf{x}_{n}$, $n=\{1, 2\}$. Larger values of $n$ led to smaller sizes of feature vectors. For 2D case, we computed six features: $\Delta \textbf{x}_{1}$, $\Delta \textbf{y}_{1}$, $d_{1}$, $\overline{d_{1,1}}$, $\overline{d_{2,1}}$ and $\theta_1$. For 3D case, 6 other features were used: $\Delta \textbf{x}_{1}$, $\Delta \textbf{y}_{1}$, $\Delta \textbf{z}_{1}$, $d_{1}$, $\overline{d_{1,1}}$, $\overline{d_{2,1}}$.

We built two similar neural network architectures for T1 and T2. Using the above-mentioned features, the output for T1 was a predicted value of $\alpha$, and the outputs for T2 were probabilities of input track belonging to one of 5 diffusive models. The architectures of both neural network were built using functions from the Keras library~\cite{chollet2015kerasSI}. In both cases, we used 3 bidirectional LSTM layers (with $2^6$, $2^5$ and $2^4$ hidden nodes, respectively), followed by 4 Dense layers (with $2^5$, $2^4$, $2^3$ and $1$ (or $5$) hidden nodes) with Dropout layers in between (with a dropout rate of $0.2$ or $0.1$). For T1, \texttt{ReLu} activation function was applied on each Dense layer, while for T2 \texttt{tanh} was applied with a \texttt{softmax} at the output layer. During the training, the models were optimized using the Adam optimizer and, as loss functions, we used the mean squared error (MSE) for T1 and categorical cross-entropy for T2.

The described networks had to be trained using trajectories with a fixed number of time points. For that, new datasets were created with the tool provided by the organizers (\url{https://github.com/AnDiChallenge/ANDI_datasets} ~\cite{andigithubSI}). To cover the variety of lengths that can be encountered in the challenge data, 4 different datasets were generated for each task, each consisting of different trajectory lengths: 50, 200, 400 or 600 time points. Thereby, each network was trained 4 times in order to create 4 distinct models. For each case (1D, 2D and 3D), we created 30000 tracks of length 50 for training and 6000 for validation (denoted 30000/6000) to keep a ratio 8:2, 7500/1500 trajectories of length 200, 3750/750 of length 400 and 2500/500 of length 600. Training and validation datasets were generated separately to ensure that all combined cases of $\alpha$ and diffusive models were present in both dataset.

The training have been carried out on a Linux system with a GPU GeForce GTX 1650 and a processor 2.60 GHz Intel 12 cores i7. An early stopping criterion was used to monitor the validation loss and prevent over-fitting. Finally, during the prediction phase and depending on the trajectory length, a combination of the different models was used to predict the outcome. Any track with a length below 100 was predicted with the model trained with 50 time points (denoted model50), any length falling between 100 and 300 with model200, between 300 and 500 with model400 and above 500 with model600. This approach would increase the accuracy of the prediction when the variety of trajectory length would be very diverse in a dataset.\\
Tasks: & T1, T2 \\
\hline
\end{tabular}
\end{table}

\begin{table}[!ht]
\linespread{1.0}\selectfont\centering
    \begin{tabular}{p{0.15\linewidth}  p{0.85\linewidth}}
    \hline
     \multicolumn{2}{l}{
Team \team{g}: \emph{HNU}}     \\ \hline\hline
Contact: & Zihan Huang \newline  School of Physics and Electronics, Hunan University \newline Changsha 410082, China \\ 
Reference: & \cite{li2021wavenetSI}\\
Method: & Just LSTM it \\ 
Platform: & Python\\ 
Open-access: & \url{https://github.com/huangzih/AnDi-Challenge} \newline \url{https://github.com/AnDiChallenge/AnDi2020_TeamG_HNU}\\
Description: & The training dataset consisting of 1D trajectories is generated at 43 specific lengths (see the open-access link for details). The total size of training dataset is about 330 GB. Each trajectory is normalized before training so that its position's average and standard deviation are 0 and 1 respectively.

A long short-term memory (LSTM)-based recursive neural network (RNN) model is used to accomplish this competition task, where the dimension of the hidden layer is 64 and the number of stacked LSTM is 3. Models for each specific length are trained separately. 80\% of training data is used for training, while the rest is used for validation. We implement the LSTM-based model by PyTorch 1.6.0. The model is trained with a batch size 512, where the loss function is the mean squared error (MSE). The optimizer is Adam with a learning rate $l=0.001$. The learning rate is changed as $l\leftarrow l/5$ if the validation loss does not decrease for 2 epochs. When the number of such changes exceeds 1, the training process is early stopped to save time and avoid overfitting. The best epoch for a specific length is determined by the lowest mean absolute error (MAE) of the validation set.

The inference of challenge data is guided by the following rule: 1) If the original length of trajectory belongs to one of the 43 specific lengths, this trajectory will be directly used for inference. 2) Otherwise, a new length of this trajectory will be set as the closest smaller specific length. For instance, the new length of a trajectory with an original length 49 should be 45. The trajectory data is subsequently transformed into 2 sequences. For clarity, we set the trajectory data as $[x_1, x_2, \cdots, x_T]$, where $T$ is the original length. We denote $T_n$ as the new length with $T_n < T$. The two sequences are $[x_1, x_2, \cdots, x_{T_n}]$ and $[x_{T-T_n+1}, x_{T-T_n+2}, \cdots, x_T]$ respectively. Such two sequences are both used for inference, with model predictions $\alpha_1$ and $\alpha_2$. The predicted exponent $\alpha$ of the original trajectory is given by $\alpha=(\alpha_1+\alpha_2)/2$.

To further improve the model performance, 5-Fold cross validation is utilized. However, due to the time limit of this competition, we only use a 3-fold average. On the other hand, by analyzing an external validation dataset containing 100000 1D trajectories, the predicted results for challenge data are multiplied by 1.011 and finally clipped to ensure reasonable predictions.

The methods for 2D and 3D tasks are both based on the solution for 1D trajectories. We separate the dimensions of the trajectories and treat the data of each dimension as 1D trajectories. Thus, we get predicted exponents $\alpha_x$, $\alpha_y$, and $\alpha_z$ for $x$, $y$, and $z$ dimensions, respectively. The final results are $\alpha_{2{\rm D}} = (\alpha_x+\alpha_y)/2$ for 2D trajectories, and $\alpha_{3{\rm D}} = (\alpha_x+\alpha_y+\alpha_z)/3$ for 3D trajectories.\\
Tasks: & T1 \\
\hline
\end{tabular}
\end{table}

\begin{table}[!ht]
\linespread{1.0}\selectfont\centering
    \begin{tabular}{p{0.15\linewidth}  p{0.85\linewidth}}
    \hline
     \multicolumn{2}{l}{
Team \team{h}:\emph{NOA}}     \\ \hline\hline
Contact: & Nicol\'{a}s Firbas \newline  Instituto Universitario de Matem\'{a}tica Pura y Aplicada, Universitat Polit\`{e}cnica de Val\`{e}ncia  \newline Valencia, Spain \\ 
Reference: & Based on Ref.~\cite{donahue2015longSI}\\
Method: & Convolutional LSTM  \\ 
Platform: & Python\\ 
Open-access: & \url{https://github.com/NicoFirbas/ConvLSTM_AnDI} \newline \url{https://github.com/AnDiChallenge/AnDi2020_TeamH_NOA}\\
Description: & 
The  convolutional long short-term memory (convLSTM) approach combines convolutional neural networks (CNN) and long short-term memory networks (LSTM), similarly as described in Ref.~\cite{donahue2015longSI}. An additional linear block placed after the LSTM uses the flattened LSTM output to predict the type of anomalous diffusion of the trajectory.

In more detail, it consists of a convolutional block (ConvBlock), a bidirectional LSTM, and a linear block (LinearOuts). The ConvBlock consists primarily of two one-dimensional convolutions with a filter size of two, each is followed by a ReLU. The first convolutional layer is more coarse and outputs 20 features, while the second layer takes the output of the first and outputs 64 features. At the end of the convolutional block, we have a dropout with dropout probability $p=0.2$, to avoid overfitting, and a one-dimensional MaxPooling layer, which cuts the output size in half by selecting the larger of two adjacent entries. The bidirectional LSTM has three layers, each layer is followed by a dropout with probability of dropout $p = 0.2$. The final Block (LinearOuts) takes the flattened (2D tensor to 1D) output of the LSTM as its input and passes it to a fully connected linear layer, which has five output units that correspond to the five models used to produce the trajectories. The first two linear layers are followed by a ReLU activation and the final layer is not, as non-linearity is handled by an instance of nn.CrossEntropyLoss, during training, called the ``criterion''.

Training of our method for the AnDi challenge was done using a hidden size of 32 and a learn rate of 0.001. However, later testing has shown that our model accuracy can be improved by increasing the hidden size to 128, while beyond that point we see a drop in accuracy.
Training was performed by merging two data sets, which were generated with the {\tt andi-datasets} package~\cite{andigithubSI}, the first of length 189810 and the second of length 150000. The resulting combined dataset was split into 75\% training data and 25\% test data. From the training data an additional 20\% was reserved for validation data to be used by our early stopping algorithm.
Our early stopping method saves the parameter state if there is an improvement in the mean validation loss, which is computed at the end of each epoch. We used 100 epochs and 10 patience for our early stopping.\\
Tasks: & T1.1D \\
\hline
\end{tabular}
\end{table}

\begin{table}[!ht]
\linespread{1.0}\selectfont\centering
    \begin{tabular}{p{0.15\linewidth}  p{0.85\linewidth}}
    \hline
     \multicolumn{2}{l}{
Team \team{i}: \emph{QuBI}}     \\ \hline\hline
Contact: & Carlo Manzo \newline  Facultat de Ci\`encies i Tecnologia, Universitat de Vic -- Universitat Central de Catalunya (UVic-UCC) \newline Vic, Spain \\ 
Reference: & \cite{manzo2021extremeSI} \\
Method: & AnDi-ELM  \\ 
Platform: & Matlab\\ 
Open-access: & \url{https://github.com/qubilab/AnDi_ELM} \newline \url{https://github.com/AnDiChallenge/AnDi2020_TeamI_QuBI}\\
Description: & Our model combines feature engineering and the use of an extreme learning machine (ELM). In brief, raw trajectories were first standardized to set their starting coordinates to zero and have a unitary standard deviation of displacements for $t_{\rm lag}=1$. For each $t_{\rm lag}=1,...,7$, two features were calculated, corresponding to
$ \frac{ {\rm log}\left< | x(t+t_{\rm lag}) -x(t) |^k \right> } { {\rm log}( t_{\rm lag} +1)}$ for $k=1,2$. In addition, the correlation of absolute displacements obtained for $t_{\rm lag}=1$ was also included, for a total of 15 features per trajectory. Features were standardized using the $z$-score over the training dataset. The mean and standard deviation obtained for each feature of the training dataset was saved and later used to standardize the validation and test datasets. For a training dataset of $n$ trajectories and $f$ features with target values ${\bf T}$, the $n \times f$ feature matrix ${\bf X}$ is fed into a ELM composed by single hidden layer feedforward network (SLFN) with $m=1000$ hidden nodes~\cite{huang2004extremeSI, huang2006extremeSI}. A  matrix of initial weights ${\bf W}$ of size $f \times m $ and a bias vector ${\bf b}$ of size $1 \times m $ are randomly initialized to connect observations to targets through:
\begin{equation*}
    f\left( {\bf X} {\bf W} + {\bf u} {\bf b^{T}}\right) {\bf B}  = {\bf H} {\bf B} = {\bf T},
\end{equation*}
where $f\left( \cdot \right)$ represents the sigmoid activation function, ${\bf u}$ is a unitary vector of size $n \times 1$, and {\bf B} is the matrix of output weight. 
The training of the SFLN is converted into solving an over-determined linear problem, whose least squares
solution corresponds to the Moore-Penrose pseudoinverse of the hidden layer matrix {\bf H}~\cite{huang2004extremeSI, huang2006extremeSI}
\begin{equation*}
    {\hat{ \bf B }}={\bf H^{\dagger}}{\bf T}.
\end{equation*}

The SFLN was trained either as a regressor or as a classifier to provide predictions for T1 and T2 for 1D trajectories. Training was performed using only the dataset provided by the organizers ($10000$ trajectories per subtask) during the Development phase of the challenge. Training took typically $5$ seconds on a MacBookPro with a 8-Core Intel Core i9 processor with 2.4GHz speed.\\
Tasks: & T1.1D, T2.1D \\
\hline
\end{tabular}
\end{table}

\begin{table}[!ht]
\linespread{1.0}\selectfont\centering
    \begin{tabular}{p{0.15\linewidth}  p{0.85\linewidth}}
    \hline
     \multicolumn{2}{l}{
Team \team{j}: \emph{FCI}}     \\ \hline\hline
Contact: & Tom Bland \newline The Francis Crick Institute \newline London, UK \\
Reference: & Based on Refs.~\cite{bai2018empiricalSI, granik2019singleSI}\\
Method: & CNN  \\ 
Platform: & Python\\ 
Open-access: & \url{https://github.com/tsmbland/andi_challenge} \newline \url{https://github.com/AnDiChallenge/AnDi2020_TeamJ_FCI}\\
Description: & We use a convolutional neural network structure adapted from the models used in Refs.~\cite{bai2018empiricalSI, granik2019singleSI}. For T1 and T2, this consists of a series of convolutional blocks, followed by a global max-pooling layer over the temporal dimension, which feeds into a dense network. For T1, the model outputs a single number representing the predicted anomalous exponent. For T2, the model outputs 5 numbers, representing a probability (from 0-1) for each diffusion type. For T3, convolutional blocks are followed by a $1 \times 1$ convolutional network, which outputs an array of size $(1, n)$, where $n$ is the number of steps in the trajectory, representing the probability of a switch at each position in the trajectory. The same network architectures can be used in 1D and higher dimensions, varying only the number of input features. Models were built using TensorFlow in Python, and code is available on Github.

Training data were generated using the {\tt andi-datasets} package~\cite{andigithubSI}. Trajectories were first preprocessed by taking the difference between successive positions, and normalized by dividing by the mean step size. For T1 and T2, a single model was simultaneously trained on trajectories of all lengths (ranging from 5-1000 steps). To permit mini-batch gradient descent with tracks of variable length, shorter tracks within each batch were padded with zeros to ensure a consistent input size (Note: padding is only necessary during training, and inference can be carried out with or without padding). For T3, training data consisted of trajectories 200-steps in length with a single changepoint, as per the challenge, but the method could be adapted to variable trajectory lengths and multiple changepoints.

For all models, training was carried out with a batch size of 32 and an Adam optimizer with a learning rate of 0.001, until a performance plateau was reached (up to a maximum of 1.28 million trajectories, with each trajectory seen by the networks only once).\\
Tasks: & T1.1D, T1.2D, T2.1D, T2.2D, T3.1D, T3.2D \\
\hline
\end{tabular}
\end{table}

\begin{table}[!ht]
\linespread{0.85}\selectfont\centering
    \begin{tabular}{p{0.15\linewidth}  p{0.85\linewidth}}
    \hline
     \multicolumn{2}{l}{
Team \team{k}: \emph{TSA}}     \\ \hline\hline
Contact: & Erez Aghion \newline Max Planck Institute for the Physics of Complex Systems \newline Dresden, Gemany \\ 
Reference: & \cite{aghion2021mosesSI}\\
Method: & Scaling analysis, and feature engineering (for T2)  \\ 
Platform: & Python \\ 
Open-access: & \url{https://github.com/ErezAgh/ANDI-challange-codes-} \newline \url{https://github.com/AnDiChallenge/AnDi2020_TeamK_TSA}\\
Description: & This approach is based on theory, as opposed to pure data-driven methods. 
Anomalous diffusion can be described via more than just the Hurst exponent. The assumptions of the central limit theorem, which leads to standard diffusion, can be violated in three distinct ways: Increment correlations (like in FBM), fat-tailed increment distribution (like in CTRW), and nonstationarity of the increments' distribution, like in SBM. Each of these three paths can be characterized by its own scaling exponent, and can be measured directly in data, using methods of time-series analysis. The exponent $J$, describing the first violation, can be measured, e.g., using detrended fluctuations analysis. The exponent $L$, for the second violation, is measured from the temporal scaling of the time-average of the squared increments of the process. Finally, The exponent $M$ is measured from the scaling of the time-average of the increments' absolute value. These exponents can be measured in any number of dimensions. Their sum leads to the Hurst exponent: $H=M+L+J-1$ \cite{chen2017anomalousSI, meyer2018anomalousSI, aghion2021mosesSI}. 

To estimate the Hurst exponent for T1, we evaluate $J$, $L$ and $M$ using methods which were specifically adapted for noise filtering. Importantly, this approach is not model-dependent, and our algorithm can be applied also to other types of data, not generated by one of the five models in the AnDi challenge. 

For T2, we construct a small set of educated questions, targeted to characterize different properties of the paths in the data set, via precise analysis of the increments of the process. 
When comparing between various models outside of the AnDi challenge, here we would need to construct a new set of questions for the new models. Some of the questions are aimed for various general relations between the three exponents described above, others, to more specific properties of the individual types of paths involved in the challenge. The answers of each question can be ``yes" ($=1$) or ``no" ($=0$) (or ``maybe" ($=2$)). An example for a question about the exponents: Is $(J-0.5)>(M-0.5)+(L-0.5)$?  Namely, is the effect of autocorrelations on the Hurst exponent stronger than the combined effect of the increment distribution?  This question separates between FBM and LW on the one side, and ATTM and SBM on the other.  An example for a question beyond the exponents, is given by the comparison of the autocorrelations of the increments of the process, versus that of their absolute value. This question is highly selective for distinguishing L\'evy walk from all the others. For each trajectory in the competition data set: we generate a set of answers using the same algorithm, and then generate an array of probabilities for this set to be either ATTM, CTRW, FBM, LW, or SBM.  This is done by counting how many times a similar line of answers appeared in the training set for each type of process, divided by the total number of occurrences. The answer is, e.g.: $[0.125; 0.025; 0.85; 0.0;...]$. The larger the training set, the more accurate is the evaluation of the probabilities. If a new set of answers is not found in the training file, a reduced number of selected questions are asked again, making the choice less selective. The selectivity of the questions, and the time-series analysis techniques used, also affect the quality of the final results. This method is similar in one and higher dimensions.\\
Tasks: & T1, T2.1D \\
\hline
\end{tabular}
\end{table}

\begin{table}[!ht]
\linespread{1.0}\selectfont\centering
    \begin{tabular}{p{0.15\linewidth}  p{0.85\linewidth}}
    \hline
     \multicolumn{2}{l}{
Team \team{l}: \emph{UCL}}     \\ \hline\hline
Contact: & Giorgio Volpe \newline Department of Chemistry, University College London \newline London, UK \\ 
Reference: & \cite{gentili2021characterizationSI}\\
Method: & CONDOR  \\ 
Platform: & Matlab \\ 
Open-access: & \url{https://github.com/sam-labUCL/CONDOR} \newline \url{https://github.com/AnDiChallenge/AnDi2020_TeamL_UCL}\\
Description: & Our method named Classifier Of aNomalous DiffusiOn tRajectories (CONDOR) relies on at first analyzing the trajectories to extract features (and their statistics) such as the trajectory length, velocity (with sign and absolute value, different sampling rates), rate of variation, Fourier Transform, Power Spectral Density, autocorrelation function, time-averaged MSD, and wavelet transform, among others. This analysis is performed on each dimension separately. 

T2: These features are the inputs for a deep feed-forward neural network (5 categories, 2 hidden layers, 20 neurons per layer, trained with a $10^5$ trajectory dataset) which classifies the model. The classification is then reprocessed in order by two similar neural networks (3 categories and 2 categories, instead of 5) that improve the precision on distinguishing among ATTM, FBM and SBM or between ATTM and CTRW, respectively. The combination of these three networks is our predictor for T2. 

T1: To estimate $\alpha$, we use the arithmetic average of the outputs of two different methods based on neural networks. Briefly, in a first method, the result of the classification (T2) is added as an input to the list of features above. These new features become the inputs for a $1 \times 4$ tree of networks (2 hidden layers, 20 neurons, trained with 3e5 trajectory datasets), where the parent network has 4 equally spaced $\alpha$ categories (in the range 0.05 to 2). Each of these categories is then branched into a different network with 5 equally spaced $\alpha$ categories in the corresponding $\alpha$ range. The (overestimated) predicted value of $\alpha$ is the average value in that category. In a second method, the result of the classification is not used as an input but is used to split the data into 5 categories each one analyzed by a different network (architecture and training as above). In particular, the networks for ATTM and CTRW have 5 $\alpha$ categories in the range 0.05 to 1. The network for LW has 5 $\alpha$ categories in the range 1 to 2. Finally, the prediction for FBM and SBM is based on a $1 \times 2$ tree of networks with the parent network having 2 equally spaced categories in the range 0.05 to 2, each then refined by a 5-category network in the corresponding range. The (underestimated) predicted value of $\alpha$ is the average value of the corresponding $\alpha$ range.\\
Tasks: & T1, T2 \\
\hline
\end{tabular}
\end{table}

\begin{table}[!ht]
\linespread{1.0}\selectfont\centering
    \begin{tabular}{p{0.15\linewidth}  p{0.85\linewidth}}
    \hline
     \multicolumn{2}{l}{
Team \team{m}: \emph{UPV-MAT}}     \\ \hline\hline
Contact: & \`{O}scar Garibo i Orts \newline  Instituto Universitario de Matem\'{a}tica Pura y Aplicada, Universitat Polit\`{e}cnica de Val\`{e}ncia  \newline Valencia, Spain \\ 
Reference: & \cite{garibo2021efficientSI} \\
Method: & Recurrent neural networks for trajectory profiling  \\ 
Platform: & Python \\ 
Open-access: & \url{https://github.com/OscarGariboiOrts/ANDI_Challenge} \newline \url{https://github.com/AnDiChallenge/AnDi2020_TeamM_UPV-MAT}\\
Description: & We have defined a recurrent neural network (RNN) architecture based on convolutional layer to feature extraction, bidirectional long short-term memory (LSTM) to learn the characteristics of the trajectory and Dense layers to smooth the signal to the final result.
For T1, we have followed the same approximation, but building up to 12 different models for trajectories of different length. We have built models for trajectories in the length intervals: $[10,20]$, $(20,30]$, $(30,40]$, $(40,50]$, $(50,100]$, $(100,200]$, $(200,300]$, $(300,400]$, $(400,500]$, $(500,600]$, $(600,800]$, and $(800,1000]$, thus checking each trajectory length and applying the proper model.\\
Tasks: & T1, T2 \\
\hline
\end{tabular}
\end{table}

\begin{table}[!ht]
\linespread{1.0}\selectfont\centering
    \begin{tabular}{p{0.15\linewidth}  p{0.85\linewidth}}
    \hline
     \multicolumn{2}{l}{
Team \team{n}: \emph{Wust ML A}}     \\ \hline\hline
Contact: &  Janusz Szwabi\'{n}ski   \newline Faculty of Pure and Applied Mathematics, Hugo Steinhaus Center, Wroc\l{}aw University of Science and Technology, \newline Wroc\l{}aw, Poland \\ 
Reference: & Based on Refs.~\cite{Lines2018SI, LeNguyen2019SI}\\
Method: & RISE for 1D - MrSEQL for 2D and 3D \\ 
Platform: & Python \\ 
Open-access: & \url{https://github.com/szwabin/ANDI-challenge/} \newline \url{https://github.com/AnDiChallenge/AnDi2020_TeamN_WustMLA}\\
Description: & RISE makes use of several series-to-series feature extraction transformers (fitted auto-regressive coefficients, estimated autocorrelation coefficients, power spectrum coefficients), which provide data to build a time series forest classifier. MrSEQL converts the numeric time series vector into strings to create multiple symbolic representations of the time series. The symbolic
representations are then used as input for a sequence learning algorithm, to select the most discriminative subsequence features for training a classifier using logistic regression.\\
Tasks: & T2 \\
\hline
\end{tabular}
\end{table}

\begin{table}[!ht]
\linespread{1.0}\selectfont\centering
    \begin{tabular}{p{0.15\linewidth}  p{0.85\linewidth}}
    \hline
     \multicolumn{2}{l}{
Team \team{o}: \emph{Wust ML B}}     \\ \hline\hline
Contact: &  Hanna Loch-Olszewska \& Patrycja Kowalek  \newline Faculty of Pure and Applied Mathematics, Hugo Steinhaus Center, Wroc\l{}aw University of Science and Technology, \newline Wroc\l{}aw, Poland \\ 
Reference: & Based on Refs.~\cite{kowalek2019classificationSI, janczura2020classificationSI, loch2020impactSI}\\
Method: & Gradient boosting regression and classification\\ 
Platform: & Python \\ 
Open-access: & \url{https://github.com/HannaLochOlszewska/ANDI_challenge} \newline  \url{https://github.com/pkowalek/ANDI-challenge} \newline \url{https://github.com/AnDiChallenge/AnDi2020_TeamO_WustMLB1} \newline \url{https://github.com/AnDiChallenge/AnDi2020_TeamO_WustMLB2}\\
Description: &  Our approach is related to the feature-based methods described in Refs.~\cite{kowalek2019classificationSI, janczura2020classificationSI, loch2020impactSI}, with an extended list of features used for extraction of the trajectories' characteristics.
We used the gradient boosting algorithm in XGBoost (T1) and Gradient Boosting (T2) architectures.
Such procedures allow us to examine trajectories with different lengths by extracting characteristics such as diffusion coefficient, anomalous diffusion exponent, fractal dimension, or gaussianity. The full set of features is listed in the Github repository.
Each task and dimension gets a different set of features, depending on the problem behind the task. 
Both algorithms (Gradient Boosting, XGBoost) belong to the class of ensemble learning, i.e., methods that generate many base classifiers/regressors (decision trees in this case) and aggregate their results. We decided to use these classifiers as the idea behind the classifiers is easy to understand and interpret. 
The training was performed on $70000$ trajectories generated using {\tt andi-datasets} package~\cite{andigithubSI} (for each task and subtask).
Each set was balanced with respect to the anomalous exponent value (T1) or the model (T2). \\
Tasks: & T1.1D, T1.2D, T2 \\
\hline
\end{tabular}
\end{table}

\clearpage
\hypertarget{sec:SI-note3_h}{ }
\section*{Supplementary note 2: Details of experiments}

\begin{table}[!ht]
\linespread{1.0}\selectfont\centering
    \begin{tabular}{p{0.35\linewidth}  p{0.65\linewidth}}
    \hline
     \multicolumn{2}{l}{
Label : \emph{GC}}     \\ \hline\hline
Reference: & \cite{golding2006physicalSI}\\
Tracer: & mRNA molecules\\
Environment: & Cytosol of {\it E. Coli } \\
Dimension: & 2D projection of a 3D movement \\
Experimental details: & The mRNA detection system consists of the bacteriophage MS2 coat protein fused to green fluorescent protein (GFP), and a reporter RNA containing 96 tandemly repeated MS2- binding sites.\\
Number of trajectories: & 54\\
Trajectory length: & 140 to 1628 frames\\
Frame rate: & 1 frame/s \\
Localization precision: & NA \\
\hline
\end{tabular}
\end{table}

\begin{table}[!ht]
\linespread{1.0}\selectfont\centering
    \begin{tabular}{p{0.35\linewidth}  p{0.65\linewidth}}
    \hline
     \multicolumn{2}{l}{
Label : \emph{W$_{\rm A}$}}     \\ \hline\hline
Reference: & \cite{stadler2017nonSI, krapf2019spectralSI}\\
Tracer: & Telomeres\\
Environment: & Nucleus of bone osteosarcoma cells (U2OS, DSMZ-No.ACC785) \\
Dimension: & 2D projection of a 3D movement \\
Experimental details: & GFP-tagged TRF-2 construct that recognizes the human telomeric sequences TTAGGG.\\
Number of trajectories: & 200\\
Trajectory length: & 500 frames\\
Frame rate: & 5 frame/s \\
Localization precision: & 18 nm \\
\hline
\end{tabular}
\end{table}

\begin{table}[!ht]
\linespread{1.0}\selectfont\centering
    \begin{tabular}{p{0.35\linewidth}  p{0.65\linewidth}}
    \hline
     \multicolumn{2}{l}{
Label : \emph{M}}     \\ \hline\hline
Reference: & \cite{manzo2015weakSI}\\
Tracer: & DC-SIGN receptor\\
Environment: & Plasma membrane of Chinese hamster ovary cells \\
Dimension: & 2D \\
Experimental details: & DC-SIGN receptors were labeled through half-antibody fragments conjugated to quantum dots.\\
Number of trajectories: & 109 \\
Trajectory length: & 182 to 2000 frames\\
Frame rate: & 60 frame/s \\
Localization precision: & $\approx20$ nm \\
\hline
\end{tabular}
\end{table}

\begin{table}[!ht]
\linespread{1.0}\selectfont\centering
    \begin{tabular}{p{0.35\linewidth}  p{0.65\linewidth}}
    \hline
     \multicolumn{2}{l}{
Label : \emph{Wi}}     \\ \hline\hline
Reference: & \cite{kindermann2017nonergodicSI}\\
Tracer: & Caesium atoms\\
Environment: &   Optical lattice\\
Dimension: & 1D \\
Experimental details: & The atoms are radially confined by a running wave optical trap. Axially the atoms are trapped within the sites of the lattice formed by two counter-propagating laser beams. During the experimental sequence, only the lattice potential is lowered, while the radial confinement is held constant at all times. This allows one to limit the diffusion of the atoms along the lattice axis, justifying an effective one-dimensional description.\\
Number of trajectories: & 3331 \\
Trajectory length: &  $\approx 10$ frames\\
Frame rate: & 2 frame/s\\
Localization precision: & 2 $\mu$m \\
\hline
\end{tabular}
\end{table}

\clearpage
\makeatletter
%\bibliography{biblio.bib}

\begin{thebibliography}{93}%
\makeatletter
\providecommand \@ifxundefined [1]{%
 \@ifx{#1\undefined}
}%
\providecommand \@ifnum [1]{%
 \ifnum #1\expandafter \@firstoftwo
 \else \expandafter \@secondoftwo
 \fi
}%
\providecommand \@ifx [1]{%
 \ifx #1\expandafter \@firstoftwo
 \else \expandafter \@secondoftwo
 \fi
}%
\providecommand \natexlab [1]{#1}%
\providecommand \enquote  [1]{``#1''}%
\providecommand \bibnamefont  [1]{#1}%
\providecommand \bibfnamefont [1]{#1}%
\providecommand \citenamefont [1]{#1}%
\providecommand \href@noop [0]{\@secondoftwo}%
\providecommand \href [0]{\begingroup \@sanitize@url \@href}%
\providecommand \@href[1]{\@@startlink{#1}\@@href}%
\providecommand \@@href[1]{\endgroup#1\@@endlink}%
\providecommand \@sanitize@url [0]{\catcode `\\12\catcode `\$12\catcode
  `\&12\catcode `\#12\catcode `\^12\catcode `\_12\catcode `\%12\relax}%
\providecommand \@@startlink[1]{}%
\providecommand \@@endlink[0]{}%
\providecommand \url  [0]{\begingroup\@sanitize@url \@url }%
\providecommand \@url [1]{\endgroup\@href {#1}{\urlprefix }}%
\providecommand \urlprefix  [0]{URL }%
\providecommand \Eprint [0]{\href }%
\providecommand \doibase [0]{https://doi.org/}%
\providecommand \selectlanguage [0]{\@gobble}%
\providecommand \bibinfo  [0]{\@secondoftwo}%
\providecommand \bibfield  [0]{\@secondoftwo}%
\providecommand \translation [1]{[#1]}%
\providecommand \BibitemOpen [0]{}%
\providecommand \bibitemStop [0]{}%
\providecommand \bibitemNoStop [0]{.\EOS\space}%
\providecommand \EOS [0]{\spacefactor3000\relax}%
\providecommand \BibitemShut  [1]{\csname bibitem#1\endcsname}%
\let\auto@bib@innerbib\@empty
%</preamble>
\bibitem [{\citenamefont {Pearson}(1905)}]{pearson1905problem}%
  \BibitemOpen
  \bibfield  {author} {\bibinfo {author} {\bibfnamefont {K.}~\bibnamefont
  {Pearson}},\ }\bibfield  {title} {\bibinfo {title} {The problem of the random
  walk},\ }\href@noop {} {\bibfield  {journal} {\bibinfo  {journal} {Nature}\
  }\textbf {\bibinfo {volume} {72}},\ \bibinfo {pages} {342} (\bibinfo {year}
  {1905})}\BibitemShut {NoStop}%
\bibitem [{\citenamefont {Klafter}\ and\ \citenamefont
  {Sokolov}(2011)}]{klafter2011first}%
  \BibitemOpen
  \bibfield  {author} {\bibinfo {author} {\bibfnamefont {J.}~\bibnamefont
  {Klafter}}\ and\ \bibinfo {author} {\bibfnamefont {I.~M.}\ \bibnamefont
  {Sokolov}},\ }\href@noop {} {\emph {\bibinfo {title} {First steps in random
  walks: from tools to applications}}}\ (\bibinfo  {publisher} {Oxford
  University Press},\ \bibinfo {year} {2011})\BibitemShut {NoStop}%
\bibitem [{\citenamefont {Hughes}\ \emph {et~al.}(1995)\citenamefont {Hughes}
  \emph {et~al.}}]{hughes1995random}%
  \BibitemOpen
  \bibfield  {author} {\bibinfo {author} {\bibfnamefont {B.~D.}\ \bibnamefont
  {Hughes}} \emph {et~al.},\ }\href@noop {} {\emph {\bibinfo {title} {Random
  walks and random environments: random walks}}},\ Vol.~\bibinfo {volume} {1}\
  (\bibinfo  {publisher} {Oxford University Press},\ \bibinfo {year}
  {1995})\BibitemShut {NoStop}%
\bibitem [{\citenamefont {Metzler}\ \emph {et~al.}(2014)\citenamefont
  {Metzler}, \citenamefont {Jeon}, \citenamefont {Cherstvy},\ and\
  \citenamefont {Barkai}}]{metzler2014anomalous}%
  \BibitemOpen
  \bibfield  {author} {\bibinfo {author} {\bibfnamefont {R.}~\bibnamefont
  {Metzler}}, \bibinfo {author} {\bibfnamefont {J.-H.}\ \bibnamefont {Jeon}},
  \bibinfo {author} {\bibfnamefont {A.~G.}\ \bibnamefont {Cherstvy}},\ and\
  \bibinfo {author} {\bibfnamefont {E.}~\bibnamefont {Barkai}},\ }\bibfield
  {title} {\bibinfo {title} {Anomalous diffusion models and their properties:
  non-stationarity, non-ergodicity, and ageing at the centenary of single
  particle tracking},\ }\href@noop {} {\bibfield  {journal} {\bibinfo
  {journal} {Physical Chemistry Chemical Physics}\ }\textbf {\bibinfo {volume}
  {16}},\ \bibinfo {pages} {24128} (\bibinfo {year} {2014})}\BibitemShut
  {NoStop}%
\bibitem [{\citenamefont {Krapf}(2015)}]{krapf2015mechanisms}%
  \BibitemOpen
  \bibfield  {author} {\bibinfo {author} {\bibfnamefont {D.}~\bibnamefont
  {Krapf}},\ }\bibfield  {title} {\bibinfo {title} {Mechanisms underlying
  anomalous diffusion in the plasma membrane},\ }\href@noop {} {\bibfield
  {journal} {\bibinfo  {journal} {Current topics in membranes}\ }\textbf
  {\bibinfo {volume} {75}},\ \bibinfo {pages} {167} (\bibinfo {year}
  {2015})}\BibitemShut {NoStop}%
\bibitem [{\citenamefont {Sabri}\ \emph {et~al.}(2020)\citenamefont {Sabri},
  \citenamefont {Xu}, \citenamefont {Krapf},\ and\ \citenamefont
  {Weiss}}]{sabri2020elucidating}%
  \BibitemOpen
  \bibfield  {author} {\bibinfo {author} {\bibfnamefont {A.}~\bibnamefont
  {Sabri}}, \bibinfo {author} {\bibfnamefont {X.}~\bibnamefont {Xu}}, \bibinfo
  {author} {\bibfnamefont {D.}~\bibnamefont {Krapf}},\ and\ \bibinfo {author}
  {\bibfnamefont {M.}~\bibnamefont {Weiss}},\ }\bibfield  {title} {\bibinfo
  {title} {Elucidating the origin of heterogeneous anomalous diffusion in the
  cytoplasm of mammalian cells},\ }\href@noop {} {\bibfield  {journal}
  {\bibinfo  {journal} {Physical Review Letters}\ }\textbf {\bibinfo {volume}
  {125}},\ \bibinfo {pages} {058101} (\bibinfo {year} {2020})}\BibitemShut
  {NoStop}%
\bibitem [{\citenamefont {Di~Pierro}\ \emph {et~al.}(2018)\citenamefont
  {Di~Pierro}, \citenamefont {Potoyan}, \citenamefont {Wolynes},\ and\
  \citenamefont {Onuchic}}]{di2018anomalous}%
  \BibitemOpen
  \bibfield  {author} {\bibinfo {author} {\bibfnamefont {M.}~\bibnamefont
  {Di~Pierro}}, \bibinfo {author} {\bibfnamefont {D.~A.}\ \bibnamefont
  {Potoyan}}, \bibinfo {author} {\bibfnamefont {P.~G.}\ \bibnamefont
  {Wolynes}},\ and\ \bibinfo {author} {\bibfnamefont {J.~N.}\ \bibnamefont
  {Onuchic}},\ }\bibfield  {title} {\bibinfo {title} {Anomalous diffusion,
  spatial coherence, and viscoelasticity from the energy landscape of human
  chromosomes},\ }\href@noop {} {\bibfield  {journal} {\bibinfo  {journal}
  {Proceedings of the National Academy of Sciences}\ }\textbf {\bibinfo
  {volume} {115}},\ \bibinfo {pages} {7753} (\bibinfo {year}
  {2018})}\BibitemShut {NoStop}%
\bibitem [{\citenamefont {Humphries}\ \emph {et~al.}(2012)\citenamefont
  {Humphries}, \citenamefont {Weimerskirch}, \citenamefont {Queiroz},
  \citenamefont {Southall},\ and\ \citenamefont
  {Sims}}]{humphries2012foraging}%
  \BibitemOpen
  \bibfield  {author} {\bibinfo {author} {\bibfnamefont {N.~E.}\ \bibnamefont
  {Humphries}}, \bibinfo {author} {\bibfnamefont {H.}~\bibnamefont
  {Weimerskirch}}, \bibinfo {author} {\bibfnamefont {N.}~\bibnamefont
  {Queiroz}}, \bibinfo {author} {\bibfnamefont {E.~J.}\ \bibnamefont
  {Southall}},\ and\ \bibinfo {author} {\bibfnamefont {D.~W.}\ \bibnamefont
  {Sims}},\ }\bibfield  {title} {\bibinfo {title} {Foraging success of
  biological L\'evy flights recorded in situ},\ }\href@noop {} {\bibfield
  {journal} {\bibinfo  {journal} {Proceedings of the National Academy of
  Sciences}\ }\textbf {\bibinfo {volume} {109}},\ \bibinfo {pages} {7169}
  (\bibinfo {year} {2012})}\BibitemShut {NoStop}%
\bibitem [{\citenamefont {Lo}\ \emph {et~al.}(2002)\citenamefont {Lo},
  \citenamefont {Amaral}, \citenamefont {Havlin}, \citenamefont {Ivanov},
  \citenamefont {Penzel}, \citenamefont {Peter},\ and\ \citenamefont
  {Stanley}}]{lo2002dynamics}%
  \BibitemOpen
  \bibfield  {author} {\bibinfo {author} {\bibfnamefont {C.-C.}\ \bibnamefont
  {Lo}}, \bibinfo {author} {\bibfnamefont {L.~N.}\ \bibnamefont {Amaral}},
  \bibinfo {author} {\bibfnamefont {S.}~\bibnamefont {Havlin}}, \bibinfo
  {author} {\bibfnamefont {P.~C.}\ \bibnamefont {Ivanov}}, \bibinfo {author}
  {\bibfnamefont {T.}~\bibnamefont {Penzel}}, \bibinfo {author} {\bibfnamefont
  {J.-H.}\ \bibnamefont {Peter}},\ and\ \bibinfo {author} {\bibfnamefont
  {H.~E.}\ \bibnamefont {Stanley}},\ }\bibfield  {title} {\bibinfo {title}
  {Dynamics of sleep-wake transitions during sleep},\ }\href@noop {} {\bibfield
  {journal} {\bibinfo  {journal} {EPL (Europhysics Letters)}\ }\textbf
  {\bibinfo {volume} {57}},\ \bibinfo {pages} {625} (\bibinfo {year}
  {2002})}\BibitemShut {NoStop}%
\bibitem [{\citenamefont {Plerou}\ \emph {et~al.}(2000)\citenamefont {Plerou},
  \citenamefont {Gopikrishnan}, \citenamefont {Amaral}, \citenamefont
  {Gabaix},\ and\ \citenamefont {Stanley}}]{plerou2000economic}%
  \BibitemOpen
  \bibfield  {author} {\bibinfo {author} {\bibfnamefont {V.}~\bibnamefont
  {Plerou}}, \bibinfo {author} {\bibfnamefont {P.}~\bibnamefont
  {Gopikrishnan}}, \bibinfo {author} {\bibfnamefont {L.~A.~N.}\ \bibnamefont
  {Amaral}}, \bibinfo {author} {\bibfnamefont {X.}~\bibnamefont {Gabaix}},\
  and\ \bibinfo {author} {\bibfnamefont {H.~E.}\ \bibnamefont {Stanley}},\
  }\bibfield  {title} {\bibinfo {title} {Economic fluctuations and anomalous
  diffusion},\ }\href@noop {} {\bibfield  {journal} {\bibinfo  {journal}
  {Physical Review E}\ }\textbf {\bibinfo {volume} {62}},\ \bibinfo {pages}
  {R3023} (\bibinfo {year} {2000})}\BibitemShut {NoStop}%
\bibitem [{\citenamefont {Scher}\ and\ \citenamefont
  {Montroll}(1975)}]{scher1975anomalous}%
  \BibitemOpen
  \bibfield  {author} {\bibinfo {author} {\bibfnamefont {H.}~\bibnamefont
  {Scher}}\ and\ \bibinfo {author} {\bibfnamefont {E.~W.}\ \bibnamefont
  {Montroll}},\ }\bibfield  {title} {\bibinfo {title} {Anomalous transit-time
  dispersion in amorphous solids},\ }\href@noop {} {\bibfield  {journal}
  {\bibinfo  {journal} {Physical Review B}\ }\textbf {\bibinfo {volume} {12}},\
  \bibinfo {pages} {2455} (\bibinfo {year} {1975})}\BibitemShut {NoStop}%
\bibitem [{\citenamefont {Mandelbrot}\ and\ \citenamefont
  {Van~Ness}(1968)}]{mandelbrot1968fractional}%
  \BibitemOpen
  \bibfield  {author} {\bibinfo {author} {\bibfnamefont {B.~B.}\ \bibnamefont
  {Mandelbrot}}\ and\ \bibinfo {author} {\bibfnamefont {J.~W.}\ \bibnamefont
  {Van~Ness}},\ }\bibfield  {title} {\bibinfo {title} {Fractional {B}rownian
  motions, fractional noises and applications},\ }\href@noop {} {\bibfield
  {journal} {\bibinfo  {journal} {SIAM Review}\ }\textbf {\bibinfo {volume}
  {10}},\ \bibinfo {pages} {422} (\bibinfo {year} {1968})}\BibitemShut
  {NoStop}%
\bibitem [{\citenamefont {Klafter}\ and\ \citenamefont
  {Zumofen}(1994)}]{klafter1994levy}%
  \BibitemOpen
  \bibfield  {author} {\bibinfo {author} {\bibfnamefont {J.}~\bibnamefont
  {Klafter}}\ and\ \bibinfo {author} {\bibfnamefont {G.}~\bibnamefont
  {Zumofen}},\ }\bibfield  {title} {\bibinfo {title} {L\'evy statistics in a
  hamiltonian system},\ }\href@noop {} {\bibfield  {journal} {\bibinfo
  {journal} {Physical Review E}\ }\textbf {\bibinfo {volume} {49}},\ \bibinfo
  {pages} {4873} (\bibinfo {year} {1994})}\BibitemShut {NoStop}%
\bibitem [{\citenamefont {Massignan}\ \emph {et~al.}(2014)\citenamefont
  {Massignan}, \citenamefont {Manzo}, \citenamefont {Torreno-Pina},
  \citenamefont {Garc{\'\i}a-Parajo}, \citenamefont {Lewenstein},\ and\
  \citenamefont {Lapeyre~Jr}}]{massignan2014nonergodic}%
  \BibitemOpen
  \bibfield  {author} {\bibinfo {author} {\bibfnamefont {P.}~\bibnamefont
  {Massignan}}, \bibinfo {author} {\bibfnamefont {C.}~\bibnamefont {Manzo}},
  \bibinfo {author} {\bibfnamefont {J.}~\bibnamefont {Torreno-Pina}}, \bibinfo
  {author} {\bibfnamefont {M.}~\bibnamefont {Garc{\'\i}a-Parajo}}, \bibinfo
  {author} {\bibfnamefont {M.}~\bibnamefont {Lewenstein}},\ and\ \bibinfo
  {author} {\bibfnamefont {G.}~\bibnamefont {Lapeyre~Jr}},\ }\bibfield  {title}
  {\bibinfo {title} {Nonergodic subdiffusion from {B}rownian motion in an
  inhomogeneous medium},\ }\href@noop {} {\bibfield  {journal} {\bibinfo
  {journal} {Physical Review Letters}\ }\textbf {\bibinfo {volume} {112}},\
  \bibinfo {pages} {150603} (\bibinfo {year} {2014})}\BibitemShut {NoStop}%
\bibitem [{\citenamefont {Lim}\ and\ \citenamefont
  {Muniandy}(2002)}]{lim2002self}%
  \BibitemOpen
  \bibfield  {author} {\bibinfo {author} {\bibfnamefont {S.}~\bibnamefont
  {Lim}}\ and\ \bibinfo {author} {\bibfnamefont {S.}~\bibnamefont {Muniandy}},\
  }\bibfield  {title} {\bibinfo {title} {Self-similar {G}aussian processes for
  modeling anomalous diffusion},\ }\href@noop {} {\bibfield  {journal}
  {\bibinfo  {journal} {Physical Review E}\ }\textbf {\bibinfo {volume} {66}},\
  \bibinfo {pages} {021114} (\bibinfo {year} {2002})}\BibitemShut {NoStop}%
\bibitem [{\citenamefont {Kepten}\ \emph {et~al.}(2015)\citenamefont {Kepten},
  \citenamefont {Weron}, \citenamefont {Sikora}, \citenamefont {Burnecki},\
  and\ \citenamefont {Garini}}]{kepten2015guidelines}%
  \BibitemOpen
  \bibfield  {author} {\bibinfo {author} {\bibfnamefont {E.}~\bibnamefont
  {Kepten}}, \bibinfo {author} {\bibfnamefont {A.}~\bibnamefont {Weron}},
  \bibinfo {author} {\bibfnamefont {G.}~\bibnamefont {Sikora}}, \bibinfo
  {author} {\bibfnamefont {K.}~\bibnamefont {Burnecki}},\ and\ \bibinfo
  {author} {\bibfnamefont {Y.}~\bibnamefont {Garini}},\ }\bibfield  {title}
  {\bibinfo {title} {Guidelines for the fitting of anomalous diffusion mean
  square displacement graphs from single particle tracking experiments},\
  }\href@noop {} {\bibfield  {journal} {\bibinfo  {journal} {PLoS One}\
  }\textbf {\bibinfo {volume} {10}},\ \bibinfo {pages} {e0117722} (\bibinfo
  {year} {2015})}\BibitemShut {NoStop}%
\bibitem [{\citenamefont {Chenouard}\ \emph {et~al.}(2014)\citenamefont
  {Chenouard}, \citenamefont {Smal}, \citenamefont {De~Chaumont}, \citenamefont
  {Ma{\v{s}}ka}, \citenamefont {Sbalzarini}, \citenamefont {Gong},
  \citenamefont {Cardinale}, \citenamefont {Carthel}, \citenamefont
  {Coraluppi}, \citenamefont {Winter} \emph {et~al.}}]{chenouard2014objective}%
  \BibitemOpen
  \bibfield  {author} {\bibinfo {author} {\bibfnamefont {N.}~\bibnamefont
  {Chenouard}}, \bibinfo {author} {\bibfnamefont {I.}~\bibnamefont {Smal}},
  \bibinfo {author} {\bibfnamefont {F.}~\bibnamefont {De~Chaumont}}, \bibinfo
  {author} {\bibfnamefont {M.}~\bibnamefont {Ma{\v{s}}ka}}, \bibinfo {author}
  {\bibfnamefont {I.~F.}\ \bibnamefont {Sbalzarini}}, \bibinfo {author}
  {\bibfnamefont {Y.}~\bibnamefont {Gong}}, \bibinfo {author} {\bibfnamefont
  {J.}~\bibnamefont {Cardinale}}, \bibinfo {author} {\bibfnamefont
  {C.}~\bibnamefont {Carthel}}, \bibinfo {author} {\bibfnamefont
  {S.}~\bibnamefont {Coraluppi}}, \bibinfo {author} {\bibfnamefont
  {M.}~\bibnamefont {Winter}}, \emph {et~al.},\ }\bibfield  {title} {\bibinfo
  {title} {Objective comparison of particle tracking methods},\ }\href@noop {}
  {\bibfield  {journal} {\bibinfo  {journal} {Nature methods}\ }\textbf
  {\bibinfo {volume} {11}},\ \bibinfo {pages} {281} (\bibinfo {year}
  {2014})}\BibitemShut {NoStop}%
\bibitem [{\citenamefont {Martin}\ \emph {et~al.}(2002)\citenamefont {Martin},
  \citenamefont {Forstner},\ and\ \citenamefont
  {K{\"a}s}}]{martin2002apparent}%
  \BibitemOpen
  \bibfield  {author} {\bibinfo {author} {\bibfnamefont {D.~S.}\ \bibnamefont
  {Martin}}, \bibinfo {author} {\bibfnamefont {M.~B.}\ \bibnamefont
  {Forstner}},\ and\ \bibinfo {author} {\bibfnamefont {J.~A.}\ \bibnamefont
  {K{\"a}s}},\ }\bibfield  {title} {\bibinfo {title} {Apparent subdiffusion
  inherent to single particle tracking},\ }\href@noop {} {\bibfield  {journal}
  {\bibinfo  {journal} {Biophysical journal}\ }\textbf {\bibinfo {volume}
  {83}},\ \bibinfo {pages} {2109} (\bibinfo {year} {2002})}\BibitemShut
  {NoStop}%
\bibitem [{\citenamefont {Weigel}\ \emph {et~al.}(2011)\citenamefont {Weigel},
  \citenamefont {Simon}, \citenamefont {Tamkun},\ and\ \citenamefont
  {Krapf}}]{weigel2011ergodic}%
  \BibitemOpen
  \bibfield  {author} {\bibinfo {author} {\bibfnamefont {A.~V.}\ \bibnamefont
  {Weigel}}, \bibinfo {author} {\bibfnamefont {B.}~\bibnamefont {Simon}},
  \bibinfo {author} {\bibfnamefont {M.~M.}\ \bibnamefont {Tamkun}},\ and\
  \bibinfo {author} {\bibfnamefont {D.}~\bibnamefont {Krapf}},\ }\bibfield
  {title} {\bibinfo {title} {Ergodic and nonergodic processes coexist in the
  plasma membrane as observed by single-molecule tracking},\ }\href@noop {}
  {\bibfield  {journal} {\bibinfo  {journal} {Proceedings of the National
  Academy of Sciences}\ }\textbf {\bibinfo {volume} {108}},\ \bibinfo {pages}
  {6438} (\bibinfo {year} {2011})}\BibitemShut {NoStop}%
\bibitem [{\citenamefont {Manzo}\ \emph {et~al.}(2015)\citenamefont {Manzo},
  \citenamefont {Torreno-Pina}, \citenamefont {Massignan}, \citenamefont
  {Lapeyre~Jr}, \citenamefont {Lewenstein},\ and\ \citenamefont
  {Parajo}}]{manzo2015weak}%
  \BibitemOpen
  \bibfield  {author} {\bibinfo {author} {\bibfnamefont {C.}~\bibnamefont
  {Manzo}}, \bibinfo {author} {\bibfnamefont {J.~A.}\ \bibnamefont
  {Torreno-Pina}}, \bibinfo {author} {\bibfnamefont {P.}~\bibnamefont
  {Massignan}}, \bibinfo {author} {\bibfnamefont {G.~J.}\ \bibnamefont
  {Lapeyre~Jr}}, \bibinfo {author} {\bibfnamefont {M.}~\bibnamefont
  {Lewenstein}},\ and\ \bibinfo {author} {\bibfnamefont {M.~F.~G.}\
  \bibnamefont {Parajo}},\ }\bibfield  {title} {\bibinfo {title} {Weak
  ergodicity breaking of receptor motion in living cells stemming from random
  diffusivity},\ }\href@noop {} {\bibfield  {journal} {\bibinfo  {journal}
  {Physical Review X}\ }\textbf {\bibinfo {volume} {5}},\ \bibinfo {pages}
  {011021} (\bibinfo {year} {2015})}\BibitemShut {NoStop}%
\bibitem [{\citenamefont {Magdziarz}\ \emph {et~al.}(2009)\citenamefont
  {Magdziarz}, \citenamefont {Weron}, \citenamefont {Burnecki},\ and\
  \citenamefont {Klafter}}]{magdziarz2009fractional}%
  \BibitemOpen
  \bibfield  {author} {\bibinfo {author} {\bibfnamefont {M.}~\bibnamefont
  {Magdziarz}}, \bibinfo {author} {\bibfnamefont {A.}~\bibnamefont {Weron}},
  \bibinfo {author} {\bibfnamefont {K.}~\bibnamefont {Burnecki}},\ and\
  \bibinfo {author} {\bibfnamefont {J.}~\bibnamefont {Klafter}},\ }\bibfield
  {title} {\bibinfo {title} {Fractional {B}rownian motion versus the
  continuous-time random walk: A simple test for subdiffusive dynamics},\
  }\href@noop {} {\bibfield  {journal} {\bibinfo  {journal} {Physical Review
  Letters}\ }\textbf {\bibinfo {volume} {103}},\ \bibinfo {pages} {180602}
  (\bibinfo {year} {2009})}\BibitemShut {NoStop}%
\bibitem [{\citenamefont {Meroz}\ \emph {et~al.}(2013)\citenamefont {Meroz},
  \citenamefont {Sokolov},\ and\ \citenamefont {Klafter}}]{meroz2013test}%
  \BibitemOpen
  \bibfield  {author} {\bibinfo {author} {\bibfnamefont {Y.}~\bibnamefont
  {Meroz}}, \bibinfo {author} {\bibfnamefont {I.~M.}\ \bibnamefont {Sokolov}},\
  and\ \bibinfo {author} {\bibfnamefont {J.}~\bibnamefont {Klafter}},\
  }\bibfield  {title} {\bibinfo {title} {Test for determining a subdiffusive
  model in ergodic systems from single trajectories},\ }\href@noop {}
  {\bibfield  {journal} {\bibinfo  {journal} {Physical Review Letters}\
  }\textbf {\bibinfo {volume} {110}},\ \bibinfo {pages} {090601} (\bibinfo
  {year} {2013})}\BibitemShut {NoStop}%
\bibitem [{\citenamefont {Chen}\ \emph {et~al.}(2017)\citenamefont {Chen},
  \citenamefont {Bassler}, \citenamefont {McCauley},\ and\ \citenamefont
  {Gunaratne}}]{chen2017anomalous}%
  \BibitemOpen
  \bibfield  {author} {\bibinfo {author} {\bibfnamefont {L.}~\bibnamefont
  {Chen}}, \bibinfo {author} {\bibfnamefont {K.~E.}\ \bibnamefont {Bassler}},
  \bibinfo {author} {\bibfnamefont {J.~L.}\ \bibnamefont {McCauley}},\ and\
  \bibinfo {author} {\bibfnamefont {G.~H.}\ \bibnamefont {Gunaratne}},\
  }\bibfield  {title} {\bibinfo {title} {Anomalous scaling of stochastic
  processes and the moses effect},\ }\href@noop {} {\bibfield  {journal}
  {\bibinfo  {journal} {Physical Review E}\ }\textbf {\bibinfo {volume} {95}},\
  \bibinfo {pages} {042141} (\bibinfo {year} {2017})}\BibitemShut {NoStop}%
\bibitem [{\citenamefont {Schwarzl}\ \emph {et~al.}(2017)\citenamefont
  {Schwarzl}, \citenamefont {Godec},\ and\ \citenamefont
  {Metzler}}]{schwarzl2017quantifying}%
  \BibitemOpen
  \bibfield  {author} {\bibinfo {author} {\bibfnamefont {M.}~\bibnamefont
  {Schwarzl}}, \bibinfo {author} {\bibfnamefont {A.}~\bibnamefont {Godec}},\
  and\ \bibinfo {author} {\bibfnamefont {R.}~\bibnamefont {Metzler}},\
  }\bibfield  {title} {\bibinfo {title} {Quantifying non-ergodicity of
  anomalous diffusion with higher order moments},\ }\href@noop {} {\bibfield
  {journal} {\bibinfo  {journal} {Scientific reports}\ }\textbf {\bibinfo
  {volume} {7}},\ \bibinfo {pages} {1} (\bibinfo {year} {2017})}\BibitemShut
  {NoStop}%
\bibitem [{\citenamefont {Weron}\ \emph {et~al.}(2017)\citenamefont {Weron},
  \citenamefont {Burnecki}, \citenamefont {Akin}, \citenamefont {Sol{\'e}},
  \citenamefont {Balcerek}, \citenamefont {Tamkun},\ and\ \citenamefont
  {Krapf}}]{weron2017ergodicity}%
  \BibitemOpen
  \bibfield  {author} {\bibinfo {author} {\bibfnamefont {A.}~\bibnamefont
  {Weron}}, \bibinfo {author} {\bibfnamefont {K.}~\bibnamefont {Burnecki}},
  \bibinfo {author} {\bibfnamefont {E.~J.}\ \bibnamefont {Akin}}, \bibinfo
  {author} {\bibfnamefont {L.}~\bibnamefont {Sol{\'e}}}, \bibinfo {author}
  {\bibfnamefont {M.}~\bibnamefont {Balcerek}}, \bibinfo {author}
  {\bibfnamefont {M.~M.}\ \bibnamefont {Tamkun}},\ and\ \bibinfo {author}
  {\bibfnamefont {D.}~\bibnamefont {Krapf}},\ }\bibfield  {title} {\bibinfo
  {title} {Ergodicity breaking on the neuronal surface emerges from random
  switching between diffusive states},\ }\href@noop {} {\bibfield  {journal}
  {\bibinfo  {journal} {Scientific reports}\ }\textbf {\bibinfo {volume} {7}},\
  \bibinfo {pages} {1} (\bibinfo {year} {2017})}\BibitemShut {NoStop}%
\bibitem [{\citenamefont {Yamamoto}\ \emph {et~al.}(2021)\citenamefont
  {Yamamoto}, \citenamefont {Akimoto}, \citenamefont {Mitsutake},\ and\
  \citenamefont {Metzler}}]{yamamoto2021universal}%
  \BibitemOpen
  \bibfield  {author} {\bibinfo {author} {\bibfnamefont {E.}~\bibnamefont
  {Yamamoto}}, \bibinfo {author} {\bibfnamefont {T.}~\bibnamefont {Akimoto}},
  \bibinfo {author} {\bibfnamefont {A.}~\bibnamefont {Mitsutake}},\ and\
  \bibinfo {author} {\bibfnamefont {R.}~\bibnamefont {Metzler}},\ }\bibfield
  {title} {\bibinfo {title} {Universal relation between instantaneous
  diffusivity and radius of gyration of proteins in aqueous solution},\
  }\href@noop {} {\bibfield  {journal} {\bibinfo  {journal} {Physical review
  letters}\ }\textbf {\bibinfo {volume} {126}},\ \bibinfo {pages} {128101}
  (\bibinfo {year} {2021})}\BibitemShut {NoStop}%
\bibitem [{\citenamefont {Truong}\ \emph {et~al.}(2020)\citenamefont {Truong},
  \citenamefont {Oudre},\ and\ \citenamefont {Vayatis}}]{truong2020selective}%
  \BibitemOpen
  \bibfield  {author} {\bibinfo {author} {\bibfnamefont {C.}~\bibnamefont
  {Truong}}, \bibinfo {author} {\bibfnamefont {L.}~\bibnamefont {Oudre}},\ and\
  \bibinfo {author} {\bibfnamefont {N.}~\bibnamefont {Vayatis}},\ }\bibfield
  {title} {\bibinfo {title} {Selective review of offline change point detection
  methods},\ }\href@noop {} {\bibfield  {journal} {\bibinfo  {journal} {Signal
  Processing}\ }\textbf {\bibinfo {volume} {167}},\ \bibinfo {pages} {107299}
  (\bibinfo {year} {2020})}\BibitemShut {NoStop}%
\bibitem [{\citenamefont {Yin}\ \emph {et~al.}(2018)\citenamefont {Yin},
  \citenamefont {Song},\ and\ \citenamefont {Yang}}]{yin2018detection}%
  \BibitemOpen
  \bibfield  {author} {\bibinfo {author} {\bibfnamefont {S.}~\bibnamefont
  {Yin}}, \bibinfo {author} {\bibfnamefont {N.}~\bibnamefont {Song}},\ and\
  \bibinfo {author} {\bibfnamefont {H.}~\bibnamefont {Yang}},\ }\bibfield
  {title} {\bibinfo {title} {Detection of velocity and diffusion coefficient
  change points in single-particle trajectories},\ }\href@noop {} {\bibfield
  {journal} {\bibinfo  {journal} {Biophysical journal}\ }\textbf {\bibinfo
  {volume} {115}},\ \bibinfo {pages} {217} (\bibinfo {year}
  {2018})}\BibitemShut {NoStop}%
\bibitem [{\citenamefont {Vega}\ \emph {et~al.}(2018)\citenamefont {Vega},
  \citenamefont {Freeman}, \citenamefont {Grinstein},\ and\ \citenamefont
  {Jaqaman}}]{vega2018multistep}%
  \BibitemOpen
  \bibfield  {author} {\bibinfo {author} {\bibfnamefont {A.~R.}\ \bibnamefont
  {Vega}}, \bibinfo {author} {\bibfnamefont {S.~A.}\ \bibnamefont {Freeman}},
  \bibinfo {author} {\bibfnamefont {S.}~\bibnamefont {Grinstein}},\ and\
  \bibinfo {author} {\bibfnamefont {K.}~\bibnamefont {Jaqaman}},\ }\bibfield
  {title} {\bibinfo {title} {Multistep track segmentation and motion
  classification for transient mobility analysis},\ }\href@noop {} {\bibfield
  {journal} {\bibinfo  {journal} {Biophysical journal}\ }\textbf {\bibinfo
  {volume} {114}},\ \bibinfo {pages} {1018} (\bibinfo {year}
  {2018})}\BibitemShut {NoStop}%
\bibitem [{\citenamefont {Akimoto}\ and\ \citenamefont
  {Yamamoto}(2017)}]{akimoto2017detection}%
  \BibitemOpen
  \bibfield  {author} {\bibinfo {author} {\bibfnamefont {T.}~\bibnamefont
  {Akimoto}}\ and\ \bibinfo {author} {\bibfnamefont {E.}~\bibnamefont
  {Yamamoto}},\ }\bibfield  {title} {\bibinfo {title} {Detection of transition
  times from single-particle-tracking trajectories},\ }\href@noop {} {\bibfield
  {journal} {\bibinfo  {journal} {Physical Review E}\ }\textbf {\bibinfo
  {volume} {96}},\ \bibinfo {pages} {052138} (\bibinfo {year}
  {2017})}\BibitemShut {NoStop}%
\bibitem [{\citenamefont {Arts}\ \emph {et~al.}(2019)\citenamefont {Arts},
  \citenamefont {Smal}, \citenamefont {Paul}, \citenamefont {Wyman},\ and\
  \citenamefont {Meijering}}]{arts2019particle}%
  \BibitemOpen
  \bibfield  {author} {\bibinfo {author} {\bibfnamefont {M.}~\bibnamefont
  {Arts}}, \bibinfo {author} {\bibfnamefont {I.}~\bibnamefont {Smal}}, \bibinfo
  {author} {\bibfnamefont {M.~W.}\ \bibnamefont {Paul}}, \bibinfo {author}
  {\bibfnamefont {C.}~\bibnamefont {Wyman}},\ and\ \bibinfo {author}
  {\bibfnamefont {E.}~\bibnamefont {Meijering}},\ }\bibfield  {title} {\bibinfo
  {title} {Particle mobility analysis using deep learning and the moment
  scaling spectrum},\ }\href@noop {} {\bibfield  {journal} {\bibinfo  {journal}
  {Scientific reports}\ }\textbf {\bibinfo {volume} {9}},\ \bibinfo {pages} {1}
  (\bibinfo {year} {2019})}\BibitemShut {NoStop}%
\bibitem [{\citenamefont {Sikora}\ \emph {et~al.}(2017)\citenamefont {Sikora},
  \citenamefont {Wy{\l}oma{\'n}ska}, \citenamefont {Gajda}, \citenamefont
  {Sol{\'e}}, \citenamefont {Akin}, \citenamefont {Tamkun},\ and\ \citenamefont
  {Krapf}}]{sikora2017elucidating}%
  \BibitemOpen
  \bibfield  {author} {\bibinfo {author} {\bibfnamefont {G.}~\bibnamefont
  {Sikora}}, \bibinfo {author} {\bibfnamefont {A.}~\bibnamefont
  {Wy{\l}oma{\'n}ska}}, \bibinfo {author} {\bibfnamefont {J.}~\bibnamefont
  {Gajda}}, \bibinfo {author} {\bibfnamefont {L.}~\bibnamefont {Sol{\'e}}},
  \bibinfo {author} {\bibfnamefont {E.~J.}\ \bibnamefont {Akin}}, \bibinfo
  {author} {\bibfnamefont {M.~M.}\ \bibnamefont {Tamkun}},\ and\ \bibinfo
  {author} {\bibfnamefont {D.}~\bibnamefont {Krapf}},\ }\bibfield  {title}
  {\bibinfo {title} {Elucidating distinct ion channel populations on the
  surface of hippocampal neurons via single-particle tracking recurrence
  analysis},\ }\href@noop {} {\bibfield  {journal} {\bibinfo  {journal}
  {Physical Review E}\ }\textbf {\bibinfo {volume} {96}},\ \bibinfo {pages}
  {062404} (\bibinfo {year} {2017})}\BibitemShut {NoStop}%
\bibitem [{\citenamefont {Bo}\ \emph {et~al.}(2019)\citenamefont {Bo},
  \citenamefont {Schmidt}, \citenamefont {Eichhorn},\ and\ \citenamefont
  {Volpe}}]{bo2019measurement}%
  \BibitemOpen
  \bibfield  {author} {\bibinfo {author} {\bibfnamefont {S.}~\bibnamefont
  {Bo}}, \bibinfo {author} {\bibfnamefont {F.}~\bibnamefont {Schmidt}},
  \bibinfo {author} {\bibfnamefont {R.}~\bibnamefont {Eichhorn}},\ and\
  \bibinfo {author} {\bibfnamefont {G.}~\bibnamefont {Volpe}},\ }\bibfield
  {title} {\bibinfo {title} {Measurement of anomalous diffusion using recurrent
  neural networks},\ }\href@noop {} {\bibfield  {journal} {\bibinfo  {journal}
  {Physical Review E}\ }\textbf {\bibinfo {volume} {100}},\ \bibinfo {pages}
  {010102} (\bibinfo {year} {2019})}\BibitemShut {NoStop}%
\bibitem [{\citenamefont {Lanoisel{\'e}e}\ and\ \citenamefont
  {Grebenkov}(2017)}]{lanoiselee2017unraveling}%
  \BibitemOpen
  \bibfield  {author} {\bibinfo {author} {\bibfnamefont {Y.}~\bibnamefont
  {Lanoisel{\'e}e}}\ and\ \bibinfo {author} {\bibfnamefont {D.~S.}\
  \bibnamefont {Grebenkov}},\ }\bibfield  {title} {\bibinfo {title} {Unraveling
  intermittent features in single-particle trajectories by a local convex hull
  method},\ }\href@noop {} {\bibfield  {journal} {\bibinfo  {journal} {Physical
  Review E}\ }\textbf {\bibinfo {volume} {96}},\ \bibinfo {pages} {022144}
  (\bibinfo {year} {2017})}\BibitemShut {NoStop}%
\bibitem [{\citenamefont {Manzo}\ and\ \citenamefont
  {Garcia-Parajo}(2015)}]{manzo2015review}%
  \BibitemOpen
  \bibfield  {author} {\bibinfo {author} {\bibfnamefont {C.}~\bibnamefont
  {Manzo}}\ and\ \bibinfo {author} {\bibfnamefont {M.~F.}\ \bibnamefont
  {Garcia-Parajo}},\ }\bibfield  {title} {\bibinfo {title} {A review of
  progress in single particle tracking: from methods to biophysical insights},\
  }\href@noop {} {\bibfield  {journal} {\bibinfo  {journal} {Reports on
  progress in physics}\ }\textbf {\bibinfo {volume} {78}},\ \bibinfo {pages}
  {124601} (\bibinfo {year} {2015})}\BibitemShut {NoStop}%
\bibitem [{\citenamefont {Thapa}\ \emph {et~al.}(2018)\citenamefont {Thapa},
  \citenamefont {Lomholt}, \citenamefont {Krog}, \citenamefont {Cherstvy},\
  and\ \citenamefont {Metzler}}]{thapa2018bayesian}%
  \BibitemOpen
  \bibfield  {author} {\bibinfo {author} {\bibfnamefont {S.}~\bibnamefont
  {Thapa}}, \bibinfo {author} {\bibfnamefont {M.~A.}\ \bibnamefont {Lomholt}},
  \bibinfo {author} {\bibfnamefont {J.}~\bibnamefont {Krog}}, \bibinfo {author}
  {\bibfnamefont {A.~G.}\ \bibnamefont {Cherstvy}},\ and\ \bibinfo {author}
  {\bibfnamefont {R.}~\bibnamefont {Metzler}},\ }\bibfield  {title} {\bibinfo
  {title} {Bayesian analysis of single-particle tracking data using the
  nested-sampling algorithm: maximum-likelihood model selection applied to
  stochastic-diffusivity data},\ }\href@noop {} {\bibfield  {journal} {\bibinfo
  {journal} {Physical Chemistry Chemical Physics}\ }\textbf {\bibinfo {volume}
  {20}},\ \bibinfo {pages} {29018} (\bibinfo {year} {2018})}\BibitemShut
  {NoStop}%
\bibitem [{\citenamefont {Burnecki}\ \emph {et~al.}(2015)\citenamefont
  {Burnecki}, \citenamefont {Kepten}, \citenamefont {Garini}, \citenamefont
  {Sikora},\ and\ \citenamefont {Weron}}]{burnecki2015estimating}%
  \BibitemOpen
  \bibfield  {author} {\bibinfo {author} {\bibfnamefont {K.}~\bibnamefont
  {Burnecki}}, \bibinfo {author} {\bibfnamefont {E.}~\bibnamefont {Kepten}},
  \bibinfo {author} {\bibfnamefont {Y.}~\bibnamefont {Garini}}, \bibinfo
  {author} {\bibfnamefont {G.}~\bibnamefont {Sikora}},\ and\ \bibinfo {author}
  {\bibfnamefont {A.}~\bibnamefont {Weron}},\ }\bibfield  {title} {\bibinfo
  {title} {Estimating the anomalous diffusion exponent for single particle
  tracking data with measurement errors-an alternative approach},\ }\href@noop
  {} {\bibfield  {journal} {\bibinfo  {journal} {Scientific Reports}\ }\textbf
  {\bibinfo {volume} {5}},\ \bibinfo {pages} {1} (\bibinfo {year}
  {2015})}\BibitemShut {NoStop}%
\bibitem [{\citenamefont {Krapf}\ \emph {et~al.}(2019)\citenamefont {Krapf},
  \citenamefont {Lukat}, \citenamefont {Marinari}, \citenamefont {Metzler},
  \citenamefont {Oshanin}, \citenamefont {Selhuber-Unkel}, \citenamefont
  {Squarcini}, \citenamefont {Stadler}, \citenamefont {Weiss},\ and\
  \citenamefont {Xu}}]{krapf2019spectral}%
  \BibitemOpen
  \bibfield  {author} {\bibinfo {author} {\bibfnamefont {D.}~\bibnamefont
  {Krapf}}, \bibinfo {author} {\bibfnamefont {N.}~\bibnamefont {Lukat}},
  \bibinfo {author} {\bibfnamefont {E.}~\bibnamefont {Marinari}}, \bibinfo
  {author} {\bibfnamefont {R.}~\bibnamefont {Metzler}}, \bibinfo {author}
  {\bibfnamefont {G.}~\bibnamefont {Oshanin}}, \bibinfo {author} {\bibfnamefont
  {C.}~\bibnamefont {Selhuber-Unkel}}, \bibinfo {author} {\bibfnamefont
  {A.}~\bibnamefont {Squarcini}}, \bibinfo {author} {\bibfnamefont
  {L.}~\bibnamefont {Stadler}}, \bibinfo {author} {\bibfnamefont
  {M.}~\bibnamefont {Weiss}},\ and\ \bibinfo {author} {\bibfnamefont
  {X.}~\bibnamefont {Xu}},\ }\bibfield  {title} {\bibinfo {title} {Spectral
  content of a single non-{B}rownian trajectory},\ }\href@noop {} {\bibfield
  {journal} {\bibinfo  {journal} {Physical Review X}\ }\textbf {\bibinfo
  {volume} {9}},\ \bibinfo {pages} {011019} (\bibinfo {year}
  {2019})}\BibitemShut {NoStop}%
\bibitem [{\citenamefont {Thapa}\ \emph {et~al.}(2020)\citenamefont {Thapa},
  \citenamefont {Wylomanska}, \citenamefont {Sikora}, \citenamefont {Wagner},
  \citenamefont {Krapf}, \citenamefont {Kantz}, \citenamefont {Chechkin},\ and\
  \citenamefont {Metzler}}]{thapa2020leveraging}%
  \BibitemOpen
  \bibfield  {author} {\bibinfo {author} {\bibfnamefont {S.}~\bibnamefont
  {Thapa}}, \bibinfo {author} {\bibfnamefont {A.}~\bibnamefont {Wylomanska}},
  \bibinfo {author} {\bibfnamefont {G.}~\bibnamefont {Sikora}}, \bibinfo
  {author} {\bibfnamefont {C.}~\bibnamefont {Wagner}}, \bibinfo {author}
  {\bibfnamefont {D.}~\bibnamefont {Krapf}}, \bibinfo {author} {\bibfnamefont
  {H.}~\bibnamefont {Kantz}}, \bibinfo {author} {\bibfnamefont
  {A.}~\bibnamefont {Chechkin}},\ and\ \bibinfo {author} {\bibfnamefont
  {R.}~\bibnamefont {Metzler}},\ }\bibfield  {title} {\bibinfo {title}
  {Leveraging large-deviationstatistics to decipher the stochastic properties
  of measured trajectories},\ }\href@noop {} {\bibfield  {journal} {\bibinfo
  {journal} {New Journal of Physics}\ } (\bibinfo {year} {2020})}\BibitemShut
  {NoStop}%
\bibitem [{\citenamefont {Cichos}\ \emph {et~al.}(2020)\citenamefont {Cichos},
  \citenamefont {Gustavsson}, \citenamefont {Mehlig},\ and\ \citenamefont
  {Volpe}}]{cichos2020machine}%
  \BibitemOpen
  \bibfield  {author} {\bibinfo {author} {\bibfnamefont {F.}~\bibnamefont
  {Cichos}}, \bibinfo {author} {\bibfnamefont {K.}~\bibnamefont {Gustavsson}},
  \bibinfo {author} {\bibfnamefont {B.}~\bibnamefont {Mehlig}},\ and\ \bibinfo
  {author} {\bibfnamefont {G.}~\bibnamefont {Volpe}},\ }\bibfield  {title}
  {\bibinfo {title} {Machine learning for active matter},\ }\href@noop {}
  {\bibfield  {journal} {\bibinfo  {journal} {Nature Machine Intelligence}\
  }\textbf {\bibinfo {volume} {2}},\ \bibinfo {pages} {94} (\bibinfo {year}
  {2020})}\BibitemShut {NoStop}%
\bibitem [{\citenamefont {Mu{\~n}oz-Gil}\ \emph
  {et~al.}(2020{\natexlab{a}})\citenamefont {Mu{\~n}oz-Gil}, \citenamefont
  {Garcia-March}, \citenamefont {Manzo}, \citenamefont {Mart{\'\i}n-Guerrero},\
  and\ \citenamefont {Lewenstein}}]{munoz2020single}%
  \BibitemOpen
  \bibfield  {author} {\bibinfo {author} {\bibfnamefont {G.}~\bibnamefont
  {Mu{\~n}oz-Gil}}, \bibinfo {author} {\bibfnamefont {M.~A.}\ \bibnamefont
  {Garcia-March}}, \bibinfo {author} {\bibfnamefont {C.}~\bibnamefont {Manzo}},
  \bibinfo {author} {\bibfnamefont {J.~D.}\ \bibnamefont
  {Mart{\'\i}n-Guerrero}},\ and\ \bibinfo {author} {\bibfnamefont
  {M.}~\bibnamefont {Lewenstein}},\ }\bibfield  {title} {\bibinfo {title}
  {Single trajectory characterization via machine learning},\ }\href@noop {}
  {\bibfield  {journal} {\bibinfo  {journal} {New Journal of Physics}\ }\textbf
  {\bibinfo {volume} {22}},\ \bibinfo {pages} {013010} (\bibinfo {year}
  {2020}{\natexlab{a}})}\BibitemShut {NoStop}%
\bibitem [{\citenamefont {Granik}\ \emph {et~al.}(2019)\citenamefont {Granik},
  \citenamefont {Weiss}, \citenamefont {Nehme}, \citenamefont {Levin},
  \citenamefont {Chein}, \citenamefont {Perlson}, \citenamefont {Roichman},\
  and\ \citenamefont {Shechtman}}]{granik2019single}%
  \BibitemOpen
  \bibfield  {author} {\bibinfo {author} {\bibfnamefont {N.}~\bibnamefont
  {Granik}}, \bibinfo {author} {\bibfnamefont {L.~E.}\ \bibnamefont {Weiss}},
  \bibinfo {author} {\bibfnamefont {E.}~\bibnamefont {Nehme}}, \bibinfo
  {author} {\bibfnamefont {M.}~\bibnamefont {Levin}}, \bibinfo {author}
  {\bibfnamefont {M.}~\bibnamefont {Chein}}, \bibinfo {author} {\bibfnamefont
  {E.}~\bibnamefont {Perlson}}, \bibinfo {author} {\bibfnamefont
  {Y.}~\bibnamefont {Roichman}},\ and\ \bibinfo {author} {\bibfnamefont
  {Y.}~\bibnamefont {Shechtman}},\ }\bibfield  {title} {\bibinfo {title}
  {Single-particle diffusion characterization by deep learning},\ }\href@noop
  {} {\bibfield  {journal} {\bibinfo  {journal} {Biophysical Journal}\ }\textbf
  {\bibinfo {volume} {117}},\ \bibinfo {pages} {185} (\bibinfo {year}
  {2019})}\BibitemShut {NoStop}%
\bibitem [{\citenamefont {Kowalek}\ \emph {et~al.}(2019)\citenamefont
  {Kowalek}, \citenamefont {Loch-Olszewska},\ and\ \citenamefont
  {Szwabi{\'n}ski}}]{kowalek2019classification}%
  \BibitemOpen
  \bibfield  {author} {\bibinfo {author} {\bibfnamefont {P.}~\bibnamefont
  {Kowalek}}, \bibinfo {author} {\bibfnamefont {H.}~\bibnamefont
  {Loch-Olszewska}},\ and\ \bibinfo {author} {\bibfnamefont {J.}~\bibnamefont
  {Szwabi{\'n}ski}},\ }\bibfield  {title} {\bibinfo {title} {Classification of
  diffusion modes in single-particle tracking data: Feature-based versus
  deep-learning approach},\ }\href@noop {} {\bibfield  {journal} {\bibinfo
  {journal} {Physical Review E}\ }\textbf {\bibinfo {volume} {100}},\ \bibinfo
  {pages} {032410} (\bibinfo {year} {2019})}\BibitemShut {NoStop}%
\bibitem [{\citenamefont {Jamali}\ \emph {et~al.}(2021)\citenamefont {Jamali},
  \citenamefont {Hargus}, \citenamefont {Ben-Moshe}, \citenamefont {Aghazadeh},
  \citenamefont {Ha}, \citenamefont {Mandadapu},\ and\ \citenamefont
  {Alivisatos}}]{jamali2021anomalous}%
  \BibitemOpen
  \bibfield  {author} {\bibinfo {author} {\bibfnamefont {V.}~\bibnamefont
  {Jamali}}, \bibinfo {author} {\bibfnamefont {C.}~\bibnamefont {Hargus}},
  \bibinfo {author} {\bibfnamefont {A.}~\bibnamefont {Ben-Moshe}}, \bibinfo
  {author} {\bibfnamefont {A.}~\bibnamefont {Aghazadeh}}, \bibinfo {author}
  {\bibfnamefont {H.~D.}\ \bibnamefont {Ha}}, \bibinfo {author} {\bibfnamefont
  {K.~K.}\ \bibnamefont {Mandadapu}},\ and\ \bibinfo {author} {\bibfnamefont
  {A.~P.}\ \bibnamefont {Alivisatos}},\ }\bibfield  {title} {\bibinfo {title}
  {Anomalous nanoparticle surface diffusion in lctem is revealed by deep
  learning-assisted analysis},\ }\bibfield  {journal} {\bibinfo  {journal}
  {Proceedings of the National Academy of Sciences}\ }\textbf {\bibinfo
  {volume} {118}},\ \href {https://doi.org/10.1073/pnas.2017616118}
  {10.1073/pnas.2017616118} (\bibinfo {year} {2021})\BibitemShut {NoStop}%
\bibitem [{\citenamefont {Mu{\~n}oz-Gil}\ \emph
  {et~al.}(2020{\natexlab{b}})\citenamefont {Mu{\~n}oz-Gil}, \citenamefont
  {Romero}, \citenamefont {Mateos}, \citenamefont {de~Llobet~Cucalon},
  \citenamefont {Beato}, \citenamefont {Lewenstein}, \citenamefont
  {Garcia-Parajo},\ and\ \citenamefont {Torreno-Pina}}]{munoz2020phase}%
  \BibitemOpen
  \bibfield  {author} {\bibinfo {author} {\bibfnamefont {G.}~\bibnamefont
  {Mu{\~n}oz-Gil}}, \bibinfo {author} {\bibfnamefont {C.}~\bibnamefont
  {Romero}}, \bibinfo {author} {\bibfnamefont {N.}~\bibnamefont {Mateos}},
  \bibinfo {author} {\bibfnamefont {L.~I.}\ \bibnamefont {de~Llobet~Cucalon}},
  \bibinfo {author} {\bibfnamefont {M.}~\bibnamefont {Beato}}, \bibinfo
  {author} {\bibfnamefont {M.}~\bibnamefont {Lewenstein}}, \bibinfo {author}
  {\bibfnamefont {M.~F.}\ \bibnamefont {Garcia-Parajo}},\ and\ \bibinfo
  {author} {\bibfnamefont {J.~A.}\ \bibnamefont {Torreno-Pina}},\ }\bibfield
  {title} {\bibinfo {title} {Phase separation of tunable biomolecular
  condensates predicted by an interacting particle model},\ }\href@noop {}
  {\bibfield  {journal} {\bibinfo  {journal} {bioRxiv}\ } (\bibinfo {year}
  {2020}{\natexlab{b}})}\BibitemShut {NoStop}%
\bibitem [{\citenamefont {Cherstvy}\ \emph {et~al.}(2019)\citenamefont
  {Cherstvy}, \citenamefont {Thapa}, \citenamefont {Wagner},\ and\
  \citenamefont {Metzler}}]{cherstvy2019non}%
  \BibitemOpen
  \bibfield  {author} {\bibinfo {author} {\bibfnamefont {A.~G.}\ \bibnamefont
  {Cherstvy}}, \bibinfo {author} {\bibfnamefont {S.}~\bibnamefont {Thapa}},
  \bibinfo {author} {\bibfnamefont {C.~E.}\ \bibnamefont {Wagner}},\ and\
  \bibinfo {author} {\bibfnamefont {R.}~\bibnamefont {Metzler}},\ }\bibfield
  {title} {\bibinfo {title} {Non-{G}aussian, non-ergodic, and non-fickian
  diffusion of tracers in mucin hydrogels},\ }\href@noop {} {\bibfield
  {journal} {\bibinfo  {journal} {Soft Matter}\ }\textbf {\bibinfo {volume}
  {15}},\ \bibinfo {pages} {2526} (\bibinfo {year} {2019})}\BibitemShut
  {NoStop}%
\bibitem [{\citenamefont {Golding}\ and\ \citenamefont
  {Cox}(2006)}]{golding2006physical}%
  \BibitemOpen
  \bibfield  {author} {\bibinfo {author} {\bibfnamefont {I.}~\bibnamefont
  {Golding}}\ and\ \bibinfo {author} {\bibfnamefont {E.~C.}\ \bibnamefont
  {Cox}},\ }\bibfield  {title} {\bibinfo {title} {Physical nature of bacterial
  cytoplasm},\ }\href@noop {} {\bibfield  {journal} {\bibinfo  {journal}
  {Physical Review Letters}\ }\textbf {\bibinfo {volume} {96}},\ \bibinfo
  {pages} {098102} (\bibinfo {year} {2006})}\BibitemShut {NoStop}%
\bibitem [{\citenamefont {Stadler}\ and\ \citenamefont
  {Weiss}(2017)}]{stadler2017non}%
  \BibitemOpen
  \bibfield  {author} {\bibinfo {author} {\bibfnamefont {L.}~\bibnamefont
  {Stadler}}\ and\ \bibinfo {author} {\bibfnamefont {M.}~\bibnamefont
  {Weiss}},\ }\bibfield  {title} {\bibinfo {title} {Non-equilibrium forces
  drive the anomalous diffusion of telomeres in the nucleus of mammalian
  cells},\ }\href@noop {} {\bibfield  {journal} {\bibinfo  {journal} {New
  Journal of Physics}\ }\textbf {\bibinfo {volume} {19}},\ \bibinfo {pages}
  {113048} (\bibinfo {year} {2017})}\BibitemShut {NoStop}%
\bibitem [{\citenamefont {Kindermann}\ \emph {et~al.}(2017)\citenamefont
  {Kindermann}, \citenamefont {Dechant}, \citenamefont {Hohmann}, \citenamefont
  {Lausch}, \citenamefont {Mayer}, \citenamefont {Schmidt}, \citenamefont
  {Lutz},\ and\ \citenamefont {Widera}}]{kindermann2017nonergodic}%
  \BibitemOpen
  \bibfield  {author} {\bibinfo {author} {\bibfnamefont {F.}~\bibnamefont
  {Kindermann}}, \bibinfo {author} {\bibfnamefont {A.}~\bibnamefont {Dechant}},
  \bibinfo {author} {\bibfnamefont {M.}~\bibnamefont {Hohmann}}, \bibinfo
  {author} {\bibfnamefont {T.}~\bibnamefont {Lausch}}, \bibinfo {author}
  {\bibfnamefont {D.}~\bibnamefont {Mayer}}, \bibinfo {author} {\bibfnamefont
  {F.}~\bibnamefont {Schmidt}}, \bibinfo {author} {\bibfnamefont
  {E.}~\bibnamefont {Lutz}},\ and\ \bibinfo {author} {\bibfnamefont
  {A.}~\bibnamefont {Widera}},\ }\bibfield  {title} {\bibinfo {title}
  {Nonergodic diffusion of single atoms in a periodic potential},\ }\href@noop
  {} {\bibfield  {journal} {\bibinfo  {journal} {Nature Physics}\ }\textbf
  {\bibinfo {volume} {13}},\ \bibinfo {pages} {137} (\bibinfo {year}
  {2017})}\BibitemShut {NoStop}%
\bibitem [{\citenamefont {Caspi}\ \emph {et~al.}(2000)\citenamefont {Caspi},
  \citenamefont {Granek},\ and\ \citenamefont {Elbaum}}]{caspi2000enhanced}%
  \BibitemOpen
  \bibfield  {author} {\bibinfo {author} {\bibfnamefont {A.}~\bibnamefont
  {Caspi}}, \bibinfo {author} {\bibfnamefont {R.}~\bibnamefont {Granek}},\ and\
  \bibinfo {author} {\bibfnamefont {M.}~\bibnamefont {Elbaum}},\ }\bibfield
  {title} {\bibinfo {title} {Enhanced diffusion in active intracellular
  transport},\ }\href@noop {} {\bibfield  {journal} {\bibinfo  {journal}
  {Physical Review Letters}\ }\textbf {\bibinfo {volume} {85}},\ \bibinfo
  {pages} {5655} (\bibinfo {year} {2000})}\BibitemShut {NoStop}%
\bibitem [{\citenamefont {He}\ \emph {et~al.}(2008)\citenamefont {He},
  \citenamefont {Burov}, \citenamefont {Metzler},\ and\ \citenamefont
  {Barkai}}]{he2008random}%
  \BibitemOpen
  \bibfield  {author} {\bibinfo {author} {\bibfnamefont {Y.}~\bibnamefont
  {He}}, \bibinfo {author} {\bibfnamefont {S.}~\bibnamefont {Burov}}, \bibinfo
  {author} {\bibfnamefont {R.}~\bibnamefont {Metzler}},\ and\ \bibinfo {author}
  {\bibfnamefont {E.}~\bibnamefont {Barkai}},\ }\bibfield  {title} {\bibinfo
  {title} {Random time-scale invariant diffusion and transport coefficients},\
  }\href@noop {} {\bibfield  {journal} {\bibinfo  {journal} {Physical Review
  Letters}\ }\textbf {\bibinfo {volume} {101}},\ \bibinfo {pages} {058101}
  (\bibinfo {year} {2008})}\BibitemShut {NoStop}%
\bibitem [{\citenamefont {Magdziarz}\ and\ \citenamefont
  {Weron}(2011)}]{magdziarz2011anomalous}%
  \BibitemOpen
  \bibfield  {author} {\bibinfo {author} {\bibfnamefont {M.}~\bibnamefont
  {Magdziarz}}\ and\ \bibinfo {author} {\bibfnamefont {A.}~\bibnamefont
  {Weron}},\ }\bibfield  {title} {\bibinfo {title} {Anomalous diffusion:
  testing ergodicity breaking in experimental data},\ }\href@noop {} {\bibfield
  {journal} {\bibinfo  {journal} {Physical Review E}\ }\textbf {\bibinfo
  {volume} {84}},\ \bibinfo {pages} {051138} (\bibinfo {year}
  {2011})}\BibitemShut {NoStop}%
\bibitem [{\citenamefont {Molina-Garc{\'\i}a}\ \emph
  {et~al.}(2016)\citenamefont {Molina-Garc{\'\i}a}, \citenamefont {Pham},
  \citenamefont {Paradisi}, \citenamefont {Manzo},\ and\ \citenamefont
  {Pagnini}}]{molina2016fractional}%
  \BibitemOpen
  \bibfield  {author} {\bibinfo {author} {\bibfnamefont {D.}~\bibnamefont
  {Molina-Garc{\'\i}a}}, \bibinfo {author} {\bibfnamefont {T.~M.}\ \bibnamefont
  {Pham}}, \bibinfo {author} {\bibfnamefont {P.}~\bibnamefont {Paradisi}},
  \bibinfo {author} {\bibfnamefont {C.}~\bibnamefont {Manzo}},\ and\ \bibinfo
  {author} {\bibfnamefont {G.}~\bibnamefont {Pagnini}},\ }\bibfield  {title}
  {\bibinfo {title} {Fractional kinetics emerging from ergodicity breaking in
  random media},\ }\href@noop {} {\bibfield  {journal} {\bibinfo  {journal}
  {Physical Review E}\ }\textbf {\bibinfo {volume} {94}},\ \bibinfo {pages}
  {052147} (\bibinfo {year} {2016})}\BibitemShut {NoStop}%
\bibitem [{\citenamefont {Lanoisel{\'e}e}\ \emph {et~al.}(2018)\citenamefont
  {Lanoisel{\'e}e}, \citenamefont {Moutal},\ and\ \citenamefont
  {Grebenkov}}]{lanoiselee2018diffusion}%
  \BibitemOpen
  \bibfield  {author} {\bibinfo {author} {\bibfnamefont {Y.}~\bibnamefont
  {Lanoisel{\'e}e}}, \bibinfo {author} {\bibfnamefont {N.}~\bibnamefont
  {Moutal}},\ and\ \bibinfo {author} {\bibfnamefont {D.~S.}\ \bibnamefont
  {Grebenkov}},\ }\bibfield  {title} {\bibinfo {title} {Diffusion-limited
  reactions in dynamic heterogeneous media},\ }\href@noop {} {\bibfield
  {journal} {\bibinfo  {journal} {Nature communications}\ }\textbf {\bibinfo
  {volume} {9}},\ \bibinfo {pages} {1} (\bibinfo {year} {2018})}\BibitemShut
  {NoStop}%
\bibitem [{\citenamefont {Dechant}\ \emph {et~al.}(2019)\citenamefont
  {Dechant}, \citenamefont {Kindermann}, \citenamefont {Widera},\ and\
  \citenamefont {Lutz}}]{dechant2019continuous}%
  \BibitemOpen
  \bibfield  {author} {\bibinfo {author} {\bibfnamefont {A.}~\bibnamefont
  {Dechant}}, \bibinfo {author} {\bibfnamefont {F.}~\bibnamefont {Kindermann}},
  \bibinfo {author} {\bibfnamefont {A.}~\bibnamefont {Widera}},\ and\ \bibinfo
  {author} {\bibfnamefont {E.}~\bibnamefont {Lutz}},\ }\bibfield  {title}
  {\bibinfo {title} {Continuous-time random walk for a particle in a periodic
  potential},\ }\href@noop {} {\bibfield  {journal} {\bibinfo  {journal}
  {Physical Review Letters}\ }\textbf {\bibinfo {volume} {123}},\ \bibinfo
  {pages} {070602} (\bibinfo {year} {2019})}\BibitemShut {NoStop}%
\bibitem [{\citenamefont {Manley}\ \emph {et~al.}(2008)\citenamefont {Manley},
  \citenamefont {Gillette}, \citenamefont {Patterson}, \citenamefont {Shroff},
  \citenamefont {Hess}, \citenamefont {Betzig},\ and\ \citenamefont
  {Lippincott-Schwartz}}]{manley2008high}%
  \BibitemOpen
  \bibfield  {author} {\bibinfo {author} {\bibfnamefont {S.}~\bibnamefont
  {Manley}}, \bibinfo {author} {\bibfnamefont {J.~M.}\ \bibnamefont
  {Gillette}}, \bibinfo {author} {\bibfnamefont {G.~H.}\ \bibnamefont
  {Patterson}}, \bibinfo {author} {\bibfnamefont {H.}~\bibnamefont {Shroff}},
  \bibinfo {author} {\bibfnamefont {H.~F.}\ \bibnamefont {Hess}}, \bibinfo
  {author} {\bibfnamefont {E.}~\bibnamefont {Betzig}},\ and\ \bibinfo {author}
  {\bibfnamefont {J.}~\bibnamefont {Lippincott-Schwartz}},\ }\bibfield  {title}
  {\bibinfo {title} {High-density mapping of single-molecule trajectories with
  photoactivated localization microscopy},\ }\href@noop {} {\bibfield
  {journal} {\bibinfo  {journal} {Nature Methods}\ }\textbf {\bibinfo {volume}
  {5}},\ \bibinfo {pages} {155} (\bibinfo {year} {2008})}\BibitemShut {NoStop}%
\bibitem [{\citenamefont {Jeon}\ \emph {et~al.}(2013)\citenamefont {Jeon},
  \citenamefont {Barkai},\ and\ \citenamefont {Metzler}}]{jeon2013noisy}%
  \BibitemOpen
  \bibfield  {author} {\bibinfo {author} {\bibfnamefont {J.-H.}\ \bibnamefont
  {Jeon}}, \bibinfo {author} {\bibfnamefont {E.}~\bibnamefont {Barkai}},\ and\
  \bibinfo {author} {\bibfnamefont {R.}~\bibnamefont {Metzler}},\ }\bibfield
  {title} {\bibinfo {title} {Noisy continuous time random walks},\ }\href@noop
  {} {\bibfield  {journal} {\bibinfo  {journal} {The Journal of Chemical
  Physics}\ }\textbf {\bibinfo {volume} {139}},\ \bibinfo {pages} {09B616\_1}
  (\bibinfo {year} {2013})}\BibitemShut {NoStop}%
\bibitem [{\citenamefont {Cherstvy}\ \emph {et~al.}(2014)\citenamefont
  {Cherstvy}, \citenamefont {Chechkin},\ and\ \citenamefont
  {Metzler}}]{cherstvy2014ageing}%
  \BibitemOpen
  \bibfield  {author} {\bibinfo {author} {\bibfnamefont {A.~G.}\ \bibnamefont
  {Cherstvy}}, \bibinfo {author} {\bibfnamefont {A.~V.}\ \bibnamefont
  {Chechkin}},\ and\ \bibinfo {author} {\bibfnamefont {R.}~\bibnamefont
  {Metzler}},\ }\bibfield  {title} {\bibinfo {title} {Ageing and confinement in
  non-ergodic heterogeneous diffusion processes},\ }\href@noop {} {\bibfield
  {journal} {\bibinfo  {journal} {Journal of Physics A: Mathematical and
  Theoretical}\ }\textbf {\bibinfo {volume} {47}},\ \bibinfo {pages} {485002}
  (\bibinfo {year} {2014})}\BibitemShut {NoStop}%
\bibitem [{\citenamefont {Mu{\~n}oz-Gil}\ \emph
  {et~al.}(2020{\natexlab{c}})\citenamefont {Mu{\~n}oz-Gil}, \citenamefont
  {Requena}, \citenamefont {Volpe}, \citenamefont {Garcia-March},\ and\
  \citenamefont {Manzo}}]{andigithub}%
  \BibitemOpen
  \bibfield  {author} {\bibinfo {author} {\bibfnamefont {G.}~\bibnamefont
  {Mu{\~n}oz-Gil}}, \bibinfo {author} {\bibfnamefont {B.}~\bibnamefont
  {Requena}}, \bibinfo {author} {\bibfnamefont {G.}~\bibnamefont {Volpe}},
  \bibinfo {author} {\bibfnamefont {M.~A.}\ \bibnamefont {Garcia-March}},\ and\
  \bibinfo {author} {\bibfnamefont {C.}~\bibnamefont {Manzo}},\ }\href
  {https://doi.org/10.5281/zenodo.4775311} {\bibinfo {title}
  {An{D}i{C}hallenge/{ANDI}\_datasets: {C}hallenge 2020 release}} (\bibinfo
  {year} {2020}{\natexlab{c}})\BibitemShut {NoStop}%
\bibitem [{\citenamefont {Bouchaud}(1992)}]{bouchaud1992weak}%
  \BibitemOpen
  \bibfield  {author} {\bibinfo {author} {\bibfnamefont {J.-P.}\ \bibnamefont
  {Bouchaud}},\ }\bibfield  {title} {\bibinfo {title} {Weak ergodicity breaking
  and aging in disordered systems},\ }\href@noop {} {\bibfield  {journal}
  {\bibinfo  {journal} {Journal de Physique I}\ }\textbf {\bibinfo {volume}
  {2}},\ \bibinfo {pages} {1705} (\bibinfo {year} {1992})}\BibitemShut
  {NoStop}%
\bibitem [{\citenamefont {Barkai}\ \emph {et~al.}(2012)\citenamefont {Barkai},
  \citenamefont {Garini},\ and\ \citenamefont {Metzler}}]{barkai2012single}%
  \BibitemOpen
  \bibfield  {author} {\bibinfo {author} {\bibfnamefont {E.}~\bibnamefont
  {Barkai}}, \bibinfo {author} {\bibfnamefont {Y.}~\bibnamefont {Garini}},\
  and\ \bibinfo {author} {\bibfnamefont {R.}~\bibnamefont {Metzler}},\
  }\bibfield  {title} {\bibinfo {title} {Strange kinetics of single molecules
  in living cells},\ }\href@noop {} {\bibfield  {journal} {\bibinfo  {journal}
  {Phys. Today}\ }\textbf {\bibinfo {volume} {65}},\ \bibinfo {pages} {29}
  (\bibinfo {year} {2012})}\BibitemShut {NoStop}%
\bibitem [{\citenamefont {Bel}\ and\ \citenamefont {Barkai}(2005)}]{2005Bel}%
  \BibitemOpen
  \bibfield  {author} {\bibinfo {author} {\bibfnamefont {G.}~\bibnamefont
  {Bel}}\ and\ \bibinfo {author} {\bibfnamefont {E.}~\bibnamefont {Barkai}},\
  }\bibfield  {title} {\bibinfo {title} {Weak ergodicity breaking in the
  continuous-time random walk},\ }\href
  {https://doi.org/10.1103/PhysRevLett.94.240602} {\bibfield  {journal}
  {\bibinfo  {journal} {Phys. Rev. Lett.}\ }\textbf {\bibinfo {volume} {94}},\
  \bibinfo {pages} {240602} (\bibinfo {year} {2005})}\BibitemShut {NoStop}%
\bibitem [{\citenamefont {Rebenshtok}\ and\ \citenamefont
  {Barkai}(2007)}]{2007Rebenshtok}%
  \BibitemOpen
  \bibfield  {author} {\bibinfo {author} {\bibfnamefont {A.}~\bibnamefont
  {Rebenshtok}}\ and\ \bibinfo {author} {\bibfnamefont {E.}~\bibnamefont
  {Barkai}},\ }\bibfield  {title} {\bibinfo {title} {Distribution of
  time-averaged observables for weak ergodicity breaking},\ }\href
  {https://doi.org/10.1103/PhysRevLett.99.210601} {\bibfield  {journal}
  {\bibinfo  {journal} {Phys. Rev. Lett.}\ }\textbf {\bibinfo {volume} {99}},\
  \bibinfo {pages} {210601} (\bibinfo {year} {2007})}\BibitemShut {NoStop}%
\bibitem [{\citenamefont {Deng}\ and\ \citenamefont
  {Barkai}(2009)}]{2009DengErgodic}%
  \BibitemOpen
  \bibfield  {author} {\bibinfo {author} {\bibfnamefont {W.}~\bibnamefont
  {Deng}}\ and\ \bibinfo {author} {\bibfnamefont {E.}~\bibnamefont {Barkai}},\
  }\bibfield  {title} {\bibinfo {title} {Ergodic properties of fractional
  {B}rownian-{L}angevin motion},\ }\href
  {https://doi.org/10.1103/PhysRevE.79.011112} {\bibfield  {journal} {\bibinfo
  {journal} {Phys. Rev. E}\ }\textbf {\bibinfo {volume} {79}},\ \bibinfo
  {pages} {011112} (\bibinfo {year} {2009})}\BibitemShut {NoStop}%
\bibitem [{\citenamefont {Godec}\ and\ \citenamefont
  {Metzler}(2013{\natexlab{a}})}]{2013Godec}%
  \BibitemOpen
  \bibfield  {author} {\bibinfo {author} {\bibfnamefont {A.}~\bibnamefont
  {Godec}}\ and\ \bibinfo {author} {\bibfnamefont {R.}~\bibnamefont
  {Metzler}},\ }\bibfield  {title} {\bibinfo {title} {Finite-time effects and
  ultraweak ergodicity breaking in superdiffusive dynamics},\ }\href
  {https://doi.org/10.1103/PhysRevLett.110.020603} {\bibfield  {journal}
  {\bibinfo  {journal} {Phys. Rev. Lett.}\ }\textbf {\bibinfo {volume} {110}},\
  \bibinfo {pages} {020603} (\bibinfo {year} {2013}{\natexlab{a}})}\BibitemShut
  {NoStop}%
\bibitem [{\citenamefont {Godec}\ and\ \citenamefont
  {Metzler}(2013{\natexlab{b}})}]{2013Godecb}%
  \BibitemOpen
  \bibfield  {author} {\bibinfo {author} {\bibfnamefont {A.}~\bibnamefont
  {Godec}}\ and\ \bibinfo {author} {\bibfnamefont {R.}~\bibnamefont
  {Metzler}},\ }\bibfield  {title} {\bibinfo {title} {Linear response,
  fluctuation-dissipation, and finite-system-size effects in superdiffusion},\
  }\href {https://doi.org/10.1103/PhysRevE.88.012116} {\bibfield  {journal}
  {\bibinfo  {journal} {Phys. Rev. E}\ }\textbf {\bibinfo {volume} {88}},\
  \bibinfo {pages} {012116} (\bibinfo {year} {2013}{\natexlab{b}})}\BibitemShut
  {NoStop}%
\bibitem [{\citenamefont {Jeon}\ and\ \citenamefont
  {Metzler}(2010)}]{jeon2010fractional}%
  \BibitemOpen
  \bibfield  {author} {\bibinfo {author} {\bibfnamefont {J.-H.}\ \bibnamefont
  {Jeon}}\ and\ \bibinfo {author} {\bibfnamefont {R.}~\bibnamefont {Metzler}},\
  }\bibfield  {title} {\bibinfo {title} {Fractional {B}rownian motion and
  motion governed by the fractional {L}angevin equation in confined
  geometries},\ }\href {https://doi.org/10.1103/PhysRevE.81.021103} {\bibfield
  {journal} {\bibinfo  {journal} {Phys. Rev. E}\ }\textbf {\bibinfo {volume}
  {81}},\ \bibinfo {pages} {021103} (\bibinfo {year} {2010})}\BibitemShut
  {NoStop}%
\bibitem [{\citenamefont {Davies}\ and\ \citenamefont
  {Harte}(1987)}]{davies1987}%
  \BibitemOpen
  \bibfield  {author} {\bibinfo {author} {\bibfnamefont {R.~B.}\ \bibnamefont
  {Davies}}\ and\ \bibinfo {author} {\bibfnamefont {D.}~\bibnamefont {Harte}},\
  }\bibfield  {title} {\bibinfo {title} {Tests for hurst effect},\ }\href@noop
  {} {\bibfield  {journal} {\bibinfo  {journal} {Biometrika}\ }\textbf
  {\bibinfo {volume} {74}},\ \bibinfo {pages} {95} (\bibinfo {year}
  {1987})}\BibitemShut {NoStop}%
\bibitem [{\citenamefont {Hosking}(1984)}]{hosking1984modeling}%
  \BibitemOpen
  \bibfield  {author} {\bibinfo {author} {\bibfnamefont {J.~R.}\ \bibnamefont
  {Hosking}},\ }\bibfield  {title} {\bibinfo {title} {Modeling persistence in
  hydrological time series using fractional differencing},\ }\href@noop {}
  {\bibfield  {journal} {\bibinfo  {journal} {Water resources research}\
  }\textbf {\bibinfo {volume} {20}},\ \bibinfo {pages} {1898} (\bibinfo {year}
  {1984})}\BibitemShut {NoStop}%
\bibitem [{\citenamefont {Michalet}(2010)}]{michalet2010mean}%
  \BibitemOpen
  \bibfield  {author} {\bibinfo {author} {\bibfnamefont {X.}~\bibnamefont
  {Michalet}},\ }\bibfield  {title} {\bibinfo {title} {Mean square displacement
  analysis of single-particle trajectories with localization error: {B}rownian
  motion in an isotropic medium},\ }\href@noop {} {\bibfield  {journal}
  {\bibinfo  {journal} {Physical Review E}\ }\textbf {\bibinfo {volume} {82}},\
  \bibinfo {pages} {041914} (\bibinfo {year} {2010})}\BibitemShut {NoStop}%
\bibitem [{\citenamefont {Ferrari}\ \emph {et~al.}(2001)\citenamefont
  {Ferrari}, \citenamefont {Manfroi},\ and\ \citenamefont
  {Young}}]{ferrari2001strongly}%
  \BibitemOpen
  \bibfield  {author} {\bibinfo {author} {\bibfnamefont {R.}~\bibnamefont
  {Ferrari}}, \bibinfo {author} {\bibfnamefont {A.}~\bibnamefont {Manfroi}},\
  and\ \bibinfo {author} {\bibfnamefont {W.}~\bibnamefont {Young}},\ }\bibfield
  {title} {\bibinfo {title} {Strongly and weakly self-similar diffusion},\
  }\href@noop {} {\bibfield  {journal} {\bibinfo  {journal} {Physica D:
  Nonlinear Phenomena}\ }\textbf {\bibinfo {volume} {154}},\ \bibinfo {pages}
  {111} (\bibinfo {year} {2001})}\BibitemShut {NoStop}%
\bibitem [{\citenamefont {Sbalzarini}\ and\ \citenamefont
  {Koumoutsakos}(2005)}]{sbalzarini2005feature}%
  \BibitemOpen
  \bibfield  {author} {\bibinfo {author} {\bibfnamefont {I.~F.}\ \bibnamefont
  {Sbalzarini}}\ and\ \bibinfo {author} {\bibfnamefont {P.}~\bibnamefont
  {Koumoutsakos}},\ }\bibfield  {title} {\bibinfo {title} {Feature point
  tracking and trajectory analysis for video imaging in cell biology},\
  }\href@noop {} {\bibfield  {journal} {\bibinfo  {journal} {Journal of
  structural biology}\ }\textbf {\bibinfo {volume} {151}},\ \bibinfo {pages}
  {182} (\bibinfo {year} {2005})}\BibitemShut {NoStop}%
\bibitem [{\citenamefont {Ariel}\ \emph {et~al.}(2015)\citenamefont {Ariel},
  \citenamefont {Rabani}, \citenamefont {Benisty}, \citenamefont {Partridge},
  \citenamefont {Harshey},\ and\ \citenamefont {Be'Er}}]{ariel2015swarming}%
  \BibitemOpen
  \bibfield  {author} {\bibinfo {author} {\bibfnamefont {G.}~\bibnamefont
  {Ariel}}, \bibinfo {author} {\bibfnamefont {A.}~\bibnamefont {Rabani}},
  \bibinfo {author} {\bibfnamefont {S.}~\bibnamefont {Benisty}}, \bibinfo
  {author} {\bibfnamefont {J.~D.}\ \bibnamefont {Partridge}}, \bibinfo {author}
  {\bibfnamefont {R.~M.}\ \bibnamefont {Harshey}},\ and\ \bibinfo {author}
  {\bibfnamefont {A.}~\bibnamefont {Be'Er}},\ }\bibfield  {title} {\bibinfo
  {title} {Swarming bacteria migrate by {L}\'{e}vy walk},\ }\href@noop {}
  {\bibfield  {journal} {\bibinfo  {journal} {Nature {C}ommunications}\
  }\textbf {\bibinfo {volume} {6}},\ \bibinfo {pages} {1} (\bibinfo {year}
  {2015})}\BibitemShut {NoStop}%
\bibitem [{\citenamefont {{\'S}l{\k{e}}zak}\ \emph {et~al.}(2019)\citenamefont
  {{\'S}l{\k{e}}zak}, \citenamefont {Metzler},\ and\ \citenamefont
  {Magdziarz}}]{slkezak2019codifference}%
  \BibitemOpen
  \bibfield  {author} {\bibinfo {author} {\bibfnamefont {J.}~\bibnamefont
  {{\'S}l{\k{e}}zak}}, \bibinfo {author} {\bibfnamefont {R.}~\bibnamefont
  {Metzler}},\ and\ \bibinfo {author} {\bibfnamefont {M.}~\bibnamefont
  {Magdziarz}},\ }\bibfield  {title} {\bibinfo {title} {Codifference can detect
  ergodicity breaking and non-{G}aussianity},\ }\href@noop {} {\bibfield
  {journal} {\bibinfo  {journal} {New Journal of Physics}\ }\textbf {\bibinfo
  {volume} {21}},\ \bibinfo {pages} {053008} (\bibinfo {year}
  {2019})}\BibitemShut {NoStop}%
\bibitem [{\citenamefont {Burov}\ \emph {et~al.}(2011)\citenamefont {Burov},
  \citenamefont {Jeon}, \citenamefont {Metzler},\ and\ \citenamefont
  {Barkai}}]{burov2011single}%
  \BibitemOpen
  \bibfield  {author} {\bibinfo {author} {\bibfnamefont {S.}~\bibnamefont
  {Burov}}, \bibinfo {author} {\bibfnamefont {J.-H.}\ \bibnamefont {Jeon}},
  \bibinfo {author} {\bibfnamefont {R.}~\bibnamefont {Metzler}},\ and\ \bibinfo
  {author} {\bibfnamefont {E.}~\bibnamefont {Barkai}},\ }\bibfield  {title}
  {\bibinfo {title} {Single particle tracking in systems showing anomalous
  diffusion: the role of weak ergodicity breaking},\ }\href@noop {} {\bibfield
  {journal} {\bibinfo  {journal} {Physical Chemistry Chemical Physics}\
  }\textbf {\bibinfo {volume} {13}},\ \bibinfo {pages} {1800} (\bibinfo {year}
  {2011})}\BibitemShut {NoStop}%
\bibitem [{\citenamefont {Wolpert}(1992)}]{wolpert1992stacked}%
  \BibitemOpen
  \bibfield  {author} {\bibinfo {author} {\bibfnamefont {D.~H.}\ \bibnamefont
  {Wolpert}},\ }\bibfield  {title} {\bibinfo {title} {Stacked generalization},\
  }\href@noop {} {\bibfield  {journal} {\bibinfo  {journal} {Neural networks}\
  }\textbf {\bibinfo {volume} {5}},\ \bibinfo {pages} {241} (\bibinfo {year}
  {1992})}\BibitemShut {NoStop}%
\bibitem [{\citenamefont {Krog}\ \emph {et~al.}(2018)\citenamefont {Krog},
  \citenamefont {Jacobsen}, \citenamefont {Lund}, \citenamefont {W{\"u}stner},\
  and\ \citenamefont {Lomholt}}]{krog2018bayesian}%
  \BibitemOpen
  \bibfield  {author} {\bibinfo {author} {\bibfnamefont {J.}~\bibnamefont
  {Krog}}, \bibinfo {author} {\bibfnamefont {L.~H.}\ \bibnamefont {Jacobsen}},
  \bibinfo {author} {\bibfnamefont {F.~W.}\ \bibnamefont {Lund}}, \bibinfo
  {author} {\bibfnamefont {D.}~\bibnamefont {W{\"u}stner}},\ and\ \bibinfo
  {author} {\bibfnamefont {M.~A.}\ \bibnamefont {Lomholt}},\ }\bibfield
  {title} {\bibinfo {title} {Bayesian model selection with fractional
  {B}rownian motion},\ }\href@noop {} {\bibfield  {journal} {\bibinfo
  {journal} {Journal of Statistical Mechanics: Theory and Experiment}\ }\textbf
  {\bibinfo {volume} {2018}},\ \bibinfo {pages} {093501} (\bibinfo {year}
  {2018})}\BibitemShut {NoStop}%
\bibitem [{\citenamefont {Park}\ \emph {et~al.}(2021)\citenamefont {Park},
  \citenamefont {Thapa}, \citenamefont {Kim}, \citenamefont {Lomholt},\ and\
  \citenamefont {Jeon}}]{park2021bayesian}%
  \BibitemOpen
  \bibfield  {author} {\bibinfo {author} {\bibfnamefont {S.}~\bibnamefont
  {Park}}, \bibinfo {author} {\bibfnamefont {S.}~\bibnamefont {Thapa}},
  \bibinfo {author} {\bibfnamefont {Y.}~\bibnamefont {Kim}}, \bibinfo {author}
  {\bibfnamefont {M.~A.}\ \bibnamefont {Lomholt}},\ and\ \bibinfo {author}
  {\bibfnamefont {J.-H.}\ \bibnamefont {Jeon}},\ }\bibfield  {title} {\bibinfo
  {title} {Bayesian inference of l$\backslash$'evy walks via hidden markov
  models},\ }\href@noop {} {\bibfield  {journal} {\bibinfo  {journal} {arXiv
  preprint arXiv:2107.05390}\ } (\bibinfo {year} {2021})}\BibitemShut {NoStop}%
\bibitem [{\citenamefont {Verdier}\ \emph {et~al.}(2021)\citenamefont
  {Verdier}, \citenamefont {Duval}, \citenamefont {Laurent}, \citenamefont
  {Cass{\'{e}}}, \citenamefont {Vestergaard},\ and\ \citenamefont
  {Masson}}]{verdier2021learning}%
  \BibitemOpen
  \bibfield  {author} {\bibinfo {author} {\bibfnamefont {H.}~\bibnamefont
  {Verdier}}, \bibinfo {author} {\bibfnamefont {M.}~\bibnamefont {Duval}},
  \bibinfo {author} {\bibfnamefont {F.}~\bibnamefont {Laurent}}, \bibinfo
  {author} {\bibfnamefont {A.}~\bibnamefont {Cass{\'{e}}}}, \bibinfo {author}
  {\bibfnamefont {C.~L.}\ \bibnamefont {Vestergaard}},\ and\ \bibinfo {author}
  {\bibfnamefont {J.-B.}\ \bibnamefont {Masson}},\ }\bibfield  {title}
  {\bibinfo {title} {Learning physical properties of anomalous random walks
  using graph neural networks},\ }\href
  {https://doi.org/10.1088/1751-8121/abfa45} {\bibfield  {journal} {\bibinfo
  {journal} {Journal of Physics A: Mathematical and Theoretical}\ }\textbf
  {\bibinfo {volume} {54}},\ \bibinfo {pages} {234001} (\bibinfo {year}
  {2021})}\BibitemShut {NoStop}%
\bibitem [{\citenamefont {He}\ \emph {et~al.}(2016)\citenamefont {He},
  \citenamefont {Zhang}, \citenamefont {Ren},\ and\ \citenamefont
  {Sun}}]{he2016deep}%
  \BibitemOpen
  \bibfield  {author} {\bibinfo {author} {\bibfnamefont {K.}~\bibnamefont
  {He}}, \bibinfo {author} {\bibfnamefont {X.}~\bibnamefont {Zhang}}, \bibinfo
  {author} {\bibfnamefont {S.}~\bibnamefont {Ren}},\ and\ \bibinfo {author}
  {\bibfnamefont {J.}~\bibnamefont {Sun}},\ }\bibfield  {title} {\bibinfo
  {title} {Deep residual learning for image recognition},\ }in\ \href@noop {}
  {\emph {\bibinfo {booktitle} {Proceedings of the IEEE conference on computer
  vision and pattern recognition}}}\ (\bibinfo {year} {2016})\ pp.\ \bibinfo
  {pages} {770--778}\BibitemShut {NoStop}%
\bibitem [{\citenamefont {Chen}\ \emph {et~al.}(2016)\citenamefont {Chen} \emph
  {et~al.}}]{chen2016guestrin}%
  \BibitemOpen
  \bibfield  {author} {\bibinfo {author} {\bibfnamefont {T.}~\bibnamefont
  {Chen}} \emph {et~al.},\ }\bibfield  {title} {\bibinfo {title} {Guestrin, c.:
  Xgboost: A scalable tree boosting system},\ }in\ \href@noop {} {\emph
  {\bibinfo {booktitle} {Proceedings of the 22nd ACM SIGKDD International
  Conference on Knowledge Discovery and Data Mining (KDD’16)}}}\ (\bibinfo
  {year} {2016})\ pp.\ \bibinfo {pages} {785--794}\BibitemShut {NoStop}%
\bibitem [{\citenamefont {Argun}\ \emph {et~al.}(2021)\citenamefont {Argun},
  \citenamefont {Volpe},\ and\ \citenamefont {Bo}}]{argun2021classification}%
  \BibitemOpen
  \bibfield  {author} {\bibinfo {author} {\bibfnamefont {A.}~\bibnamefont
  {Argun}}, \bibinfo {author} {\bibfnamefont {G.}~\bibnamefont {Volpe}},\ and\
  \bibinfo {author} {\bibfnamefont {S.}~\bibnamefont {Bo}},\ }\bibfield
  {title} {\bibinfo {title} {Classification, inference and segmentation of
  anomalous diffusion with recurrent neural networks},\ }\href
  {https://doi.org/10.1088/1751-8121/ac070a} {\bibfield  {journal} {\bibinfo
  {journal} {Journal of Physics A: Mathematical and Theoretical}\ }\textbf
  {\bibinfo {volume} {54}},\ \bibinfo {pages} {294003} (\bibinfo {year}
  {2021})}\BibitemShut {NoStop}%
\bibitem [{\citenamefont {Li}\ \emph {et~al.}(2021)\citenamefont {Li},
  \citenamefont {Yao},\ and\ \citenamefont {Huang}}]{li2021wavenet}%
  \BibitemOpen
  \bibfield  {author} {\bibinfo {author} {\bibfnamefont {D.}~\bibnamefont
  {Li}}, \bibinfo {author} {\bibfnamefont {Q.}~\bibnamefont {Yao}},\ and\
  \bibinfo {author} {\bibfnamefont {Z.}~\bibnamefont {Huang}},\ }\bibfield
  {title} {\bibinfo {title} {Wave{N}et-based deep neural networks for the
  characterization of anomalous diffusion ({WADN}et)},\ }\bibfield  {journal}
  {\bibinfo  {journal} {Journal of Physics A: Mathematical and Theoretical}\
  }\href {https://doi.org/10.1088/1751-8121/ac219c} {10.1088/1751-8121/ac219c}
  (\bibinfo {year} {2021})\BibitemShut {NoStop}%
\bibitem [{\citenamefont {Donahue}\ \emph {et~al.}(2015)\citenamefont
  {Donahue}, \citenamefont {Anne~Hendricks}, \citenamefont {Guadarrama},
  \citenamefont {Rohrbach}, \citenamefont {Venugopalan}, \citenamefont
  {Saenko},\ and\ \citenamefont {Darrell}}]{donahue2015long}%
  \BibitemOpen
  \bibfield  {author} {\bibinfo {author} {\bibfnamefont {J.}~\bibnamefont
  {Donahue}}, \bibinfo {author} {\bibfnamefont {L.}~\bibnamefont
  {Anne~Hendricks}}, \bibinfo {author} {\bibfnamefont {S.}~\bibnamefont
  {Guadarrama}}, \bibinfo {author} {\bibfnamefont {M.}~\bibnamefont
  {Rohrbach}}, \bibinfo {author} {\bibfnamefont {S.}~\bibnamefont
  {Venugopalan}}, \bibinfo {author} {\bibfnamefont {K.}~\bibnamefont
  {Saenko}},\ and\ \bibinfo {author} {\bibfnamefont {T.}~\bibnamefont
  {Darrell}},\ }\bibfield  {title} {\bibinfo {title} {Long-term recurrent
  convolutional networks for visual recognition and description},\ }in\
  \href@noop {} {\emph {\bibinfo {booktitle} {Proceedings of the IEEE
  conference on computer vision and pattern recognition}}}\ (\bibinfo {year}
  {2015})\ pp.\ \bibinfo {pages} {2625--2634}\BibitemShut {NoStop}%
\bibitem [{\citenamefont {Manzo}(2021)}]{manzo2021extreme}%
  \BibitemOpen
  \bibfield  {author} {\bibinfo {author} {\bibfnamefont {C.}~\bibnamefont
  {Manzo}},\ }\bibfield  {title} {\bibinfo {title} {Extreme learning machine
  for the characterization of anomalous diﬀusion from single trajectories
  (andi-{ELM})},\ }\bibfield  {journal} {\bibinfo  {journal} {Journal of
  Physics A: Mathematical and Theoretical}\ }\href
  {https://doi.org/10.1088/1751-8121/ac13dd} {10.1088/1751-8121/ac13dd}
  (\bibinfo {year} {2021})\BibitemShut {NoStop}%
\bibitem [{\citenamefont {Bai}\ \emph {et~al.}(2018)\citenamefont {Bai},
  \citenamefont {Kolter},\ and\ \citenamefont {Koltun}}]{bai2018empirical}%
  \BibitemOpen
  \bibfield  {author} {\bibinfo {author} {\bibfnamefont {S.}~\bibnamefont
  {Bai}}, \bibinfo {author} {\bibfnamefont {J.~Z.}\ \bibnamefont {Kolter}},\
  and\ \bibinfo {author} {\bibfnamefont {V.}~\bibnamefont {Koltun}},\
  }\bibfield  {title} {\bibinfo {title} {An empirical evaluation of generic
  convolutional and recurrent networks for sequence modeling},\ }\href@noop {}
  {\bibfield  {journal} {\bibinfo  {journal} {arXiv preprint arXiv:1803.01271}\
  } (\bibinfo {year} {2018})}\BibitemShut {NoStop}%
\bibitem [{\citenamefont {Aghion}\ \emph {et~al.}(2021)\citenamefont {Aghion},
  \citenamefont {Meyer}, \citenamefont {Adlakha}, \citenamefont {Kantz},\ and\
  \citenamefont {Bassler}}]{aghion2021moses}%
  \BibitemOpen
  \bibfield  {author} {\bibinfo {author} {\bibfnamefont {E.}~\bibnamefont
  {Aghion}}, \bibinfo {author} {\bibfnamefont {P.~G.}\ \bibnamefont {Meyer}},
  \bibinfo {author} {\bibfnamefont {V.}~\bibnamefont {Adlakha}}, \bibinfo
  {author} {\bibfnamefont {H.}~\bibnamefont {Kantz}},\ and\ \bibinfo {author}
  {\bibfnamefont {K.~E.}\ \bibnamefont {Bassler}},\ }\bibfield  {title}
  {\bibinfo {title} {{M}oses, {N}oah and {J}oseph effects in l\'{e}vy walks},\
  }\href@noop {} {\bibfield  {journal} {\bibinfo  {journal} {New Journal of
  Physics}\ }\textbf {\bibinfo {volume} {23}} (\bibinfo {year}
  {2021})}\BibitemShut {NoStop}%
\bibitem [{\citenamefont {Gentili}\ and\ \citenamefont
  {Volpe}(2021)}]{gentili2021characterization}%
  \BibitemOpen
  \bibfield  {author} {\bibinfo {author} {\bibfnamefont {A.}~\bibnamefont
  {Gentili}}\ and\ \bibinfo {author} {\bibfnamefont {G.}~\bibnamefont
  {Volpe}},\ }\bibfield  {title} {\bibinfo {title} {Characterization of
  anomalous diffusion classical statistics powered by deep learning
  ({CONDOR})},\ }\href {https://doi.org/10.1088/1751-8121/ac0c5d} {\bibfield
  {journal} {\bibinfo  {journal} {Journal of Physics A: Mathematical and
  Theoretical}\ }\textbf {\bibinfo {volume} {54}},\ \bibinfo {pages} {314003}
  (\bibinfo {year} {2021})}\BibitemShut {NoStop}%
\bibitem [{\citenamefont {Garibo~i Orts}\ \emph {et~al.}(2021)\citenamefont
  {Garibo~i Orts}, \citenamefont {Garcia-March},\ and\ \citenamefont
  {Conejero}}]{garibo2021efficient}%
  \BibitemOpen
  \bibfield  {author} {\bibinfo {author} {\bibfnamefont {O.}~\bibnamefont
  {Garibo~i Orts}}, \bibinfo {author} {\bibfnamefont {M.~A.}\ \bibnamefont
  {Garcia-March}},\ and\ \bibinfo {author} {\bibfnamefont {J.~A.}\ \bibnamefont
  {Conejero}},\ }\bibfield  {title} {\bibinfo {title} {Efficient recurrent
  neural network methods for anomalously diffusing single-particle short and
  noisy trajectories},\ }\href@noop {} {\bibfield  {journal} {\bibinfo
  {journal} {arXiv preprint arXiv:2108.02834}\ } (\bibinfo {year}
  {2021})}\BibitemShut {NoStop}%
\bibitem [{\citenamefont {Lines}\ \emph {et~al.}(2018)\citenamefont {Lines},
  \citenamefont {Taylor},\ and\ \citenamefont {Bagnall}}]{Lines2018}%
  \BibitemOpen
  \bibfield  {author} {\bibinfo {author} {\bibfnamefont {J.}~\bibnamefont
  {Lines}}, \bibinfo {author} {\bibfnamefont {S.}~\bibnamefont {Taylor}},\ and\
  \bibinfo {author} {\bibfnamefont {A.}~\bibnamefont {Bagnall}},\ }\bibfield
  {title} {\bibinfo {title} {Time series classification with hive-cote: The
  hierarchical vote collective of transformation-based ensembles},\ }\bibfield
  {journal} {\bibinfo  {journal} {ACM Trans. Knowl. Discov. Data}\ }\textbf
  {\bibinfo {volume} {12}},\ \href {https://doi.org/10.1145/3182382}
  {10.1145/3182382} (\bibinfo {year} {2018})\BibitemShut {NoStop}%
\bibitem [{\citenamefont {Le~Nguyen}\ \emph {et~al.}(2019)\citenamefont
  {Le~Nguyen}, \citenamefont {Gsponer}, \citenamefont {Ilie}, \citenamefont
  {O'Reilly},\ and\ \citenamefont {Ifrim}}]{LeNguyen2019}%
  \BibitemOpen
  \bibfield  {author} {\bibinfo {author} {\bibfnamefont {T.}~\bibnamefont
  {Le~Nguyen}}, \bibinfo {author} {\bibfnamefont {S.}~\bibnamefont {Gsponer}},
  \bibinfo {author} {\bibfnamefont {I.}~\bibnamefont {Ilie}}, \bibinfo {author}
  {\bibfnamefont {M.}~\bibnamefont {O'Reilly}},\ and\ \bibinfo {author}
  {\bibfnamefont {G.}~\bibnamefont {Ifrim}},\ }\bibfield  {title} {\bibinfo
  {title} {Interpretable time series classification using linear models and
  multi-resolution multi-domain symbolic representations},\ }\href
  {https://doi.org/10.1007/s10618-019-00633-3} {\bibfield  {journal} {\bibinfo
  {journal} {Data Mining and Knowledge Discovery}\ }\textbf {\bibinfo {volume}
  {33}},\ \bibinfo {pages} {1183} (\bibinfo {year} {2019})}\BibitemShut
  {NoStop}%
\bibitem [{\citenamefont {Janczura}\ \emph {et~al.}(2020)\citenamefont
  {Janczura}, \citenamefont {Kowalek}, \citenamefont {Loch-Olszewska},
  \citenamefont {Szwabi{\'n}ski},\ and\ \citenamefont
  {Weron}}]{janczura2020classification}%
  \BibitemOpen
  \bibfield  {author} {\bibinfo {author} {\bibfnamefont {J.}~\bibnamefont
  {Janczura}}, \bibinfo {author} {\bibfnamefont {P.}~\bibnamefont {Kowalek}},
  \bibinfo {author} {\bibfnamefont {H.}~\bibnamefont {Loch-Olszewska}},
  \bibinfo {author} {\bibfnamefont {J.}~\bibnamefont {Szwabi{\'n}ski}},\ and\
  \bibinfo {author} {\bibfnamefont {A.}~\bibnamefont {Weron}},\ }\bibfield
  {title} {\bibinfo {title} {Classification of particle trajectories in living
  cells: Machine learning versus statistical testing hypothesis for fractional
  anomalous diffusion},\ }\href@noop {} {\bibfield  {journal} {\bibinfo
  {journal} {Physical Review E}\ }\textbf {\bibinfo {volume} {102}},\ \bibinfo
  {pages} {032402} (\bibinfo {year} {2020})}\BibitemShut {NoStop}%
\bibitem [{\citenamefont {Loch-Olszewska}\ and\ \citenamefont
  {Szwabi{\'n}ski}(2020)}]{loch2020impact}%
  \BibitemOpen
  \bibfield  {author} {\bibinfo {author} {\bibfnamefont {H.}~\bibnamefont
  {Loch-Olszewska}}\ and\ \bibinfo {author} {\bibfnamefont {J.}~\bibnamefont
  {Szwabi{\'n}ski}},\ }\bibfield  {title} {\bibinfo {title} {Impact of feature
  choice on machine learning classification of fractional anomalous
  diffusion},\ }\href@noop {} {\bibfield  {journal} {\bibinfo  {journal}
  {Entropy}\ }\textbf {\bibinfo {volume} {22}},\ \bibinfo {pages} {1436}
  (\bibinfo {year} {2020})}\BibitemShut {NoStop}%
\end{thebibliography}

\begin{thebibliography}{34}%
\makeatletter
\providecommand \@ifxundefined [1]{%
 \@ifx{#1\undefined}
}%
\providecommand \@ifnum [1]{%
 \ifnum #1\expandafter \@firstoftwo
 \else \expandafter \@secondoftwo
 \fi
}%
\providecommand \@ifx [1]{%
 \ifx #1\expandafter \@firstoftwo
 \else \expandafter \@secondoftwo
 \fi
}%
\providecommand \natexlab [1]{#1}%
\providecommand \enquote  [1]{``#1''}%
\providecommand \bibnamefont  [1]{#1}%
\providecommand \bibfnamefont [1]{#1}%
\providecommand \citenamefont [1]{#1}%
\providecommand \href@noop [0]{\@secondoftwo}%
\providecommand \href [0]{\begingroup \@sanitize@url \@href}%
\providecommand \@href[1]{\@@startlink{#1}\@@href}%
\providecommand \@@href[1]{\endgroup#1\@@endlink}%
\providecommand \@sanitize@url [0]{\catcode `\\12\catcode `\$12\catcode
  `\&12\catcode `\#12\catcode `\^12\catcode `\_12\catcode `\%12\relax}%
\providecommand \@@startlink[1]{}%
\providecommand \@@endlink[0]{}%
\providecommand \url  [0]{\begingroup\@sanitize@url \@url }%
\providecommand \@url [1]{\endgroup\@href {#1}{\urlprefix }}%
\providecommand \urlprefix  [0]{URL }%
\providecommand \Eprint [0]{\href }%
\providecommand \doibase [0]{https://doi.org/}%
\providecommand \selectlanguage [0]{\@gobble}%
\providecommand \bibinfo  [0]{\@secondoftwo}%
\providecommand \bibfield  [0]{\@secondoftwo}%
\providecommand \translation [1]{[#1]}%
\providecommand \BibitemOpen [0]{}%
\providecommand \bibitemStop [0]{}%
\providecommand \bibitemNoStop [0]{.\EOS\space}%
\providecommand \EOS [0]{\spacefactor3000\relax}%
\providecommand \BibitemShut  [1]{\csname bibitem#1\endcsname}%
\let\auto@bib@innerbib\@empty
%</preamble>
\bibitem [{\citenamefont {Golding}\ and\ \citenamefont
  {Cox}(2006)}]{golding2006physicalSI}%
  \BibitemOpen
  \bibfield  {author} {\bibinfo {author} {\bibfnamefont {I.}~\bibnamefont
  {Golding}}\ and\ \bibinfo {author} {\bibfnamefont {E.~C.}\ \bibnamefont
  {Cox}},\ }\bibfield  {title} {\bibinfo {title} {Physical nature of bacterial
  cytoplasm},\ }\href@noop {} {\bibfield  {journal} {\bibinfo  {journal}
  {Physical Review Letters}\ }\textbf {\bibinfo {volume} {96}},\ \bibinfo
  {pages} {098102} (\bibinfo {year} {2006})}\BibitemShut {NoStop}%
\bibitem [{\citenamefont {Stadler}\ and\ \citenamefont
  {Weiss}(2017)}]{stadler2017nonSI}%
  \BibitemOpen
  \bibfield  {author} {\bibinfo {author} {\bibfnamefont {L.}~\bibnamefont
  {Stadler}}\ and\ \bibinfo {author} {\bibfnamefont {M.}~\bibnamefont
  {Weiss}},\ }\bibfield  {title} {\bibinfo {title} {Non-equilibrium forces
  drive the anomalous diffusion of telomeres in the nucleus of mammalian
  cells},\ }\href@noop {} {\bibfield  {journal} {\bibinfo  {journal} {New
  Journal of Physics}\ }\textbf {\bibinfo {volume} {19}},\ \bibinfo {pages}
  {113048} (\bibinfo {year} {2017})}\BibitemShut {NoStop}%
\bibitem [{\citenamefont {Krapf}\ \emph {et~al.}(2019)\citenamefont {Krapf},
  \citenamefont {Lukat}, \citenamefont {Marinari}, \citenamefont {Metzler},
  \citenamefont {Oshanin}, \citenamefont {Selhuber-Unkel}, \citenamefont
  {Squarcini}, \citenamefont {Stadler}, \citenamefont {Weiss},\ and\
  \citenamefont {Xu}}]{krapf2019spectralSI}%
  \BibitemOpen
  \bibfield  {author} {\bibinfo {author} {\bibfnamefont {D.}~\bibnamefont
  {Krapf}}, \bibinfo {author} {\bibfnamefont {N.}~\bibnamefont {Lukat}},
  \bibinfo {author} {\bibfnamefont {E.}~\bibnamefont {Marinari}}, \bibinfo
  {author} {\bibfnamefont {R.}~\bibnamefont {Metzler}}, \bibinfo {author}
  {\bibfnamefont {G.}~\bibnamefont {Oshanin}}, \bibinfo {author} {\bibfnamefont
  {C.}~\bibnamefont {Selhuber-Unkel}}, \bibinfo {author} {\bibfnamefont
  {A.}~\bibnamefont {Squarcini}}, \bibinfo {author} {\bibfnamefont
  {L.}~\bibnamefont {Stadler}}, \bibinfo {author} {\bibfnamefont
  {M.}~\bibnamefont {Weiss}},\ and\ \bibinfo {author} {\bibfnamefont
  {X.}~\bibnamefont {Xu}},\ }\bibfield  {title} {\bibinfo {title} {Spectral
  content of a single non-{B}rownian trajectory},\ }\href@noop {} {\bibfield
  {journal} {\bibinfo  {journal} {Physical Review X}\ }\textbf {\bibinfo
  {volume} {9}},\ \bibinfo {pages} {011019} (\bibinfo {year}
  {2019})}\BibitemShut {NoStop}%
\bibitem [{\citenamefont {Manzo}\ \emph {et~al.}(2015)\citenamefont {Manzo},
  \citenamefont {Torreno-Pina}, \citenamefont {Massignan}, \citenamefont
  {Lapeyre~Jr}, \citenamefont {Lewenstein},\ and\ \citenamefont
  {Parajo}}]{manzo2015weakSI}%
  \BibitemOpen
  \bibfield  {author} {\bibinfo {author} {\bibfnamefont {C.}~\bibnamefont
  {Manzo}}, \bibinfo {author} {\bibfnamefont {J.~A.}\ \bibnamefont
  {Torreno-Pina}}, \bibinfo {author} {\bibfnamefont {P.}~\bibnamefont
  {Massignan}}, \bibinfo {author} {\bibfnamefont {G.~J.}\ \bibnamefont
  {Lapeyre~Jr}}, \bibinfo {author} {\bibfnamefont {M.}~\bibnamefont
  {Lewenstein}},\ and\ \bibinfo {author} {\bibfnamefont {M.~F.~G.}\
  \bibnamefont {Parajo}},\ }\bibfield  {title} {\bibinfo {title} {Weak
  ergodicity breaking of receptor motion in living cells stemming from random
  diffusivity},\ }\href@noop {} {\bibfield  {journal} {\bibinfo  {journal}
  {Physical Review X}\ }\textbf {\bibinfo {volume} {5}},\ \bibinfo {pages}
  {011021} (\bibinfo {year} {2015})}\BibitemShut {NoStop}%
\bibitem [{\citenamefont {Kindermann}\ \emph {et~al.}(2017)\citenamefont
  {Kindermann}, \citenamefont {Dechant}, \citenamefont {Hohmann}, \citenamefont
  {Lausch}, \citenamefont {Mayer}, \citenamefont {Schmidt}, \citenamefont
  {Lutz},\ and\ \citenamefont {Widera}}]{kindermann2017nonergodicSI}%
  \BibitemOpen
  \bibfield  {author} {\bibinfo {author} {\bibfnamefont {F.}~\bibnamefont
  {Kindermann}}, \bibinfo {author} {\bibfnamefont {A.}~\bibnamefont {Dechant}},
  \bibinfo {author} {\bibfnamefont {M.}~\bibnamefont {Hohmann}}, \bibinfo
  {author} {\bibfnamefont {T.}~\bibnamefont {Lausch}}, \bibinfo {author}
  {\bibfnamefont {D.}~\bibnamefont {Mayer}}, \bibinfo {author} {\bibfnamefont
  {F.}~\bibnamefont {Schmidt}}, \bibinfo {author} {\bibfnamefont
  {E.}~\bibnamefont {Lutz}},\ and\ \bibinfo {author} {\bibfnamefont
  {A.}~\bibnamefont {Widera}},\ }\bibfield  {title} {\bibinfo {title}
  {Nonergodic diffusion of single atoms in a periodic potential},\ }\href@noop
  {} {\bibfield  {journal} {\bibinfo  {journal} {Nature Physics}\ }\textbf
  {\bibinfo {volume} {13}},\ \bibinfo {pages} {137} (\bibinfo {year}
  {2017})}\BibitemShut {NoStop}%
\bibitem [{\citenamefont {Mu{\~n}oz-Gil}\ \emph
  {et~al.}(2020{\natexlab{a}})\citenamefont {Mu{\~n}oz-Gil}, \citenamefont
  {Romero}, \citenamefont {Mateos}, \citenamefont {de~Llobet~Cucalon},
  \citenamefont {Beato}, \citenamefont {Lewenstein}, \citenamefont
  {Garcia-Parajo},\ and\ \citenamefont {Torreno-Pina}}]{munoz2020phaseSI}%
  \BibitemOpen
  \bibfield  {author} {\bibinfo {author} {\bibfnamefont {G.}~\bibnamefont
  {Mu{\~n}oz-Gil}}, \bibinfo {author} {\bibfnamefont {C.}~\bibnamefont
  {Romero}}, \bibinfo {author} {\bibfnamefont {N.}~\bibnamefont {Mateos}},
  \bibinfo {author} {\bibfnamefont {L.~I.}\ \bibnamefont {de~Llobet~Cucalon}},
  \bibinfo {author} {\bibfnamefont {M.}~\bibnamefont {Beato}}, \bibinfo
  {author} {\bibfnamefont {M.}~\bibnamefont {Lewenstein}}, \bibinfo {author}
  {\bibfnamefont {M.~F.}\ \bibnamefont {Garcia-Parajo}},\ and\ \bibinfo
  {author} {\bibfnamefont {J.~A.}\ \bibnamefont {Torreno-Pina}},\ }\bibfield
  {title} {\bibinfo {title} {Phase separation of tunable biomolecular
  condensates predicted by an interacting particle model},\ }\href@noop {}
  {\bibfield  {journal} {\bibinfo  {journal} {bioRxiv}\ } (\bibinfo {year}
  {2020}{\natexlab{a}})}\BibitemShut {NoStop}%
\bibitem [{\citenamefont {Wolpert}(1992)}]{wolpert1992stackedSI}%
  \BibitemOpen
  \bibfield  {author} {\bibinfo {author} {\bibfnamefont {D.~H.}\ \bibnamefont
  {Wolpert}},\ }\bibfield  {title} {\bibinfo {title} {Stacked generalization},\
  }\href@noop {} {\bibfield  {journal} {\bibinfo  {journal} {Neural networks}\
  }\textbf {\bibinfo {volume} {5}},\ \bibinfo {pages} {241} (\bibinfo {year}
  {1992})}\BibitemShut {NoStop}%
\bibitem [{\citenamefont {Krog}\ \emph {et~al.}(2018)\citenamefont {Krog},
  \citenamefont {Jacobsen}, \citenamefont {Lund}, \citenamefont {W{\"u}stner},\
  and\ \citenamefont {Lomholt}}]{krog2018bayesianSI}%
  \BibitemOpen
  \bibfield  {author} {\bibinfo {author} {\bibfnamefont {J.}~\bibnamefont
  {Krog}}, \bibinfo {author} {\bibfnamefont {L.~H.}\ \bibnamefont {Jacobsen}},
  \bibinfo {author} {\bibfnamefont {F.~W.}\ \bibnamefont {Lund}}, \bibinfo
  {author} {\bibfnamefont {D.}~\bibnamefont {W{\"u}stner}},\ and\ \bibinfo
  {author} {\bibfnamefont {M.~A.}\ \bibnamefont {Lomholt}},\ }\bibfield
  {title} {\bibinfo {title} {Bayesian model selection with fractional
  {B}rownian motion},\ }\href@noop {} {\bibfield  {journal} {\bibinfo
  {journal} {Journal of Statistical Mechanics: Theory and Experiment}\ }\textbf
  {\bibinfo {volume} {2018}},\ \bibinfo {pages} {093501} (\bibinfo {year}
  {2018})}\BibitemShut {NoStop}%
\bibitem [{\citenamefont {Park}\ \emph {et~al.}(2021)\citenamefont {Park},
  \citenamefont {Thapa}, \citenamefont {Kim}, \citenamefont {Lomholt},\ and\
  \citenamefont {Jeon}}]{park2021bayesianSI}%
  \BibitemOpen
  \bibfield  {author} {\bibinfo {author} {\bibfnamefont {S.}~\bibnamefont
  {Park}}, \bibinfo {author} {\bibfnamefont {S.}~\bibnamefont {Thapa}},
  \bibinfo {author} {\bibfnamefont {Y.}~\bibnamefont {Kim}}, \bibinfo {author}
  {\bibfnamefont {M.~A.}\ \bibnamefont {Lomholt}},\ and\ \bibinfo {author}
  {\bibfnamefont {J.-H.}\ \bibnamefont {Jeon}},\ }\bibfield  {title} {\bibinfo
  {title} {Bayesian inference of {L}\'evy walks via hidden {M}arkov
  models},\ }\href@noop {} {\bibfield  {journal} {\bibinfo  {journal} {arXiv
  preprint arXiv:2107.05390}\ } (\bibinfo {year} {2021})}\BibitemShut {NoStop}%
\bibitem [{\citenamefont {Verdier}\ \emph {et~al.}(2021)\citenamefont
  {Verdier}, \citenamefont {Duval}, \citenamefont {Laurent}, \citenamefont
  {Cass{\'{e}}}, \citenamefont {Vestergaard},\ and\ \citenamefont
  {Masson}}]{verdier2021learningSI}%
  \BibitemOpen
  \bibfield  {author} {\bibinfo {author} {\bibfnamefont {H.}~\bibnamefont
  {Verdier}}, \bibinfo {author} {\bibfnamefont {M.}~\bibnamefont {Duval}},
  \bibinfo {author} {\bibfnamefont {F.}~\bibnamefont {Laurent}}, \bibinfo
  {author} {\bibfnamefont {A.}~\bibnamefont {Cass{\'{e}}}}, \bibinfo {author}
  {\bibfnamefont {C.~L.}\ \bibnamefont {Vestergaard}},\ and\ \bibinfo {author}
  {\bibfnamefont {J.-B.}\ \bibnamefont {Masson}},\ }\bibfield  {title}
  {\bibinfo {title} {Learning physical properties of anomalous random walks
  using graph neural networks},\ }\href
  {https://doi.org/10.1088/1751-8121/abfa45} {\bibfield  {journal} {\bibinfo
  {journal} {Journal of Physics A: Mathematical and Theoretical}\ }\textbf
  {\bibinfo {volume} {54}},\ \bibinfo {pages} {234001} (\bibinfo {year}
  {2021})}\BibitemShut {NoStop}%
\bibitem [{\citenamefont {He}\ \emph {et~al.}(2016)\citenamefont {He},
  \citenamefont {Zhang}, \citenamefont {Ren},\ and\ \citenamefont
  {Sun}}]{he2016deepSI}%
  \BibitemOpen
  \bibfield  {author} {\bibinfo {author} {\bibfnamefont {K.}~\bibnamefont
  {He}}, \bibinfo {author} {\bibfnamefont {X.}~\bibnamefont {Zhang}}, \bibinfo
  {author} {\bibfnamefont {S.}~\bibnamefont {Ren}},\ and\ \bibinfo {author}
  {\bibfnamefont {J.}~\bibnamefont {Sun}},\ }\bibfield  {title} {\bibinfo
  {title} {Deep residual learning for image recognition},\ }in\ \href@noop {}
  {\emph {\bibinfo {booktitle} {Proceedings of the IEEE conference on computer
  vision and pattern recognition}}}\ (\bibinfo {year} {2016})\ pp.\ \bibinfo
  {pages} {770--778}\BibitemShut {NoStop}%
\bibitem [{\citenamefont {Chen}\ \emph {et~al.}(2016)\citenamefont {Chen} \emph
  {et~al.}}]{chen2016guestrinSI}%
  \BibitemOpen
  \bibfield  {author} {\bibinfo {author} {\bibfnamefont {T.}~\bibnamefont
  {Chen}} \emph {et~al.},\ }\bibfield  {title} {\bibinfo {title} {Guestrin, c.:
  Xgboost: A scalable tree boosting system},\ }in\ \href@noop {} {\emph
  {\bibinfo {booktitle} {Proceedings of the 22nd ACM SIGKDD International
  Conference on Knowledge Discovery and Data Mining (KDD’16)}}}\ (\bibinfo
  {year} {2016})\ pp.\ \bibinfo {pages} {785--794}\BibitemShut {NoStop}%
\bibitem [{\citenamefont {Argun}\ \emph {et~al.}(2021)\citenamefont {Argun},
  \citenamefont {Volpe},\ and\ \citenamefont {Bo}}]{argun2021classificationSI}%
  \BibitemOpen
  \bibfield  {author} {\bibinfo {author} {\bibfnamefont {A.}~\bibnamefont
  {Argun}}, \bibinfo {author} {\bibfnamefont {G.}~\bibnamefont {Volpe}},\ and\
  \bibinfo {author} {\bibfnamefont {S.}~\bibnamefont {Bo}},\ }\bibfield
  {title} {\bibinfo {title} {Classification, inference and segmentation of
  anomalous diffusion with recurrent neural networks},\ }\href
  {https://doi.org/10.1088/1751-8121/ac070a} {\bibfield  {journal} {\bibinfo
  {journal} {Journal of Physics A: Mathematical and Theoretical}\ }\textbf
  {\bibinfo {volume} {54}},\ \bibinfo {pages} {294003} (\bibinfo {year}
  {2021})}\BibitemShut {NoStop}%
\bibitem [{\citenamefont {Mu{\~n}oz-Gil}\ \emph
  {et~al.}(2020{\natexlab{b}})\citenamefont {Mu{\~n}oz-Gil}, \citenamefont
  {Requena}, \citenamefont {Volpe}, \citenamefont {Garcia-March},\ and\
  \citenamefont {Manzo}}]{andigithubSI}%
  \BibitemOpen
  \bibfield  {author} {\bibinfo {author} {\bibfnamefont {G.}~\bibnamefont
  {Mu{\~n}oz-Gil}}, \bibinfo {author} {\bibfnamefont {B.}~\bibnamefont
  {Requena}}, \bibinfo {author} {\bibfnamefont {G.}~\bibnamefont {Volpe}},
  \bibinfo {author} {\bibfnamefont {M.~A.}\ \bibnamefont {Garcia-March}},\ and\
  \bibinfo {author} {\bibfnamefont {C.}~\bibnamefont {Manzo}},\ }\href
  {https://doi.org/10.5281/zenodo.4775311} {\bibinfo {title}
  {An{D}i{C}hallenge/{ANDI}\_datasets: {C}hallenge 2020 release}} (\bibinfo
  {year} {2020}{\natexlab{b}})\BibitemShut {NoStop}%
\bibitem [{\citenamefont {Arts}\ \emph {et~al.}(2019)\citenamefont {Arts},
  \citenamefont {Smal}, \citenamefont {Paul}, \citenamefont {Wyman},\ and\
  \citenamefont {Meijering}}]{arts2019particleSI}%
  \BibitemOpen
  \bibfield  {author} {\bibinfo {author} {\bibfnamefont {M.}~\bibnamefont
  {Arts}}, \bibinfo {author} {\bibfnamefont {I.}~\bibnamefont {Smal}}, \bibinfo
  {author} {\bibfnamefont {M.~W.}\ \bibnamefont {Paul}}, \bibinfo {author}
  {\bibfnamefont {C.}~\bibnamefont {Wyman}},\ and\ \bibinfo {author}
  {\bibfnamefont {E.}~\bibnamefont {Meijering}},\ }\bibfield  {title} {\bibinfo
  {title} {Particle mobility analysis using deep learning and the moment
  scaling spectrum},\ }\href@noop {} {\bibfield  {journal} {\bibinfo  {journal}
  {Scientific reports}\ }\textbf {\bibinfo {volume} {9}},\ \bibinfo {pages} {1}
  (\bibinfo {year} {2019})}\BibitemShut {NoStop}%
\bibitem [{\citenamefont {Goodfellow}\ \emph {et~al.}(2016)\citenamefont
  {Goodfellow}, \citenamefont {Bengio},\ and\ \citenamefont
  {Courville}}]{goodfellow2016deepSI}%
  \BibitemOpen
  \bibfield  {author} {\bibinfo {author} {\bibfnamefont {I.}~\bibnamefont
  {Goodfellow}}, \bibinfo {author} {\bibfnamefont {Y.}~\bibnamefont {Bengio}},\
  and\ \bibinfo {author} {\bibfnamefont {A.}~\bibnamefont {Courville}},\
  }\href@noop {} {\emph {\bibinfo {title} {Deep Learning}}}\ (\bibinfo
  {publisher} {MIT Press},\ \bibinfo {year} {2016})\ \bibinfo {note}
  {\url{http://www.deeplearningbook.org}}\BibitemShut {NoStop}%
\bibitem [{\citenamefont {Chollet}\ \emph {et~al.}(2015)\citenamefont {Chollet}
  \emph {et~al.}}]{chollet2015kerasSI}%
  \BibitemOpen
  \bibfield  {author} {\bibinfo {author} {\bibfnamefont {F.}~\bibnamefont
  {Chollet}} \emph {et~al.},\ }\href@noop {} {\bibinfo {title} {Keras}},\
  \bibinfo {howpublished} {\url{https://keras.io}} (\bibinfo {year}
  {2015})\BibitemShut {NoStop}%
\bibitem [{\citenamefont {Li}\ \emph {et~al.}(2021)\citenamefont {Li},
  \citenamefont {Yao},\ and\ \citenamefont {Huang}}]{li2021wavenetSI}%
  \BibitemOpen
  \bibfield  {author} {\bibinfo {author} {\bibfnamefont {D.}~\bibnamefont
  {Li}}, \bibinfo {author} {\bibfnamefont {Q.}~\bibnamefont {Yao}},\ and\
  \bibinfo {author} {\bibfnamefont {Z.}~\bibnamefont {Huang}},\ }\bibfield
  {title} {\bibinfo {title} {Wave{N}et-based deep neural networks for the
  characterization of anomalous diffusion ({WADN}et)},\ }\bibfield  {journal}
  {\bibinfo  {journal} {Journal of Physics A: Mathematical and Theoretical}\
  }\href {https://doi.org/10.1088/1751-8121/ac219c} {10.1088/1751-8121/ac219c}
  (\bibinfo {year} {2021})\BibitemShut {NoStop}%
\bibitem [{\citenamefont {Donahue}\ \emph {et~al.}(2015)\citenamefont
  {Donahue}, \citenamefont {Anne~Hendricks}, \citenamefont {Guadarrama},
  \citenamefont {Rohrbach}, \citenamefont {Venugopalan}, \citenamefont
  {Saenko},\ and\ \citenamefont {Darrell}}]{donahue2015longSI}%
  \BibitemOpen
  \bibfield  {author} {\bibinfo {author} {\bibfnamefont {J.}~\bibnamefont
  {Donahue}}, \bibinfo {author} {\bibfnamefont {L.}~\bibnamefont
  {Anne~Hendricks}}, \bibinfo {author} {\bibfnamefont {S.}~\bibnamefont
  {Guadarrama}}, \bibinfo {author} {\bibfnamefont {M.}~\bibnamefont
  {Rohrbach}}, \bibinfo {author} {\bibfnamefont {S.}~\bibnamefont
  {Venugopalan}}, \bibinfo {author} {\bibfnamefont {K.}~\bibnamefont
  {Saenko}},\ and\ \bibinfo {author} {\bibfnamefont {T.}~\bibnamefont
  {Darrell}},\ }\bibfield  {title} {\bibinfo {title} {Long-term recurrent
  convolutional networks for visual recognition and description},\ }in\
  \href@noop {} {\emph {\bibinfo {booktitle} {Proceedings of the IEEE
  conference on computer vision and pattern recognition}}}\ (\bibinfo {year}
  {2015})\ pp.\ \bibinfo {pages} {2625--2634}\BibitemShut {NoStop}%
\bibitem [{\citenamefont {Manzo}(2021)}]{manzo2021extremeSI}%
  \BibitemOpen
  \bibfield  {author} {\bibinfo {author} {\bibfnamefont {C.}~\bibnamefont
  {Manzo}},\ }\bibfield  {title} {\bibinfo {title} {Extreme learning machine
  for the characterization of anomalous diﬀusion from single trajectories
  (andi-{ELM})},\ }\bibfield  {journal} {\bibinfo  {journal} {Journal of
  Physics A: Mathematical and Theoretical}\ }\href
  {https://doi.org/10.1088/1751-8121/ac13dd} {10.1088/1751-8121/ac13dd}
  (\bibinfo {year} {2021})\BibitemShut {NoStop}%
\bibitem [{\citenamefont {Huang}\ \emph {et~al.}(2004)\citenamefont {Huang},
  \citenamefont {Zhu},\ and\ \citenamefont {Siew}}]{huang2004extremeSI}%
  \BibitemOpen
  \bibfield  {author} {\bibinfo {author} {\bibfnamefont {G.-B.}\ \bibnamefont
  {Huang}}, \bibinfo {author} {\bibfnamefont {Q.-Y.}\ \bibnamefont {Zhu}},\
  and\ \bibinfo {author} {\bibfnamefont {C.-K.}\ \bibnamefont {Siew}},\
  }\bibfield  {title} {\bibinfo {title} {Extreme learning machine: a new
  learning scheme of feedforward neural networks},\ }in\ \href@noop {} {\emph
  {\bibinfo {booktitle} {2004 IEEE international joint conference on neural
  networks (IEEE Cat. No. 04CH37541)}}},\ Vol.~\bibinfo {volume} {2}\ (\bibinfo
  {organization} {IEEE},\ \bibinfo {year} {2004})\ pp.\ \bibinfo {pages}
  {985--990}\BibitemShut {NoStop}%
\bibitem [{\citenamefont {Huang}\ \emph {et~al.}(2006)\citenamefont {Huang},
  \citenamefont {Zhu},\ and\ \citenamefont {Siew}}]{huang2006extremeSI}%
  \BibitemOpen
  \bibfield  {author} {\bibinfo {author} {\bibfnamefont {G.-B.}\ \bibnamefont
  {Huang}}, \bibinfo {author} {\bibfnamefont {Q.-Y.}\ \bibnamefont {Zhu}},\
  and\ \bibinfo {author} {\bibfnamefont {C.-K.}\ \bibnamefont {Siew}},\
  }\bibfield  {title} {\bibinfo {title} {Extreme learning machine: theory and
  applications},\ }\href@noop {} {\bibfield  {journal} {\bibinfo  {journal}
  {Neurocomputing}\ }\textbf {\bibinfo {volume} {70}},\ \bibinfo {pages} {489}
  (\bibinfo {year} {2006})}\BibitemShut {NoStop}%
\bibitem [{\citenamefont {Bai}\ \emph {et~al.}(2018)\citenamefont {Bai},
  \citenamefont {Kolter},\ and\ \citenamefont {Koltun}}]{bai2018empiricalSI}%
  \BibitemOpen
  \bibfield  {author} {\bibinfo {author} {\bibfnamefont {S.}~\bibnamefont
  {Bai}}, \bibinfo {author} {\bibfnamefont {J.~Z.}\ \bibnamefont {Kolter}},\
  and\ \bibinfo {author} {\bibfnamefont {V.}~\bibnamefont {Koltun}},\
  }\bibfield  {title} {\bibinfo {title} {An empirical evaluation of generic
  convolutional and recurrent networks for sequence modeling},\ }\href@noop {}
  {\bibfield  {journal} {\bibinfo  {journal} {arXiv preprint arXiv:1803.01271}\
  } (\bibinfo {year} {2018})}\BibitemShut {NoStop}%
\bibitem [{\citenamefont {Granik}\ \emph {et~al.}(2019)\citenamefont {Granik},
  \citenamefont {Weiss}, \citenamefont {Nehme}, \citenamefont {Levin},
  \citenamefont {Chein}, \citenamefont {Perlson}, \citenamefont {Roichman},\
  and\ \citenamefont {Shechtman}}]{granik2019singleSI}%
  \BibitemOpen
  \bibfield  {author} {\bibinfo {author} {\bibfnamefont {N.}~\bibnamefont
  {Granik}}, \bibinfo {author} {\bibfnamefont {L.~E.}\ \bibnamefont {Weiss}},
  \bibinfo {author} {\bibfnamefont {E.}~\bibnamefont {Nehme}}, \bibinfo
  {author} {\bibfnamefont {M.}~\bibnamefont {Levin}}, \bibinfo {author}
  {\bibfnamefont {M.}~\bibnamefont {Chein}}, \bibinfo {author} {\bibfnamefont
  {E.}~\bibnamefont {Perlson}}, \bibinfo {author} {\bibfnamefont
  {Y.}~\bibnamefont {Roichman}},\ and\ \bibinfo {author} {\bibfnamefont
  {Y.}~\bibnamefont {Shechtman}},\ }\bibfield  {title} {\bibinfo {title}
  {Single-particle diffusion characterization by deep learning},\ }\href@noop
  {} {\bibfield  {journal} {\bibinfo  {journal} {Biophysical Journal}\ }\textbf
  {\bibinfo {volume} {117}},\ \bibinfo {pages} {185} (\bibinfo {year}
  {2019})}\BibitemShut {NoStop}%
\bibitem [{\citenamefont {Aghion}\ \emph {et~al.}(2021)\citenamefont {Aghion},
  \citenamefont {Meyer}, \citenamefont {Adlakha}, \citenamefont {Kantz},\ and\
  \citenamefont {Bassler}}]{aghion2021mosesSI}%
  \BibitemOpen
  \bibfield  {author} {\bibinfo {author} {\bibfnamefont {E.}~\bibnamefont
  {Aghion}}, \bibinfo {author} {\bibfnamefont {P.~G.}\ \bibnamefont {Meyer}},
  \bibinfo {author} {\bibfnamefont {V.}~\bibnamefont {Adlakha}}, \bibinfo
  {author} {\bibfnamefont {H.}~\bibnamefont {Kantz}},\ and\ \bibinfo {author}
  {\bibfnamefont {K.~E.}\ \bibnamefont {Bassler}},\ }\bibfield  {title}
  {\bibinfo {title} {{M}oses, {N}oah and {J}oseph effects in l\'{e}vy walks},\
  }\href@noop {} {\bibfield  {journal} {\bibinfo  {journal} {New Journal of
  Physics}\ }\textbf {\bibinfo {volume} {23}} (\bibinfo {year}
  {2021})}\BibitemShut {NoStop}%
\bibitem [{\citenamefont {Chen}\ \emph {et~al.}(2017)\citenamefont {Chen},
  \citenamefont {Bassler}, \citenamefont {McCauley},\ and\ \citenamefont
  {Gunaratne}}]{chen2017anomalousSI}%
  \BibitemOpen
  \bibfield  {author} {\bibinfo {author} {\bibfnamefont {L.}~\bibnamefont
  {Chen}}, \bibinfo {author} {\bibfnamefont {K.~E.}\ \bibnamefont {Bassler}},
  \bibinfo {author} {\bibfnamefont {J.~L.}\ \bibnamefont {McCauley}},\ and\
  \bibinfo {author} {\bibfnamefont {G.~H.}\ \bibnamefont {Gunaratne}},\
  }\bibfield  {title} {\bibinfo {title} {Anomalous scaling of stochastic
  processes and the moses effect},\ }\href@noop {} {\bibfield  {journal}
  {\bibinfo  {journal} {Physical Review E}\ }\textbf {\bibinfo {volume} {95}},\
  \bibinfo {pages} {042141} (\bibinfo {year} {2017})}\BibitemShut {NoStop}%
\bibitem [{\citenamefont {Meyer}\ \emph {et~al.}(2018)\citenamefont {Meyer},
  \citenamefont {Adlakha}, \citenamefont {Kantz},\ and\ \citenamefont
  {Bassler}}]{meyer2018anomalousSI}%
  \BibitemOpen
  \bibfield  {author} {\bibinfo {author} {\bibfnamefont {P.~G.}\ \bibnamefont
  {Meyer}}, \bibinfo {author} {\bibfnamefont {V.}~\bibnamefont {Adlakha}},
  \bibinfo {author} {\bibfnamefont {H.}~\bibnamefont {Kantz}},\ and\ \bibinfo
  {author} {\bibfnamefont {K.~E.}\ \bibnamefont {Bassler}},\ }\bibfield
  {title} {\bibinfo {title} {Anomalous diffusion and the {M}oses effect in an
  aging deterministic model},\ }\href@noop {} {\bibfield  {journal} {\bibinfo
  {journal} {New Journal of Physics}\ }\textbf {\bibinfo {volume} {20}},\
  \bibinfo {pages} {113033} (\bibinfo {year} {2018})}\BibitemShut {NoStop}%
\bibitem [{\citenamefont {Gentili}\ and\ \citenamefont
  {Volpe}(2021)}]{gentili2021characterizationSI}%
  \BibitemOpen
  \bibfield  {author} {\bibinfo {author} {\bibfnamefont {A.}~\bibnamefont
  {Gentili}}\ and\ \bibinfo {author} {\bibfnamefont {G.}~\bibnamefont
  {Volpe}},\ }\bibfield  {title} {\bibinfo {title} {Characterization of
  anomalous diffusion classical statistics powered by deep learning
  ({CONDOR})},\ }\href {https://doi.org/10.1088/1751-8121/ac0c5d} {\bibfield
  {journal} {\bibinfo  {journal} {Journal of Physics A: Mathematical and
  Theoretical}\ }\textbf {\bibinfo {volume} {54}},\ \bibinfo {pages} {314003}
  (\bibinfo {year} {2021})}\BibitemShut {NoStop}%
\bibitem [{\citenamefont {Garibo~i Orts}\ \emph {et~al.}(2021)\citenamefont
  {Garibo~i Orts}, \citenamefont {Garcia-March},\ and\ \citenamefont
  {Conejero}}]{garibo2021efficientSI}%
  \BibitemOpen
  \bibfield  {author} {\bibinfo {author} {\bibfnamefont {O.}~\bibnamefont
  {Garibo~i Orts}}, \bibinfo {author} {\bibfnamefont {M.~A.}\ \bibnamefont
  {Garcia-March}},\ and\ \bibinfo {author} {\bibfnamefont {J.~A.}\ \bibnamefont
  {Conejero}},\ }\bibfield  {title} {\bibinfo {title} {Efficient recurrent
  neural network methods for anomalously diffusing single-particle short and
  noisy trajectories},\ }\href@noop {} {\bibfield  {journal} {\bibinfo
  {journal} {arXiv preprint arXiv:2108.02834}\ } (\bibinfo {year}
  {2021})}\BibitemShut {NoStop}%
\bibitem [{\citenamefont {Lines}\ \emph {et~al.}(2018)\citenamefont {Lines},
  \citenamefont {Taylor},\ and\ \citenamefont {Bagnall}}]{Lines2018SI}%
  \BibitemOpen
  \bibfield  {author} {\bibinfo {author} {\bibfnamefont {J.}~\bibnamefont
  {Lines}}, \bibinfo {author} {\bibfnamefont {S.}~\bibnamefont {Taylor}},\ and\
  \bibinfo {author} {\bibfnamefont {A.}~\bibnamefont {Bagnall}},\ }\bibfield
  {title} {\bibinfo {title} {Time series classification with hive-cote: The
  hierarchical vote collective of transformation-based ensembles},\ }\bibfield
  {journal} {\bibinfo  {journal} {ACM Trans. Knowl. Discov. Data}\ }\textbf
  {\bibinfo {volume} {12}},\ \href {https://doi.org/10.1145/3182382}
  {10.1145/3182382} (\bibinfo {year} {2018})\BibitemShut {NoStop}%
\bibitem [{\citenamefont {Le~Nguyen}\ \emph {et~al.}(2019)\citenamefont
  {Le~Nguyen}, \citenamefont {Gsponer}, \citenamefont {Ilie}, \citenamefont
  {O'Reilly},\ and\ \citenamefont {Ifrim}}]{LeNguyen2019SI}%
  \BibitemOpen
  \bibfield  {author} {\bibinfo {author} {\bibfnamefont {T.}~\bibnamefont
  {Le~Nguyen}}, \bibinfo {author} {\bibfnamefont {S.}~\bibnamefont {Gsponer}},
  \bibinfo {author} {\bibfnamefont {I.}~\bibnamefont {Ilie}}, \bibinfo {author}
  {\bibfnamefont {M.}~\bibnamefont {O'Reilly}},\ and\ \bibinfo {author}
  {\bibfnamefont {G.}~\bibnamefont {Ifrim}},\ }\bibfield  {title} {\bibinfo
  {title} {Interpretable time series classification using linear models and
  multi-resolution multi-domain symbolic representations},\ }\href
  {https://doi.org/10.1007/s10618-019-00633-3} {\bibfield  {journal} {\bibinfo
  {journal} {Data Mining and Knowledge Discovery}\ }\textbf {\bibinfo {volume}
  {33}},\ \bibinfo {pages} {1183} (\bibinfo {year} {2019})}\BibitemShut
  {NoStop}%
\bibitem [{\citenamefont {Kowalek}\ \emph {et~al.}(2019)\citenamefont
  {Kowalek}, \citenamefont {Loch-Olszewska},\ and\ \citenamefont
  {Szwabi{\'n}ski}}]{kowalek2019classificationSI}%
  \BibitemOpen
  \bibfield  {author} {\bibinfo {author} {\bibfnamefont {P.}~\bibnamefont
  {Kowalek}}, \bibinfo {author} {\bibfnamefont {H.}~\bibnamefont
  {Loch-Olszewska}},\ and\ \bibinfo {author} {\bibfnamefont {J.}~\bibnamefont
  {Szwabi{\'n}ski}},\ }\bibfield  {title} {\bibinfo {title} {Classification of
  diffusion modes in single-particle tracking data: Feature-based versus
  deep-learning approach},\ }\href@noop {} {\bibfield  {journal} {\bibinfo
  {journal} {Physical Review E}\ }\textbf {\bibinfo {volume} {100}},\ \bibinfo
  {pages} {032410} (\bibinfo {year} {2019})}\BibitemShut {NoStop}%
\bibitem [{\citenamefont {Janczura}\ \emph {et~al.}(2020)\citenamefont
  {Janczura}, \citenamefont {Kowalek}, \citenamefont {Loch-Olszewska},
  \citenamefont {Szwabi{\'n}ski},\ and\ \citenamefont
  {Weron}}]{janczura2020classificationSI}%
  \BibitemOpen
  \bibfield  {author} {\bibinfo {author} {\bibfnamefont {J.}~\bibnamefont
  {Janczura}}, \bibinfo {author} {\bibfnamefont {P.}~\bibnamefont {Kowalek}},
  \bibinfo {author} {\bibfnamefont {H.}~\bibnamefont {Loch-Olszewska}},
  \bibinfo {author} {\bibfnamefont {J.}~\bibnamefont {Szwabi{\'n}ski}},\ and\
  \bibinfo {author} {\bibfnamefont {A.}~\bibnamefont {Weron}},\ }\bibfield
  {title} {\bibinfo {title} {Classification of particle trajectories in living
  cells: Machine learning versus statistical testing hypothesis for fractional
  anomalous diffusion},\ }\href@noop {} {\bibfield  {journal} {\bibinfo
  {journal} {Physical Review E}\ }\textbf {\bibinfo {volume} {102}},\ \bibinfo
  {pages} {032402} (\bibinfo {year} {2020})}\BibitemShut {NoStop}%
\bibitem [{\citenamefont {Loch-Olszewska}\ and\ \citenamefont
  {Szwabi{\'n}ski}(2020)}]{loch2020impactSI}%
  \BibitemOpen
  \bibfield  {author} {\bibinfo {author} {\bibfnamefont {H.}~\bibnamefont
  {Loch-Olszewska}}\ and\ \bibinfo {author} {\bibfnamefont {J.}~\bibnamefont
  {Szwabi{\'n}ski}},\ }\bibfield  {title} {\bibinfo {title} {Impact of feature
  choice on machine learning classification of fractional anomalous
  diffusion},\ }\href@noop {} {\bibfield  {journal} {\bibinfo  {journal}
  {Entropy}\ }\textbf {\bibinfo {volume} {22}},\ \bibinfo {pages} {1436}
  (\bibinfo {year} {2020})}\BibitemShut {NoStop}%
\end{thebibliography}
%
\makeatother

\end{document}